\documentclass[10pt]{report}
\usepackage{graphicx}
\usepackage{url}
\usepackage{subcaption}
\usepackage{float}
\usepackage{fancyhdr}
\usepackage{tocloft}
\usepackage{setspace}
\usepackage{lastpage}
\usepackage{xcolor}
\usepackage{titlesec}
\usepackage{wrapfig}
\usepackage{rotating}
\usepackage{changepage}

\usepackage[most]{tcolorbox}
\usepackage[hidelinks]{hyperref}

\usepackage{tabularray}
\usepackage{siunitx}
\usepackage{lipsum}
\usepackage{tabularx}
\usepackage{tikz}
\usepackage[utf8]{inputenc}
\usepackage{glossaries}
\usepackage{multicol}
\usepackage{multirow}
\usepackage{cancel}
\usepackage{amssymb}
\usepackage{placeins}
\usepackage{caption}
\usepackage{booktabs}
\usepackage{listings}
\usepackage[top=2.5cm, bottom=2.5cm, left=3cm, right=2cm]{geometry}
\usepackage{parskip}
\usepackage{amsmath}
\usepackage{cleveref}
\usepackage{enumitem}
\usepackage{xargs}
\usepackage[colorinlistoftodos,prependcaption,textsize=tiny]{todonotes}
\usepackage{balance}

\floatstyle{plain}
\newfloat{listing}{tbp}{lol}   
\crefname{lstlisting}{Listing}{Listings}

\newtcolorbox{custombox}{
  colframe=customframe,
  colback=customback,
  boxrule=0.5mm,
  arc=0mm,
  boxsep=1mm,
  left=1mm,
  right=1mm,
  top=1mm,
  bottom=1mm,
  width=\textwidth,
  enhanced,
  drop shadow={black!50!white},
}

\definecolor{background}{rgb}{0.95,0.95,0.95}
\definecolor{string}{rgb}{0.6,0,0}
\definecolor{key}{rgb}{0,0,0.6}
\definecolor{value}{rgb}{0,0,0}
\definecolor{customframe}{HTML}{2D5C8A}
\definecolor{customback}{HTML}{E1E8F0}
\definecolor{CodGray}{HTML}{151515}

\newcommand{\boxelement}[2]{
  \noindent \begin{tabular}{@{}p{1.5cm}p{\dimexpr\linewidth-2cm}@{}}
  \textbf{#1} & #2
  \end{tabular} \par \vspace{0.5em}
}

\newcounter{mf}
\crefname{mf}{}{}                

\newcounter{finding}
\renewcommand{\thefinding}{MF\arabic{finding}}
\crefname{finding}{(MF)}{MFs}

\newcommand{\finding}[2]{
  \refstepcounter{finding}\label{#1}%
  \boxelement{\textbf{\thefinding}}{#2}%
}

\newcommand{\blankpage}{%
  \newpage            
  \thispagestyle{empty}
  \mbox{}             
  \newpage            
}

\lstdefinelanguage{json}{
    basicstyle=\ttfamily,
    morestring=[b]",
    literate=
     *{0}{{{\color{value}0}}}{1}
      {1}{{{\color{value}1}}}{1}
      {2}{{{\color{value}2}}}{1}
      {3}{{{\color{value}3}}}{1}
      {4}{{{\color{value}4}}}{1}
      {5}{{{\color{value}5}}}{1}
      {6}{{{\color{value}6}}}{1}
      {7}{{{\color{value}7}}}{1}
      {8}{{{\color{value}8}}}{1}
      {9}{{{\color{value}9}}}{1}
      {:}{{{\color{value}:}}}{1}
      {,}{{{\color{value},}}}{1}
      {\{}{{{\color{key}\{}}}{1}
      {\}}{{{\color{key}\}}}}{1}
      {[}{{{\color{key}[}}}{1}
      {]}{{{\color{key}]}}}{1},
    stringstyle=\color{string},
    keywordstyle=\color{key},
    commentstyle=\color{gray},
    backgroundcolor=\color{background},
    showstringspaces=false
}

\lstdefinelanguage{python}{
    basicstyle=\ttfamily,
    morestring=[b]",
    morestring=[b]',
    morecomment=[l]{\#},
    literate=
     *{0}{{{\color{value}0}}}{1}
      {1}{{{\color{value}1}}}{1}
      {2}{{{\color{value}2}}}{1}
      {3}{{{\color{value}3}}}{1}
      {4}{{{\color{value}4}}}{1}
      {5}{{{\color{value}5}}}{1}
      {6}{{{\color{value}6}}}{1}
      {7}{{{\color{value}7}}}{1}
      {8}{{{\color{value}8}}}{1}
      {9}{{{\color{value}9}}}{1}
      {:}{{{\color{value}:}}}{1}
      {,}{{{\color{value},}}}{1}
      {(}{{{\color{key}(}}}{1}
      {)}{{{\color{key})}}}{1}
      {[}{{{\color{key}[}}}{1}
      {]}{{{\color{key}]}}}{1}
      {\{}{{{\color{key}\{}}}{1}
      {\}}{{{\color{key}\}}}}{1}
      {=}{{{\color{key}=}}}{1}
      {+}{{{\color{key}+}}}{1}
      {-}{{{\color{key}-}}}{1}
      {*}{{{\color{key}*}}}{1}
      {/}{{{\color{key}/}}}{1}
      {<}{{{\color{key}<}}}{1}
      {>}{{{\color{key}>}}}{1}
      {.}{{{\color{key}.}}}{1},
    keywords=[1]{def, class, if, else, elif, for, while, try, except, finally, 
                 with, as, import, from, return, yield, break, continue, pass, 
                 lambda, and, or, not, in, is, True, False, None},
    keywords=[2]{int, float, str, list, dict, tuple, set, bool, len, range, 
                 print, input, open, close, read, write, append, pop, insert, 
                 remove, sort, reverse, join, split, strip, replace, find, 
                 startswith, endswith, upper, lower, capitalize},
    stringstyle=\color{string},
    keywordstyle=[1]\color{key}\bfseries,
    keywordstyle=[2]\color{blue}\bfseries,
    commentstyle=\color{gray},
    backgroundcolor=\color{background},
    showstringspaces=false,
    breaklines=true,
    breakatwhitespace=true,
    tabsize=4,
    showspaces=false,
    showtabs=false
}

\lstdefinelanguage{bash}{
  basicstyle=\ttfamily,
  morecomment=[l]{\#},
  morestring=[b]",
  literate=
   *{:}{{{\color{value}:}}}{1}
    {-}{{{\color{value}-}}}{1}
    {_}{{{\color{value}\_}}}{1}
    {=}{{{\color{value}=}}}{1}
    {(}{{{\color{key}(}}}{1}
    {)}{{{\color{key})}}}{1},
  keywords=[1]{cd,ls,cp,mv,rm,mkdir,rmdir,chmod,chown,grep,awk,sed,echo,
              find,tar,zip,unzip,ssh,scp,git,docker,srun},
  keywordstyle=[1]\color{key}\bfseries,
  stringstyle=\color{string},
  commentstyle=\color{gray},
  backgroundcolor=\color{background},
  showstringspaces=false,
  breaklines=true,
  breakatwhitespace=true,
  tabsize=4
}

\lstdefinestyle{commands}{
  language=bash,
  basicstyle=\footnotesize\ttfamily,
  numbers=left,
  numberstyle=\tiny\color{gray},
  captionpos=b,
  backgroundcolor=\color{background},
  breaklines=true,
  breakatwhitespace=true
}

\lstdefinestyle{promptinteraction}{
  basicstyle=\footnotesize\ttfamily,          
  breaklines=false,                           
  columns=fullflexible,                       
  numbers=left,                               
  numberstyle=\tiny\color{gray},              
  captionpos=b,                               
  commentstyle=\color{gray},
  morekeywords=[1]{Step,Task},           
  morekeywords=[2]{Compute, Not, cached},                  
  morekeywords=[3]{(Partially), Cached, Retrieve, Cache},
  keywordstyle=[1]\color{blue}\bfseries,             
  keywordstyle=[2]\color{red}\bfseries,              
  keywordstyle=[3]\color{green!60!black}\bfseries,   
  keywordstyle=[4]\color{orange!80!black}\bfseries   
}

\titleformat{\chapter}[block]
  {\normalfont\huge\bfseries}
  {\flushright\resizebox{!}{7ex}{\thechapter}}
  {4em}
  {\flushleft\parbox[t]{\dimexpr\textwidth-20ex-4em}{\raggedright\Large}}

\titlespacing*{\chapter}
  {0pt}{20pt plus 1ex minus .5ex}{20pt plus 2pt minus 2pt}

\newcommandx{\unsure}[2][1=]{\todo[linecolor=red,backgroundcolor=red!25,bordercolor=red,#1]{#2}}
\newcommandx{\change}[2][1=]{\todo[linecolor=blue,backgroundcolor=blue!25,bordercolor=blue,#1]{#2}}
\newcommandx{\info}[2][1=]{\todo[linecolor=OliveGreen,backgroundcolor=OliveGreen!25,bordercolor=OliveGreen,#1]{#2}}
\newcommandx{\improvement}[2][1=]{\todo[linecolor=Plum,backgroundcolor=Plum!25,bordercolor=Plum,#1]{#2}}
\newcommandx{\thiswillnotshow}[2][1=]{\todo[disable,#1]{#2}}

\DeclareRobustCommand{\circled}[1]{%
    \tikz[baseline=(char.base)]{
        \node[shape=circle, fill=black, text=white, inner sep=1pt] (char) {#1};
    }%
}

\DeclareRobustCommand{\circledd}[1]{%
    \tikz[baseline=(char.base)]{
        \node[shape=circle, fill=black, text=white, inner sep=0pt] (char) {#1};
    }%
}

\usepackage{relsize}       

\newcommandx{\todolarge}[2][1=]{\todo[inline,size=\large,linecolor=blue,backgroundcolor=cyan,bordercolor=blue,#1]{#2}}

\begin{document}
\pagestyle{fancy}
\fancyhf{} 

\fancyhead[L]{\leftmark} 

\fancyfoot[C]{\thepage} 

\renewcommand{\chaptermark}[1]{\markboth{Chapter \thechapter: #1}{}} 
\renewcommand{\sectionmark}[1]{}
\renewcommand{\subsectionmark}[1]{\markright{\thesubsection\ #1}} 

\renewcommand{\headrulewidth}{0.4pt} 
\renewcommand{\footrulewidth}{0pt} 

\fancypagestyle{noheader}{
  \fancyhf{}  
  \fancyfoot[C]{\thepage}  
  \renewcommand{\headrulewidth}{0pt}  
}

\newgeometry{top=4cm, bottom=2cm, left=2cm, right=2cm}

\begin{center}
\thispagestyle{empty}
\textit{Vrije Universiteit Amsterdam}
\vspace{2mm}

\hfill
\includegraphics[height=17mm]{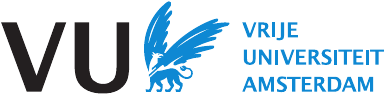}
\hfill
\vspace{0.5cm}

{\large Bachelor Thesis}
\vspace{0.5cm}

\rule{\linewidth}{.16pt}\\[0.4cm] 
{\large
\textbf{Kavier: Exploring Performance, Sustainability, and Efficiency of LLM Ecosystems under Inference through Cache-Aware Discrete-Event Simulation}
\vspace{2mm}
}

\rule{\linewidth}{.6pt}\\[1.5cm]

{\large
\begin{tabular}{lll}
{\bf Author:} ~~Radu Nicolae ~~ (2760443) \\
\end{tabular}
}
\\ {r.nicolae@vu.nl}

\vspace*{1.5cm}

\begin{tabular}{lll}
{\it 1st supervisor:}   & ~~ Prof. Dr. Ir. Cav. Alexandru Iosup & ~~ (VU Amsterdam) \\
{\it Daily supervisor:} & ~~ Dr. Animesh Trivedi & ~~ (IBM Research Europe) \\
{\it 2nd reader:} & ~~ Dr. Ir. Jesse Donkervliet & ~~ (VU Amsterdam)
\end{tabular}

\vspace*{2cm}

\textit{A thesis submitted in fulfillment of the requirements for the \\ VU Bachelor of Science degree in Computer Science.}

{July 15, 2025}


{{\textit{Many thanks to AtLarge team, our partners, and people involved in this research.}}}

\begin{tabular}{ccccc}
\includegraphics[height=10mm]{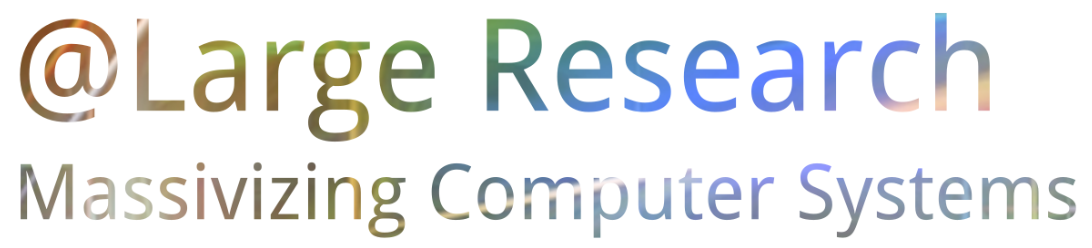} &
\shortstack{{IBM Research} \\\vspace{0.1cm} {Europe/Zurich}} \hspace{0.3cm}
\includegraphics[height=10mm]{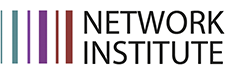} &
\includegraphics[height=10mm]{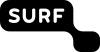} &
\includegraphics[height=10mm]{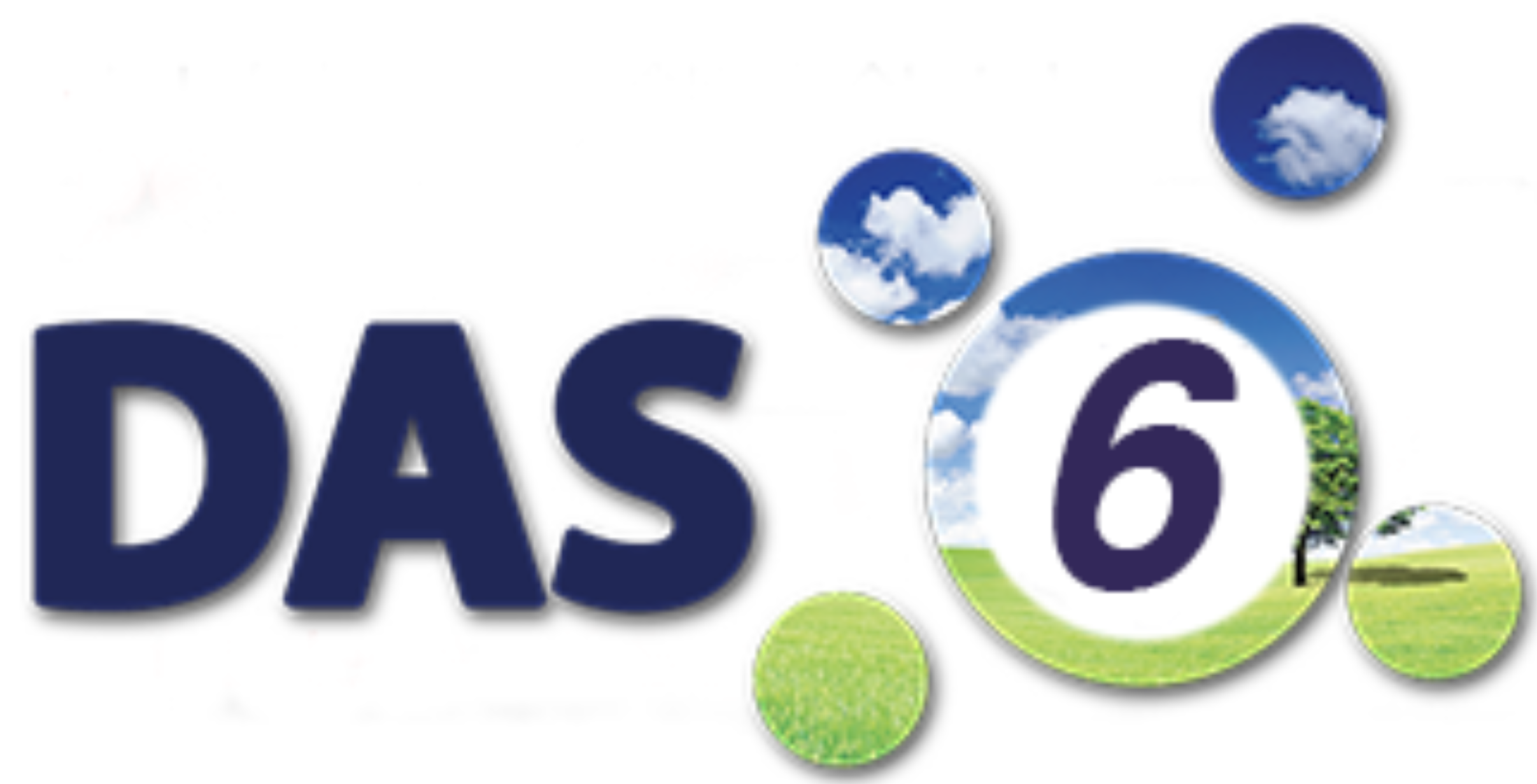} &
\end{tabular}

\end{center}
\newpage
\restoregeometry

\blankpage

\thispagestyle{empty} 
\chapter*{Abstract}\label{sec:abstract}

Large Language Models (LLMs) are widely used by our increasingly digitalized society, but raise sustainability, performance, and financial concerns, especially as inference workloads grow. 
To improve the design and operation of LLM ecosystems, we envision simulators and simulation-based digital twins becoming primary decision-making tools. LLM ecosystems leverage many heterogeneous components, making simulation a non-trivial, yet critical operation. 
The simulation challenge is exacerbated by the absence of a comprehensive reference architecture of LLM ecosystems; the lack of such a conceptual model can be costly and could misguide the designers and engineers. Without a reference architecture, even the most experienced stakeholders could tinker in researching, engineering, or maintaining LLM ecosystems.
In this work, we bring a three-fold contribution to the scientific community. 
Firstly, we synthesize, propose, and validate a reference architecture (RA) of LLM ecosystems under inference. 
Then, adhering to the reference architecture, we design \underline{K}a\underline{v}ier, the first simulation instrument able to predict the performance, sustainability, and efficiency of LLM ecosystems under inference, through discrete-event and cache-aware simulation, focusing on \underline{K}ey-\underline{V}alue-(KV-)Caching and prompt prefix caching policies. 

Through experiments with a Kavier prototype and real-world traces, (i) we measure the accuracy of Kavier and its performance in massive-scale simulations, (ii) we compare the performance of different KV-Caching policies, and (iii) we analyze the performance, sustainability, and efficiency of LLM ecosystems under various prefix caching policies. 
Through experiment (i), we demonstrate that Kavier can simulate hundreds of GPU hours in a matter of seconds, at second granularity, and with error rates of less than 10\%. 
Through experiments (ii) and (iii), we identify and quantify operational aspects of caching in the context of LLM inference. 
Specifically, in experiment (ii), we quantify improvements of 2-3 orders of magnitude on performance when LLM ecosystems adopt KV-Caching. In experiment (iii), we identify that prefix caching can reduce latency by up to 65\%, with cascading improvements also in environmental and financial costs.
Overall, we show that Kavier enables operators, researchers, and engineers to predict LLM ecosystems in a time, performance, and cost-efficient way.

\subsection*{Keywords} \label{sec:keywords}
LLMs, LLM ecosystems, KV-Cache, discrete-event simulation, performance, sustainability, efficiency, energy utilization, OpenDC 
 \newpage

\blankpage

\newpage

\chapter*{}
\vspace{-14.5em}

{
    \huge \bfseries
    Acknowledgments

    \vspace{0.1em}
}

\textit{``Massivizing Computer Systems, from the Metaverse to the Continuum"} -- this was my very first lecture at Vrije~Universiteit~Amsterdam, on September 6th, 2022. Back then, I already had a few years of experience in software engineering, and I was in love with computers and programming (turns out, years later, I still am!), but I had no idea what science was. I remember the excitement the lecturer had towards computer science and towards researching distributed computer ecosystems with impact on millions. That passion was quite contagious, and suddenly I was also excited about the massive-scale computer (eco)systems. I wanted to learn more. The conversation afterwards with the lecturer sparked a large interest in the field. 

Almost three years later, I am taking the last steps of my journey towards my Bachelor's degree. During this time, I had unique research and teaching opportunities. Through my enrollment in the Honours Programme, I had the opportunity to lead an ambitious scientific research project on datacenter multi-model simulation through which I broke new ground. Still on the scientific side, I had the pleasure to give numerous invited talks, write a scientific article for peer-reviewed publication, and co-supervise students in their first scientific steps. On the teaching side, I had the opportunity to run large-enrollment courses, give lectures, and contribute to shaping the computer science and honours programme curricula of the VU. I was deeply honoured to be selected as the student of the year by the Faculty of Science and to be awarded the Student~Talent~Award by the Faculty. I owe much to all the people who facilitated all of these.

For the remainder of this (too) limited space, I would like to offer both nominal and non-nominal gratitude to all the people who have been with me through this exciting journey, either always or episodically. 

Alexandru, \textit{mulțumesc}. You have been my guide through science and taught me almost everything I know about ``compsys," research methodology, and scientific reasoning. You believed in me from the very beginning and invited me to be part of \textit{the} leading computer systems research group in the country.
You've offered me high-tier, high-responsibility opportunities, and we've built some big things together. I'm looking forward to seeing how they'll concretize in the future (see DT :D). Thank you for the long and varied conversations, for mentoring me when I was unsure, and for the encouragements when I felt low (the \textit{``capul sus"} always made the situation at least n\% better). From \textit{``fishing"} hours, 6 am, to late-evening meetings, you've always inspired me to learn more, challenge myself, and seek the highest level of depth. For all, I'm grateful.

Animesh, it's been a pleasure to work together and learn more about your research and teaching style. I am grateful for all the shared knowledge and insightful papers, and especially thankful for the constructive feedback and healthy harshness. Jesse, thank you for teaching me research methodology, computer networking, metaverse, and online gaming systems, and I'm deeply grateful for our conversations on both professional and personal topics. I also want to thank the entire AtLarge team for always being so helpful, approachable, and supportive; especially, many thanks (alphabetically), Daniele, Dante, Hexiang, Krijn, Matthijs, Sacheen, Tiziano, Vincent, and Xiaoyu.

Thank you to the CS Department for your support and opportunities. Ivano, I had a lot of fun running the Network Institute together; it was very insightful to explore the endless world of interdisciplinary science. Thank you, Sara, for your help in so many things - I'd probably double the size of this thesis if I were to mention all of them. Thilo, it's been a pleasure to run computer programming together - thank you for your trust and support. Thank you, Mojca, \textit{``the mother of the CS department"}, and your always good energy during our sometimes too long conversations; Mojca, you never failed to put a smile on my face.

All these would not have been possible without my amazing friends and colleagues, who helped me disconnect from scientific fun when I needed a break. Thank you, Daniel, Cristi, Sofia, Ana, Lara, Maja, Clara, Cătălin, Traian (Finu), Lennart, Dovydas (Dovy), Sorin (Sorinel), Matei, Isidora, Nader, Ștefan (Fane), Alexis $...$ . A special thank you to Fadime -- \textit{sağ ol}, Fa. Thank you to my riding friends who constantly reminded me that two wheels and some miles are the best therapy: Rareș, Voshon, Fifi (wherever you'd be, bud), and Daniel.

I would like to deeply thank my family for all their support, of all kinds, and at all times. This goes beyond my Bachelor's and beyond my career -- thank you for always being here and e-here. Thank you for supporting me in my passion and in moving to the other side of the continent to follow this passion. Regardless of the proximity, parts of my heart will always be in Bucharest, Victoria, and Hațeg. Lastly and firstly, thank you, D.S. and A.B.

\hspace{40.99em}
\textit{Radu Nicolae}
 \newpage

\setcounter{tocdepth}{1}
\tableofcontents
\thispagestyle{empty}  

\clearpage
\pagestyle{fancy}
 \newpage
\clearpage

\thispagestyle{noheader}
\chapter{Introduction}\label{sec:intro}

LLM ecosystems are being adopted at an unprecedented scale~\cite{DBLP:journals/corr/abs-2206-03259, DBLP:conf/sosp/ZhangDLKMWLYLLZ25}, and are hosted on massive-scale, intensely used ICT infrastructure, hence raising concerns about performance, sustainability, and efficiency~\cite{DBLP:journals/cacm/Chien23a, DBLP:conf/mlsys/WuRGAAMCBHBGGOM22, DBLP:journals/corr/abs-2405-21015, simon2024llm-mooreslaw}. To understand ICT infrastructure, numerous simulators have been proposed in the past decades, such as OpenDC~\cite{DBLP:conf/ccgrid/MastenbroekAJLB21}, DCSim~\cite{gupta2011gdcsim}, GDCSim~\cite{gupta2011gdcsim}, or CloudSim~\cite{DBLP:journals/spe/CalheirosRBRB11}. However, no simulator supports the prediction of LLM ecosystems under inference and cache-awarely. For LLM ecosystems, the caching system is crucial, especially Key-Value Caching (KV-Caching) and prompt prefix caching~\cite{vaswani2017attention,  DBLP:conf/sosp/ZhangDLKMWLYLLZ25, openai_prompt_caching}, and proven to have significant impacts on performance, sustainability, and efficiency, sometimes of orders of magnitude. Although simulators for predicting LLMs have been proposed (e.g., Vidur~\cite{agrawal2024vidur}, LLMServingSim~\cite{cho2024llmservingsim}), none of them can cache-awarely simulate the performance, sustainability, and efficiency of LLM ecosystems under inference. Exacerbating this challenge, the absence of a reference architecture of LLM ecosystems under inference prevents rigorously designing and implementing a scientific simulation instrument; adhering to state-of-the-art methodology~\cite{DBLP:conf/icdcs/IosupVTETBFMT19}, followed by top-tier publications~\cite{DBLP:conf/sc/AndreadisVMI18, DBLP:conf/ccgrid/JansenAPTI23, DBLP:journals/tpds/AndreadisMBI22, DBLP:conf/ccgrid/MastenbroekAJLB21}, simulators should be designed and implemented as mapped to validated reference architectures. Furthermore, the absence of a reference architecture can misguide even the advanced groups of engineers, researchers, and operators, which could overlook crucial aspects of the ecosystem (e.g., prompt prefix caching). Identifying the absence of a conceptual model and of a simulation instrument leads to the main research question: \textit{(MRQ) How to enable analysis of LLM Ecosystems under inference through discrete-event simulation?} In this work, we address the MRQ and propose a dual main contribution, first, a detailed and comprehensive reference architecture to guide how LLM-inference systems are designed, deployed, and analyzed; second, the Kavier tool to simulate and analyze LLM-inference systems based on the reference architecture. Kavier facilitates the community to explore real-world LLM ecosystems in a time and cost efficiency way, through simulation-driven experimentation. Having a better understanding of massive-scale computer ecosystems, especially LLM ecosystems, can lead to significant improvements in these systems' performance, sustainability, and efficiency overall~\cite{DBLP:journals/corr/abs-2206-03259}.

Our society and economy are increasingly dependent on AI services, especially on Large Language Models (LLMs)~\cite{DBLP:journals/corr/abs-2206-03259, DBLP:conf/sosp/ZhangDLKMWLYLLZ25, DBLP:conf/mlsys/WuRGAAMCBHBGGOM22}. Correspondingly, LLMs are becoming more accessible to the public at large, while leveraging more complex architectures and consuming massive computational resources~\cite{DBLP:journals/cacm/Chien23a, DBLP:conf/mlsys/WuRGAAMCBHBGGOM22}. Since the launch of GPT-3 in November 2022, LLMs have been and are being increasingly embedded in operational processes across industry, government, and academia~\cite{Agrawal2025EnergyEL, DBLP:journals/corr/abs-2408-07326}. In industry, as of 2025\footnote{This thesis has been written and submitted in 2025, including information and data which might change over time. Although apparently ephemeral, the presented numbers highlight the massive scale of LLM services and their societal impact, which is projected to grow in the upcoming decades.}, tech giants integrate LLMs into their search engines (e.g., Gemini, Copilot), and customer service platforms, processing billions of queries daily~\cite{DBLP:journals/corr/abs-2408-07326, DBLP:journals/corr/abs-2501-11006}; in academia, LLMs are widely used in research, especially in field as computer science, medicine, chemistry, or biology, in processes of scientific writing, data analysis, or programming assistance~\cite{alzaabi2023chatgpt, kim2023chatgpt, meyer2023chatgpt, liang2024mapping}; governments deploy LLMs for public service automation~\cite{qin2024study} and policy analysis~\cite{safaei2024end, dai2024applications}. The wide use of these services is reflected in various costs -- LLMs are trained and run at massive sustainability, financial, and performance costs~\cite{DBLP:conf/sosp/ZhangDLKMWLYLLZ25, DBLP:journals/corr/abs-2206-03259}.

The environmental footprint of LLMs is massive and is only expected to grow, further exacerbating the already concerning global challenges in resource allocation, energy consumption, and CO2 emissions \cite{DBLP:journals/cacm/Chien23a}. This footprint begins as soon as the production of the hardware on which the LLM ecosystem is deployed~\cite{DBLP:journals/corr/abs-2408-07326}. For example, manufacturing a single NVIDIA H100 GPU, widely used in AI training and inference infrastructure, generates hundreds of kgCO2 (estimated between 200-500kg CO2 per GPU unit)~\cite{nvidia_h100, DBLP:journals/cacm/Chien23a}; equipping a hyperscale datacenter with 10,000 NVIDIA H100 GPUs results in over 20,000-50,000 tCO2 emissions before even deploying the LLM ecosystem and without considering other components of such a datacenter~\cite{iea2023netherlands, DBLP:journals/corr/abs-2501-11006}. The footprint of the training process is significant even for GPT-3, a small LLM by 2025 standards, which consumed 1.28~GWh of energy, and emitted 553 metric tons of CO2~\cite{patterson2022carbon}, equivalent to 123 gasoline cars driven for a year~\cite{epa2023emissions}. Larger LLMs, such as Google's PaLM-2 (340B parameters), require 3.4x more energy than GPT-3 during training, while models like Antropic's Claude 3 (500B parameters) consume over 5~GWh per training run~\cite{Agrawal2025EnergyEL}. Lastly, inference exacerbates exponentially this climate footprint - ChatGPT service, at peak usage, consumes over 1~GWh daily~\cite{DBLP:journals/corr/abs-2408-07326}, on par with approximately 40,000 Dutch houses~\cite{iea2023netherlands}. However, the CO2 intensity varies by the energy source: LLMs run on coal-based power grid can emit up to 2-3 orders of magnitude more CO2 than LLMs run on renewable or nuclear-powered datacenters~\cite{masanet2020recalibrating, DBLP:journals/jmlr/LuccioniVL23}. Cumulatively, in 2025, LLMs are estimated to account for approximately 3\% of the electricity consumption of the global datacenters, and this proportion is expected to exponentially grow~\cite{andrae2015global}.

The financial costs of training and running LLM are increasing at an unprecedented pace, with an estimated growth of an order of magnitude per year in compute costs~\cite{DBLP:journals/corr/abs-2405-21015}. \textit{Cottier et al.} note the magnitude growth in the regression mean for training frontier AI models, from approximately \$10k in 2016, to approximately \$8M (\$0.08B) in 2024, and expected to overtake \$1B by 2027~\cite{DBLP:journals/corr/abs-2405-21015}. Inference costs are equally staggering~\cite{shoham2024longcontext, DBLP:conf/wosp/ChowTLRB24, long_context2025}: ChatGPT costs \$700,000 per day to operate, and using GPT-4 to support customer service can cost a small business \$21,000 a month~\cite{DBLP:conf/wosp/ChowTLRB24}. The financial costs can grow exponentially for million-scale token contexts (e.g., Google's Gemini 1.5), which increases GPU memory usage by 4-8x, and proportionally the cloud hosting fees~\cite{long_context2025}; some cloud providers estimate a single 100K-token prompt at \$0.50 scale in cloud compute fees, while a small prompt at only fractions of a cent \cite{shoham2024longcontext,long_context2025}. Still, it is essential to note that many factors could influence the price of LLM inference, and, although the number of tokens influences the price, there is no correlation between these two metrics; for example, the measured cost of running Jamba 1.5 Large (256k parameters) on Amazon Bedrock was of \$0.32 per 100k tokens of prompt (equivalent to \$0.81 per prompt), while the cost of running Clause 3.5 Sonnet (200k parameters) on the same infrastructure was of \$0.54 per 100k tokens of prompt (equivalent to \$1.08 per prompt)~\cite{shoham2024longcontext}. Scaled to an audience of tens/hundreds of millions of active, intense users, these costs become unsustainable.

The performance costs of LLMs are reflected into a ``modern-day Moore's law"~\cite{simon2024llm-mooreslaw}, where available infrastructure fails to meet the ultra-high demands of ultra-large LLMs. GPT-4 was trained in 5,000 - 10,000 GPU years\footnote{GPT-4 is estimated to have used 25,000 NVIDIA A100 GPUs run for 3-4 months~\cite{achiam2023gpt, xu2024hethub, DBLP:journals/corr/abs-2405-21015}. Official numbers are undisclosed by OpenAI, for undisclosed reasons.}; the Ice Age terminated $\approx$11,700 (human) years ago, if we sequentialize the GPUs and assume no external factors, the training of GPT-4 would have started a few millennia before pyramids were built, just after the end of the Ice Age~\cite{achiam2023gpt, xu2024hethub, DBLP:journals/corr/abs-2405-21015, cheng2009ice, spence2000ancient}. Albeit the massive scale of the training stage, the training becomes the smaller sibling of the inference, which increases proportionally with the exponentially growing number of users \textit{and} size of models; Google estimates the ratio between training and inference as 40 to 60, thus showing how "modern"-day AI spends most of its lifetime in the inference stage~\cite{Cho2023AICarbon, Patterson2022GoodNews}. \textit{Chien et al.} analyze ChatGPT and Google AI services and observe that annual inference needs 25x, respectively 1,386x more compute resources, than were needed to train GPT-3~\cite{Chien2023Reducing}.

LLMs run on LLM ecosystems, which are \textit{``non-trivially heterogenous groups of computer systems, distributed in nature"}~\cite{book-distributed-systems}, and spanning across all three layers of the Compute Continuum: endpoint, edge, and cloud~\cite{DBLP:conf/ccgrid/JansenAPTI23}.
We argue that predicting LLM ecosystems is a society-critical yet non-trivial simulation problem that could lead to significant service improvements, cost savings, and greener LLMs towards a better sustainability of worldwide digital services. Simulation enables large-scale and fine-grained exploration, analysis, and comparison of systems technologies~\cite{DBLP:conf/ccgrid/MastenbroekAJLB21, radunicolae-hp-m3sa}. The constant accelerating, increasing rate and demand for computing power, further exacerbated by the public-wide availability of LLM, has led to a substantial expansion of datacenter infrastructure, especially in scale and complexity, making datacenter simulation essential from economical, performance, and environmental perspectives~\cite{DBLP:conf/ccgrid/MastenbroekAJLB21, DBLP:journals/fgcs/MastenbroekMBI25}. The climate impact of experimentation through simulation of a datacenter configuration under workload, compared to the climate impact of experimenting with a real-life building, configuration, and running the workload, is 8 to 12 orders of magnitude lower, assuming the simulation is accurate and correct~\cite{iosup2021massivizingkeynote, DBLP:conf/ccgrid/MastenbroekAJLB21,radunicolae-hp-m3sa}; for example, an analysis conducted by Mastenbroek~et~al. estimate a ratio of 1:116,000,000,000 in energy consumed to conduct simulations over the equivalent real-world experiments~\cite{DBLP:conf/ccgrid/MastenbroekAJLB21}. Although the high importance of simulating ICT infrastructure which hosts highly resource-hungry systems, the current state-of-the-art simulators can predict only individual components e.g., memory, CPUs, GPUs, networking, or only shallowly integrated; it has never been proposed a simulator, nor a comprehensive reference architecture of such an instrument, able to simulate performance, sustainability, and efficiency of LLM ecosystems run on large-scale ICT infrastructure. 

Although such a simulation instrument does not exist, the development and hosting of LLM ecosystems is only accelerating and considered to be a modern-day Moore's law~\cite{bommasani2021opportunities, simon2024llm-mooreslaw}, with number of parameters growing exponentially and increasing the performance gap between LLM size and hardware performance \cite{lazuka2024llm, narayanan2021efficient, ramponi2024llm}. While most LLM-oriented research is focused on improving the performance of LLM training, towards higher accuracy, higher throughput, and lower latency, the high-level picture of the LLM ecosystem~\cite{lazuka2024llm}, highly integrated and heavily distributed, is yet blurry and remains unexplored~\cite{DBLP:conf/hpec/SamsiZMLMJBKTG23}. One approach to meet the performance-hardware gap, and host the increasingly heavy LLM ecosystems, is scaling up the hardware used for training and inference~\cite{bommasani2021opportunities, Agrawal2025EnergyEL}; this is the current approach adopted by AI giants~\cite{microsoft-openai-10billion, threemileisland}, yet projected to reach soon the so-called ``modern-day Moore's law"~\cite{lazuka2024llm, simon2024llm-mooreslaw}. Another approach to bridge this gap is by carefully anticipating datacenters, through accurate and reliable simulation processes that predict ICT infrastructure under real-world workloads and systems (e.g., LLM ecosystems); such anticipation can happen both before building datacenters, or after building, during operation, helping in dynamic adjustments and resource allocation~\cite{DBLP:journals/tpds/AndreadisMBI22, DBLP:journals/fgcs/MastenbroekMBI25, DBLP:conf/ccgrid/MastenbroekAJLB21}. Many vetted and community-wide tested simulators already exist~\cite{DBLP:conf/hpec/SamsiZMLMJBKTG23, DBLP:journals/corr/abs-2408-13386, DBLP:journals/simpra/JawaddiI24, DBLP:conf/ccgrid/MastenbroekAJLB21, DBLP:journals/tomacs/GuptaBAVJFGM14}. However, none support the simulation of performance, sustainability, and efficiency of LLM ecosystems. Exacerbating the simulation challenge, there is no comprehensive reference architecture for LLM ecosystems, which could be further integrated into simulation processes.

In this work, we identify and address the major sustainability concerns raised by LLMs, the increasing performance gap between LLM ecosystems and ICT infrastructure, and the efficiency concerns of LLM ecosystems. We identify the lack of understanding of how these ecosystems operate and how their internal (eco)systems interact. Addressing these community-knowledge gaps, we propose a high-level conceptual model of the LLM continuum, which we design and validate across real-world ecosystems. We refer to this conceptual model as \textit{reference architecture.} Then, we propose Kavier, a scientific instrument for simulating the LLM continuum; our instrument adopts \textit{discrete-event simulation}, where the operation of a system is represented as a sequence of events over time, with the assumption that no changes occur in-between events~\cite{banks2005discrete, DBLP:conf/ccgrid/MastenbroekAJLB21}. We design a Kavier capable of predicting the performance, sustainability, and efficiency of LLM ecosystems under inference, as well as cache-aware simulation. We then implement a prototype of this design, which we release as open science. We identify the absence of traces showing the relationship between the amount of prefill/decode tokens and their impact on performance. To address this challenge, we deploy LLM ecosystems on real-world infrastructure and conduct measurements. Lastly, through trace-based experimentation, we successfully validate the engineered prototype and demonstrate the superiority of simulation-driven experiments over real-world infrastructure experimentation. We then evaluate the impact of different caching policies on performance, sustainability, and efficiency by first analyzing prefix caching and subsequently examining KV-Caching in autoregressive transformer LLMs.

\section{Problem Statement}\label{sec:intro:probem-statement}


Our society and economy are increasingly dependent on AI services, especially on LLMs, which are being increasingly embedded in operational processes across industry~\cite{DBLP:journals/corr/abs-2408-07326, DBLP:journals/corr/abs-2501-11006}, academia~\cite{alzaabi2023chatgpt, kim2023chatgpt, meyer2023chatgpt, liang2024mapping}, and government~\cite{qin2024study, safaei2024end, dai2024applications}. However, LLMs are not free to train and host, and surely not cheap: inference of ChatGPT consumes energy, daily, on par with approximately 40,000 Dutch houses~\cite{DBLP:journals/corr/abs-2408-07326, iea2023netherlands}, GPT-4 training time is estimated at 5k-10k GPU years~\cite{achiam2023gpt, xu2024hethub, DBLP:journals/corr/abs-2405-21015}, and the training process of GPT-3, an already ``small" model by nowadays standards, consumed 553 metric tons of CO2~\cite{patterson2022carbon}. AI services, including LLM ecosystems, are trained and hosted on large-scale ICT infrastructure~\cite{DBLP:journals/corr/abs-2206-03259, DBLP:conf/sosp/ZhangDLKMWLYLLZ25}; therefore, the performance and sustainability of LLM services is directly dependent on the performance and sustainability of the datacenter on which the systems are run.

Proven to be critical instruments in datacenter designing, scaling, maintaining, and building, many datacenter simulators have been developed and used worldwide~\cite{DBLP:conf/hpec/SamsiZMLMJBKTG23, DBLP:journals/corr/abs-2408-13386, DBLP:journals/simpra/JawaddiI24, DBLP:conf/ccgrid/MastenbroekAJLB21, DBLP:journals/tomacs/GuptaBAVJFGM14}. Simulation is a powerful tool that can help stakeholders anticipate ICT infrastructure at a fraction of a cost; \textit{Mastenbroek et al.} estimate a ratio of 1:116,000,000,000 in energy consumed to conduct simulations over the equivalent real-world experiments~\cite{DBLP:journals/fgcs/MastenbroekMBI25, iosup2021massivizingkeynote}. 

An alternative to simulation is scaling up the infrastructure~\cite{threemileisland, microsoft-openai-10billion} (``scaling-by-credit-card"~\cite{iosup2021massivizingkeynote}). This approach, albeit currently functional, is unsustainable in the long run, leading to a projected ``modern-day Moore's law"~\cite{bommasani2021opportunities, simon2024llm-mooreslaw}, where the number of parameters grows exponentially, and the hardware performance logarithmically, at best linearly~\cite{lazuka2024llm, narayanan2021efficient, ramponi2024llm}. To cover this gap, we propose carefully and responsibly anticipating datacenters hosting LLM ecosystems, through accurate and reliable simulation processes. Therefore, we identify \textbf{PS1}:

\begin{enumerate}[label=\textbf{PS1}]
    \item \label{sec:intro:thesisContributions:item1} Although robustly simulating ICT infrastructure is highly important for the community, especially simulating LLM ecosystems, no simulator currently is capable of predicting performance, sustainability, and efficiency of LLM ecosystems under inference. Without such a scientific instrument, exploration of LLM ecosystems, timely and efficiently, can be hindered.
\end{enumerate}

To rigorously design a datacenter simulator, tailored to predicting LLM ecosystems, the state-of-the-art is materializing a reference architecture into a simulation tool or instrument~\cite{DBLP:journals/tpds/AndreadisMBI22, DBLP:conf/ccgrid/MastenbroekAJLB21}. However the high importance of simulation and, thus, the high importance of a reference architecture, currently, there is no comprehensive reference architecture for inference of LLM ecosystems, hence exacerbating PS1. Although progress has been made in this direction, proposed reference architectures are incomplete~\cite{lu2023towards}, non-inference oriented~\cite{lu2023towards, bucaioni2025functional, mahr2024reference}, assume a (too) high degree of homogeneity of LLM ecosystems~\cite{bucaioni2025functional}, vetted, following state-of-the-art approaches in distributed systems, but too universal~\cite{DBLP:conf/ccgrid/JansenAPTI23}, or do not follow a distributed systems approach~\cite{bucaioni2025functional}. The existing reference architectures are not necessarily wrong, yet they are unsuitable for materializing within a unitary simulation entity. 

We identify various ecosystems for serving LLM inference in practice, such as the IBM inference ecosystem, highly homogeneous and self-contained, the Databricks inference ecosystem, highly heterogeneous and self-contained, and the inference ecosystem envisioned by Ubicloud, highly heterogeneous and distributed. We identify the main requirement of proposing a reference architecutre abstract enough to model these distinct natures of these ecosystems, while still being specific enough to model intricate behaviour of LLM inference as opposed to traditional, more homogeneous computer ecosystems (e.g., storage, video streaming).

Moreover, we identify very visible effects of the lack of a reference architecture over the compute continuum of LLM ecosystems under inference. The lack of comprehension of this conceptual model leads to incomparable designs, inability to discuss practical shortcomings of existing ecosystems, and difficulties in operating and expanding existing ecosystems.

In this work, we propose the first reference architecture of LLM ecosystems under inference and map our reference architecture to numerous vetted, standard, and universal reference architecture, thus generalizing our model to even more applications that can leverage LLM pipelines. In other words, we are generalizing over the continuum and tailoring for LLM inference pipelines. This raises \textbf{PS2}:

\begin{enumerate}[label=\textbf{PS2}]
    \item \label{sec:intro:thesisContributions:item1} Currently, there is no comprehensive reference architecture for inference of LLM ecosystems.
\end{enumerate}

Recent advances in KV-Caching optimization focus on local improvements: PyramidInfer achieves double throughput and halves GPU memory reduction in KV-Cache, using layer-wise dynamic allocation~\cite{DBLP:conf/acl/YangHGHZ024}; AIBrix proposes a distributed KV-Cache approach, which boosts token reuse across nodes, leading to a 50\% increase in throughput and a 70\% reduction in inference latency~\cite{team2025aibrix}; Jenga introduces a two-level memory allocator that reduces fragmentation by 80\% through LCM-based page sizing and request-aware allocation, thus improving throughput by 1.8x in heterogenous LLMs. However, these approaches neglect system-wide impacts, primarily focusing on KV-Cache-GPU interaction, without modeling datacenter-scale performance, energy usage, or CO2 emissions. This leads to a critical gap: current KV-Caching research focuses on isolated operational layers of the Compute Continuum, mainly KV-GPU, while this work pioneers vertical simulation of the KV-Caching system, as integrated within an LLM ecosystem. Therefore, we identify \textbf{PS3}:

\begin{enumerate}[label=\textbf{PS3}]
    \item \label{sec:intro:thesisContributions:item1} Current research and experimentation focuses on isolated operational layers of the Compute Continuum, mainly on the interaction between KV-Caches and GPUs, and fails to provide a vertical simulation of the KV-Caching system as integrated within a unitary, highly-distributed, and heterogeneous LLM ecosystem.
\end{enumerate}

\section{Research Questions} \label{sec:intro:research-questions}
To address the aforementioned challenges, we raise the main research question \textbf{(MRQ)}, from which we refine a sequence of four research questions \textbf{(RQ)}.

\noindent
\begin{tabular}{@{}p{1cm} p{14cm}@{}}
    \textbf{MRQ} & How to enable analysis of LLM ecosystems, through discrete-event simulation?
\end{tabular}

\section*{Research Question 1} \label{sec:intro:rq1}
LLM ecosystems have a massive societal impact when run at a worldwide scale, and raise significant sustainability, performance, and economic concerns, especially when workload grows~\cite{DBLP:conf/sosp/ZhangDLKMWLYLLZ25, DBLP:journals/corr/abs-2206-03259, DBLP:journals/cacm/Chien23a}. LLM ecosystems are run on large-scale infrastructure, resource-hungry, and with an increasing gap between the LLM needed resources and infrastructure capabilities~\cite{lazuka2024llm, simon2024llm-mooreslaw}; scaling up datacenters is only a temporary fix~\cite{microsoft-openai-10billion, threemileisland, simon2024llm-mooreslaw}. As the size and demand of ICT infrastructure grow, we envision simulators and simulation-based digital twins becoming primary decision-making tools, helping to meet Service Level Objectives (SLOs) without trading sustainability. LLM ecosystems become increasingly heterogeneous~\cite{DBLP:conf/sosp/ZhangDLKMWLYLLZ25, lazuka2024llm}, making simulation a non-trivial, yet critical operation. The simulation challenge is exacerbated by the absence of a comprehensive reference architecture of LLM ecosystems.

Toward answering RQ1, we propose a high-level abstraction of LLM ecosystems, leveraging the entire process from user's input to the system output. We identify the main components of such a system tailored with industry-standard technologies, and individually describe each component, its purpose, and its interaction with other components. Proposing a high-level and comprehensive abstraction of LLM ecosystems raises the research question:

\noindent
\begin{tabular}{@{}p{1cm} p{14cm}@{}}
    \textbf{RQ1} & How to synthesize and validate a reference architecture of LLM ecosystems? \label{RQ1}
\end{tabular}

\section*{Research Question 2} 
\label{sec:intro:rq2}
It has never been proposed a scientific instrument for simulating the performance, sustainability, and efficiency of LLM ecosystems under inference, following a cache-aware and discrete-event simulation model. We identify the main challenge of proposing a design of such a scientific instrument which ensures not only meeting the functional requirements, but also simulating with a close-to-reality accuracy, lightning-fast performance, and seamless integration with a peer-reviewed and community-vetted datacenter simulator. 

Toward answering RQ2, we propose \underline{K}a\underline{v}ier, a scientific instrument for (\underline{KV})-cache-aware simulation of the continuum of LLM ecosystems under inference. Following AtLarge Design Methodology~\cite{DBLP:conf/icdcs/IosupVTETBFMT19}, we establish a set of functional and non-functional requirements to guide our design process. Then, also adhering to the reference architecture obtained after answering RQ1, we analyze multiple design choices and select the best-aligned decisions with the established requirements. Then, we focus on each core simulation component (e.g., performance, sustainability, and efficiency) and individually detail the simulation approach for each such model. We also detail the cache-aware simulation component and how caches affect prefill and decode time. The design process is a critical and non-trivial step in the research process, which raises the main research question:

\begin{tabular}{@{}p{1cm} p{14cm}@{}}

    \textbf{RQ2} & How to design Kavier, a scientific instrument for cache-aware simulation analysis of the performance, sustainability, and efficiency of LLM ecosystems under inference? \label{RQ2}
    \\
\end{tabular}

\section*{Research Question 3} 
\label{sec:intro:rq2}
Proposing a (successful) novel simulation concept and scientific instrument is a rarity in our field and represents a potential massive-scale contribution if widely adopted, especially in widely used technologies such as LLM ecosystems, heavily reliant on KV-Caches, with potential improvements of orders of magnitude~\cite{DBLP:conf/sosp/ZhangDLKMWLYLLZ25, radunicolae-hp-m3sa}. The main challenge is demonstrating the ability to rigorously implement and integrate Kavier into a vetted simulator. This raises three sub-challenges. Firstly, we identify the challenge of materializing the design proposed in RQ2 into an engineered prototype with minimal redundancy, maximized accuracy, performance, integration, and abstraction, following industry-standard, state-of-the-art techniques. Secondly, we identify the challenge of integrating the engineered prototype with a top-tier, peer-reviewed simulator while respecting the functional and non-functional requirements. Thirdly, we identify the challenge of integration towards further development and research processes, following principles of open-source and open-science, and allowing for further steps towards simulating LLM ecosystems, following the reference architecture proposed in RQ1.

Toward answering RQ3, we propose Kavier, a scientific instrument able to simulate LLM ecosystems and strictly adhere to the design obtained by answering RQ2. To answer RQ3, we identify the main requirement of integrating Kavier within a large-scale, top-tier datacenter simulator. We engineer Kavier as able to predict by leveraging multiple simulation models, following principles of Multi-Model~\cite{radunicolae-hp-m3sa, harrison2018brief, myhre2017multi} and Meta-Model simulation~\cite{nicolae2025m3sa, radunicolae-hp-m3sa}. We integrate Kavier within a top-tier datacenter simulator, and leverage its peer-reviewed capabilities of simulating sustainability. The implementation and component raise the research question:

\begin{tabular}{@{}p{1cm} p{14cm}@{}}
    \textbf{RQ3} & How to implement and integrate Kavier within a peer-reviewed, discrete-event datacenter simulator? \label{RQ3}
\end{tabular}

\section*{Research Question 4} 
\label{sec:intro:rq4}

Reiterating the statement above, proposing a (successful) novel simulation concept and scientific instrument is a rarity, yet with society-wide impact~\cite{nicolae2025m3sa, DBLP:conf/ccgrid/MastenbroekAJLB21}. Towards evaluating Kavier, we design three experiments. The experiments focus both on quantifying and evaluating Kavier, against well-defined criteria (i.e., functional and non-functional requirements), and on exploring real-world scenarios using Kavier (e.g., exploring sustainability - CO2 emissions, energy consumption - of LLM ecosystems, evaluating workload performance).

To answer RQ4, we establish the experiment setup and synthesize an overview of the experiments. State-of-the-art methodology in the field adopts simulation to check various operational properties (both functional-, and especially non-functional requirements) for distributed systems and ecosystems~\cite{DBLP:conf/icdcs/IosupVTETBFMT19, Mastenbroek2023RADICE, DBLP:journals/tpds/AndreadisMBI22, dniewenhuis_hotcloud_footprinter, DBLP:conf/ccgrid/MastenbroekAJLB21}. We firstly quantify Kavier's accuracy and its performance during simulation. We further explore two real-world scenarios using the just-designed and prototyped instrument, and summarize the main findings. The experimentation journey raises the research question:

\begin{tabular}{@{}p{1cm} p{14cm}@{}}
    \textbf{RQ4} & How to evaluate a Kavier prototype with trace-based realistic scenarios? \label{RQ4}
\end{tabular}

\section{Approach} \label{sec:intro:approach}
Throughout the research and engineering process, we approach the problem statement and the subsequent research questions with a distributed systems approach,\textit{``a combination of conceptual, technical, and experimental work"} \cite{DBLP:journals/fgcs/MastenbroekMBI25}, guided by the state-of-the-art AtLarge Design Process \cite{DBLP:conf/icdcs/IosupVTETBFMT19}.

To answer RQ1, we analyze existing systems and technical documentation and discuss with selected experts in the field various variants of reference architectures of LLM ecosystems, and we contrast these with a set of selected peer-reviewed articles in the community. We summarize, per architecture, the positives and negatives. We present this overview in \Cref{sec:background}. Further, in \Cref{sec:refarch}, we propose a high-level abstraction of LLM ecosystems, leveraging the entire process, from the user's input to the system's output. We follow a distributed systems approach in the modeling process and model multiple layers of abstraction, focusing on how different components connect and interact. We describe both the reference architecture, from a high-level perspective, and individually describe the functionality and scope of each component of the distributed LLM ecosystem. We then detail the KV-Caching system and provide an extensive description of its integration within the system. Lastly, we validate the proposed reference architecture against LLM inference ecosystems from the industry and against a peer-reviewed reference architecture of the ICT Compute Continuum.

To answer RQ2, we propose Kavier, a first-of-its-kind discrete-event and cache-aware simulation instrument for predicting performance, sustainability, and efficiency of LLM ecosystems under inference. Following the vetted AtLarge Vision on the Design of Distributed Systems and Ecosystems~\cite{DBLP:conf/icdcs/IosupVTETBFMT19}, we center RQ2 towards researching a rigorous design of Kavier. We synthesize functional and non-functional requirements of the simulation system, with a focus on accuracy, performance, and system-wide embed of Kavier; we integrate the latest, state-of-the-art simulation techniques, such as simulation based on multi- and meta-models\footnote{Multi-Model, or simulation using multiple models, is a novel simulation technique which uses multiple models for predicting datacenter infrastructure under workloads. These individual models are run in parallel, without interfering, and their predictions are further leveraged within a unitary prediction system, towards providing the user with a better explanation of the simulation results. The Meta-Model is an aggregation model that predicts using other models' predictions.}. Then, we propose a high-level design for a simulation instrument and detail design choices, analysis, and simulation components and models of simulator.

To answer RQ3, we implement the design from RQ2 of Kavier, and produce an engineered prototype, following state-of-the-art software engineering technologies, methods, and principles~\cite{DBLP:journals/sigsoft/Herzog15}. We integrate Kavier within a top-tier, vetted simulator to simulate the inference process of LLM ecosystems. We detail the engineering and integration process in \Cref{sec:prototype}.

To answer RQ4, we evaluate the developed prototype of Kavier against the established functional and non-functional requirements using real-world scenarios and data. After answering RQ4, thus at the end of \Cref{sec:evaluation}, also corroborated with analysis from previous chapters, we successfully validate all the functional and non-functional requirements of Kavier. We run trace-based experiments through a built prototype and analyze the impacts of various configurations of  and workloads on metrics as performance, sustainability, and efficiency. 

\section{Contributions}\label{sec:intro:thesisContributions}
This paper will impact the scientific community by providing a reference architecture of the most rapidly growing technology of the 2020s: LLMs and LLM ecosystems. A comprehensive reference architecture of the inference process for LLM ecosystems would be beneficial in generating a base of knowledge on how to design and build LLM ecosystems, as well as in easing the understanding of how the components of LLM ecosystems interact, communicate, and function together. Furthermore, such an abstraction of LLM ecosystems could facilitate further research in various directions, such as system simulation or simulation of individual components (e.g., KV-Cache). A reference architecture and simulation of the Compute Continuum behind LLM ecosystems is critical, especially with the rapid growth of these services; further exploration of these techniques could aim to responsibly \textit{massivize} LLM ecosystems.


With this research, our key contributions are:
\begin{enumerate}[label=\textbf{C\arabic*}]
    \item \label{sec:intro:thesisContributions:item1} We conduct an unsystematic literature study and detailed technology analysis to identify, synthesize, and characterize existing reference architectures of, preferably, but not restricted to, inference. We review existing system models and analyze the pros and cons of each selected model. We design a reference architecture for LLM ecosystems under inference, then validate this conceptual model against industry-leading LLM ecosystems, scientific community standards, and through structured and non-structured discussions with experts. We thus address \textbf{PS2}.
    
    \item \label{sec:intro:thesisContributions:item2} We design Kavier, a simulator for predicting LLM ecosystems as modular and integrable with datacenter simulators. Kavier is a first-of-its-kind tool, a simulator able to predict performance, sustainability, and efficiency of LLM ecosystems under inference, following a discrete-event and cache-aware simulation model. We thus address \textbf{PS1} and contribute to addressing \textbf{PS3}.
    
    \item \label{sec:intro:thesisContributions:item3} We prototype Kavier, following state-of-the-art software engineering technologies and principles. We further integrate Kavier into a peer-reviewed, top-tier datacenter simulator. We use OpenDC, an open-source platform for cloud datacenter simulation, built through 8+ years of development and operations, and vetted across numerous venues~\cite{DBLP:conf/ccgrid/MastenbroekAJLB21, DBLP:journals/fgcs/MastenbroekMBI25, DBLP:journals/tpds/AndreadisMBI22}. Following principles of open science, we release the open-source integration Kavier-OpenDC. Engineering this prototype, thus, addresses \textbf{PS1} and contributes to addressing \textbf{PS3} through simulation-driven experimentation.

    \item \label{sec:intro:thesisContributions:item4} We evaluate Kavier-OpenDC integration using real-world experimentation and traces. We seek traces that show the impact of the prefill/decode length on performance metrics, yet find no such traces available; we thus deploy and trace LLM ecosystems; we release these traces as open science. We also release the validation of Kavier's accuracy and performance as open science. Lastly, we analyze, with Kavier and trace-based experiments, the caching impacts on LLM performance, sustainability, and efficiency. We thus address \textbf{PS3}.

    \item \label{sec:intro:openScience} We release all the artifacts in this work as FAIR~\cite{GOFAIR_FAIRPrinciples} datasets and software. The artefacts from this thesis have been peer-reviewed by members of AtLarge Research Group and are available via: \\ \url{https://github.com/atlarge-research/On-Simulating-LLM-Ecosystems-under-Inference}.
\end{enumerate}

This work also represents the culmination of over eight years of contact with the computer science field, out of which three years were invested in intense academic activities, research, and piles of accumulated knowledge. This paper has a significant impact on my personal development as an independent researcher, with contributions to the computer-science community and worldwide society, towards \textit{Massivizing Computer Systems}. Many thanks to Alexandru, Animesh, and team AtLarge (more in \S Acknowledgements).

\section{Impact on Society and Computer Systems Community}\label{sec:intro:societal_impcat}

Through our contributions, we envision a significant impact on society and the computer systems community.

\begin{enumerate}[label=\textbf{I\arabic*}]
    \item \label{sec:intro:impact:item1} We anticipate our proposed reference architecture to be beneficial in generating base knowledge on how to design, engineer, operate, and expand LLM Ecosystems. Such a conceptual model could be a major step towards standardisation and allow stakeholders to compare existing ecosystems, identify strong and weak points, and take better-informed decisions on improving efficiency. This would, thus, help to address current efficiency concerns related to LLMs, such as low performance, high energy, or high CO2 footprint. 

    \item \label{sec:intro:impact:item2} We anticipate Kavier as a simulator which could be adopted by large datacenter providers, to anticipate how various ICT configurations would function under various large- and massive-scale workloads of LLM inference. We envision a multi-step approach, starting from our collaborators as a proof-of-concept (e.g., IBM, Solvinity), and internationally scaling to the largest LLM providers (e.g., Google, Meta, OpenAI). A large-scale adoption of simulation-driven experimentation would allow operators and C-level stakeholders to better reason about their infrastructure and products, and potentially alleviate the concerning resource over-exploitation.

    \item \label{sec:intro:impact:item3} We anticipate the LLM Trace Archive introduced in this work, a FAIR dataset~\cite{GOFAIR_FAIRPrinciples}, to be highly beneficial for researchers and students exploring LLM inference in the future. A unified dataset would thus alleviate efforts otherwise spent on data collection and would instead allow scientists to focus on other research processes at a higher depth. Furthermore, the LLM Trace Archive contains traces unique in the community, and we are the first to FAIRly release measurements on the relationship between the amount of prefill and decode tokens and the system performance. Releasing these traces thus offers researchers access to information otherwise inaccessible (inexistent), difficult to obtain (e.g., via tracing), or not possible to obtain in case of lack of access to such infrastructure.

    \item \label{sec:intro:impact:item4} We plan to develop educational material around simulating LLM ecosystems, aided by Kavier, and deliver as a series of interactive workshops, seminars, and assignments to educate groups of various academic ages. Furthermore, thanks to the FAIR nature of all our contributions, such material can be developed both by us and by other researchers and educators from the community, and can be in-depth explored by students who would engage in these educational activities. We envision expanding the Modern Distributed Systems MOOC course on edX\footnote{\url{https://www.edx.org/learn/computer-science/delft-university-of-technology-modern-distributed-systems}}, which uses a form of OpenDC that leverages some of these concepts and will include an exercise based on Kavier in the next edition.
\end{enumerate}

\section{Plagiarism Declaration}\label{sec:intro:plagiarismDeclaration}

I confirm that this thesis is my own work, is not copied from any source (person, Internet, or machine), and has not been submitted elsewhere for assessment. The work, findings, and formulations that do not represent my contribution are given explicit recognition via citations. The plagiarism declaration excepts \Cref{sec:intro:plagiarismDeclaration}, which is \textit{ad litteram} copied from the template report :).
\section{Thesis Structure}\label{sec:intro:thesisStructure}

\begin{figure}[t]
    \centering
    \includegraphics[width=\linewidth]{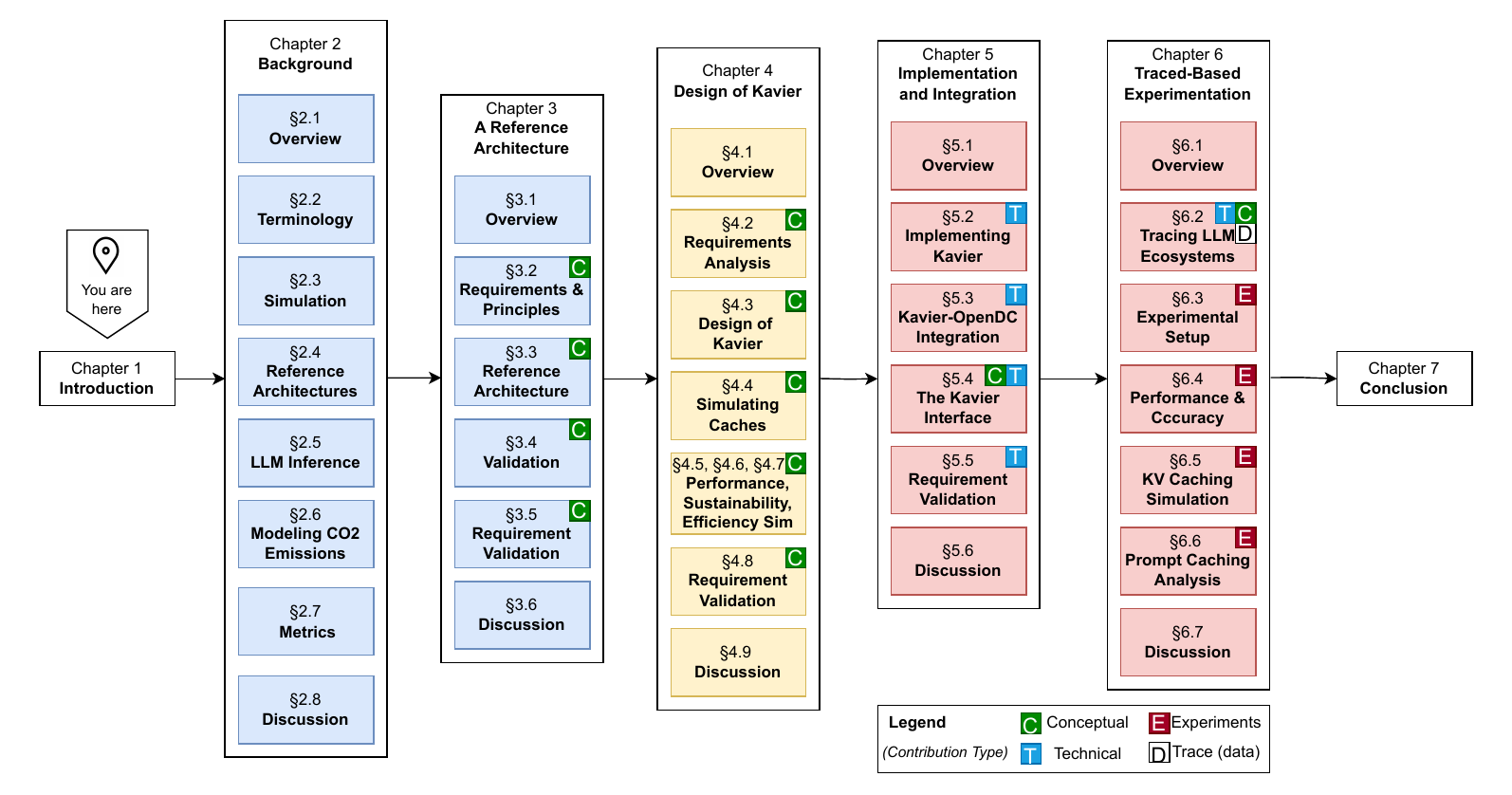}
    \caption{The structure of this thesis.}
    \label{fig:intro:thesis-structure}
\end{figure}

\Cref{fig:intro:thesis-structure} visually represents the structure of this work, and highlights the four main types of contributions of this work, to the scientific community: \underline{c}onceptual (C), \underline{t}echnical (software) (T), contributions from measurements of real-world LLM ecosystems synthesized in \underline{d}ata-traces (D), and contributions from trace-based \underline{e}xperiments (E).

In \Cref{sec:background}, we describe relevant background information on LLM ecosystems, simulators, and latest simulation innovations in the community. 
In \Cref{sec:refarch}, we propose a Reference Architecture of LLM ecosystems under inference and validate it by aligning this conceptual model with industry ecosystems and peer-reviewed reference architectures of the Compute Continuum~\cite{DBLP:conf/ccgrid/JansenAPTI23}. 
In \Cref{sec:design}, we design Kavier following vetted design processes in the computer systems community~\cite{DBLP:conf/icdcs/IosupVTETBFMT19, DBLP:journals/corr/abs-2206-03259}.
In \Cref{sec:prototype}, we implement Kavier and integrate the resulted prototype within OpenDC, a state-of-the-art and peer-reviewed datacenter simulator with over 8 years of development~\cite{DBLP:conf/ccgrid/MastenbroekAJLB21}.
With the Kavier-OpenDC integration, in \Cref{sec:experiments} we evaluate Kavier-prototype with trace-based realistic scenarios, following experimentation processes introduced by AtLarge~\cite{DBLP:conf/ccgrid/MastenbroekAJLB21, DBLP:journals/tpds/AndreadisMBI22, DBLP:journals/fgcs/MastenbroekMBI25, nicolae2025m3sa}, currently standards in the computer systems community.
 
 \newpage
\thispagestyle{noheader}
\chapter{Background}\label{sec:background}

In this chapter, we present a comprehensive, yet not exhaustive, background on subjects relevant to reference architectures, simulation, and LLM inference. Sections \ref{sec:background:terminology:simulation}, \ref{sec:background:dc-simulators}, 
\ref{sec:background:carbonModels}, and
\ref{sec:background:energyMetrics}, are adapted from my honours programme thesis \textit{"M3SA: Exploring the Performance and Climate Impact of Datacenters by Multi-Model Simulation and Analysis"}~\cite{radunicolae-hp-m3sa}, authored by Radu Nicolae (myself), and supervised by Prof.~Dr.~Ir.~Cav.~Alexandru~Iosup and Dante Niewenhuis.

\section{Overview}\label{sec:refarch:overview}

We now present an overview of this chapter through a top-down presentation. In this chapter, our contribution is six-fold:

\begin{enumerate}
    \item We firstly present, in \Cref{sec:background:terminology}, terminology used in operating LLMs and ICT infrastructure and introduce concepts such as \textit{token}, \textit{prefill/decode}, \textit{workload}, \textit{trace}, or \textit{model}, further used for the rest of this work.

    \item In \Cref{sec:background:dc-simulators}, we provide background on datacenter simulation, and expand on a peer-reviewed, community-vetted, open-source datacenter simulator.

    \item In \Cref{sec:background:llm_arch} we conduct an analysis of existing reference architectures of LLM ecosystems under inference, and present for each architecture advantages and drawbacks. This is a crucial foundation for \Cref{sec:refarch}, where we propose a comprehensive, state-of-the-art, following a distributed-systems approach reference architecture.

    \item We then provide background and how our community addresses, through simulation, the emerging concern of CO2 emissions from massive-scale ICT infrastructure under workload~(\Cref{sec:background:carbonModels}).

    \item We then present background on KV-Caching, with a top-down approach, starting from what KV-Caching is, what its main functions are, and delving into the latest scientific discoveries on KV-Caching and its potential magnitude-scale impact on system throughput and latency (\Cref{sec:background:kv_caching}).

    \item Then, in \Cref{sec:background:energyMetrics}, we offer an overview of metrics used in our community to quantify ecosystems (e.g., power usage effectiveness, sustainability, performance, efficiency) and metrics to quantify simulation accuracy (e.g., MAPE).

\end{enumerate}

\section{Terminology} \label{subsec:simulator-terminology}\label{sec:background:terminology}

We now present terminology used in this work. We begin by defining terminology related to AI and LLM ecosystems in \Cref{sec:background:terminology:llm-ecosystems}, then provide terminology related to ICT simulation in \Cref{sec:background:terminology:simulation}.

\subsection{A background on  terminology}\label{sec:background:terminology:llm-ecosystems}

AI Inference is \textit{"the ability of trained AI models to recognize patterns and draw conclusions from information that they haven’t seen before"}~\cite{flinders2024ai}.

\textit{"A token is a collection of characters that has semantic meaning for a model. Tokenization is the process of converting the words in your prompt into tokens"}~\cite{ibm_tokens}. In this work, we consider one token to be one word.

Prefill is \textit{"the stage in which the model processes the prompt tokens of a new request. It computes the transformer attention and stores the Key (K) Value (V) tensors of the attention for each token and each layer into the KV cache blocks"}~\cite{li2025life}.

Decode is \textit{"the stage in which the model generates the next output token or the output tensor for intermediate layers repeatedly for ongoing requests. It reuses the stored KV-Cache of all preceding tokens and computes the Query (Q) tensors based on the most recent token"}~\cite{li2025life}.

KV-Cache is \textit{"the GPU memory region used to store the transformer attention keys and values for each token in a request. It is managed globally across all requests and devices in vLLM"}~\cite{li2025life}. We provide more background on KV-Cache(ing) in \Cref{sec:background:kv_caching}.

A distributed ecosystem is \textit{"a non-trivially heterogeneous group of computer systems distributed in nature, collectively called constituents. Constituents are autonomous, but often in competition and even antagonistic with each other. The ecosystem structure and organization ensure its collective responsibility: completing functions with humans in the loop, providing desirable non-functional properties that go beyond traditional performance, subject to agreements with clients. Ecosystems experience short- and long-term dynamics: operating well although challenging, possibly changing conditions external to the control of the ecosystem"}~\cite{book-distributed-systems}.

\subsection{A background on ICT simulation terminology}\label{sec:background:terminology:simulation}

\textit{``Simulation is defined as the imitation of the operation of a system or real-world process over time, and in many cases, manufacturing provides one of the most important applications of simulation"~\cite{DBLP:conf/wsc/LeeMS03}}. Simulation provides datacenter stakeholders with operational insights into how the ICT infrastructure behaves under different configurations, workloads, and operational phenomena (e.g., infrastructure failures).

\textit{Workloads} contain tasks operated on physical machines, virtual machines (VM), or containers~\cite{nicolae2025m3sa}.

\textit{Traces} are fine-grained recordings of real-world events, capturing detailed operational data of infrastructure under different workload(s); traces provide a granular view of resource usage, essential for driving simulations or replaying real-world scenarios~\cite{nicolae2025m3sa}. In this work, traces are monitored at a constant time granularity to provide details on the computational demand over time, and are crucial in replaying real-world scenarios to predict system behavior, energy consumption, and CO2 emissions.

\textit{Predictive models} in large-scale computer systems are empirical prediction systems that analyze, combine, and compute various input elements to produce fine-grained output predictions~\cite{nicolae2025m3sa, modsim:book/ZaraiN15:orig}. In other words, \textit{predictive models} are empirical prediction systems that analyze, combine, and compute atomic input elements to produce a comprehensive, sometimes exhaustive output. We use models to predict real-world workloads run on ICT infrastructure with various specifications modeled by users. Models help understand and optimize resource allocation, workload management, and monitoring of overall performance metrics, such as energy consumption and CO2 emissions.

\textit{The export rate} of the simulator represents the granularity at which the instrument samples and exports simulation data. For example, an export rate of 30 seconds will lead to 2 exported samples per minute. In this work, we address simulations with different sample rates, towards analyzing various metrics (e.g., performance) of the researched and developed tools.

\textit{Multi-Model} proposes leveraging multiple predictive models, run in parallel, without interference, into a unified tool. The \textit{Meta-Model} simulation vision proposes aggregating multiple predictive models into a unified model, the Meta-Model, able to predict based on other models' predictions~\cite{nicolae2025m3sa, radunicolae-hp-m3sa}. OpenDC supports Multi- and Meta-Model simulation~\cite{DBLP:conf/ccgrid/MastenbroekAJLB21, nicolae2025m3sa}.

\section{The Main Analytical Tool: Simulation}\label{sec:background:dc-simulators}

Datacenters serve as vital cloud infrastructure, playing a crucial role in the digital society by serving stakeholders from industry, government, and academia~\cite{DBLP:conf/sc/AndreadisVMI18, DBLP:journals/tpds/AndreadisMBI22, DBLP:conf/ccgrid/MastenbroekAJLB21, DBLP:conf/ispdc/IosupABBENOTVV17}. Extensive research has been conducted in this field, including analyzing and predicting data traffic evolution, developing datacenter simulators, and proposing novel scheduling techniques. In this section, we present the data traffic trends (Section \ref{sec:background:dc-simulators:dataTraffic}), which increased by one order of magnitude within the last decade; this data, is handled, created, transferred, and reproduced via massive-scale ICT infrastructure, which is in a continuous expansion and growth~\cite{iosup2021massivizingkeynote, DBLP:conf/ispdc/IosupABBENOTVV17, IEADataCentresNetworks}. The current state-of-the-art consists of simulating before building; we further expand on datacenter simulation frameworks in Section \ref{sec:background:dc-simulators:openDC}.

\subsection{Data Traffic Trends} \label{sec:background:dc-simulators:dataTraffic}
\textit{Reinsel et al.} analyzed data traffic trends, estimating an one-order-of-magnitude increase to 163ZB reached in 2025, compared to just 16ZB in 2016. These data include 25ZB of critical information and 4ZB of hypercritical data, directly impacting users' health, life, commercial air travel, military security, and numerous other situations. In addition, as of 2025, approximately 75\% of the global population is estimated to be connected to the Internet. The research carried out by Reinsel et al., as part of an IDC White Paper sponsored by Seagate, underscores the vital importance of establishing reliable and efficient datacenters, scalable to billions of people and tens of billions of devices~\cite{seagateData}.

\subsection{Simulation with OpenDC} \label{sec:background:dc-simulators:openDC}
\textit{``Simulation is defined as the imitation of the operation of a system or real-world process over time, and in many cases, manufacturing provides one of the most important applications of simulation"~\cite{DBLP:conf/wsc/LeeMS03}}. Simulation provides datacenter stakeholders with operational insights into how the ICT infrastructure behaves under different configurations, workloads, and operational phenomena (e.g., infrastructure failures).

\textit{Iosup et al.} analyzed existing datacenter simulators, highlighted, and addressed simulation challenges by introducing OpenDC 1.0~\cite{DBLP:conf/ispdc/IosupABBENOTVV17}, succeeded by OpenDC 2.0~\cite{DBLP:conf/ccgrid/MastenbroekAJLB21}, introduced by \textit{Mastenbroek et al.} OpenDC is an open-source platform for modeling, simulation, and experimentation with cloud datacenters. OpenDC 2.0 addresses multiple key challenges:
\begin{enumerate}
    \item contains models for emerging technologies, such as serverless computing and machine learning workloads running in datacenters;
    \item contains models for CO2 emission predictions and models for energy usage predictions, calibrated with real-life data;
    \item provides an intuitive interface with enhanced visualization and interaction tools, supporting various input/output formats and metrics. OpenDC 2.0 facilitates the process of designing and sharing (parts of) complex datacenters;
    \item provides both a GUI and JSON interfaces towards accommodating a wide range of stakeholders, including experts and general users.
\end{enumerate}

OpenDC 2.0 is a pioneering and re-engineered iteration of the 1.0 prototype, becoming the first simulator to integrate serverless and machine-learning execution while leveraging discrete-event simulation. This simulator integrates a model for the TensorFlow ecosystem and primarily employs Kotlin as the main programming language for the codebase. The authors compare the developed datacenter simulation concepts and architecture with \textit{i) Mathematical Analysis}, albeit faster, too high-level for the processes from a datacenter and with \textit{ii) Real-world experimentation}, which yields accurate results, is non-trivial to run at a large scale due to high energy footprint and extensive waiting times. 

With highly precise and accurate simulations, open-source nature, and a wide variety of distinct models used in simulations, OpenDC has proven results through multiple peer-reviewed, award-winning, top-tier publications~\cite{dniewenhuis_hotcloud_footprinter, hongyuhe2021hpreport, DBLP:conf/ispdc/IosupABBENOTVV17, iosup2021massivizingkeynote, DBLP:conf/ccgrid/MastenbroekAJLB21, Mastenbroek2023RADICE, DBLP:journals/fgcs/MastenbroekMBI25, DBLP:journals/tpds/AndreadisMBI22, DBLP:journals/tpds/AndreadisMBI22}. We identify the OpenDC simulation framework and the related work as highly relevant for this research. 


\section{The Root of Every Systematic Simulation - the Reference Architecture}\label{sec:background:llm_arch}


Prior to conducting this scientific research, we searched for relevant literature proposing reference architectures (RAs) for LLM ecosystems. 
We define a set of keywords: \textit{"reference architecture LLM ecosystems,"} and explore existing literature on 
ACM Digital Library\footnote{\url{https://dl.acm.org/action/doSearch?AllField=reference\%20architecture\%20llm\%20systems}},
Google Scholar\footnote{\url{https://scholar.google.com/scholar?hl=en&as_sdt=0,5&q=reference+architecture+llm+systems}},
and DBLP\footnote{\url{https://dblp.org/search?q=reference\%20architecture\%20llm\%20systems}}.
We discover one relevant architecture for LLM ecosystems proposed by \textit{Bucaioni et al.} (\Cref{fig:background:ra:bucaioni}) and a community-vetted reference architecture for the Compute Continuum, proposed by \textit{Jansen et al.} (\Cref{sec:background:llm_arch:continuum}). Alongside the aforementioned RAs, we discover RAs with various purposes (e.g., for deploying LLMs \cite{mahr2024reference}), albeit valuable, are not aligned with the scope of this research. 
However, although not in direct alignment with the scope of this work, in \Cref{sec:background:llm_arch:foundation}, we present the reference architecture proposed by \textit{Lu et al.} for designing foundational model-based systems, which we envision as a highly relevant envision of the LLM evolution over the next decade(s).

In this section, we present two community-vetted reference architectures for LLM ecosystems and a reference architecture on the evolution of foundational models. For each RA, we present an overview and highlight present and absent key points. We regard this (sub)-section as critical for \Cref{sec:refarch}, in which we envision a comprehensive reference architecture for the inference process of LLM ecosystems, following a distributed systems approach, and mapped to the Compute Continuum.

\subsection{A Functional Software RA}\label{sec:background:llm_arch:bucaioni}

\textbf{Overview:} \textit{Bucaioni et al.} propose a \textit{``preliminary functional reference architecture as a conceptual framework"} which addresses the lack of systematic reasoning about the design and quality attributes of LLM software systems~\cite{bucaioni2025functional}.

\begin{figure}[t]
        \centering
        \includegraphics[width=0.75\linewidth]{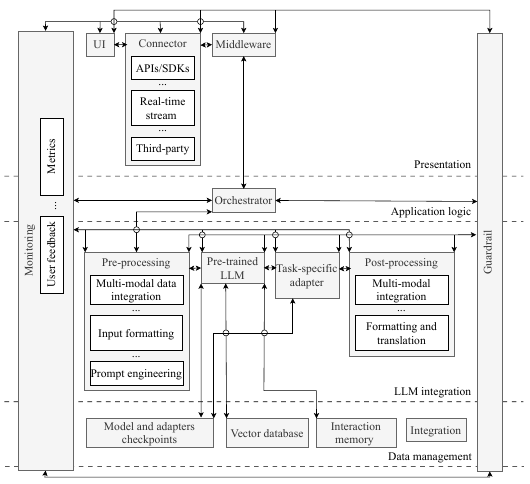}
        \caption{Existing reference architecture for LLM-integrated ecosystems, from~\cite{bucaioni2025functional}.}
        \label{fig:background:ra:bucaioni}
\end{figure}

The authors conduct a literature survey and identify architectural concerns for ``large language model-integrated systems" (LLM ecosystems) and propose a four-layer functional reference architecture emphasizing modular service decomposition and cross-layer monitoring (\S III~\cite{bucaioni2025functional}), which we illustrate in \Cref{fig:background:ra:bucaioni}.

\textit{(i) Presentation layer:} handles multimodal user interactions through a user interface (UI), linked to a connector component, and linked to the other layers via an orchestrator.

\textit{(ii) Application logic layer:} contains a single, non-shared component, the orchestrator, which dynamically determines workflows based on user input. The orchestrator bridges the (i) presentation layer with the (iii) LLM integration layer. To address scalability concerns, the orchestrator supports asynchronous, event-driven workflows.

\textit{(iii) LLM integration layer:} is the system's core and handles input-processing, detailing from the very first input formatting, called \textit{pre-processing} (e.g., for single reasoning model workflows) and prompt engineering (e.g., for cascading model workflows, where the user prompt generates multiple, in-LLM prompts), to the \textit{post-processing steps}, such as multi-modal integration, or formatting and translation. Although the authors include the \textit{multi-modal} element in the RA, they do not further expand on this topic.

\textit{(iv) Data management layer:} "ensures efficient data handling"~\cite{bucaioni2025functional}. This layer includes multiple elements for performance enhancement, such as a vector database for retrieval-augmented generation for knowledge-grounded outputs, thus improving accuracy, following the model of Pinecone's integration with Notion AI~\cite{bucaioni2025functional}, or the integration of a memory component, which maintains context across sessions.  

\textit{All layers:} The authors propose two elements that cover the entire continuum and span over all four layers. The monitoring component proposes collecting performance metrics (e.g., latency, throughput), and user feedback. The guardrail component ensures security and privacy, ensuring law-compliance over all the layers; however, this component is still blurry, without a detailed description in the paper. 

\textbf{Advantages of this RA:} \textit{Bucaioni et al.}'s architecture employs layered interoperability, rather than a strict stack, which allows for both vertical flow (layers (i)-(iv)) and horizontal flow, with monitoring and guardrail elements covering each layer. This RA allows for modularity through a high-level approach, while maintaining end-to-end workflow cohesion. 

\textbf{Weak points:} While \textit{Bucaioni et al.} propose a valid, community-wide recognized, and thus peer-reviewed, reference architecture, we argue this RA exhibits two critical gaps, making it insufficient for abstracting inference of LLM ecosystems operating across the Compute Continuum. While these elements don't invalidate the correctness of the RA proposed by \textit{Bucaioni et al.}, it makes the RA insufficient for the scope of our work.

\textit{G1: High homogeneity:} We argue this reference architecture proposes a too-high degree of homogeneity, especially in the memory systems and computing infrastructure. The reference architecture contains an \textit{Interaction memory} component, which, however, fails to reflect the real-world degree of heterogeneity, especially hierarchical layers of the memory component. Furthermore, the computing infrastructure is disregarded from this architecture, making it unclear where the heavyweight computation is connected to. 

\textit{G2: Blurry memory component:} While this reference architecture abstracts (some of) the components of the LLM ecosystems under inference workload, it fails to provide an in-depth model of the memory component, which, in reality, is highly hierarchical, and contains elements which can impact performance by orders of magnitude. One such component, the KV-Caching system (background provided in \Cref{sec:background:kv_caching}), is a key component of nowadays massive-scale LLM ecosystems, and a main focus of this paper. The reference architecture proposed by \textit{Bucaioni et al.}, fails to present how the KV-Caching system is integrated with the memory hierarchy and with the Compute Continuum of the system

\subsection{The Compute Continuum}\label{sec:background:llm_arch:continuum}

\begin{figure}[t]
  \centering
  \begin{minipage}[b]{0.49\linewidth}
    \centering
    \includegraphics[width=\linewidth]{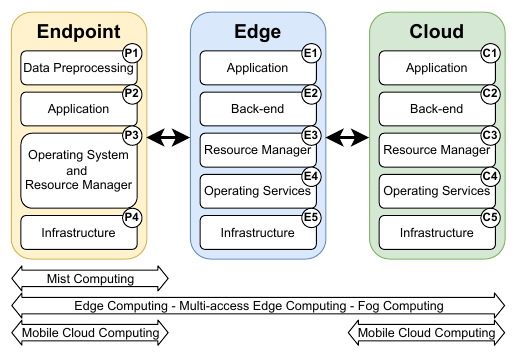}
    \caption{Reference architecture for the Compute Continuum taken from \cite{DBLP:conf/ccgrid/JansenAPTI23}.}
    \label{fig:background-refarch-continuum-2}
  \end{minipage}\hfill%
  \begin{minipage}[b]{0.49\linewidth}
    \centering
    \includegraphics[width=\linewidth]{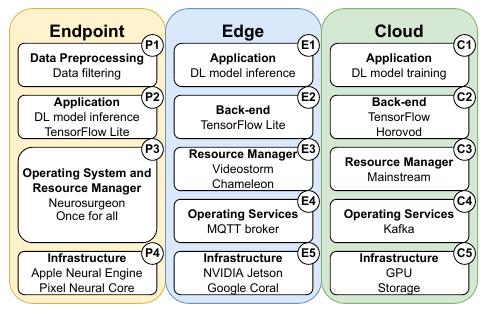}
    \caption{Deep learning architecture applied on the reference architecture proposed and taken from \cite{DBLP:conf/ccgrid/JansenAPTI23}.}
    \label{fig:background-refarch-continuum-3}
  \end{minipage}
\end{figure}

\textit{Jansen et al.} \cite{DBLP:conf/ccgrid/JansenAPTI23} propose a three-tier reference architecture of a Compute Continuum, as depicted in \Cref{fig:background-refarch-continuum-2}, comprising cloud, edge, and endpoint, and addressing the fragmentation of 17 existing computing models, via systematic synthesis of common characteristics and design patterns, following the community-vetted AtLarge Design Process~\cite{DBLP:conf/icdcs/IosupVTETBFMT19}. 

\textbf{Overview:} \textit{Jansen et al.} have a five-fold contribution, out of which contributions (ii) and (iii) are highly representative for this work. In~\cite{DBLP:conf/ccgrid/JansenAPTI23}, the authors:
(i) conduct a literature survey on computing models, synthesize properties, and identify opportunities for unification;
(ii) propose a unified reference architecture, the first in the community to consider the entire edge-cloud Compute Continuum;
(iii) synthesize two domain-specific architectures, one for deep learning, illustrated in \Cref{fig:background-refarch-continuum-3}, highly relevant for the scope of our work, and one for industrial IoT;
(iv) offer an open-source workload deployment and benchmarking framework;
(v) formulate analytical performance models for exploring workload deployment scenarios in the continuum.

The authors propose the SPEG-RG reference architecture illustrated in \Cref{fig:background-refarch-continuum-2}. The architecture comprises three tiers of systems: cloud, edge, and endpoint, where each component is further expanded in \S III~\cite{DBLP:conf/ccgrid/JansenAPTI23}. 

Endpoints are \textit{``the last hop of processing and connectivity to users"}~\cite{DBLP:conf/ccgrid/JansenAPTI23}, typically single-tenant~\cite{DBLP:conf/wmcsa/SatyanarayananG19}, resource and energy-constrained (e.g., smartphones, cameras, sensors), with four main responsibilities: (P1) pre-processing data before pushing to the edge and cloud servers, (P2) running user-defined logic to process incoming data and make decisions, (P3) OS-level resource managers and multiplexers for workload management, and (P4) the physical computing available to the operating systems (e.g., processing units, memory, network, storage). 

Edge and cloud are presented as sharing the same high-level design, both able to run multi-tenant workloads on shared infrastructure~\cite{DBLP:conf/ccgrid/JansenAPTI23}. Edge and cloud mainly differ by the resources and energy constraints, where cloud operates at a higher scale. Unlike cloud and uniquely at the edge, there should be support for application offloading both vertically (cloud to edge and back) and horizontally (from one edge system to another)~\cite{DBLP:conf/ccgrid/JansenAPTI23, DBLP:conf/edge/MortazaviSGPL17}. The authors propose five elements in the edge and cloud: (E1) Applications are the first step from endpoint to cloud and are in the best position to make decisions regarding placements, offloading, scheduling, or conduct other user-defined decision-making processes, towards meeting workload-specific objectives, (E2) the backend represents more general-purpose application execution frameworks, usually light(er)weight in the edge than in the cloud, (E3) Resource Managers manage systems' application-independent physical and virtual resources, such as virtual machines and containers, (E4) Operating services are described as providing "support to build distributed applications, and their responsibilities include (but are not limited to) communication, metadata management, consensus services, monitoring, storage services, etc."~\cite{DBLP:conf/ccgrid/JansenAPTI23}, and (E5), similarly to endpoints, yet at orders-of-magnitude higher scale, infrastructure contains compute, memory, networking, and storage resources; however, unlike endpoints, infrastructure contains resources split into physical and virtual resources.

\textbf{Deep Learning architecture mapped to the Compute Continuum:} \textit{Jansen et al.} create architectures for deep learning and industrial IoT; we regard the architecture of deep learning systems as highly aligned with the scope of this research and further provide background. In \Cref{fig:background-refarch-continuum-3}, we illustrate the deep learning architecture proposed in \S III, \cite{DBLP:conf/ccgrid/JansenAPTI23}. As the authors mention, an important trend for deep learning in the continuum is that model training and inference tasks are split across all the tiers of devices (i.e., endpoint, edge, cloud). 

\textbf{Advantages of this RA:} The SPEC-RG reference architecture poses several advantages. This is the first peer-reviewed RA in the community to consider the entire edge-cloud Compute Continuum and further synthesize the proposals in a unitary architecture, thus leveraging previously fragmented computing models under a cohesive framework. Furthermore, this RA proves its validity through systematic research methodology and evaluation, following state-of-the-art AtLarge Design Process~\cite{DBLP:conf/icdcs/IosupVTETBFMT19}. Lastly, the authors contribute to open science by providing an engineered prototype of an open-source framework, released as open-source.

\textbf{Weak points:} While comprehensive and, at present, regarded as the state-of-the-art for the computer (eco)systems community, this reference architecture exhibits some limitations in regards to the scope of this work, limitations which, however, do not undermine the validity of the Compute Continuum. The RA focuses primarily on a higher-level design, yet does not present technical depth on how real-world large-scale, highly heterogeneous systems would map to this architecture (e.g., highly distributed LLM ecosystems). Furthermore, the RA proposed by \textit{Jansen et al.} presents a high-level abstraction of e.g., Application components (P1, E1, C1), which, in this work, alongside other components, we expand and detail.

\subsection{RA for Foundation Model Based Systems}\label{sec:background:llm_arch:foundation}

\begin{figure}
    \centering
    \includegraphics[width=0.9\linewidth]{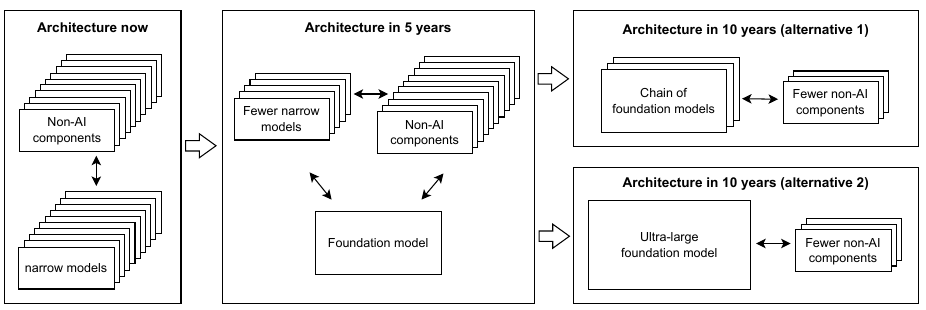}
    \caption{From ``foundation-model-as-a-connector" to ``foundation-model-as-a-monolithic-architecture," taken from~\cite{lu2023towards}.}
    \label{fig:background-refarch-foundational}
    \vspace*{-0.5cm}
\end{figure}

\textit{Lu et al.} identify a broad consensus that "foundational models will be the fundamental building blocks for future AI systems"~\cite{lu2023towards}; however, there is a lack of systematic guidance on the architecture design. The authors present an architecture evolution of AI systems (\Cref{fig:background-refarch-foundational}), then propose a pattern-oriented reference architecture for designing responsible foundation-model-based systems. 

\textbf{Overview:} In this section, we detail \Cref{fig:background-refarch-foundational}, as \textit{Lu et al.} present architectures in three time periods. 
The authors identify the architecture now as containing "many narrow AI models and many non-AI components", yet this is prone to change over the next decade. 
The predicted transition in the next five years regards foundational models (FM) as a connector, with one FM, a few narrow models, and many non-AI components. 

The authors envision two evolution alternatives of LLM FMs within the next 10 years. 
\textit{Alternative one} proposes a chain of FMs and only a few non-AI components, an alternative in which most of the software components could be absorbed into the FMs, chained together, without requiring additional training or fine-tuning. 
Those FMs would be connected via APIs, with external non-AI components that offer additional functionalities (e.g., robotic systems or web search engines).
\textit{Alternative two} proposes one ultra-large FM and only a few non-AI components, following a monolithic architecture. The ultra-large FM would be unitary, massive, and capable of performing a variety of tasks by incorporating different types of tasks, from searching, reasoning, self-inputting (self-prompt engineering), and outputting in various shapes and forms. The non-AI component may include context engineering components (e.g., multimodal context injection), prompt engineering components (e.g., AI-powered prompt optimizer), and responsible AI components, ensuring privacy, security, and a continuous risk assessment.


\section[The Main Components LLM Inference -- Self-Attention Mechanism and KV-caching]{The Main Components of the LLM Inference Ecosystem -- Self-Attention Mechanism and KV-caching}\label{sec:background:kv_caching}

\textit{"Attention is all you need"}~\cite{vaswani2017attention}. In 2017, \textit{Vaswani et al.} proposed the Transformer, \textit{"a model architecture eschewing recurrent and instead relying entirely on an attention mechanism"}, a state-of-the-art approach allowing better parallelization, with a working prototype trained for a limited time on limited resources ~\cite{vaswani2017attention}. With approximately 180,000 citations as of 2025, this work revolutionized sequence modeling by replacing recurrence with self-attention, a pivotal contribution in enabling parallelized training and inference towards high throughput and low latency. Although this paper does not explicitly introduce \textbf{KV}-caching, the authors adopt an autoregressive approach involving \textbf{K}ey and \textbf{V}value matrices across time steps to avoid recomputation.

\textit{Zhang et al.} introduce LLMs as \textit{"autoregressive models that generate tokens iteratively, one at a time"}, with the inference engine storing \textit{"KV caches-intermediate tensors produced by attention layers."} \textit{"The computation of a new token depends on interactions between its embedding and the previously stored intermediate KV cache tensors~\cite{DBLP:conf/sosp/ZhangDLKMWLYLLZ25}"}.

HuggingFace presents KV as follows: KV vectors are used to calculate attention scores; KV scores are calculated depending on the previous tokens~\cite{huggingface-kv-cache}. However, this means that \textit{"each prediction depends on the previous tokens"}, which is equivalent to the model performing \textit{"the same computation each time"}~\cite{huggingface-kv-cache}. The key-value (KV) vectors are used to calculate attention scores. For autoregressive models, KV scores are calculated every time because the model predicts one token at a time. Each prediction depends on the previous tokens, which means the model performs the same computations each time.

\subsection{Self-Attention Mechanism} \label{sec:background:kv_caching:self_attention}
To better comprehend KV-Caching, we first focus on the self-attention mechanism. 
\textit{Vaswani et al.} describe an attention function as \textit{"mapping a query and a set of key-value pairs to an output, where the query, keys, values, and output are all vectors. The output is computed as a weighted sum of the values, where a compatibility function of the query with the corresponding key computes the weight assigned to each value."} They propose the following equation, a standard in the nowadays community:

\begin{align}
    \quad Attention(Q,K,V) = softmax(\frac{QK^T}{\sqrt{d_k}})V \label{eq:attention} 
\end{align}
\begin{small}
    \textit{where $Q$ is the query token for which the model is computing attention, $K$ represents all tokens (\textbf{k}eys) in the sequences used to determine the relevance to the query, $V$ contains the information to be aggregated (\textbf{v})alues, weighted by the attention scores, and $d_k$ denotes the dimension of the key vectors.}
\end{small}

\textit{Inputs:} Q, K, V are each matrices derived from the input embeddings (i.e., the previous layer outputs) via the learned linear transformation. Considering the input sequence of length n and hidden size d, Q and K have the shape of $(n, d_k)$, and V has the shape $(n, d_v)$. 

\textit{Outputs:} The attention function outputs a matrix of shape $(n, d_v)$, where each row is a contextually weighted sum of the value vectors and corresponds to each input position. The attention weights can be interpreted as how much each input token attends to every other token~\cite{DBLP:journals/corr/abs-2102-08606}.

\begin{figure}[t]
    \centering
    \includegraphics[width=\linewidth]{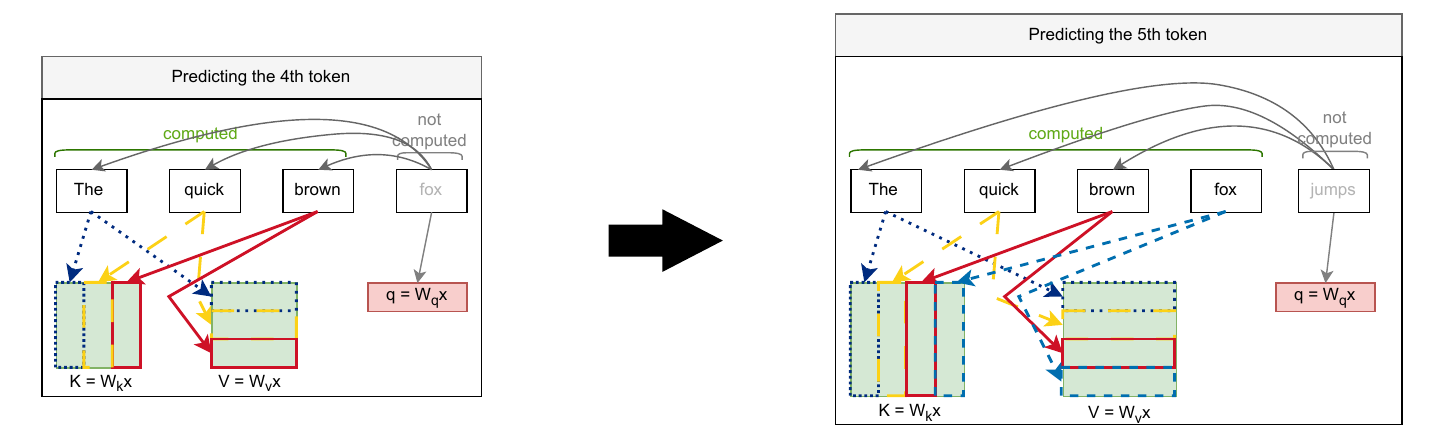}
    \caption{LLMs predicting without KV-Caching. $O(n^2)$ time complexity.}
    \label{fig:background:without-kv}
\end{figure}

\begin{figure}[t]
    \centering
    \includegraphics[width=\linewidth]{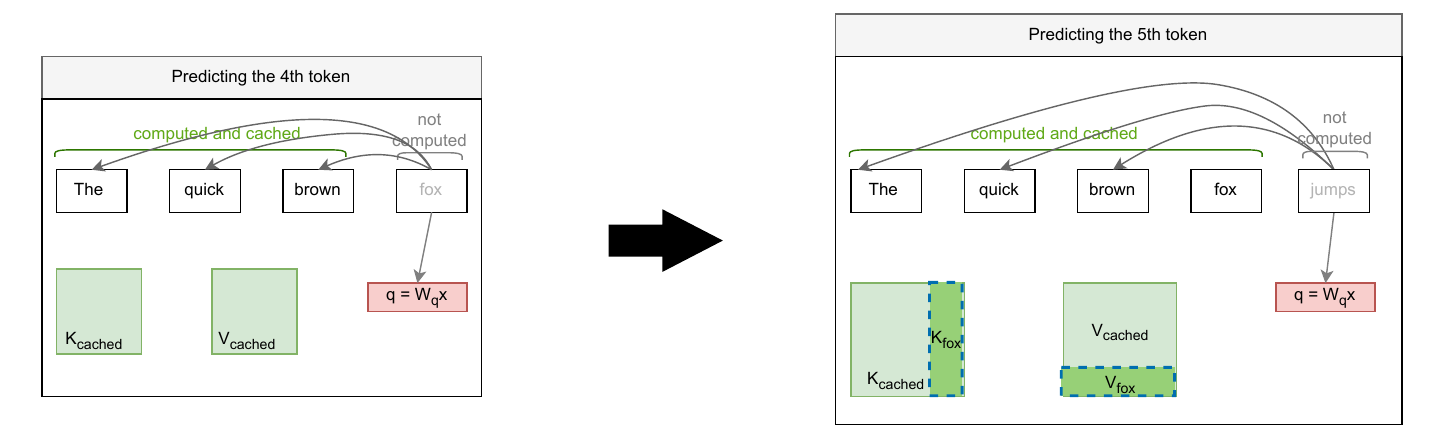}
    \caption{LLMs predicting with KV-Caching. $O(n)$ time complexity.}
    \label{fig:background:with-kv}
\end{figure}

In \Cref{fig:background:without-kv}, we present an example of an LLM predicting without KV-Caching. For the sake of simplicity, we illustrate a scenario where the LLM contains a short, visually comprehensive sequence.

In the left part of \Cref{fig:background:without-kv}, the LLM contains the sequence \textit{"the quick brown"}, where each word is a stored token. We generate one word at a time, following the mechanism employed by autoregressive encoding. The K and V matrices contain information about the entire sequence, while the query vector Q contains information only about the last token (i.e., \textit{"fox"}). 

In the right part of \Cref{fig:background:without-kv}, the LLM contains the sequence of \textit{"the quick brown fox"} and predicts the fifth token. We observe that matrices K and V don't change much, but rather receive an additional column and row for each additional token. However, the model still needs to perform the heavy, yet redundant, computational work of computing the key and value vectors for each word.

This approach results in $O(n^2)$ time complexity for total generation, where n represents the number of tokens per input. While, in this example simplified for visual comprehension, $O(n^2)$ is not computationally heavy, the absence of KV-Caching quickly increases the number of computations needed for predicting large phrases.

\subsection{Key-Value Caching}

A KV-Cache stores the calculations from \Cref{sec:background:kv_caching:self_attention} so they can be reused without recomputing them. Efficient caching is crucial for optimizing model performance because it reduces computation time and improves response rates~\cite{huggingface-kv-cache}.

The KV-Caching proposes a simple, yet powerful technique: when the model reads a new token, it generates the query vector ($Q$), similarly to the approach presented in \Cref{sec:background:kv_caching:self_attention}, yet the system caches the values already computed for the previous tokens to reduce redundancy and multiple calculations for each token. Instead, the model only computes a new column and a new row for the key-value matrix.

In \Cref{fig:background:with-kv}, we illustrate the KV-Caching approach, using the same example as in \Cref{fig:background:without-kv}. With this design, for the sequence \textit{"The quick brown"}, the memory holds two 3-by-3 matrices, one for keys, one for values, and a query vector, similarly to the no-KV-approach. However, when predicting the fourth token, \textit{"fox"}, the system computes only one new column for the key matrix and one new row for the value matrix, instead of recomputing the entire matrix. 

\subsection{KV-Caching and GPU-memory}

The cached key and value matrices (i.e., K and V, from \Cref{fig:background:with-kv}) need to be stored in the GPU's memory, such that the already-computed matrices can be reused when predicting the next token, thus preventing the otherwise redundant computation introduced by the no-KV alternative.

The KV-Cache memory usage can be computed with the formula:

\begin{align}
    memory &= 2 \times L \times H \times d \times N \times sizeof(type)  \label{eq:kv-memory}
\end{align}
\begin{small}
    \textit{where $L$ is the number of transformer layers in the model, $H$ is the number of attention heads, $d$ is the dimension per head, $N$ is the number of tokens in the sequence, and $sizeof(type)$ represents the size of the data type in bytes (e.g., float16 represents 2 bytes, float32 4 bytes). The factor of 2 represents storing two matrices, one for keys and one for values.}
\end{small}

Bai Li, through Efficient NLP~\cite{efficientnlp2023kv}, computes the memory needed for OPT-30B~\cite{zhang2022opt}, a 30-billion-parameter model, a small to medium-sized model. OTP-30B runs inference in 16 bits (i.e., 2 bytes), contains 48 layers, and has 7168 dimensions. In this example, Bai proposed a sequence length of 1024 tokens and a batch size of 128.

\begin{align}
    memory_{kv} &= 2 \times L \times H \times d \times N \times sizeof(type)  \label{eq:kv-memory} \\
    memory_{kv} &= 2 \times 48 \times 128 \times 7168 \times 1024 \times 2  \label{eq:kv-memory} \\
    memory_{kv} &= 176,160,768B \approx 176GB \\
    memory_{model} &= 2 \times 30B = 60GB
\end{align}

    

We thus observe that KV-Caches use 2.9x more memory than the model itself, a common effect for inference scenarios of LLMs, where the KV-Caching is a dominant factor~\cite{DBLP:conf/acl/YangHGHZ024, DBLP:conf/sosp/ZhangDLKMWLYLLZ25, efficientnlp2023kv}. However, thanks to KV-Caching, the time complexity decreases from $O(n^2)$ to $O(n)$ since the KV-Caching approach prevents the redundant recomputation of previous token scores, and caches these computations instead.
\section{An Emerging Concern: Modelling CO2 Emissions}\label{sec:background:carbonModels}
\textit{Niewenhuis et al.} proposed FootPrinter, a \textit{``first-of-its-kind tool that supports datacenter designers and operators in assessing the environmental impact of their datacenter"}~\cite{dniewenhuis_hotcloud_footprinter}. As part of their research, engineering, and evaluation, matching the state-of-the-art AtLarge Design Process~\cite{DBLP:conf/icdcs/IosupVTETBFMT19}, the authors proposed and integrated into OpenDC a model able to predict the CO2 emissions of ICT infrastructure under workload.

FootPrinter simulation process leveragies the energy module of OpenDC, which simulates the amount of power draw and energy usage for infrastructure under workloads. Then, using the FootPrinter module, the simulator computes the amount of CO2 emissions based on the carbon intensity at the time and the power draw at the time

\textit{Niewenhuis et al.} calibrated the researched CO2-emissions model with data from the ENTSO-E Transparency Platform~\cite{site:entso-e}. The efforts have been concertized in a research paper, part of a top-tier conference, which hence confirms the validity of the research methodology, and accuracy and validity of the designed and engineered tool.







\section{Community Agreement on What to Measure: Metrics}\label{sec:background:energyMetrics}

In this section, we present metrics stakeholders use to quantify datacenter sustainability and efficiency, and metrics the community uses to quantify simulation accuracy. In \Cref{sec:background:metrics:energy}, we present metrics that quantify the energy efficiency of ICT infrastructure and emphasize the need for efficient ICT operation. In \Cref{sec:background:metrics:carbon}, we present carbon metrics which we use to simulate and quantify the sustainability of LLM ecosystems. In \Cref{sec:background:metrics:performance} we present metrics, such as throughput and latency, which we use to quantify performance of LLM ecosystems. In \Cref{sec:background:metrics:efficiency} we present metrics we use to quantify system's financial and sustainability efficiency. In \Cref{sec:background:metrics:accuracy}, we present the Mean Absolute Percentage Error (MAPE) ratio, a widely used metric to quantify the accuracy of simulation models and divergence with real-world measurements.

\subsection{Metrics on quantifying the energy effectiveness of the systems}\label{sec:background:metrics:energy}

Designing, building, operating, and expanding energy-efficient datacenters is becoming an increasingly concerning challenge for our increasingly digitalized society. The wide adoption of LLM ecosystems deepens the gap between LLM needs and hardware efficiency, which results in high energy usage. To quantify the efficiency of existing ICT infrastructure, we identify Power Usage Effectiveness (PUE) and Datacenter Performance Efficiency (DCPE) as core metrics.

\subsubsection{Power Usage Effectiveness (PUE)} \label{sec:background:energyMetrics:pue}
Introduced in 2006 by \textit{Malone et al.}~\cite{malone2006proceedings-pue, hongyuhe2021hpreport}, Power Usage Effectiveness (PUE) is an end-user tool consisting of a metric \textit{``for understanding how well a datacenter is delivering energy to its information technology equipment"}~\cite{avelar2012pue}. PUE is the ratio of the total energy and the energy that is used for the actual computation.

\begin{center}
\begin{equation} \label{eq:pue}
\displaystyle PUE = \frac{E_{T}}{E_{IT}}
\end{equation}
\end{center}
\textit{where $E_{T}$ denotes the total energy used by the datacenter and $E_{IT}$ denotes the energy used by the IT components of the datacenter.
}

Equation \eqref{eq:pue}~\cite{avelar2012pue} provides a high-level mathematical equation to compute the Power Usage Effectiveness factor of an ICT infrastructure. PUE can take values between a minimum of 1.0 and an infinite maximum. Lower values of PUE are better, and the aim is to get as close to 1.0 as possible. A PUE of 1.0 means that the IT equipment uses all the energy received by the datacenter, yet it is impossible to achieve due to the laws of physics. 

\begin{figure}[t]
    \centering
    \includegraphics[width=0.75\textwidth]{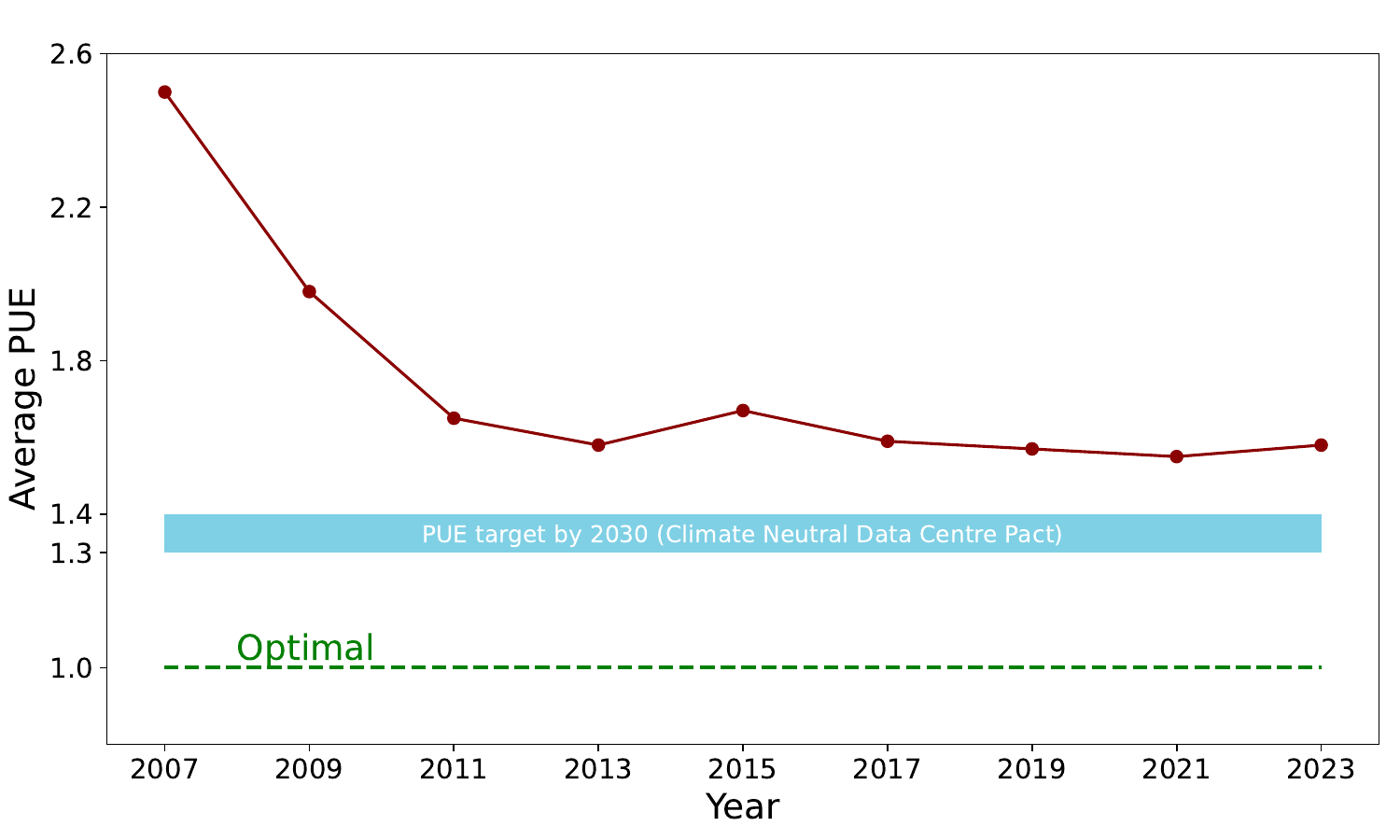}
    \caption{PUE Evolution Between 2007 and 2023~\cite{statistaPUE2023}.}
    \label{fig:pue_evolution_2007_2023}
\end{figure}

The Climate Neutral Data Centre Pact~\cite{ClimateNeutralDataCentre} mandates that, by 2030, all datacenters must meet the PUE target of 1.3 in cool climates and 1.4 in warm climates.~\cite{ClimateNeutralDataCentre}. Although significant improvements in PUE occurred, from an average of 2.6 in 2007 to 1.6 in 2015, the decline has stagnated in recent years (Figure \ref{fig:pue_evolution_2007_2023}), while the overall energy usage is alarmingly increasing~\cite{dniewenhuis_hotcloud_footprinter}. Although Google achieved an average annual PUE of 1.1 in 2023~\cite{google2023efficiency}, and BTDC (Sweden) set a PUE record of 1.014 in 2021~\cite{lowestPUE-article, lowestPUE-website}, the global average PUE remains worryingly high, at 1.58 in 2023~\cite{statistaPUE2023}. Besides environmental concerns, the sharp increase in energy prices in 2022 has a significant economic impact on administrators of datacenters with a bad (high) PUE factor~\cite{Mastenbroek2023RADICE, DBLP:journals/fgcs/MastenbroekMBI25}.

To transpose the numbers above to real examples, we will consider a hyperscale datacenter, which aims to improve the average annual PUE. In our hypothesis, we consider the annual PUE of the datacenter equal to 1.58, denoted as \textit{$x_1$}. This value represents the average PUE of datacenters worldwide in 2023~\cite{statistaPUE2023}. To meet and improve beyond the Pact-mandated metrics, the administrators want to improve and achieve an average annual PUE of 1.25 (i.e., the target PUE), denoted as \textit{$x_2$}. We assume that the datacenter consumes 100~GWh per year (i.e., total facility energy) and is denoted as \textit{y}. We denote the energy used for computation (i.e., IT Equipment Energy) as \textit{$z_1$}, for the current PUE, and as \textit{$z_2$}, for the target PUE. We assume the average price per GWh is approximately €350,000 (i.e., the average price per GWh in 2024, in the Netherlands~\cite{StatistaElectricityPrices2024}), and denote as \textit{p}.

\begin{align}
    p &\approx350,000\,EUR && \text{(approx. price per GWh, Netherlands, 2024~\cite{StatistaElectricityPrices2024})} \\
x_1 &= 1.58 && \text{(current PUE of the datacenter)} \label{eq:current_pue} \\
x_2 &= 1.25 && \text{(target PUE of the datacenter)} \label{eq:target_pue} \\
y &= 100\,GWh && \text{(total yearly consumption)} \label{eq:total_energy} \\
z_1 &= \frac{y}{x_1} \approx 63.29\,GWh && \text{(IT components yearly consumption with the $x_1$)} \label{eq:energy_current_pue} \\
z_2 &= \frac{y}{x_2} = 80.00\,GWh && \text{(IT components yearly consumption with the $x_2$)} \label{eq:energy_target_pue}
\end{align}

\begin{align}
i &= \frac{\lvert z_1 - z_2 \lvert}{z_2} \approx \frac{\lvert x_1 - x_2 \lvert}{x_2} && \approx 20.89\% && \text{(energy saved)} \\
\Delta z &\approx z_2 - z_1 &&\approx 16.71\,GWh && \text{(energy saved yearly)} \\
\Delta p &\approx \Delta z \cdot p &&\approx 5,848,500\,EUR &&\text{(money saved yearly)}
\end{align}

Under the aforementioned hypothesis, we identify a 20.89\% improvement in energy consumption, resulting in approximately 16.71~GWh saved per year, equivalent to savings of approximately 5,848,500 EUR.

\subsubsection{Datacenter Performance Efficiency (DCPE)}\label{sec:background:energyMetrics:dcpe}
Derived from PUE, Datacenter Performance Efficiency (DCPE), also referred to as Compute Power Efficiency (CPE), is a metric used to measure the computational efficiency of datacenters. DCPE was introduced by \textit{Malone et al.} and used to capture the fraction of energy used for computation. 

\begin{center}
\begin{equation} \label{eq:cpe}
DCPE = \frac{U_{IT}}{P} = \frac{U_{IT} \cdot E_{IT}}{E_{T}}
\end{equation}
\end{center}
\textit{$U_{IT}$ is the IT Equipment Utilization, $P$ is PUE, $E_{IT}$ is the energy used by the IT components of the datacenter, $E_{T}$ is the total energy used by the datacenter.}

We observe that even slight changes in the Power Usage Effectiveness metrics significantly impact the datacenter Performance Efficiency factor. To illustrate this, we use the example of the hyperscale datacenter presented in Section \ref{sec:background:energyMetrics:pue}. We analyze the increase in the DCPE factor between the PUE of $x_{1} = 1.58$ and $x_{2} = 1.25$. We determine a 26.98\% improvement in the DCPE factor.

\begin{align}
U_{IT}   &                                        && \text{(IT Equipemnt Utilization)} \\
d_1 &= \frac{U_{IT}}{x_1} = \frac{1}{1.58} = 0.63 && \text{(DCPE for PUE of $x_1=1.58$)} \\
d_2 &= \frac{U_{IT}}{x_2} = \frac{1}{1.25} = 0.80  && \text{(DCPE for PUE of $x_2=1.25$)} \\
p_i &= \frac{\lvert d_1 - d_2 \lvert}{d_1} = 26.98\%        && \text{(Performance Improvement)}
\end{align}


\subsection{Metrics on quantifying CO2 footprint of the system}\label{sec:background:metrics:carbon}
PUE is an excellent metric to quantify ICT infrastructure's performance and energy efficiency. However, PUE does not consider the energy efficiency of applications and workloads~\cite{zhou2013axpue} and overlooks the type of energy used~\cite{dniewenhuis_hotcloud_footprinter}.  While there is a correlation between a datacenter's power draw (i.e., the energy consumed) and the CO2 emissions, several other factors influence the amount of CO2 emitted. Determining the CO2 footprint of the datacenter under a specific workload is an environment-critical, yet not trivial, challenge. Many datacenters use energy from the grid, generated through various sources, with various environmental impacts (e.g., solar, wind, coal). In some cases, energy used from renewable sources, such as wind or solar, can emit up to 20x less CO2 compared to traditional energy sources, such as coal~\cite{dniewenhuis_hotcloud_footprinter, gupta2011gdcsim}.

\textit{Niewenhuis et al.} presents two types of CO2 emissions in the datacenters: \textit{i) the embodied} carbon footprint and \textit{ii) the operational} carbon footprint. Embodied carbon denotes the manufacturing and production results emissions. Operational Carbon Footprint is the CO2 emissions caused by energy usage during datacenter operations. This work proposes a simulation-based solution to alleviating the concerning and deepening environmental problem of CO2 emissions, focusing on the operational carbon footprint~\cite{dniewenhuis_hotcloud_footprinter}.

\subsubsection{Carbon Intensity}\label{sec:background:carbonMetrics:carbonIntensity}
The \textit{Carbon Intensity} of an energy source defines the amount of CO2 emitted per unit of energy used~\cite{dniewenhuis_hotcloud_footprinter}. The measurement unit in the international system is ${[gCO_2 / kWh]}_{S.I.}$. Datacenters utilize energy from the grid~\cite{dniewenhuis_hotcloud_footprinter}; the energy is often provided by multiple sources with distinct Carbon Intensities~\cite{dniewenhuis_hotcloud_footprinter}. Therefore, the Carbon Intensity of the grid is calculated by adding up the Carbon Intensity of each source, proportional to the amount of energy consumed (Equation \ref{eq:carbon_intensity_ci}).

\begin{align}
\text{CI}_{g} = \sum_{s \in S} \text{CI}_{\text{s}} \cdot \frac{E_{\text{s}}}{E_{\text{g}}} && {[gCO_2 / kWh]}_{S.I.} \label{eq:carbon_intensity_ci}
\end{align}
\textit{where $CI_g$ is the Carbon Intensity of the grid, $CI_s$ is the Carbon Intensity of the source, $E_s$ is the energy from a specific source, $E_g$ is the total grid consumption, $s$ is the selected source, $S$ the set of all available energy sources~\cite{dniewenhuis_hotcloud_footprinter}.}

\subsubsection{Carbon Emissions}\label{sec:background:carbonMetrics:carbonIntensity}

\begin{figure}[t]
    \centering
    \includegraphics[width=0.60\linewidth]{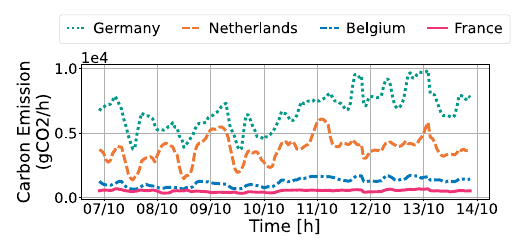}
    \caption{CO2 emission fluctuation, location-dependent. Taken with permission from~\cite{dniewenhuis_hotcloud_footprinter}.}
    \label{fig:co2_location_fluctuation}
\end{figure}

\begin{figure}[t]
    \centering
    \includegraphics[width=0.60\linewidth]{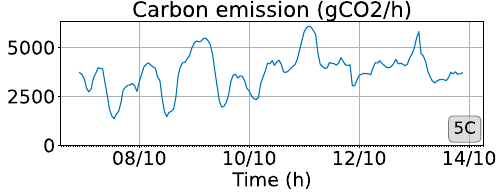}
    \caption{CO2 emission fluctuation, over time. Taken with permission from~\cite{dniewenhuis_hotcloud_footprinter}.}
    \label{fig:co2_time_fluctuation}
\end{figure}

The \textit{Carbon Emissions} of the grid fluctuate depending on the geographical location (Figure \ref{fig:co2_location_fluctuation}), time of the day (Figure \ref{fig:co2_time_fluctuation}), temperature, weather conditions, et cetera. The amount of green energy delivered peaks during the day, while during the night, energy from "grey" sources (e.g., coal) is predominantly used~\cite{eea_energy_consumption}.

\subsubsection{Operational Carbon Footprint}\label{sec:background:carbonMetrics:carbonIntensity}

The \textit{Operational Carbon Footprint} denotes the CO2 emitted when the system is running. The \textit{Operational Carbon Footprint} can be computed using Equation \eqref{eq:operational_carbon_footprint_cop}.

\begin{align}
    \text{C}_{op} = \text{CI}_{d} \cdot E_{op} && {[gCO_2]}_{S.I.} \label{eq:operational_carbon_footprint_cop}
\end{align}

\textit{where $C_{op}$ is the Operational Carbon Footprint, $\text{CI}_d$ is the Carbon Intensity of the datacenter [${gCO_2 / kWh}]_{S.I.}$, \text{$E_{op}$} is the operational energy of the datacenter $[kWh]_{S.I.}$~\cite{dniewenhuis_hotcloud_footprinter}}.

In this work, we employ simulation based on multiple models to predict the \textit{Operational Carbon Footprint} of various configurations of datacenters, under distinct workloads and scenarios.

\subsection{Metrics on quantifying the performance of the system}\label{sec:background:metrics:performance}

We identify performance as an increasingly concerning problem of nowadays LLM ecosystems, which is projected to have lead to an already-starting ``modern-day Moore's Law"~\cite{simon2024llm-mooreslaw}, which identifies a growing gap between the performance needs of LLMs and the actual performance of available physical infrastructure.

\subsubsection{Latency}\label{sec:background:metrics:latency}
\textit{Latency} is defined as the time delay between a cause and its observed effect, and measures \textit{how long} an operation takes~\cite{arpaci2018operating}.

In this work, we quantify latency of LLM ecosystems by the amount of time required to process one token, or unit-scaled (e.g., million tokens). Thus, we measure latency in seconds.

\subsubsection{Throughput}\label{sec:background:metrics:performance:throughput}

\textit{Throughput} is defined as the rate at which a system completes operations, and measures \textit{"how many"} operations (\textit{"how much"} work) the system delivers within a given timeframe~\cite{arpaci2018operating}.

In this work, we quantify throughput of LLM ecosystems by measuring the amount of tokens processed in a per unit timeframe. Thus, we measure latency in tokens per second.

\subsection{Metrics on quantifying the efficiency of the system}\label{sec:background:metrics:efficiency}

We identify efficiency metrics as crucial metrics for datacenter and LLM operators in making informed decisions about potential deployments, as this component offers a homogeneous comparison metric for each type of efficiency, directly comparable, simple to understand, represent, and explain.

\subsubsection{Financial efficiency}\label{sec:background:metrics:efficiency:financial}

We express financial efficiency as the cost per token per second, essentially for the monetary aspect of running LLM ecosystems at scale, in profit-driven processes. The financial efficiency is, thus, represented in \textit{currency per token per second} e.g., \textit{€/t/s}, \textit{LEU/t/s}. Financial efficiency is computed as exemplified in Equation \eqref{eq:efficiency:bg-financial}.

\begin{align}
    E_{f} = \frac{C}{T} = \frac{C}{\frac{T_p + T_d}{\Delta T_{P} + \Delta T_{D}}} = \frac{C \times (\Delta T_{P} + \Delta T_{D})}{T_P + T_D}
    \label{eq:efficiency:bg-financial}
\end{align}
\begin{small}
    \textit{Where $E_{f}$ represents the financial efficiency, $C$ represents the operational cost, $T_P$ and $T_D$ represent the amount of prefill and decode tokens, respectively, and $\Delta T_{P}$ and $\Delta T_{D}$ represent the total inference time for prefill and decode stages, respectively.}
\end{small}

\subsubsection{Sustainability efficiency}\label{sec:background:metrics:efficiency:sustainability}

We express sustainability efficiency as the sustainability cost (e.g., energy, CO2 emissions) per token per second, essential to quantify and compare LLM ecosystems across the increasingly concerning problem of environmental sustainability. The sustainability efficiency is, thus, represented in \textit{metric per token per second} e.g., \textit{Wh/t/s}, \textit{CO2/t/s}. Sustainability efficiency is computed as exemplified in Equation \eqref{eq:efficiency:bg-sustainability}.

\begin{align}
    E_{s} = \frac{S}{T} = \frac{S}{\frac{T_p + T_d}{\Delta T_{P} + \Delta T_{D}}} = \frac{S \times (\Delta T_{P} + \Delta T_{D})}{T_P + T_D}
    \label{eq:efficiency:bg-sustainability}
\end{align}
\begin{small}
    \textit{Where $E_{s}$ represents the sustainability efficiency, $S$ represents the sustainability cost, $T_P$ and $T_D$ represent the amount of prefill and decode tokens, respectively, and $\Delta T_{P}$ and $\Delta T_{D}$ represent the total inference time for prefill and decode stages, respectively.}
\end{small}

\subsection{Metrics on quantifying the accuracy of the simulation}\label{sec:background:metrics:accuracy}

We quantify accuracy using the Mean Absolute Percentage Error (MAPE) ratio, also known as Mean Absolute Percentage Deviation (MAPD), and widely used in the field~\cite{nicolae2025m3sa, dniewenhuis_hotcloud_footprinter, de2016mean, radunicolae-hp-m3sa, oracle2014mape, moreno2013using-mape}. MAPE equally penalizes positive and negative errors and is calculated using Equation \eqref{eq:mape}, where \textit{n} is the number of samples, \textit{R} is real-world data, \textit{S} is simulation data, \textit{i} is the sample index~\cite{nicolae2025m3sa}:

\begin{equation} \label{eq:mape}
    MAPE~[\%] = \frac{1}{n} \sum_{i=0}^{n} \left| \frac{R_{i} - S_{i}}{R_{i}} \right| \times 100
\end{equation}

\section{Discussion}\label{sec:background:discussion}

In this chapter, we introduces the most important elements for this work. We adopted a top to bottom, conceptual to practical overview, starting from terminology and the conceptual concepts of simulation and reference architectures, continued with specifics of LLM ecosystems, caching, and simulating sustainability of ICT infrastructure, and concluded with community agreements on measuring various operational aspects of ICT infrastructure, LLM ecosystems, and simulation, through community standard metrics. 

We identify several other important aspects, such as in-depth details of LLM inference, validation, experimental setup, and scientific methodology. However, we argue that, albeit overall relevant for this work, these topics are outside the scope of this chapter.
 \newpage
\thispagestyle{noheader}
\chapter{A Reference Architecture for LLM ecosystems}\label{sec:refarch}

The lack of a reference architecture or a community-wide underlying conceptual model can be costly; conceptually, stakeholders (e.g., infrastructure operators, researchers) could overlook essential components, and even capable teams of researchers and engineers could tinker, leading to architectural and deployment challenges~\cite{DBLP:conf/sc/AndreadisVMI18}. 
Furthermore, without a comprehensive reference architecture of LLM ecosystems, simulators of such infrastructure cannot be rigorously designed; later in this work (\Cref{sec:design} - \Cref{sec:evaluation}), we design, implement, integrate, and engineer a simulation instrument for KV-Caching system of LLM ecosystems under inference.

To rigorously design an ICT simulator, tailored to predicting LLM ecosystems, the state-of-the-art consists of materialising a reference architecture of the respective infrastructure, into a simulation tool or instrument, following scientific methodologies vetted and well-followed by the computer systems community~\cite{DBLP:journals/tpds/AndreadisMBI22,DBLP:conf/icdcs/IosupVTETBFMT19, DBLP:conf/ccgrid/MastenbroekAJLB21, DBLP:conf/sc/AndreadisVMI18}.
However, we argue that, currently, it doesn't exist no comprehensive reference architecture of LLM ecosystems under inference.
This raises the research question: \textit{(RQ1) How to synthesize and validate a reference architecture of LLM ecosystems?}

In this chapter, we address RQ1 by proposing the first reference architecture for LLM ecosystems under inference workloads, following community-vetted scientific processes for design and validation. 

\section{Overview}\label{sec:refarch:overview}

We propose the first reference architecture for LLM ecosystems under inference, following a distributed systems approach. Throughout this chapter, we match the state-of-the-art AtLarge Design Process \cite{DBLP:conf/icdcs/IosupVTETBFMT19}. Our contribution in this chapter is six-fold:

\begin{enumerate}
    \item We define and establish design requirements and principles which guide in proposing the reference architecture (\Cref{sec:refarch:requirements}).

    \item We synthesize a comprehensive reference architecture for LLM ecosystems under inference in \Cref{sec:refarch:high_level}, following the requirements and principles defined in \Cref{sec:refarch:requirements}. We model the entire interaction loop, from user input to system output, as a high-level picture, then we detail, individually, both the input and output process. Lastly, we detail the feedback loop, essential for processes of Reinforcement Learning from Human Feedback~(RLHF).

    \item We conduct a per-component analysis and discuss integration with other components of the ecosystem, providing real-life examples of employed technologies (\Cref{sec:refarch:high_level}).

    \item We propose a detailed design for the KV-Caching system in \Cref{sec:refarch:high_level:prompt}. Albeit optional in LLM ecosystems, the KV-Caching sub-system can enhance, by orders of magnitude, the performance (e.g., throughput, latency) of the LLM ecosystems under inference.

    \item We validate the reference architecture in \Cref{sec:refarch:validation}; 
    firstly, we validate the proposed architecture by aligning with the community-vetted, peer-reviewed Compute Continuum~\cite{DBLP:conf/ccgrid/JansenAPTI23}; then, we align our proposed architecture with a domain-specific, industry existing LLM-inference ecosystem from OpenAI; then, we align the architecture with another state-of-the-art ecosystem, used by IBM for LLM inference; many thanks to IBM Research Europe for their many-fold contributions to open-science. Lastly, we present a high-level overview of how our proposed reference architecture aligns with real-world LLM ecosystems.

    \item We address, in \Cref{sec:refarch:requirements-adressal} each design requirement and principle presented in \Cref{sec:refarch:requirements}.
\end{enumerate}

\section{Design Requirements and Principles}\label{sec:refarch:requirements}

In this section, we discuss the main design requirements and principles that guided our design process. We design to fulfill a set of requirements, corresponding to stakeholders of this reference architecture; we envision the main stakeholders of our work to be the researchers in the field of AI and LLMs, datacenter operators, C-level decision-making stakeholders, and students. 

In~\cite{DBLP:conf/sc/AndreadisVMI18}, \textit{Andreadis et al.} present a reference architecture for datacenter scheduling, a well-recognized scientific contribution published in SC18, the International Conference for High Performance Computing, Networking, Storage, and Analysis 2018. They define two main design requirements, \textit{validity} and \textit{usefulness}, also relevant for our reference architecture. Below, we expand on these \underline{\textbf{D}}esign \underline{\textbf{R}}equirements.

\begin{enumerate}[label=(\textbf{DR\arabic*}), leftmargin=*, align=left, labelwidth=1cm]
    \item \textbf{Ensure the \textit{validity} of the reference architecture.}\label{DR1}\label{DR:validity} \\
    Validity \textit{``is the property of the proposed model to accurately represent the field of"} LLM systems~\cite{DBLP:conf/sc/AndreadisVMI18}. 
    The reference architecture should cover, for each component, the state-of-the-art from both industry and academia. 
    Albeit fundamentally a subjective task quantifying the validity, we argue that mapping existing LLM systems to this reference architecture, and mapping the proposed reference architecture to the Compute Continuum prove the validity of this reference architecture to both abstract and align with real-world examples (i.e., real world LLM systems under inference) and align with higher level representations of the distributed ecosystems (i.e., the Compute Continuum).

    \item \textbf{Ensure the \textit{usefulness} of the reference architecture}\label{DR2}\label{DR:usefulness} \\
    Usefulness \textit{``gives the reference architecture a real-world purpose which motivates its creation"}\cite{DBLP:conf/sc/AndreadisVMI18}. 
    Alike \textit{validity}, \textit{usefulness} is a fundamentally subjective design requirement, yet we argue that usefulness can be evaluated by demonstrating the ability of the reference architecture to enable stakeholders to better reason about LLM system design in practice.
\end{enumerate}

We further derive seven \underline{\textbf{D}}esign \underline{\textbf{P}}rinciples the reference architecture should follow to ensure a comprehensive, actionable, and future-proof architecture, contributing to an end goal of simulating and digitally twinning, holistically, accurately and robustly, the Compute Continuum.

\begin{enumerate}[label=(\textbf{DP\arabic*}), leftmargin=*, align=left, labelwidth=1cm]
    \item \textbf{Design components with clear distinct responsibilities.}\label{DP:distinct-components}\\
    We regard distinct-responsibility components as essential for designing a state-of-the-art reference architectures, especially following community standards~\cite{DBLP:conf/sc/AndreadisVMI18, rozanski2012software}. 
    Each system component should have its own set of responsibilities, defined boundaries of those responsibilities, and a set of interfaces which define its services to other components~\cite{DBLP:conf/sc/AndreadisVMI18}. 
    Albeit included in the reference architecture, we acknowledge that not all the components need to be used by every real-world stakeholder of LLM systems; e.g., while essential in system performance, KV-Caching can be omitted in LLM systems, making the system still functional, yet non-performant.

    \item \textbf{Group related components.}\label{DP:grouping-components}\\
    Corresponding to best practices for packaging components~\cite{rozanski2012software}, and following vetted design processes of reference architectures~\cite{DBLP:conf/sc/AndreadisVMI18}, related components should be grouped according to their responsibility. 
    We acknowledge this introduces a degree of subjectivity and, thus, we regard also other reference architectures for LLM systems, although potentially not matching the same structure as the RA from this work proposes.

    \item \textbf{Aim for extensibility and modularity}\label{DP:modularity}\\
    The architecture should be modular, corresponding to the best practices of designing software systems~\cite{rozanski2012software}, thus allowing for independent extension, detailing, analysis, or replacement of components. 
    We regard this as a critical design principle to ensure a future-proof property of the proposed reference architecture. 
    Furthermore, such extensibility and modularity allow the reference architecture to seamlessly integrate and integrate with emerging technologies that will emerge, without redesigning the whole.

    \item \textbf{Separate mechanisms from policies and goals.}\label{DP:mechanism-policy}\\
    Each architectural component should clearly distinguish between mechanisms (how it operates), policies (how decisions are made), and the metrics used in measurement and evaluation (e.g., latency, throughput, power draw, CO2 emissions). 
    To provide the necessary~\cite{DBLP:conf/sc/AndreadisVMI18} level of abstraction, the reference architecture should follow a qualitative model~\cite{rozanski2012software}, without mandating specific policies~\cite{DBLP:conf/sc/AndreadisVMI18}.

    \item \textbf{Cover end-to-end prompt-to-response LLM workflow.}\label{DP:end-to-end}\\
    The architecture should model the entire inference workflow, from the user input to the system output, and model all the intermediate stages, with the necessary level of abstraction. 
    This end-to-end exhaustive view ensures a continuous interaction-feedback loop, and ensures that, e.g., bottlenecks, cross-component effects, and trade-offs can be analyzed and eventually simulated.

    \item \textbf{Support multiple users in the ecosystem.}\label{DP6}\label{DP:multi-user}\\
    The architecture should support multiple users operating simultaneously, each giving workloads (i.e., prompts) to the LLM ecosystem, and each receiving responses, following the end-to-end prompt-to-response format proposed in \ref{DP:end-to-end}.

    \item \textbf{Model components responsible for decision processes across the system.}\label{DP:decision-processes}\label{DP:decision-processes}\\
    The architecture should represent decision-making processes that occur across each layer of the continuum. 
    This includes (pre)processing, workload and resource allocation, inference management, and overall orchestration and monitoring.

    \item \textbf{Model different types of prompts execution workflow.}\label{DP:flexible-prompt}\\
    The architecture should model different types of prompt-workflow and consider at least workflows with prompt reasoning, where each prompt generates a subsequent prompt, and workflows with branching prompts.
    This design principle allows for modelling a wide range of prompts, tailored to today's standards in LLMs and with future developments.
\end{enumerate}
\begin{figure}[t]
    \centering
    \includegraphics[width=\linewidth]{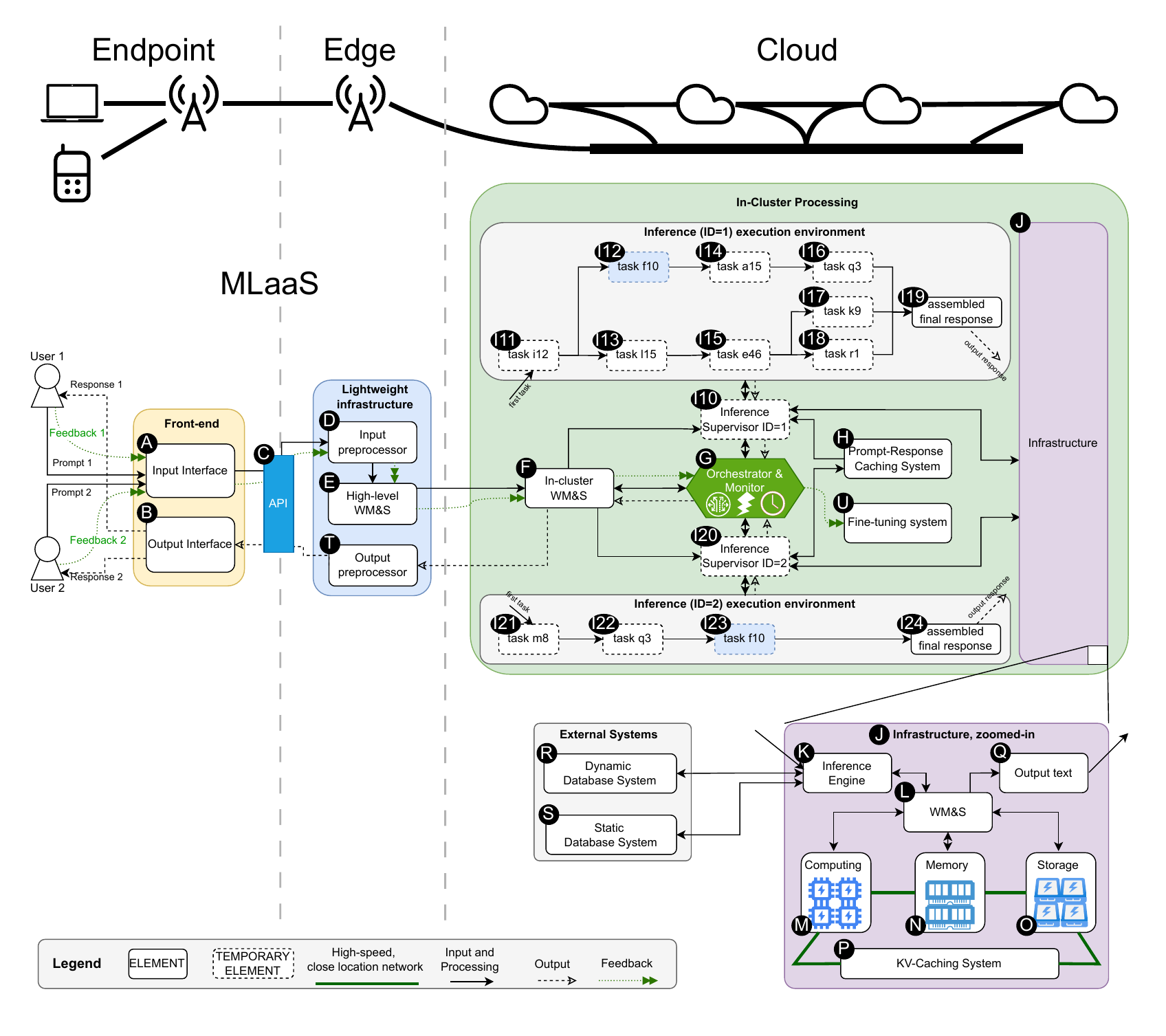}
    \caption{Reference Architecture of LLM ecosystems under multi-user inference workload.}
    \label{fig:reference-architecture}
\end{figure}

\section{Overview of the Reference Architecture}\label{sec:refarch:high_level}

In this section, we present an overview of the proposed reference architecture, present the overall design and workflow, which covers an end-to-end prompt-response-feedback loop, and the main responsibilities of each tier, component, and mechanism. \Cref{fig:reference-architecture} presents the proposed reference architecture.

Following design principle~\ref{DP:distinct-components}, we propose a reference architecture of an ecosystem which leverages sets of components, distributed by tier of operation and grouped where related~\ref{DP:grouping-components}, with each layer and component abstracted as modular and extensible~\ref{DP:modularity}. We model all three tiers of the continuum~\cite{DBLP:conf/ccgrid/JansenAPTI23}, namely \textit{endpoint}, \textit{edge}, and \textit{cloud}, and attach related components to the specific tier~\ref{DP:distinct-components},~\ref{DP:grouping-components}. 
\\
\\
\textit{Endpoint:} The proposed reference architecture begins with the user's prompt and finalizes with the ecosystem's response, following an end-to-end prompt-to-response workflow~\ref{DP:end-to-end}; optionally, the user can offer feedback on the LLM's response, feedback further used for fine-tuning the LLM and for better tailoring responses to the user's preferences.

\textit{Edge:} We propose a multi-user ecosystem, where each user-given workloads (i.e., prompts) are preprocessed in the edge, and forwarded to the most suitable cloud infrastructure, following decision processes~\ref{DP:decision-processes}, and the Machine Learning as a Service (MLaaS) operational model.

\textit{Cloud:} Once in the cloud, each prompt is preprocessed and managed by an in-cluster workload manager and scheduler, a cloud decision-making component which ensures fair resource allocation per prompt, per user~\ref{DP:mechanism-policy}~\ref{DP:multi-user}~\ref{DP:decision-processes}.  
We represent two inference processes, run in parallel, one with a complex, branching, reasoning workflow (inference id 1), and the other with a sequential reasoning workflow (inference id 2), both coupled to a shared caching system of prompts and responses~\ref{DP:flexible-prompt}. 
All processes executing in the cloud are orchestrated and monitored for performance, energy consumption, and failure detection, ensuring quality of service (QoS) and meeting service level objectives (SLOs)~\ref{DR:validity}-\ref{DR:usefulness}, \ref{DP:distinct-components}-\ref{DP:flexible-prompt}.

\textit{Infrastructure:} We now focus on component~\circled{J}. The workload enters the infrastructure through the \textit{Inference Engine}~\circled{K}, which communicates with the\textit{ WM\&S}~\circled{L} and orchestrates the computation, the backbone of the LLM inference process. We abstract the computation process as shared between three types of physical infrastructure, namely \textit{Computing Units}~\circled{M}, \textit{Memory Racks}~\circled{N}, and \textit{Storage Units}~\circled{O}, all of them employing techniques and linked to a \textit{ System}~\circled{P}; the physical infrastructure is interconnected by a high-speed, close-location network (e.g., InfiniBand).

\textit{External Systems:} The infrastructure is linked with \textit{External Systems}, which expand the LLM capabilities beyond the isolated functionality an independent and disconnected LLM could give. Component~\circled{S}, the \textit{Static Database System}, allows the LLM to access local, in-cloud databases, which store, index, and manage embedding vectors used for similarity search and retrieval augmented generation (RAG)~\cite{ibm_milvus_2024}. Component~\circled{R}, the \textit{Dynamic Database System} enables the LLM to conduct web searches~\cite{ibm_elasticsearch_2024}. Both the Dynamic and the Static database systems are orchestrated by the \textit{Inference Engine}~\circled{K}.

\textit{Workflows:} We identify and detail three main workflows~\ref{DP:end-to-end}:
\textit{(i) in \Cref{sec:refarch:high_level:prompt}}, we detail the prompt-execution workflow, containing the steps between the user's input and the response of the LLM;
\textit{(ii) in \Cref{sec:refarch:high_level:response}}, we detail the response workflow, containing the processes between the LLM response and the display of the response on the user's interface;
\textit{(iii) in \Cref{sec:refarch:high_level:feedback}}, we detail the optional feedback workflow, through which users can review the LLM's response; further feedback is passed to the ecosystem for LLM fine-tuning and prompt enhancement. 

\begin{custombox}
    \boxelement{Note}{In the following sections, we detail one workflow per section, detailing specifics of each workflow, and de-focusing non-relevant components for better visual comprehension. In \Cref{sec:refarch:high_level:prompt} we detail the prompt execution workflow, in \Cref{sec:refarch:high_level:response} we detail the prompt response workflow, and in \Cref{sec:refarch:high_level:feedback} we detail the feedback workflow.}
\end{custombox}

\subsection{Detailed Design of Prompt Execution Workflow}\label{sec:refarch:high_level:prompt}

\begin{figure}[t]
    \centering
    \includegraphics[width=\linewidth]{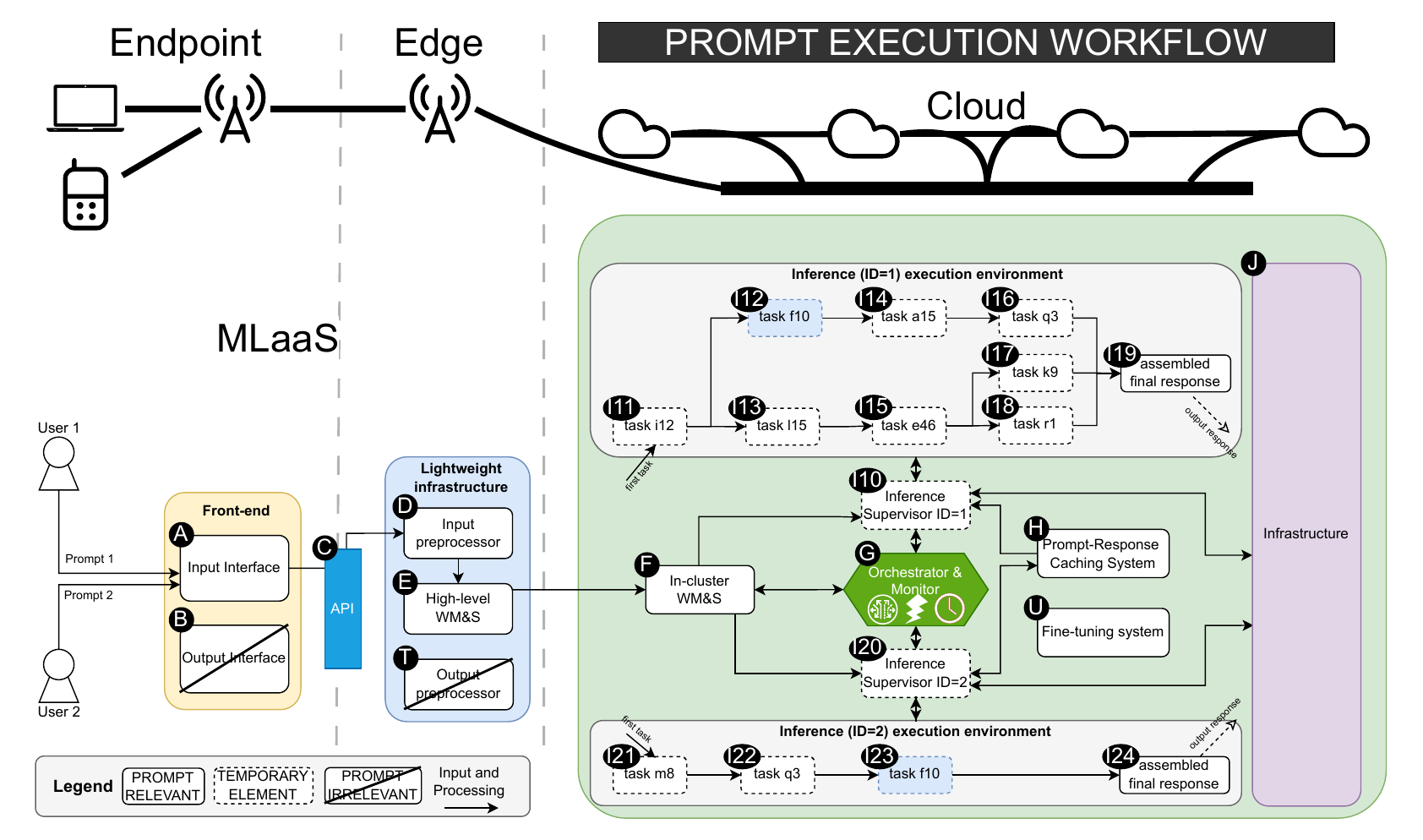}
    \caption{Reference Architecture of LLM ecosystems detailing the prompt execution workflow. The prompt execution workflow focuses on components from \circled{A} to \circled{J}.}
    \label{fig:reference-architecture:workflow:execution}
\end{figure}

In this section, we detail the \textit{prompt execution workflow}, which starts with the user prompt and ends with the final leveraged response. \Cref{fig:reference-architecture:workflow:execution} illustrates this process.

\subsubsection{Front-end (endpoint) tier}

The prompt execution workflow begins with the users, who provide input (prompts), visually represented in the left-most part of the reference architecture from \Cref{fig:reference-architecture:workflow:execution}. 
Although we present only two users for visual purposes, the proposed RA can scale indefinitely from a design perspective, yet is upper-bounded by limitations of physical resources.

Users interact with the ecosystem via an \textit{input interface}~\circled{A}, as part of the \textit{front-end}, which is often through a web interface or through a mobile application, but could also be through an API call via e.g., a command line interface (CLI) environment. 
The \textit{front-end} tier is the equivalent to the \textit{endpoint} tier from the RA of the Compute Continuum~\cite{DBLP:conf/ccgrid/JansenAPTI23}; we further expand and align our RA with the Compute Continuum in~\Cref{sec:refarch:validation:continuum}.

\subsubsection{Lightweight infrastructure (edge) tier}

Following the MLaaS operational model, the user's prompt is transferred to the \textit{cloud} where the heavyweight computation happens; however, in this transfer process, the edge plays a crucial role.

The user's prompt is transferred from the \textit{front-end} to the \textit{lightweight infrastructure} via an \textit{application programming interface (API)}~\circled{C} and is further parsed by an \textit{input preprocessor}~\circled{D}. We argue that the overall process of data preprocessing (mainly happening in~\circled{C}) is critical in reducing the amount of data transferred from the edge to the cloud, and in restricting the execution prompts to only desired (e.g., policy-adherent) prompts.

\subsubsection{A detailed design on the Prompt-Response Caching System}

\begin{figure}[t]
    \centering
    \includegraphics[width=0.95\linewidth]{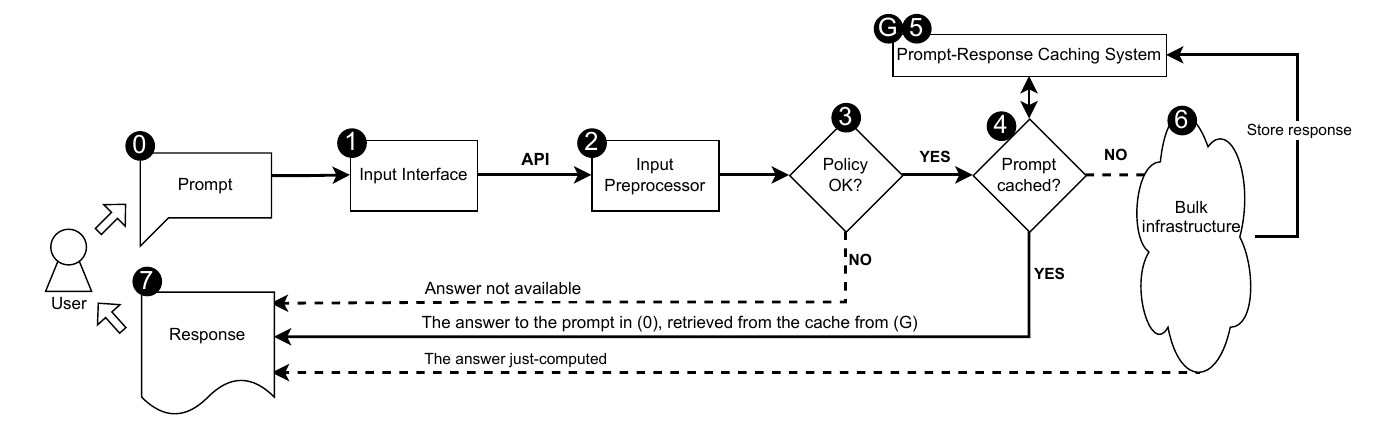}
    \caption{Prompt-response workflow, preventing redundant computation of already-generated responses to prompts, employing an inference, prompt-level caching system. }
    \label{fig:prompt-response-caching}
\end{figure}

In \Cref{fig:prompt-response-caching}, we present a prompt-response workflow that prevents the re-computation of prompts the ecosystem has already responded to, thus minimizing redundancy. 
Although not essential for the system's base functionality, the illustrated caching approach offers a theoretical advantage by restricting the execution of prompt workflows to only when encountering a non-cached prompt. OpenAI employs a prompt-caching system that considers only prompts exceeding 1,024 tokens~\cite{openai_prompt_caching}. OpenAI claims to perform only \textit{``halfway prompt caching,"} where they store prefill weights in the caching system and always run the decode stage for each cache hit, for each user: \textit{"the actual response is computed anew each time based on the cached prompt."}~\cite{openai_prompt_caching}. The OpenAI codebase is closed-source.

In \Cref{fig:prompt-response-caching}, aligned with the overall RA presented in \Cref{fig:reference-architecture} and \Cref{fig:reference-architecture:workflow:execution}, the process begins with the \textit{user's prompt}~\circled{0}, through an \textit{input interface}~\circled{1}, and further parsed through an \textit{input preprocessor}~\circled{2}. 
In the preprocessing step, the system checks if the prompt matches the company's policy and laws in the country of operation (e.g., prompt content, lawfulness)~\circled{3}. 
If the prompt complies, the system checks whether the prompt is already cached~\circled{4} in a \textit{prompt-response caching system}~\circled{5}; the \textit{prompt-response caching system} is equivalent to element~\circled{G} from \Cref{fig:reference-architecture}. This process is mainly handled by 
(i) the \textit{High-Level WM\&S} from the \textit{lightweight infrastructure} (\circled{D}, \Cref{fig:reference-architecture:workflow:execution}),
(ii) the \textit{In-Cluster WM\&S} (\circled{G}, \Cref{fig:reference-architecture:workflow:execution}), and 
(iii) the \textit{Orchestrator and Monitor} (\circled{F}, \Cref{fig:reference-architecture:workflow:execution}), all decision-making components of their respective layer~\ref{DP:decision-processes}.

If the prompt is cached, the already generated and cached response is retrieved and delivered to the user. 
We envision this approach as improving performance, boosting the ecosystem's throughput, reducing the system's latency, and overall reducing the amount of resources consumed by the infrastructure (e.g., power draw), with the caveat that the caching system is well-designed and efficient. The response retrieval process doesn't consume more resources than generating the response itself.
If the prompt-response pair is not found in the caching system, the system forwards the workload to a bulk infrastructure which generates a response; this response is ultimately stored in the \textit{prompt-response caching} system ~\circled{G}/\circled{5}, and offered to the user~\circled{7}.

In practice, OpenAI employs a similar technique; however, they claim to store only prefill weights in a caching system, and for each cache hit, the decode stage is rerun for each user. With this halfway caching technique, OpenAI claims to have OpenAI claims this technique to have reduced latency \textit{``by up to 80\% and cost by 50\% for long prompts"}~\cite{openai_prompt_caching}.

\subsubsection{In-cluster (cloud) tier}
\textit{Orchestrator \& Monitor:} The bulk computation and storage occur in the cloud, a massive-scale, highly heterogeneous, and distributed infrastructure, where a core, central component supervises the overall process, the \textit{Orchestrator \& Monitor}~\circled{G}.
Component \circled{G} serves as a core decision-making component and manages, using data obtained from monitoring the ecosystem, other supervisors of the ecosystem, e.g., \textit{inference supervisor}~\circledd{I10},~\circledd{I20}, \textit{in-cluster WM\&S}~\circled{F}. The monitoring responsibility involves measuring performance metrics (e.g., throughput, latency), sustainability metrics (e.g., the amount of hourly emitted CO2, energy consumption), and system failures (e.g., uptime, amount of jobs completed/failed). Based on the metrics gathered from the monitoring process, \textit{Orchestrator and Monitor}~\circled{G} analyses and predicts further behaviour (e.g., using an ecosystem simulator), and makes decisions in orchestrating inference workloads. Furthermore, the \textit{Orchestrator and Monitor}~\circled{G} has management and supervision access over the part of the cloud dedicated to the ecosystem.

\textit{Edge-cloud WM\&S communication:} The \textit{high-level WM\&S}~\circled{E}, from the edge, forwards prompts to the \textit{in-cluster WM\&S}~\circled{F}, which manages prompts and ensures fair, policy-compliant responses to users' requests. 
\circled{F} generates inference supervisors for each prompt received; in \Cref{fig:reference-architecture:workflow:execution}, we exemplify using two prompts, each with its own inference supervisor, namely supervisors \circledd{I10}, \circledd{I20}.

\textit{The inference supervisor:} serves as a middleware between the workload tasks and the overall datacenter orchestrator \& monitor~\circled{F}. For example, \textit{inference supervisor 1}, represented in the upper half of \Cref{fig:reference-architecture:workflow:execution}, handles the inference process for prompt 1, from user 1. 

\textit{Complex, branching, prompt:}
We exemplify a prompt that requires a reasoning process that branches, where each task generates one or more new tasks, until a final response is obtained. 
In the illustrated reference architecture, the \textit{inference supervisor} firstly generates a starting task \circledd{I11}, which, after completion, generates two new tasks \circledd{I12} and \circledd{I13}. This recurrent process recurrently repeats and, depending on the task, one or more tasks are generated until a \textit{final response} is assembled, in our example \circledd{I19}. The \textit{inference supervisor} determines when a response is final.

\textit{Simple, sequential, prompt:}
However, prompts can also generate simpler, sequential, and non-branching tasks, such as prompt 2, assigned inference supervisor 2 and the tasks \circledd{I21}-\circledd{I24}.

\textit{Prompt-response caching system:}
For each generated task (\circledd{I11}-\circledd{I19}, \circledd{I21}-\circledd{I24}), the workload is managed by the corresponding inference supervisor, which checks the \textit{prompt-response caching system} \circled{G}. 
If the response to the specific task\footnote{in nowadays ecosystems, e.g., OpenAI's ChatGPT~\cite{openaichatgpt} or Google's Gemini~\cite{google2023efficiency}, each answered task is temporary saved and also phrased as a prompt for the future task(s).} is found in the \textit{prompt-response caching system} (cache hit), then the \textit{inference supervisor} retrieves and uses the response as the response to the specific task. 
Otherwise, when the response to the prompt is not already cached (cache miss), the inference supervisor forwards the workload to the infrastructure~\circled{J}. The infrastructure~\circled{J} computes and redirects the response to the caching system~\circled{H}; here, if the caching policy is matched, the response is saved. Further, the response is forwarded to the inference process, either as an intermediate response or as an assembled final response (e.g., \circledd{I19}, \circledd{I24}).

\textit{Identical tasks:}
In \Cref{fig:reference-architecture:workflow:execution}, we exemplify two prompts with an identical intermediate task, namely \textit{task f10} (\circledd{I12}, \circledd{I23}). 
This could happen in user prompts with similar tasks, and, thus, an identical intermediate task to achieve the response to a certain task. 
We exemplify below with two analogies, one with a Mathematical analogy and one with a Path(Route)-Finding example. 

\textit{Analogies of identical tasks:} In the \textit{Mathematical Analogy (\Cref{lst:refarch:analogy1})}, we showcase a scenario where two users give two distinct prompts, yet with an identical sub-task (i.e., computing 10 factorial, equivalent to $10!$, where $10! = 10 \times 9 \times 8 \dots \times 2 \times 1$); this task happens only once, for the first encountered prompt, and is retrieved for the second prompt, instead of redundantly re-computed. 

In the \textit{Path-Finding Analogy~(\Cref{lst:refarch:analogy2})}, similarly, two users give two distinct prompts. The first user requests a path from Amsterdam to Bucharest, and the LLM ecosystem, unable to identify the response to such a prompt in the cache, computes, generates, and caches the response to each intermediate task. The second user requests a path from Amsterdam to Bratislava; the LLM finds this path as cached, and only retrieves the response from the \textit{prompt-response caching system}, without redundant (re)computation. Then, the LLM ecosystem only computes the rest of the response.

\begin{lstlisting}[style=promptinteraction,caption={A Mathematical analogy of the LLM inference and prompt-response caching process.}, label={lst:refarch:analogy1}]
 Task 1: Calculate 12 x 6 x 1980 x 10!
 Task 2: Calculate 23 x 2 x 2004 x 10!
 
 LLM approach:
 Step 1.1: Determine 12 x 6 x 1980. Not cached. Compute. Cache. 
 Step 1.2: Determine 10!. Not cached. Compute. Cache.
 Step 1.3: Solve final task 1. Not cached. Compute. Cache.
 
 Step 2.1: Determine 23 x 2 x 2004 and cache. Not cached. Compute. Cache.
 Step 2.2: Determine 10!. Cached! Retrieve!
 Step 2.3: Solve final task 2. Not cached. Compute. Cache.
\end{lstlisting} \label{lst:llm-mathematical}


\begin{lstlisting}[style=promptinteraction,caption={A Path-Finding analogy of the LLM inference and prompt-response caching process.}, label={lst:refarch:analogy2}]
 Task 1: Find a path from Amsterdam to Bucharest for a motorbike drive.
 Task 2: Find a path from Amsterdam to Bratislava for a motorbike drive.
 
 LLM approach:
 Step 1.1: Find intermediate checkpoints for the most time-efficient route between Amsterdam and Bucharest 
 (e.g., LLM finds Amsterdam, Leipzig, Bratislava, Arad, Bucharest). Not cached. Compute. Cache.
 Step 1.2: Find the most efficient route Amsterdam - Leipzig. Not cached. Compute. Cache.
 Step 1.3: Find the most efficient route Leipzig - Bratislava. Not cached. Compute. Cache.
 Step 1.4: Find the most efficient route Brastislava - Arad. Not cached. Compute. Cache.
 Step 1.5: Find the most efficient route Arad-Bucharest. Not cached. Compute. Cache.
 Step 1.6: Generate and export GPX file to user. Not cached. Compute. Cache.
 
 Step 2.1: Find intermediate checkpoints for the most time-efficient route between Amsterdam and Bratislava 
 (e.g., LLM finds Amsterdam, Leipzig, Bratislava). Not cached. Compute. Cache.
 Step 2.2: Find the most efficient route Amsterdam - Leipzig. Cached! Retrieve!
 Step 2.3: Find the most efficient route Leipig - Bratislava. Cached! Retrieve!
 Step 2.4: Generate and export GPX file to user. (Partially) Cached! Retrieve! Compute the rest. Cache.
\end{lstlisting} \label{lst:llm-gps}

\subsection{Detailed Design of the LLM Response Workflow}\label{sec:refarch:high_level:response}

\begin{figure}[t]
    \centering
    \includegraphics[width=\linewidth]{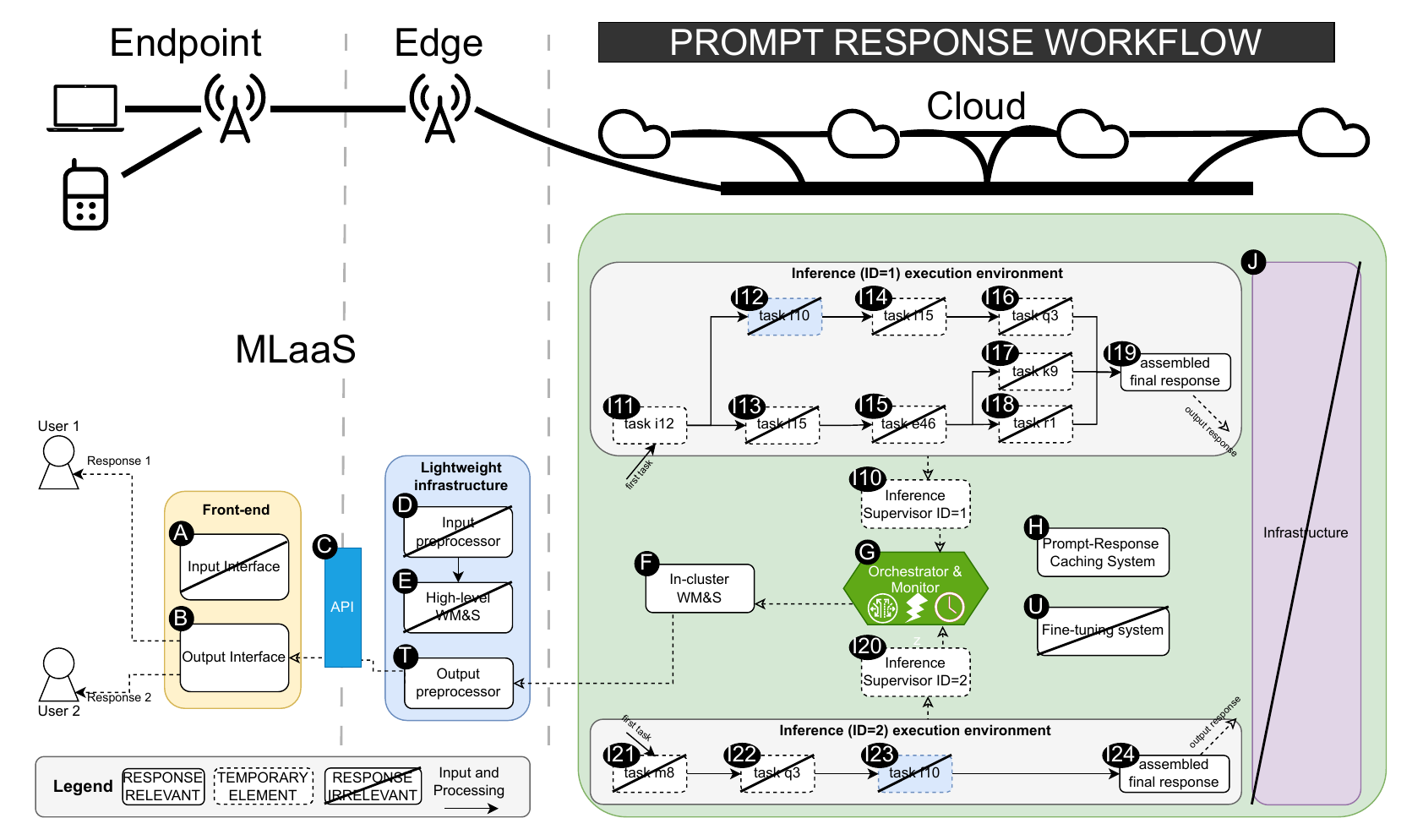}
    \caption{Reference Architecture of LLM ecosystems detailing the LLM response processing workflow. The prompt-response workflow focuses on components \circledd{I19}, \circledd{I24},  \circledd{I10}, \circledd{I20}, \circled{H}, \circled{G}, \circled{F}, \circled{T}, \circled{C}, and \circled{B}.}
    \label{fig:reference-architecture:workflow:response}
\end{figure}

In this section, we detail the \textit{prompt-response workflow}, which begins once the final response is assembled and finalised, and ends once the response is displayed on the user's interface \Cref{fig:reference-architecture:workflow:response} illustrates this process.

\textit{Cloud: }The response output process begins from \circledd{I19}, for inference with ID=1, and from \circledd{I24}, for inference with ID=2. The output response is further transferred to the corresponding \textit{inference supervisor}, further transferred to the \textit{in-cluster WM\&S}~\circled{F}. The entire workflow executed in the cluster is constantly monitored by the \textit{orchestrator and monitored component}~\circled{G}.

\textit{Edge:} The leveraged response, now located in \circled{F}, the WM\&S of the cloud, is transferred to the \textit{output processor}~\circled{T} from the \textit{edge} (i.e., \textit{lightweight infrastructure}). 

\textit{Endpoint:} Lastly, the response is transferred from the \textit{edge}, via the API, to the \textit{endpoint} and displayed on the \textit{output interface}~\circled{B} of the LLM ecosystem, belonging to the \textit{front-end}.

\subsection{Detailed Design of Feedback Workflow}\label{sec:refarch:high_level:feedback}

\begin{figure}[t]
    \centering
    \includegraphics[width=\linewidth]{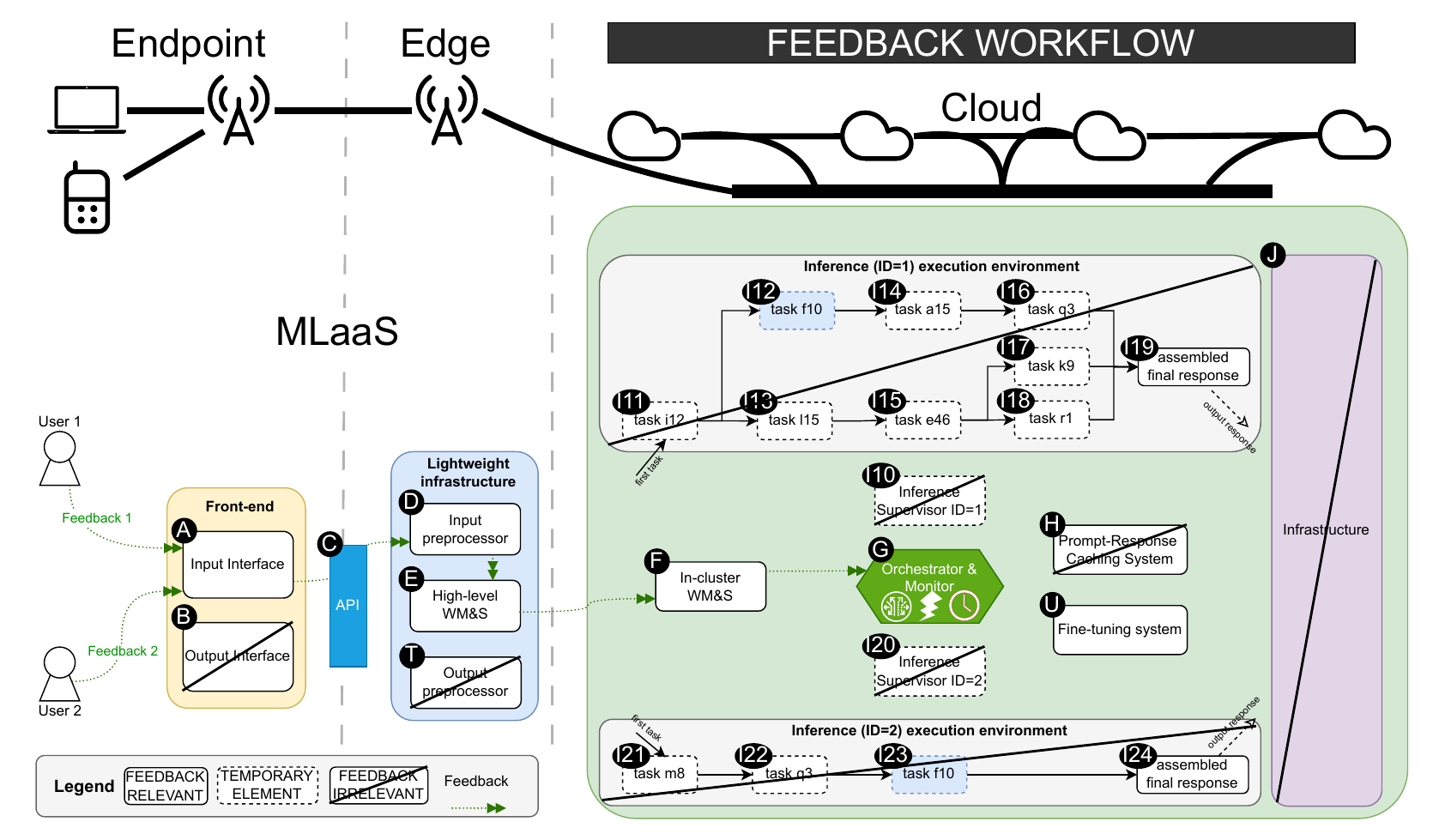}
    \caption{Reference Architecture of LLM ecosystems detailing the feedback processing workflow. The feedback workflow focuses on components \circled{A}, \circled{C},  \circled{D}, \circled{E}, \circled{F}, \circled{G}, and \circled{U}.}
    \label{fig:reference-architecture:workflow:feedback}
\end{figure}

In this section, we detail the \textit{feedback processing workflow}, an optional workflow of the inference process which users often skip, yet is critical for LLM finetuning and tailoring responses to users' needs and preferences~\cite{dai2023safe, ouyang2022training}. This workflow starts with the user's feedback, through the input interface, and ends with the system receiving and optionally adopting this feedback. \Cref{fig:reference-architecture:workflow:feedback} illustrates this process.

\textit{Why feedback matters: } Feedback is crucial for refining LLM responses and improving performance through Reinforcement Learning from Human Feedback (RLHF). For example, OpenAI researchers identify even small RLHF-trained models, of 1.3B parameters, as better-preferred, higher-accurate, than larger models such as 175B GPT-3, although having 100x fewer parameters~\cite{ouyang2022training}. While limited reports exist from large LLM providers, RLHF is a widely used technique, cheap to scale compared to traditional LLM finetuning or extended training, preventing and reducing financial, performance, and sustainability costs~\cite{achiam2023gpt, dong2024rlhf, dai2023safe, ouyang2022training}.

\textit{Endpoint:} We model a scenario in which both user 1 and user 2 evaluate the LLM's prompt. The feedback detail depends on the platform, and can vary from e.g., selecting between positive and negative to e.g., giving detailed feedback on multiple categories, with written components. Users' feedbacks are inputted via component \circled{A}, the \textit{Input Interface}, and transferred to the cloud via the endpoint, similarly to users' prompts.

\textit{Edge}: The \textit{API}~\circled{C} links the \textit{Endpoint} to the \textit{Edge} (and vice versa), transferring the user's feedback to an \textit{Input preprocessor}~\circled{D}. Depending on the platform design, \circled{D} can filter users' text feedback and check policy adherence; if the feedback doesn't involve a text component, step \circled{D} may be skipped. Further, a \textit{High-level WM\&S}~\circled{E} component forwards the feedback to the cloud.

\textit{Cloud:} The feedback is processed by an \textit{in-cluster WM\&S}~\circled{F}, which forwards user's feedback to an \textit{Orchestrator \& Monitor}~\circled{G} component, a centric element of the cloud with monitoring, analysis, and decision capabilities over the datacenter part reserved for the inference process of the LLM ecosystem. The feedback is forwarded to a \textit{Fine-Tuning System}~\circled{U} which handles the feedback.

\textit{Feedback policies:} How the ecosystem handles users' feedback is dependent on the provider's policies and regulations. For example, an LLM provider can choose to tailor LLM's responses only for the conversation in progress, without keeping cached feedback for other conversations, and without using users' feedback for fine-tuning the global LLM for other users. However, a provider with less strict privacy policies can use feedback for fine-tuning the model for all users, not only for the specific user who gave the feedback. To ensure generality and universality of the proposed reference architecture, we abstract the feedback and fine-tuning system into a unitary component \circled{U}.

\begin{custombox}
    \boxelement{Note}{The process described throughout this section (\Cref{sec:refarch:high_level}), and the last few pages, executes within (milli)seconds in real-world ecosystems~\cite{lazuka2024llm, simon2024llm-mooreslaw, DBLP:conf/sosp/ZhangDLKMWLYLLZ25}.}
\end{custombox}
\section[Mapping Real-World LLM Inference Ecosystems to the Reference Architecture]{Mapping Real-World LLM Inference Ecosystems to the \\Reference Architecture}\label{sec:refarch:validation}

Reference architectures are most useful when they accurately depict real-world instances~\cite{DBLP:conf/sc/AndreadisVMI18}. In this section, we align our reference architecture with industry-leading LLM ecosystems and with a peer-reviewed, community-standard reference architecture of the Compute Continuum.

We identify some components as non-disclosed; non-disclosed components are components that are likely to exist in real-world (deployed) LLM-inference systems, but their presence is not disclosed by some designers and operators, and only inferred by communities of practice, e.g., on sites such as Hacker News and Reddit, or disclosed by other designers and operators. For example, OpenAI does not disclose the usage of input preprocessor, but the Ubicloud-envisioned ecosystem and the Databricks ecosystem disclose they use Llama Guard~\cite{Erdogan2024}, and Databricks Guardrails~\cite{databricks_guardrails_2025}, respectively, to serve this component.

\subsection{Alignment with OpenAI LLM inference ecosystem}\label{sec:refarch:validation:openai}

\begin{figure}[t]
    \centering
    \includegraphics[width=\linewidth]{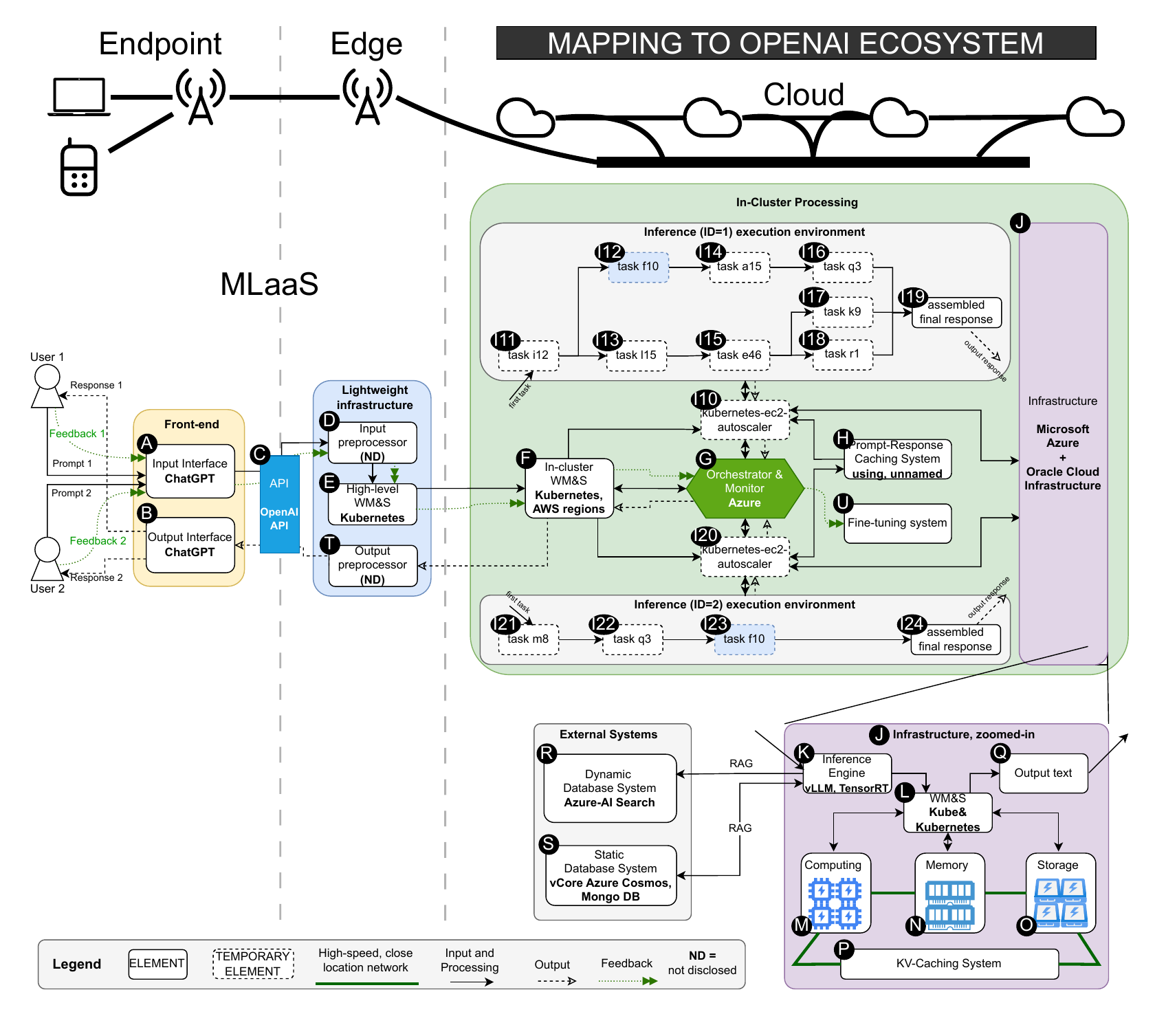}
    \caption{OpenAI LLM Inference Ecosystem mapped to the reference architecture.}
    \label{fig:refarch:validation:openai}
\end{figure}

In this section, we validate the proposed reference architecture by mapping to it the OpenAI Ecosystem for LLM inference; we present the alignment in~\Cref{fig:refarch:validation:openai}. While OpenAI doesn't explicitly present a reference architecture of its ecosystem, as of May 2025, the company discloses technologies through publicly released web articles. However, several components remain non-disclosed, represented in~\Cref{fig:refarch:validation:openai} as \textit{ND}.

\textit{Endpoint:} Users interact with the ecosystem via a front-end component with an input~\circled{A} and output~\circled{B} interface, such as the ChatGPT mobile application or website. 

\textit{Edge:} The \textit{endpoint} communicates with the \textit{edge} via the OpenAI API~\cite{openai_api}~\circled{C}. We note that OpenAI does not disclose information on the \textit{input preprocessor} component~\circled{D}. OpenAI utilizes Kubernetes~\circled{E} at the edge as a high-level workload manager and scheduler, which redirects workloads to the appropriate cloud~\cite{openai_deep_learning_infrastructure}.

\textit{Cloud:} Overall, OpenAI is highly reliant on Microsoft Azure services for system monitoring, orchestration, and management~\circled{G}, physical infrastructure~\circled{J}, and external systems~\circled{R}~\cite{microsoft_azure_openai_data}. OpenAI's Kubernetes implementation, Kubernetes-ec2-autoscaler, addresses bursty and unpredictable workloads that can scale from single-machine operations to hundreds of cores. The Kubernetes-ec2-autoscaler is a batch-optimized scaling manager and maps to~\circledd {I10} and~\circledd{I20} in our reference architecture~\cite{openai_deep_learning_infrastructure}. The resource autoscaling approach has been explored in recent literature and has proven its effectiveness in improving performance and SLO adherence; for example, Chiron is a hierarchical autoscaler for LLM-inference, which can enhance SLO attainment by 90\% and GPU efficiency by up to 70\% compared to its absence~\cite{patke2025hierarchical}.

\textit{Prompt-Response Caching:} OpenAI employs prompt caching, which is claimed to reduce latency by 80\% and cost by 50\% for long (more than 1,024 tokens) prompts~\cite{openai_prompt_caching}, \circled{H}. The system checks if the prefix of the prompt is stored in the cache, and if a matching prefix is found, the system uses the cache's result. Alternatively, the system processes the full prompt. OpenAI keeps caches active for 5-10 minutes and up to 1 hour during off-peak periods~\cite{openai_prompt_caching}. 

\textit{Infrastructure:} The infrastructure, component~\circled{J}, orchestrates computational, memory, and storage for LLM inference in the \textit{Cloud} tier. In 2016, OpenAI was mostly using \textit{``TensorFlow (or Theano) for GPU computing; for CPU, we }(note: OpenAI) \textit{use those or Numpy"}~\cite{openai_deep_learning_infrastructure}. While the exact technologies used by the inference engine are undisclosed in 2025, we argue that OpenAI follows the community standard of employing vLLM, TensorRT, or similar technologies\cite{kwon2023efficient, nvidia_tensorrt_2023, DBLP:journals/corr/abs-2404-14294, DBLP:journals/tmc/HeFYL24}. For \textit{workload management and scheduling}~\circled{L}, OpenAI uses Kubernetes customized for heavy ML workloads and scaled to managing thousands of nodes: in 2018, OpenAI was running 2,500 nodes, while in 2021, OpenAI was running 7,500 nodes~\cite{openai2021kubernetes}; exact numbers are undisclosed for 2025.

Critical to performance, OpenAI employs KV-Caching~\circled{P}, which prevents the redundant computation of the attention mechanism. Although the KV-Cache implementation remains undisclosed, OpenAI's approach proves to reduce latency by 80\% and half the costs. Similarly, the exact infrastructure of OpenAI is undisclosed in 2025; we expect major hardware, middleware, and software advances in the upcoming years as a response to the significant funding announced for OpenAI (e.g., potential \$500 billion from Stargate~\cite{openai2025stargate}, \$40 billion from Softbank~\cite{openai2025funding}).

\textit{External systems:} To access information available online (e.g., news) without retraining the model at a financially, computationally, and environmentally unsustainable granularity, OpenAI uses Azure AI Search, formerly Azure Cognitive Search, an \textit{``information retrieval system for your heterogenous content"}~\cite{microsoft_azure_openai_data}, which we map to component~\circled{R}, the \textit{Dynamic Database System}. As a \textit{Static Database System}, OpenAI leverages Azure Cosmos DB, which allows for retrieval-augmented generation capabilities and stores frequently requested information as cached responses. \circled{S} helps in reducing latency and costs by minimizing real-time web access~\circled{R} through pre-indexed content~\cite{microsoft_azure_cosmosdb_2024}.

\subsection{Alignment with IBM LLM inference ecosystem}\label{sec:refarch:validation:ibm} 

\begin{figure}[t]
    \centering
    \includegraphics[width=\linewidth]{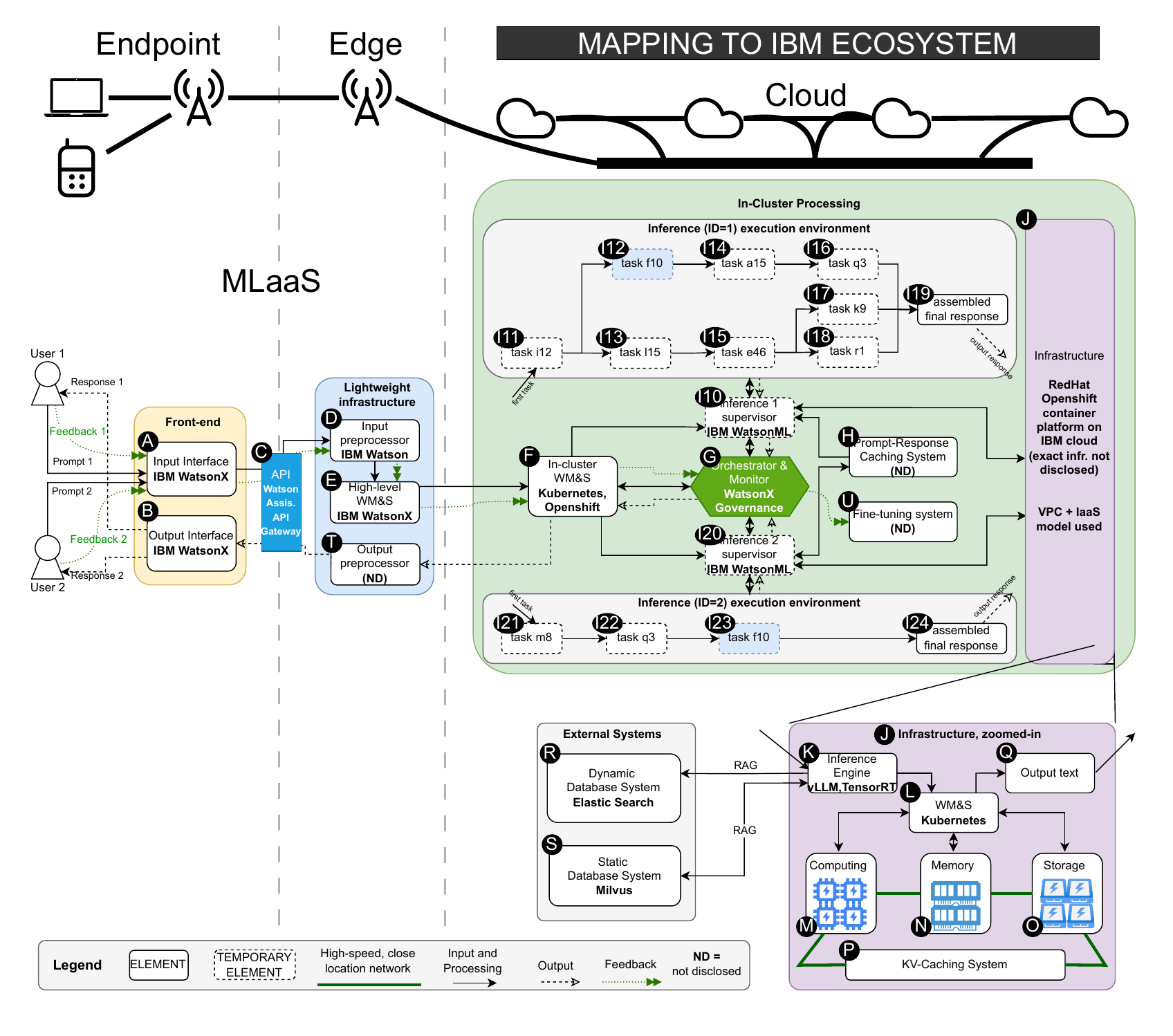}
    \caption{Reference architecture aligned with IBM LLM inference ecosystems.}
    \label{fig:refarch:validation:ibm}
\end{figure}

In this section, we validate the proposed reference architecture by mapping it to the IBM Ecosystem for LLM inference and present the alignment in~\Cref{fig:refarch:validation:ibm}. Although IBM does not exhaustively disclose technologies used in its reference architecture, as of 2025, IBM releases to the public and open science significantly more information than OpenAI. We validate the alignment of our reference architecture with OpenAI's ecosystem in \Cref{sec:refarch:validation:openai}.

IBM inference stack widely employs WatsonX, ``\textit{a portfolio of AI products that accelerates the impact of generative AI in core workflows to drive productivity}"\cite{ibm-watson}.

\textit{Endpoint:} IBM implements the endpoint component through Watson Assistant, with a simplistic and performant interface for input~\circled{A} and output~\circled{B}~\cite{ibm_watson_assistant_getting_started}.

\textit{Edge:} Following the MLaaS model, the \textit{endpoint} connects to the \textit{edge} through an API component; Watsonx API enables this functionality as a core component of the Watson stack, and maps to component~\circled{C} from the proposed~RA~\cite{ibm_watson_assistant_getting_started}. Once the user's input reaches the \textit{edge}, Watson Assistant's natural language understanding processes and filters the prompt for policy compliance, formatting, and enhancement~\circled{D}. This component, together with the \textit{High-Level WM\&S}~\circled{E}, decides to route the customer's request to \textit{``the appropriate resolution mechanism, which might be an action or a search of existing content"}~\cite{ibm_watsonx_apis}. 

\textit{Cloud:} IBM relies on RedHat's OpenShift, which provisions and manages container images, workloads, and inference processes underlying Kubernetes~\circled{F}. In direct communication with OpenShift, Watson Governance handles the overall system orchestration and monitoring for performance and cost~\circled{G}~\cite{ibm_watsonx_governance}. Component~\circled{I} maps to Watson Machine Learning services, which handle and supervise inference processes~\cite{ibm_watson_machine_learning}.

\textit{Prompt-Response Caching:} While not explicitly disclosed, we argue that, similarly to OpenAI, IBM employs a \textit{prompt-response caching system}~\circled{H}, enabling prompt caching through Watson Assistant's conversation memory and action-based storage mechanisms~\cite{ibm_watson_assistant_getting_started}. 

\textit{Infrastructure:} IBM implements infrastructure through IBM Virtual Private Cloud (VPC)~\cite{ibm_genai_rag_2024}. Watsonx.ai handles the core LLM inference process; the Inference Engine~\circled{K} follows a multi-framework approach and supports TensorFlow, PyTorch, as well as vLLM for KV-Caching~\circled{P}~\cite{ibm_genai_rag_2024}; IBM's infrastructure adopts RedHat's Openshift Container Platform for workload management and determining the \textit{``the optimal node in the cluster is for each pod to run on"}~\cite{ibm_workload_placement_2025}.

\textit{External systems:} IBM implements component \circled{R}, \textit{Dynamic Database System}, through Elasticsearch, an IBM-provided service, coupled with Watsonx Assistant, with RAG capabilities and the ability to access web resources~\cite{ibm_elastic_partnership_2024}. For component \circled{S}, \textit{Static Database System}, IBM uses Milvus within WatsonX Data to store precomputed embeddings, enable efficient similarity searches, and thus reduce real-time query load by up to 40\%~\cite{ibm_watsonx_milvus_2025, ibm_think_milvus_2025}. The external systems are linked to the \textit{Inference Engine}~\circled{K} and to the \textit{Orchestrator and Monitor}~\circled{G}.

\subsection{Alignment with the Compute Continuum}\label{sec:refarch:validation:continuum}

\begin{figure}[t]
    \centering
    \begin{minipage}[b]{0.495\textwidth}
        \centering
        \includegraphics[width=\textwidth]{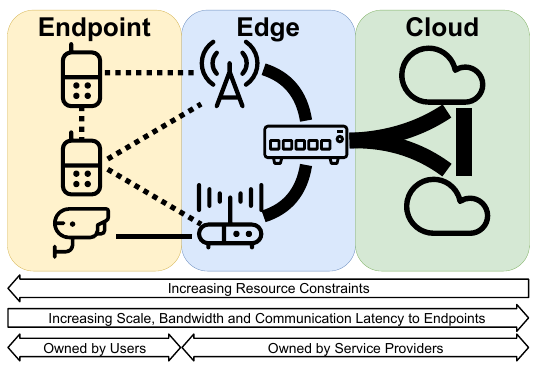}
        \caption{A high-level taken with permission from the Compute Continuum~\cite{DBLP:conf/ccgrid/JansenAPTI23}.}
        \label{fig:refarch:continuum-high-level}
    \end{minipage}
    \hfill
    \begin{minipage}[b]{0.495\textwidth}
        \centering
        \includegraphics[width=\textwidth]{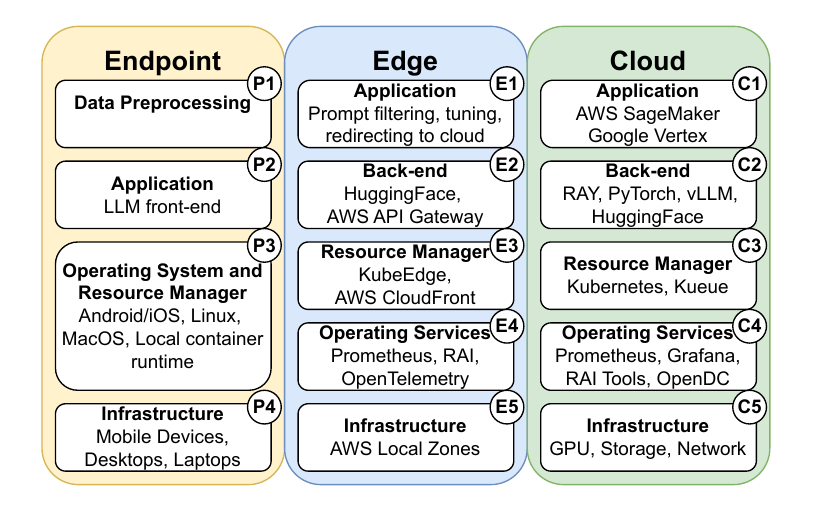}
        \caption{LLM ecosystems architecture for inference workflows aligned with the Compute Continuum.}
        \label{fig:refarch:continuum-alignment}
    \end{minipage}
\end{figure}

We align our RA with the Compute Continuum, a peer-reviewed reference architecture proposed by \textit{Jansen et al.} and a community standard. In \Cref{fig:refarch:continuum-high-level}, we present a high-level view of the continuum, comprising three main tiers: the endpoint, the edge, and the cloud. These pivotal elements align with the tiers outlined in the reference architecture we propose in~\Cref{sec:refarch:overview}. 

\textit{Endpoint:} The LLM inference process begins from the endpoint, where users give prompts via the LLM front-end \circledd{P2}. The main endpoint infrastructure is represented by regular user devices, such as mobile devices, desktops, or laptops \circledd{P4}, leveraging the operating system and resource manager of the device itself \circledd{P3}, yet the endpoint could also be accessed via APIs (e.g., OpenAI API). The endpoint infrastructure redirects the workload to the edge.

\textit{Edge:} Running \textit{large} language models \textit{at scale} on the edge becomes unmanageable when demand grows~\cite{DBLP:journals/corr/abs-2404-14294, yan2025we}; to address the growing demand of running large-scale LLMs (over 10B parameters, \cite{DBLP:journals/corr/abs-2404-14294}) at nowadays massive-scale public of users, LLM services use the edge for redirecting the workload to the most suitable datacenter capable of providing the best combination of performance, financial, sustainability metrics~\cite{DBLP:journals/corr/abs-2501-14205, DBLP:journals/tmc/HeFYL24,DBLP:journals/corr/abs-2404-14294}. The \textit{edge tier} is responsible for prompt processing at the \textit{application level} \circledd{E1}, where prompts are filtered for matching the service's policy, optionally system-enhanced (prompt tuning), and further redirected to the cloud. The \textit{back-end component} \circledd{E2}, e.g., HuggingFace, provides lightweight AI analysis and contributes to AI-powered decision-making processes (e.g., routing to datacenters, NLP for analyzing prompts). At the same time, AWS API Gateway handles secure communication with both the \textit{endpoint tier} and the \textit{cloud tier}. While Edge doesn't conduct the core inference process, it still contains \textit{resource managers} \circledd{E3} for orchestrating edge devices (e.g., KubeEdge) and for in-edge workload distribution (e.g., AWS CloudFront). \textit{Operating Services} \circledd{E4} monitor, collect performance metrics (Prometheus), ensure responsible AI governance (RAI), and collect telemetry data for troubleshooting, debugging, and application management (OpenTelemetry). 

\textit{Cloud:} Once the \textit{edge back-end} \circledd{E2} redirects the prompt to the \textit{cloud tier}, potentially in the form of a workload for execution, the \textit{application} layer \circledd{C1} enables inference processes under the available resources, using technologies such as AWS SageMaker or Google Vertex. \circledd{C3}, the \textit{resource manager}, could employ the Kubernetes-Kueue tuple, where Kubernetes orchestrates containerized workload across the infrastructure, and Kueue ensures fair workload scheduling. Similarly to the \textit{edge tier}, yet on a larger scale, \textit{operating services} (e.g., \circledd{C4}) collect (e.g., Prometheus) and visualize (e.g., Grafana) datacenter metrics and ensure responsible AI governance (e.g., RAI). We envision future research in the field of digital twinning, where state-of-the-art simulators (e.g., OpenDC) serve as key decision-making tools in a simulation-infrastructure adjustment-simulation loop. The \textit{back-end}~\circledd{C2} enables distributed computing across the infrastructure (with tools such as Ray), offer deep learning capabilities (with PyTorch as the state-of-the-art as of 2025 standards), and employs KV-Caching tools and techniques for enhanced performance (e.g., vLLM). The infrastructure is primarily composed of GPUs, typically A100/H100, storage systems, which are generally multi-layered for caching, temporary, and bulk storage purposes, and high-bandwidth networking, such as InfiniBand.

\textit{Cloud - Edge - Endpoint:} Once the final task response is obtained, the \textit{cloud tier} forwards the response to the \textit{edge tier}, which forwards to the \textit{endpoint tier}, to which the user has access via the LLM front-end. Depending on the LLM ecosystem and service, this process can happen in a single step, where the entire response is offered to the user at once (e.g., Google's Gemini)~\cite{googlegemini}, or in multiple steps, where the response is sequentially offered to the user in intermediate steps (e.g., OpenAI's ChatGPT)~\cite{openaichatgpt}.

\subsection{Multi-Ecosystem validation}

\hspace*{-2cm}
\begin{table}[t]
\scriptsize
\centering
\caption{Overview of the proposed reference architecture against real-world ecosystems for serving LLM inference. U/U = using, but the component is unnamed, N/D = not disclosed.}

\begin{adjustwidth}{-1.3cm}{-1.5cm}
\begin{tblr}{
    width = \textwidth,                 
  colspec = {c l                     
             p{3.3cm}                
             p{3.3cm}                
             p{3.3cm}                
             p{3.3cm}},              
  row{1} = {l},
  cell{2}{1} = {l},
  cell{3}{1} = {l},
  cell{4}{1} = {l},
  cell{5}{1} = {l},
  cell{6}{1} = {l},
  cell{7}{1} = {l},
  cell{8}{1} = {l},
  cell{9}{1} = {l},
  cell{10}{1} = {l},
  cell{11}{1} = {l},
  cell{12}{1} = {l},
  cell{13}{1} = {l},
  cell{14}{1} = {l},
  cell{15}{1} = {l},
  cell{16}{1} = {l},
  cell{17}{1} = {l},
  cell{18}{1} = {l},
  cell{19}{1} = {l},
  cell{20}{1} = {l},
  cell{21}{1} = {l},
  hline{1-2,6,10,14,18,22} = {-}{},
}
ID & Component Name                    & IBM Ecosystem         & OpenAI Ecosystem                      & Ubicloud                                    & Databricks                                 \\
A  & Input Interface                   & WatsonX~\cite{ibm_watson_assistant_getting_started}               & ChatGPT~\cite{openaichatgpt}             & EuroGPT~\cite{Erdogan2024}                       & {Databricks \\ Notebooks~\cite{databricks_notebooks_doc}}       \\
B  & Output Interface                  & WatsonX~\cite{ibm_watson_assistant_getting_started}               & ChatGPT~\cite{openaichatgpt}             & EuroGPT~\cite{Erdogan2024}                       & {Databricks \\ Notebooks~\cite{databricks_notebooks_doc}}       \\
C  & API                               & WatsonAssistant~\cite{ibm_watsonx_apis}       & OpenAI~API~\cite{openai_api}          & Ubicloud~API~\cite{li2025life}     & MosaicAI Serving~\cite{databricks_mosaic_serving_2025}          \\
D  & Input Preprocessor                & Watson~\cite{ibm_watsonx_apis}                & {N/D}                                    & Llama~Guard~\cite{Cubukcu2024}                         & {Databricks Guardrails~\cite{databricks_guardrails_2025}}                                          \\
E  & High-level WMS                    & WatsonX~\cite{ibm_watsonx_apis}               & Kubernetes~\cite{openai_deep_learning_infrastructure}                            & Scheduler,unnamed~\cite{Li2025}                             & {N/D}                                          \\
F  & In-cluster WMS                    & Kubernetes, Openshift                         & Kubernetes, AWS~\cite{openai_deep_learning_infrastructure}                       & Scheduler,unnamed~\cite{Li2025}                           & Kubernetes~\cite{databricks_k8s_upgrade_2024} \\
G  & {Orchestrator \\ Monitor}         & WatsonX Governance~\cite{ibm_watsonx_governance}    & Azure~\cite{microsoft_azure_openai_data}                                 & EngineCore~\cite{Li2025}                                     & {MosaicAI~\cite{databricks_mosaic_serving_2025}}                                          \\
H  & {Prompt-Response\\Caching System} & {N/D}                                            & U/U~\cite{openai_prompt_caching}                       & {N/D}                                           & {N/D}                                          \\
I  & Inference Supervisor              & WatsonML~\cite{ibm_watson_machine_learning}          & {kubernetes-ec2-\\autoscaler~\cite{openai_deep_learning_infrastructure}}                                           & AsyncLLM~\cite{li2025life}                                     & {MosaicAI~\cite{databricks_mosaic_serving_2025}}                                          \\
J  & Infrastructure                    & Openshift~\cite{ibm_workload_placement_2025}             & {Azure, Oracle Cloud\\Infrastructure}~\cite{microsoft_azure_openai_data} & Ubicloud~\cite{li2025life}                 & {Databricks (U/U)}              \\
K  & Inference Engine                  & vLLM, TensorRT~\cite{ibm_genai_rag_2024}      & {{N/D}, LLM, TensorRT\\~\cite{kwon2023efficient,nvidia_tensorrt_2023,DBLP:journals/corr/abs-2404-14294,DBLP:journals/tmc/HeFYL24}} & vLLM~\cite{li2025life}                     & {TensorRT, \\ TensorFlow~\cite{databricks_resnet_tensorrt_2025}}             \\
L  & Infrastructure WMS                & Kubernetes                                    & Kubernetes~\cite{openai2021kubernetes}                      & Cloud Hypervisor~\cite{Erdogan2024}                           & Kubernetes~\cite{databricks_k8s_upgrade_2024} \\
M  & Computing Unit                    & U/U                                & U/U                        & U/U                              & U/U                              \\
N  & Memory Racks                      & U/U                                & U/U                        & U/U                              & U/U                              \\
O  & Storage Racks                     & U/U                                & U/U                        & U/U                              & U/U                              \\
P  & KV-Caching                        & U/U                                & U/U                        & U/U                              & U/U                              \\
Q  & Output text                       & U/U                                & U/U                        & U/U                              & U/U                              \\
R  & Dynamic Database                  & Elastic Search~\cite{ibm_elastic_partnership_2024}        & Azure AI Search~\cite{microsoft_azure_openai_data}                       & Lantern~\cite{Cubukcu2024}                               & {Databricks\\ Vector Search~\cite{databricks_vector_search_2025}}                \\
S  & Static Database                   & Milvus~\cite{ibm_watsonx_milvus_2025,ibm_think_milvus_2025}          & {vCore Azure Cosmos,\\Mongo DB}~\cite{microsoft_azure_cosmosdb_2024}       & PostgreSQL~\cite{Cubukcu2024}                   & Delta Lake~\cite{databricks_docs_2025}                          \\
T  & Output Preprocessor               & {N/D}                                            & {N/D}                                    & Llama Guard~\cite{Cubukcu2024}                           & {Databricks Guardrails~\cite{databricks_guardrails_2025}}                                          \\                         
\end{tblr}
\label{table:refarchi:validation-overview}
\end{adjustwidth}
\end{table}

In this section, we present a high-level validation overview and compare our proposed reference architecture with real-world ecosystems for serving LLM inference.

\textit{OpenAI:} OpenAI is one of the largest (perhaps, the largest) LLM providers as of 2025. In~\Cref{sec:refarch:validation:openai}, we validated our proposed reference architecture against the ecosystem OpenAI uses to serve LLM inference. Although OpenAI does not disclose some of the components we include in our reference architecture, OpenAI still releases sufficient information for our validation. We identify that OpenAI mainly uses ChatGPT, Azure, Oracle, and vLLM; we also identify that components of our reference architecture closely match the components OpenAI use in their LLM inference ecosystem. 

\textit{IBM:} IBM is one of the largest LLM providers as of 2025. In \Cref{sec:refarch:validation:ibm}, we validated our proposed reference architecture against the ecosystem IBM uses to serve LLM inference; IBM releases to the public large amounts of information on how they deploy and operate LLM ecosystems, although not fully disclosing all components (e.g., prompt-response caching system). We identify that IBM mainly relies on Watson, OpenShift, and vLLM, and the components of our reference architecture closely match the components IBM discloses as used for LLM inference.

\textit{Ubicloud:} Ubicloud offers an open-source alternative to cloud providers like AWS, Azure, and Google Cloud~\cite{li2025life}. Ubicloud envisions an LLM Inference stack and publishes details of this ecosystem through their engineering blog~\cite{li2025life}. Ubicloud LLM service is EuroGPT~\cite{Erdogan2024}, which runs Meta's Llama 3.1 405B model on European infrastructure and uses Llama Guard for prompt moderation~\cite{Cubukcu2024}. In the overview of LLM ecosystems from~\Cref{table:refarchi:validation-overview}, we provide a high-level overview of LLM inference technologies from Ubicloud~\cite{li2025life}. We identify the Ubicloud ecosystem as mainly leveraging EuroGPT, Llama-3.1 405B~\cite{meta_llama31_2024}, vLLM, and Kubernetes; our reference architecture closely matches the components of the ecosystem Ubicloud proposes.

\textit{Databricks:} Databricks is one of the largest (perhaps the largest) AI and data lakehouse platforms as of 2025. Databricks offers services for serving AI, including serving LLM inference, and provides extensive information about their AI/LLM inference stack through their online documentation~\cite{databricks_resnet_tensorrt_2025, databricks_mosaic_serving_2025, databricks_guardrails_2025, databricks_vector_search_2025}. In the overview of LLM ecosystems from~\Cref{table:refarchi:validation-overview}, we align the components from our reference architecture with real-world components from the Databricks ecosystems. We identify that Databricks serves LLM inference mainly through MosaicAI, a \textit{``platform for building, evaluating, deploying, and monitoring generative AI applications~(gen~AI~apps)"}~\cite{databricks_mosaic_capabilities_2025}, coupled with NVIDIA's TensorRT and TensorFlow for serving inference~\cite{databricks_resnet_tensorrt_2025}. Moreover, Databricks uses their in-house build system for input/output filtering, Databricks Guardrails~\cite{databricks_guardrails_2025}, and Databricks Vector Search and Delta Lake for dynamic and static database systems~\cite{databricks_vector_search_2025, databricks_docs_2025}. Overall, we identify that the components from our reference architecture closely match (abstractise) the real-world components from the Databricks ecosystem.

\section{Requirement Validation}\label{sec:refarch:requirements-adressal}

In this subsection, we present how the proposed reference architecture addresses each design requirement and principle detailed in \Cref{sec:refarch:requirements}.

\begin{enumerate}[label=(\textbf{DR\arabic*}), leftmargin=*, align=left, labelwidth=1cm]
    \item \textbf{Ensure the \textit{validity} of the reference architecture.} \\
    {We regard our reference architecture as situated in the middle of the spectrum, "low-high-level.", as a mid-level conceptual model.

    Firstly, we align with \textit{two mid-low-level} LLM ecosystems, used by the largest LLM providers of 2025, OpenAI and IBM. We validate our proposed reference architecture by mapping each component to elements of the real-world equivalent ecosystems. We identify that some components may not be disclosed by one provider, but are disclosed by the other provider (e.g., the prompt-response caching system). Similarly, we identify that some components are not recognized or distinguished as components. Still, the released documentation acknowledges the existence of such functionality and system (e.g., the input preprocessor from the edge).

    Secondly, we align our reference architecture with a \textit{high-level conceptual model} of the ICT Compute Continuum, a scientific, peer-reviewed, and community-standard model for conceptualizing the ICT field. 

    Thus, we align our reference architecture across three gradual steps on the low-to-high-level spectrum, encompassing both the industry and academic worlds. We, therefore, argue that we have proven the validity of our proposed reference architecture. }

    \item \textbf{Ensure the \textit{usefulness} of the reference architecture.} \\
    {We identify and address the design requirement of \textit{usefulness}, defined and subjectively quantified by the \textit{``real-world purpose which motivates its }(note: the reference architecture's) \textit{creation."} We motivate the need for a reference architecture which would aid various groups of stakeholders, especially decision-making LLM/infrastructure operators, researchers, and students, in offering a high-level picture of the ecosystem, with high-level components which otherwise could be omitted or misinterpreted even by experienced groups~\cite{2023-icpewip-scheduler-apis}. }
\end{enumerate}

\begin{enumerate}[label=(\textbf{DP\arabic*}), leftmargin=*, align=left, labelwidth=1cm]
    \item \textbf{Design components with clear distinct responsibilities.} \\
    {We identify and address the design principle of clearly distinguishing components with distinct responsibilities and, thus, adhering to the community standards for defining reference architectures. We distinguish components by two layers of abstraction.
    From a high-level perspective, we identify three main distinct large-components (\textit{tiers}): the \textit{front-end}, the \textit{lightweight infrastructure}, and the \textit{datacenter(s)} (heavyweight, massive-scale infrastructure). In other words, we identify the \textit{endpoint}, the \textit{edge}, and the \textit{cloud}.
    From a lower-level perspective, yet still high enough to offer abstraction, we identify components for each tier, each with its own specific responsibility and scope. We detail in \Cref{sec:refarch:high_level} how components interact and make the inference process of the LLM ecosystem tick. }

    \item \textbf{Group related components.} \\
    {We identify and address the design principle of grouping related components according to their responsibility. We identify the main tiers, namely (i) front-end, (ii) lightweight infrastructure, and (iii) the datacenter, and identify the main components for each tier.
    (i) For the \textit{front-end tier}, we identify the input and output interface, with which the user interacts and we, thus, group together. 
    (ii) In the \textit{lightweight infrastructure}, we identify components for preprocessing prompts, conducting workload management and scheduling, and outputting the response to the user; we identify a main common responsibility for each element in the lightweight infrastructure (edge) of being a middleware between the endpoint and the cloud. 
    (iii) For the tier of \textit{in-cluster processing} (i.e., cloud, datacenter), we group elements directly contributing to inference management, execution, and supervision. Furthermore, for each inference process, we group the inference supervisor with the corresponding set of (sub)tasks it executes (e.g., \circledd{I10} supervisor grouped with \circledd{I11}-\circledd{I15}).}

    \item \textbf{Aim for extensibility and modularity.}\\
    {We identify the design principle of extensibility and modularity by designing the RA in accordance with community-vetted standards for RAs~\cite{DBLP:conf/ccgrid/JansenAPTI23, DBLP:conf/sc/AndreadisVMI18}, adapted from software architecture practices~\cite{rozanski2012software}. Each component of the architecture contains a degree of abstraction sufficient to both understand the purpose of a specific component (e.g., infrastructure, orchestrator and monitor) and to be replaced or extended. For example, component \circled{G}, the \textit{Prompt-Response Caching System}, allows future stakeholders to extend, detail, or even skip entirely, based on individual needs and visions on the LLM ecosystem. This design principle is crucial for the lifespan and adoption of the proposed reference architecture.}

    \item \textbf{Separate mechanisms from policies and goals.}\\
    {We identify and address the design principle of distinguishing between mechanisms, policies, and metrics. Corroborated with \ref{DP:distinct-components}, we design each component as single-purpose, and separate mechanism elements, answering \textit{how it operates}, (e.g., \textit{input interface}, \textit{prompt-response caching system}), from policy elements \textit{how decisions are made} (e.g., \textit{in-cluster WM\&S}), from goals/metrics elements (e.g., \textit{orchestrator and monitor}). Consequently, systems such as the \textit{infrastructure} or \textit{prompt-response caching system} contain both mechanisms and policies, yet without diminishing the validity and adherence to \ref{DP:mechanism-policy} or \ref{DP:distinct-components}.}

    \item \textbf{Cover end-to-end prompt-to-response LLM workflow.}\\
    {We identify and address the design principle of modeling the complete feedback loop, from the user input to the system output. Specifically, this is represented by an outgoing arrow from the user to the \textit{input interface}, and an incoming arrow from the \textit{output interface} to the user. In \Cref{sec:refarch:high_level:prompt}, we detail the execution phase, spanning the workflow from the \textit{user's prompt} to the \textit{assembled final response}, and in \Cref{sec:refarch:high_level:response}, we detail the response phase, spanning the workflow from \textit{the assembled final response} to its delivery on the \textit{output interface}.}

    \item \textbf{Support multiple users in the ecosystem.}\\
    {We identify and address the LLM ecosystem-specific design principle of designing a reference architecture which accommodates multiple users interacting simultaneously with the ecosystem. For better comprehension purposes, we visually represent only two users, yet present and detail the inference process for a many-user interaction (\Cref{sec:refarch:overview}). We emphasize the role of each component responsible for the multi-user operability, especially the role of the \textit{high-level WM\&S} from the lightweight infrastructure, the \textit{in-cluster WM\&S}, the \textit{prompt-response caching system}, and the \textit{orchestrator and monitor}.}

    \item \textbf{Model components responsible for decision processes across the system.}\\
    {We identify and address the LLM ecosystem-specific design principle of employing components responsible for decision-making processes, such as the workload manager and schedules \circled{D}, \circled{E}, the \textit{orchestrator and monitor} \circled{F}, central to the cloud infrastructure, the and the \textit{inference supervisors} \circledd{I10}, \circledd{I20}. These decision-making components span both \textit{lightweight infrastructure} and \textit{datacenter} tiers, responsible for management and supervision across different layers of the ecosystem.
    }

    \item \textbf{Model different types of prompts execution workflow.}\\
    {
    We identify and address the design principle of modelling flexible prompts by presenting two different inference execution processes, one of which contains branching prompts that ultimately aggregate in a final response, and one which contains no branching, but only reasoning. For simplicity and visual comprehension, we model only two examples, yet still covering the distinct execution behaviour inference processes have.. 
    }
    
\end{enumerate}

\section{Discussion}\label{sec:refarch:discussion}

We summarize contributions of this chapter, envision future research, and discuss potential threats to validity.

\textit{Summary:} In this chapter, we propose and validate the first comprehensive reference architecture for LLM ecosystems under inference, following the vetted AtLarge Design Process~\cite{DBLP:conf/icdcs/IosupVTETBFMT19}, and addressing RQ1. We validate this reference architecture by aligning with the Compute Continuum~\cite{DBLP:conf/ccgrid/JansenAPTI23}, with a state-of-the-art ecosystem from IBM, and with an LLM inference ecosystem from OpenAI. The proposed reference architecture, combined with a design focused on long lifespan, enables LLM operators, researchers, and students to gain a better understanding of the ecosystem and, where applicable, make more informed decisions.

\textit{Future Research:} We envision future research in modeling LLM ecosystems. LLM training is a resource-very-hungry~\cite{DBLP:journals/cacm/Chien23a, achiam2023gpt}, computationally intensive~\cite{achiam2023gpt, xu2024hethub, DBLP:journals/corr/abs-2405-21015}, financially expensive~\cite{DBLP:journals/corr/abs-2405-21015, shoham2024longcontext}, and sustainability-concerning~\cite{DBLP:conf/mlsys/KorthikantiCLMA23, Agrawal2025EnergyEL, DBLP:journals/corr/abs-2408-07326}; similarly, to the inference process, currently there is no comprehensive reference architecture of LLM ecosystems under the training phase. We also envision future work in modeling and simulating the inference process with different types of prompts, e.g., prompts with no intermediate tasks, deep research prompts~\cite{openai2025deepresearch}, and prompts with intermediate tasks (reasoning).

\textit{Threats to Validity:} The reference architecture proposed in this chapter has been designed in accordance with a well-established set of design requirements and principles, utilizing state-of-the-art design and validation methodologies, and leveraging resources and knowledge from the open-source and open-science communities.

While comprehensive and universal, we cannot guarantee full-spectrum validity regarding alignment with existing closed-source ecosystems. However, we expect those ecosystems to follow a similar (if not identical) reference architecture. Offering an analogy from physics, we can completely validate or invalidate only theories and principles applicable to the known universe; there is no theory or principle of physics for which we can guarantee it holds in the unknown universe. Similarly, there is no reference architecture for which one can guarantee its validity in the \textit{``unknown universe."}

 \newpage
\thispagestyle{noheader}
\chapter{Design of Kavier: a tool for simulating LLM inference and KV-Caching}\label{sec:design}

LLM ecosystems are becoming increasingly large, distributed, and heterogeneous, and raise performance, sustainability, and efficiency concerns~\cite{vaswani2017attention, efficientnlp2023kv, ramponi2024llm, DBLP:journals/cacm/Chien23a}. It is crucial to understand how LLM ecosystems, and the (eco)systems orchestrated by LLM ecosystems, operate and behave at scale. Addressing this concern, in \Cref{sec:refarch}, we proposed the first comprehensive reference architectures of LLM ecosystems under inference, which provide a vital conceptual overview of the LLM continuum. We envision simulation as a natural next step for systematically anticipating how LLM ecosystems would behave under different workloads and configurations; simulation enables experimentation and prediction of performance, sustainability, and efficiency in a time and cost-efficient way~\cite{DBLP:conf/ccgrid/MastenbroekAJLB21, DBLP:journals/corr/abs-2206-03259}.

Designing a simulator capable of cache-awarely predicting the performance, sustainability, and efficiency of LLM ecosystems under inference, using discrete-event simulation, is a critical yet non-trivial scientific challenge. Currently, there is no such scientific instrument. This raises the research question: \textit{(RQ2) How to design Kavier, a scientific instrument for cache-aware simulation analysis of the
performance, sustainability, and efficiency of LLM ecosystems under inference?}

In this chapter, we address the RQ2 by designing Kavier, a first-of-its-kind scientific instrument for cache-aware simulation of the performance, sustainability, and efficiency of LLM ecosystems under inference. In \Cref{sec:prototype}, we leverage this design to create an implemented prototype of the simulator, further integrated with a top-tier, community-vetted, and peer-reviewed data center simulator. In \Cref{sec:evaluation}, we validate our design through real-world, trace-based experimentation.

\section{Overview} \label{sec:design:overview}
We design Kavier matching the state-of-the-art AtLarge design process of designing computer systems and ecosystems~\cite{DBLP:conf/icdcs/IosupVTETBFMT19}. Our contribution in this chapter is seven-fold:

\begin{enumerate}
    \item We define and establish functional and non-functional requirements for Kavier in \Cref{sec:design:requirements-establish}.
    \item We propose a high-level design for the architecture of Kavier in \Cref{sec:design:kavier:high-level}.
    \item We present models Kavier uses to simulate the performance of LLM inference in \Cref{sec:design:performance}. These models simulate LLMs under various caching policies, both for KV-Caching and prefix matching.
    \item We present models Kavier uses for predicting sustainability in \Cref{sec:design:sustainability}.
    \item We present models Kavier uses for computing efficiency metrics, namely performance-cost and sustainability-cost in \Cref{sec:design:efficiency_simulation}.
    \item We address each functional and non-functional requirement in \Cref{sec:design:requirement-addressal}. 
    \item Lastly, we reflect on our design and envision future work in designing simulation instruments for LLM ecosystems \Cref{sec:design:discussion}.
\end{enumerate}

\section{Requirements Analysis}\label{sec:design:requirements-establish}
In this section, we establish a set of functional requirements (FRs) and non-functional requirements (NFRs) that guide the design process of Kavier, a tool for simulating LLM ecosystems under inference, with a focus on the caching component. This matches stage 1 of the AtLarge Design Process on Distributed Systems and Ecosystems~\cite{DBLP:conf/icdcs/IosupVTETBFMT19}.


\textbf{Main Functional Requirement (MFR):} Simulate performance, sustainability, and efficiency of LLM ecosystems under inference.

\subsection{Functional Requirements}
We identify a set of six functional requirements which guide our design process and tell \textit{``what the system should be able to do"}~\cite{book-distributed-systems}.

\begin{enumerate}[label=(\textbf{FR\arabic*}), leftmargin=*, align=left, labelwidth=1cm]
    \item \textbf{Support holistic simulation of the LLM inference process.} \label{fr:discrete} \label{sec:design:fr1}\\
    The simulator should model the entire LLM inference process executed in the cloud tier, both the \textit{prefill} and the \textit{decode} stage. Kavier should support splitting the inference process between these two stages and tailor it accordingly to the different performance characteristics of each stage. Furthermore, Kavier should be a discrete-event simulator with a user-configurable prediction granularity. Without \ref{fr:discrete}, Kavier would omit the distinct behaviours of prefill and decode and, thus, cannot accurately simulate the performance of LLM workloads.
    
    \item \textbf{Simulate with cache awareness.}\label{fr:caching} \label{sec:design:fr2}\\
    Kavier, as a cache-oriented simulator for LLM inference, should support the simulation of the key-value caching (KV-Caching) mechanism used in transformer models. Kavier should allow for enabling or disabling KV-Caching for the simulation scenario, thus enabling comprehensive modelling of real-world LLM processes and facilitating versatile experimentation. Without \ref{sec:design:fr2}, the tool cannot explore impacts of caching policies on performance \ref{sec:design:fr3}, sustainability \ref{sec:design:fr4}, and efficiency \ref{sec:design:fr5}.

    \item \textbf{Predict the performance of the LLM ecosystem under workload.} \label{fr:performance} \label{sec:design:fr3}\\
    The simulator should predict ecosystem performance, specifically, predict latency and throughput. We identify latency as the amount of time required to answer a prompt; we identify throughput as the number of tokens that can be executed per second. We consider a sequential execution of prompts and identify prompt parallelisation as an area of future research in the simulation of LLM inference. Results should be recorded in a structured trace format, both as a task-based trace, containing cumulated trace details, and as a fragment-based trace, detailing snapshots of each task, snapshots taken at a user-established granularity. Without \ref{sec:design:fr3}, Kavier would not provide insight into performance metrics, thus limiting a further accurate evaluation of sustainability and efficiency metrics.

    \item \textbf{Predict the sustainability of LLM ecosystems under workload.}  \label{fr:sustainability} \label{sec:design:fr4}\\
    The simulator should predict the ecosystem's sustainability, using models for estimating the energy consumption of the GPU infrastructure and the resulting CO2 emissions for the simulated workload. Kavier, coupled with a peer-reviewed simulator, should predict power usage over time, following the user-established granularity, and the total energy consumption run by a batch of LLM inference workloads. Addressing the increasingly concerning CO2 emissions for large-scale and massive infrastructure, Kavier should predict the carbon footprint of LLM inference workloads, using real-world CO2 traces. Without \ref{sec:design:fr4}, the simulator cannot assess the sustainability impact of LLM inference, a pivotal concern in a digital world with increasingly overexploited resources.

    \item \textbf{Predict the efficiency of LLM ecosystems under workload.} \label{fr:efficiency} \label{sec:design:fr5}\\
    The simulator should predict efficiency metrics of the ecosystem, using a configurable financial model and the performance and sustainability metrics expanded in \ref{sec:design:fr3} and \ref{sec:design:fr4}, respectively. Kavier should allow for performance-cost metrics, estimating financial cost per token per second. Kavier should also allow for performance-sustainability metrics, estimating sustainability cost per token per second. \ref{sec:design:fr5} thus enables a clear and direct comparison between ecosystems, aiding stakeholders in making informed decisions when deploying, maintaining, and expanding LLM ecosystems.

    \item \textbf{Design Kavier compatible with other simulators and extensible.} \label{fr:modularity} \label{sec:design:fr6}\\
    The simulator should allow simple integration with a peer-reviewed datacenter simulation framework, and be designed following modularity principles. Kavier's output should align with the datacenter simulator's input formats (e.g., input traces, experiment setup). Kavier should also be designed as modular, and further engineered strictly following this design approach, thus aligning with state-of-the-art software architecture and design principles~\cite{rozanski2012software}; this functional requirement is crucial for ensuring long software life and allowing for adding future functionality. Without \ref{sec:design:fr6}, Kavier would have limited usefulness as a universally applicable simulator, hindering its adoption and evolution as part of a datacenter simulator.
\end{enumerate}

\subsection{Non-Functional Requirements}
In addition to the set of functional requirements aforementioned, we identify four non-functional requirements, which guide the design and engineering process of Kavier, and tell us \textit{``how well the features should work"}~\cite{book-distributed-systems}. We address non-functional requirements at implementation and integration time, in ~\Cref{sec:prototype:requirements_addressal}.

\begin{enumerate}[label=(\textbf{NFR\arabic*}), leftmargin=*, align=left, labelwidth=1cm]
    \item \textbf{Provide in-meeting, near-interactive, same-day simulation results.} \label{nfr:speed} \label{sec:design:nfr1} \\
    Cloud infrastructure currently operates at an unprecedented scale~\cite{Mastenbroek2023RADICE, radunicolae-hp-m3sa}. The system should run efficiently, output the simulation results promptly, and support predictions of very large-scale batches of tasks. Simulating system performance should take less than 1\% of the actual run of the experiment, for prompts with prefill and decode times larger than 10 seconds cumulatively. For example, if a batch of 1,000 prompts, each 10 seconds long, would take 10,000 seconds in total, Kavier should offer predictions in a matter of 1-2 minutes. Similarly, we identify the requirement of selecting a fast and efficient, peer-reviewed datacenter simulator that can predict system sustainability rapidly. However, although relevant for the speed of the overall system's performance, optimizing external simulators (e.g., the datacenter simulator) is beyond the scope of this paper. Without \ref{sec:design:nfr1}, the Kavier cannot be reasonably used in interactive settings or for large-scale batches of LLM workloads.

    \item \textbf{Aim to provide adequate simulation accuracy.}\label{nfr:accuracy} \label{sec:design:nfr2}\\
    The predictions produced by Kavier should be on par with reality and within a Mean Absolute Error Ratio (MAPE) margin of 10\%. MAPE penalizes overestimates and underestimates equally throughout a series of predictions, making it suitable for quantifying the accuracy of discrete-event simulations. The timing predictions should be calibrated against empirical data traced from real-world systems. Without \ref{sec:design:nfr2}, the instrument would give unreliable insights. We later validate Kavier's accuracy of prefill and decode time, through trace-based experiments in \Cref{sec:evaluation}.

    \item \textbf{Facilitate reproducibility and open science.}\label{nfr:reproducibility} \label{sec:design:nfr3}\\
    The results produced by Kavier should be fully reproducible, and Kavier should be built and released in accordance with open-science principles. The code, configuration, and experiment traces should be made available, thus adhering to principles of open source and open science. Simulation involving randomness should be controllable via seeds to ensure perfect reproducibility of the experiment. Kavier should be released with rigorous documentation and tutorials for usage. We regard \ref{sec:design:nfr3} as a critical requirement for Kavier to be considered a real and valuable contribution to the scientific community.
    
    \item \textbf{Adhere to modern software design and development standards.} \label{nfr:software-art} \label{sec:design:nfr4} \\
    The simulator's codebase must be maintainable and adaptable for future changes in LLM systems. The codebase should contain clean code and adherence to software engineering best practices, such as modularity, clarity, and tests. The system should not only be to this work but should integrate with a peer-reviewed datacenter simulator, evolve, and adapt to future engineering. Without \ref{sec:design:nfr4}, Kavier's future development and maintenance would be unsustainable in the long run.
\end{enumerate}

By meeting the above functional and non-functional requirements, Kavier would serve as a KV-Cache-aware LLM inference simulator, providing accuracy, efficiency, speed, and utility for various stakeholder groups, with a long software lifecycle and simplicity in expansion by future contributors.

\section{Overview of Kavier} \label{sec:design:kavier:high-level}

In this section, we present a high-level overview of Kavier as coupled with OpenDC, a state-of-the-art, peer-reviewed simulator. The holistic simulation infrastructure follows a discrete-event simulation model~\ref{fr:discrete}, and predicts performance~\ref{fr:performance}, sustainability~\ref{fr:sustainability}, and efficiency~\ref{fr:efficiency} of both small and massive-scale batches of LLM inference workloads. Kavier integrates with OpenDC, an open-source, peer-reviewed, and state-of-the-art simulation framework for datacenters, with over 8 years of development, operation, and constant contributions to the scientific community~\cite{DBLP:conf/ccgrid/MastenbroekAJLB21, DBLP:conf/ispdc/IosupABBENOTVV17, nicolae2025m3sa, dniewenhuis_hotcloud_footprinter, Mastenbroek2023RADICE, DBLP:journals/tpds/AndreadisMBI22}. Following the AtLarge Design Process~\cite{DBLP:conf/icdcs/IosupVTETBFMT19}, we design, implement, and validate Kavier iteratively; this process begins with bootstrapping the creative process (stage 3), then focusing on the high-level and low-level design (stage 4)~\cite{Mastenbroek2023RADICE, DBLP:conf/icdcs/IosupVTETBFMT19}.

\begin{figure}[t]
    \centering
    \includegraphics[width=0.95\linewidth]{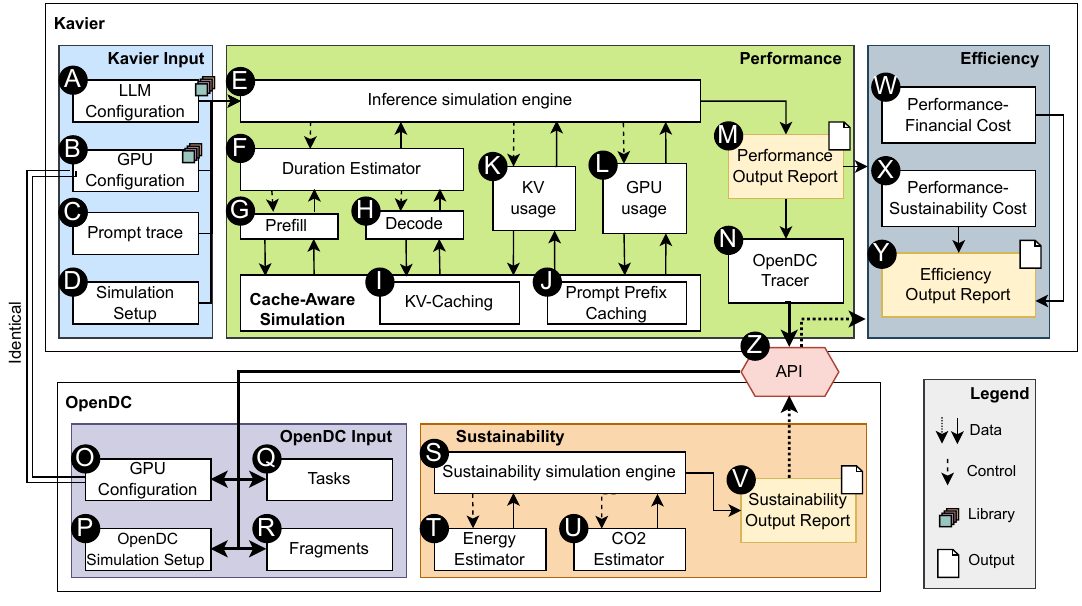}
    \caption{Overview of the high-level architecture of Kavier and OpenDC.}
    \label{fig:kavier-high-level-overview}
\end{figure}

\subsection{Design Choices} \label{sec:design:choices}
In this section, we analyze design choices of the high-level architecture of Kavier. We identify three main classes of analysis: type of simulation (e.g., discrete-event, continuous, or total), integration with other simulation tools, and simulation pipeline. We now discuss each alternative.

\textit{(DC1) Discrete-event simulation:} We identify three main simulation models: discrete-event simulation, continous-simulation, and total simulation. We identify discrete-event simulation as the most suitable for addressing the \textbf{MFR}, as this simulation model enables both post- and during-prediction analysis, unlike total simulation, which offers only overall results without providing insights into how the ecosystem evolves~\cite{DBLP:conf/ccgrid/MastenbroekAJLB21, banks2005discrete}. We also identify continuous simulation model which, however, does not align with LLM inference operational model, which involves discrete-events (e.g., token generation, cache hits/misses) occuring at a specific granularity. In contrast, discrete-event simulation matches LLM inference by matching the operational model of LLM inference, and simulating at a user-established granularity.

\textit{(DC2) Integration with other simulators:} We identify three main simulation scopes (later modules) for Kavier: performance \ref{fr:performance}, sustainability \ref{fr:sustainability}, and efficiency \ref{fr:efficiency}. For each scope, we identify two design choices: we can either leverage existing, peer-reviewed work or design and implement from scratch. We analyzed, per scope, peer-reviewed literature and identified instruments for predicting the sustainability of (LLM) Ecosystems under workload, but no instrument capable of predicting performance or efficiency of LLM ecosystems, both cache-awarely and discrete-event. We thus choose to leverage a simulator with peer-reviewed capabilities for predicting sustainability, instead of building our own sustainability simulation module. We argue that, although adapting Kavier to an external simulator increases the engineering complexity compared to creating an in-Kavier module dedicated to simulating sustainability, adapting and using peer-reviewed functionality is more important. We thus choose to design and engineer only the modules that have never been explored before by our community (i.e., performance and efficiency), and leverage the peer-reviewed capabilities of a simulator for predicting sustainability.

\textit{(DC3) Simulation pipeline:} We identify the MFR of predicting performance, sustainability, and efficiency of LLM ecosystems under inference. We identify four main pipeline architectures: 1) sequential pipeline (first performance simulation, then sustainability simulation, and lastly efficiency calculation), 2) parallel pipeline, where all modules run simultaneously with a final aggregation, 3) integrated pipeline, with a single monolithic simulator handling all the simulation and calculation aspects, and 4) hierarchical pipeline, adopting a multi-level simulation with different granularities. We identify a hybrid pipeline between the aforementioned pipelines, basing on a sequential design, where firstly the simulation system predicts performance~\ref{fr:performance}, then sustainability~\ref{fr:sustainability}, then efficiency~\ref{fr:efficiency}. This pipeline enables per-module validation, which also allows for individual module adoption (thus, for leveraging the sustainability module from a peer-reviewed simulator, DC2). This pipeline, unlike the others, allows for a human-in-the-loop setup, who can manual verify and analyze between stages, orchestrate stages based on their needs, and make adjustments. Moreover, this architecture allows for failure tolerance, if a module fails (e.g., financial efficiency is functional without sustainability predictions, performance simulation is functional without sustainability simulation) \ref{fr:modularity} and simplified debugging. Lastly, this pipeline maximized modularity and validation rigor, while adhering to principles of software design and architecture~\cite{rozanski2012software, softwareDesignPhilosophy}, and addressing \ref{sec:design:nfr4}. 

Throughout the design process, also matching the community-standard methodology on designing computer systems and ecosystems~\cite{DBLP:conf/icdcs/IosupVTETBFMT19}, we identify and analyze multiple design choices and select the option (usually, the tradeoff) that best aligns with the established requirements. For example, in \Cref{sec:design:caching}, we analyze various designs of a system able to simulate performance with cache-awareness. In \Cref{sec:design:performance}, we analyze two main design choices in simulating GPU performance, and compare simulation leveraging empirical-measurements and simulation leveraging mathematical and statistical approaches.

\subsection{Kavier Input} \label{sec:kavier:overview:kavier-input}
The Kavier process begins in the input stage, where Kavier receives the experiment setup. 
Through \textit{LLM Configuration}~\circled{A}, the user can either select a prefab from the LLM Library or build their own by offering to the system the configuration parameters.
Similarly, through \textit{GPU Configuration}~\circled{B}, the user can either select a prefab from the GPU Library or build their own. 
The \textit{Prompt trace} contains two mandatory columns, the amount of input tokens and the amount of output tokens, and two optional columns, the tokenized input and the tokenized output; while the latter columns are optional for the overall simulation process, their presence allows for simulating with \textit{Prompt Prefix Caching} policies~\circled{J}. 
The \textit{Simulation Setup}~\circled{D} allows the user to configure and customize the simulation based on their own needs and preferences, and provides configurable options such as snapshot granularity, simulation models, and output preferences.

\subsection{Performance Simulator} \label{sec:kavier:overview:performance}
Once the simulation setup is finalized, the simulation process begins. 
The \textit{Performance Simulation Engine}~\circled{E} orchestrates and manages simulation processes which predict throughput and latency metrics~\ref{fr:performance}. 
Firstly, the simulator predicts the duration of the prefill and decode stage, following a cache-aware simulation approach~\circled{I},\circled{J}~\ref{fr:caching}. We identify caching as a central component of Kavier, with major and various impacts on the simulation time, highly dependent on the simulation policy. 

Secondly, once Kavier predicted the amount of time per prefill and decode stage, it breaks the simulation time in simulation snapshots, based on the simulation granularity the user selected~\ref{fr:discrete}, following the formula $N_{i} = \lceil (T_{p} + T_{d}) / T_{i} \rceil $, where $N_{i}$ is the number of intervals at which snapshots occur, $T_{p}$ is the prefill time, $T_{d}$ is the decode time, and $T_{i}$ is the user-selected granularity at which the snapshoting occurs. For example, if the prefill time is 1.1, decode time is 9.0 seconds, and user selected a snapshotting interval is of 1 second, this would result in a total of 11 snapshots ($\lceil1.1 + 9.0\rceil \ 10 = \lceil10.1 / 1\rceil = \lceil10.1\rceil = 11$ snapshots). 

Thirdly, for each monitoring snapshot, Kavier simulates \textit{KV usage}~\circled{K} and \textit{GPU usage}~\circled{L}, following trace-based simulation models which allow for versatile, stage-specific predictions~\ref{sec:design:fr1}, thus capturing the specific and distinct behaviours of prefill stage and decode stage. Both components \circled{K} and \circled{L} are linked to the caching system, which naturally influences the usage of the infrastructure based on the presence or absence of caching, or based on the caching policy~\ref{fr:caching}.

Lastly, once the \textit{Inference Simulation Engine}~\circled{E} finishes the simulation process, it transfers data to component~\circled{M}, where a \textit{Performance Report} is generated. This report contains Kavier's predictions on inference latency per prompt and system throughput~\ref{fr:performance}. This data is ultimately transferred to the efficiency module ad is essential to compute the \textit{Performance-Financial Cost}~\circled{W}, and to ~\circled{N}, where Kavier's predictions are adjusted and made compatible with OpenDC input requirements. This results in a Kavier-output, OpenDC-input file. The transfer between Kavier and OpenDC happens through an internal \textit{API} between the systems~\circled{Z}.

\subsection{OpenDC Input} \label{sec:kavier:overview:opendc-input}
OpenDC input consists of specifications of hardware infrastructure, simulation setups, and workload traces. The \textit{GPU configuration} from \circled{B} coincides with the configuration from \circled{O}; this configuration can either be manually set up by the user or can be selected from a list of prefabs. Component~\circled{P} represents the simulation setups of OpenDC, partially coinciding yet not exhaustively with the simulation setup of Kavier. Component~\circled{Q} is the workload trace OpenDC uses to predict energy consumption, while component~\circled{R} is the CO2 trace OpenDC uses to predict the amount of CO2 emitted for running the batch of inference tasks in a real-world environment. All the inputs are forwarded to the \textit{Sustainability simulation engine}~\circled{S}.

\subsection{Sustainability Simulator} \label{sec:kavier:overview:sustainability}
Following a similar approach of discrete-event simulation, OpenDC's simulation process is orchestrated by a \textit{Sustainability simulation engine}~\circled{S}. This firstly predicts the amount of energy the GPU infrastructure would consume in a real-life setup, through the \textit{Energy Estimator}~\circled{T}~\ref{fr:sustainability}. 

Once energy predictions are completed, the \textit{Sustainability simulation engine} redirects results for a \textit{CO2 estimator}~\circled{U}, a tool which leverages the given CO2 trace and the predicted amount of energy consumption and predicts the amount of CO2 consumed at every timestamp, following the granularity selected by the user~\ref{fr:sustainability},~\ref{fr:discrete}. The CO2 estimator component of OpenDC is proposed and detailed in depth by \textit{Niewenhuis et al.} in \cite{dniewenhuis_hotcloud_footprinter}. 

Lastly, the sustainability predictions are aggregated into a \textit{Sustainability Report}~\circled{V}~\ref{fr:sustainability}. The sustainability report is transferred through the API interface~\circled{Z} to Kavier, into the efficiency module, and is essential in computing \textit{Performance-Sustainability Cost}~\circled{X}.

\subsection{Efficiency Simulator} \label{sec:kavier:overview:efficiency}
Addressing~\ref{fr:efficiency}, the efficiency component computes \textit{Performance-Financial Cost}~\circled{W}, represented in price per token per second, and the \textit{Performance-Sustainability Cost}~\circled{X}, expressed in watts per token per second. The \textit{Performance-Financial Cost}, component~\circled{W}, simulates economic efficiency by combining a predefined, yet simple to modify and expand~\ref{fr:modularity}, financial model with the Kavier-simulated performance. Specifically, the simulator computes the cost of serving LLM inference under given hardware price, amortized over the number of tokens generated per second. Similarly, the \textit{Performance-Sustainability Cost}, component~\circled{X}, quantifies environmental costs. Environmental efficiency by combining power usage with token throughput. We identify power usage as a more robust metric than CO2 emissions, as CO2 emissions are location-dependent (different locations, especially countries, emit varying amounts of CO2 for energy production), whereas power usage is location-independent. We expand both efficiency models in \Cref{sec:design:efficiency_simulation}.

\section[Kavier Components for Simulating Key-Value and Prompt-Prefix Caching]{Kavier Components for Simulating Key-Value and \\ Prompt-Prefix Caching}\label{sec:design:caching}

In this section, we expand the caching abilities of Kavier, able to simulate key-value caching mechanisms in LLM inference~\ref{fr:caching}. KV-Caching is a technique widely used in large-scale LLM deployments, which improves performance by reducing redundancy computation during the autoregressive decode phase~\cite{ kwon2023efficient, vaswani2017attention, efficientnlp2023kv, huggingface-kv-cache}; we expand KV-Caching simulation in \Cref{sec:design:caching:kv-caching}. We also model prompt prefix caching, a technique which stores previously computed results in a caching system for a given period, and, if prompted again within the period, the system retrieves from memory instead of recomputing; we expand this caching technique in \Cref{sec:design:caching:prompt}.

\subsection{KV-Caching Simulation} \label{sec:design:caching:kv-caching}
\textit{Design Choices:} We identify two main design choices for facilitating (KV-)cache-aware simulation~\ref{fr:caching}. First, we consider predicting the average KV-Caching usage for the entire inference \textit{(total-simulation model)}. Second, we consider simulating KV-Caching at each timestamp at a user-selected granularity \textit{(discrete-event simulation model)}. We identify the latter approach to be better suited for Kavier, as it provides precise estimations of the KV-Cache usage at each timestamp and at adjustable granularity. Thus, discrete-event simulation model allows operators to analyze the ecosystem in finer detail and analyze how caching usage evolves over time. However, this discrete-event simulation approach comes with a higher performance cost ($O(n)$), compared to the total-simulation model, where KV-Caching is estimated only once, as an average over the simulation ($O(1)$).

We simulate following a discrete-event simulation approach~\cite{DBLP:conf/ccgrid/MastenbroekAJLB21}. During autoregressive generation, at each timestamp, the model takes as input the new token and the past keys/values from previous tokens' attention layers; then, instead of recomputing, the model caches the already computed states while generating tokens. For example, the model computes the attention layer for the first token from scratch. Then, for token 2, the model recomputes the attention only for token 2, and retrieves from the cache the layer for token 1. Then, for token 3, the model reuses the result for tokens 1 and 2. This process runs recursively until the last token is decoded. 

KV-Caching reduces the time complexity from quadratic to linear. Specifically, KV-Caching reduces the time complexity from $O(n^2)$, where the model would process all $n$ previous tokens for each of the $n$ tokens, to $O(n)$, where the model only processes the new token and retrieves the past $n$ computations from the cache. In our simulator, we assume KV-Caching as enabled by default, reflecting the current state-of-the-art in nowadays LLM serving frameworks (e.g., vLLM 0.9.1~\cite{kwon2023efficient}). However, the current design also allows disabling KV-Caching, improving Kavier's versatility for various scenario simulations.

The memory used by KV-Caching for each prompt is simulated using the community-vetted formula represented in Equation~\eqref{eq:kv-memory:design}.

\begin{align}
    \quad \text{KV}_\text{usage} = 2 \times L \times H \times d \times N \times sizeof(type)  \label{eq:kv-memory:design} 
\end{align}
\begin{small}
    \textit{where $L$ is the number of transformer layers in the model, $H$ is the number of attention heads, $d$ is the dimension per head, $N$ is the number of tokens in the sequence, and $sizeof(type)$ represents the size of the data type in bytes (e.g., float16 represents 2 bytes, float32 4 bytes). The factor of 2 represents storing two matrices, one for keys and one for values.}
\end{small}

The logic of computing decoding time reflects the versatility of decoding with and without using KV-Caching, thereby illustrating the linear and quadratic behavior of the decoding stage. We represent this functionality in \Cref{lst:design:kv}, where $n_\text{out}$ represents the number of output (decode) tokens.

\begin{lstlisting}[language=python, caption={KV-Caching, enabled and disabled.}, float, label={lst:design:kv}]
def get_decode_time(...):
    ...
    if kv_cache:
        return n_out * time_per_token
    else:
        return (n_out * (n_out + 1) / 2) * time_per_token
\end{lstlisting}

\subsection{Prompt Prefix Caching} \label{sec:design:caching:prompt}
Prefix caching is a system-level technique for performance improvement that caches new, unseen queries for a fixed amount of time. If a new query arrives with a matching prefix (e.g., the first 256 tokens), the results of the previous computations are retrieved from the cache instead of being recomputed~\cite{openai_prompt_caching}. While, to the best of our knowledge, OpenAI is the only company as of July 2025 to acknowledge using a similar caching technique, we believe it is an industry standard. However, it is still hidden under the curtains of closed-source codebases and inference pipelines. 

\begin{figure}[t]
    \centering
    \includegraphics[width=0.75\linewidth]{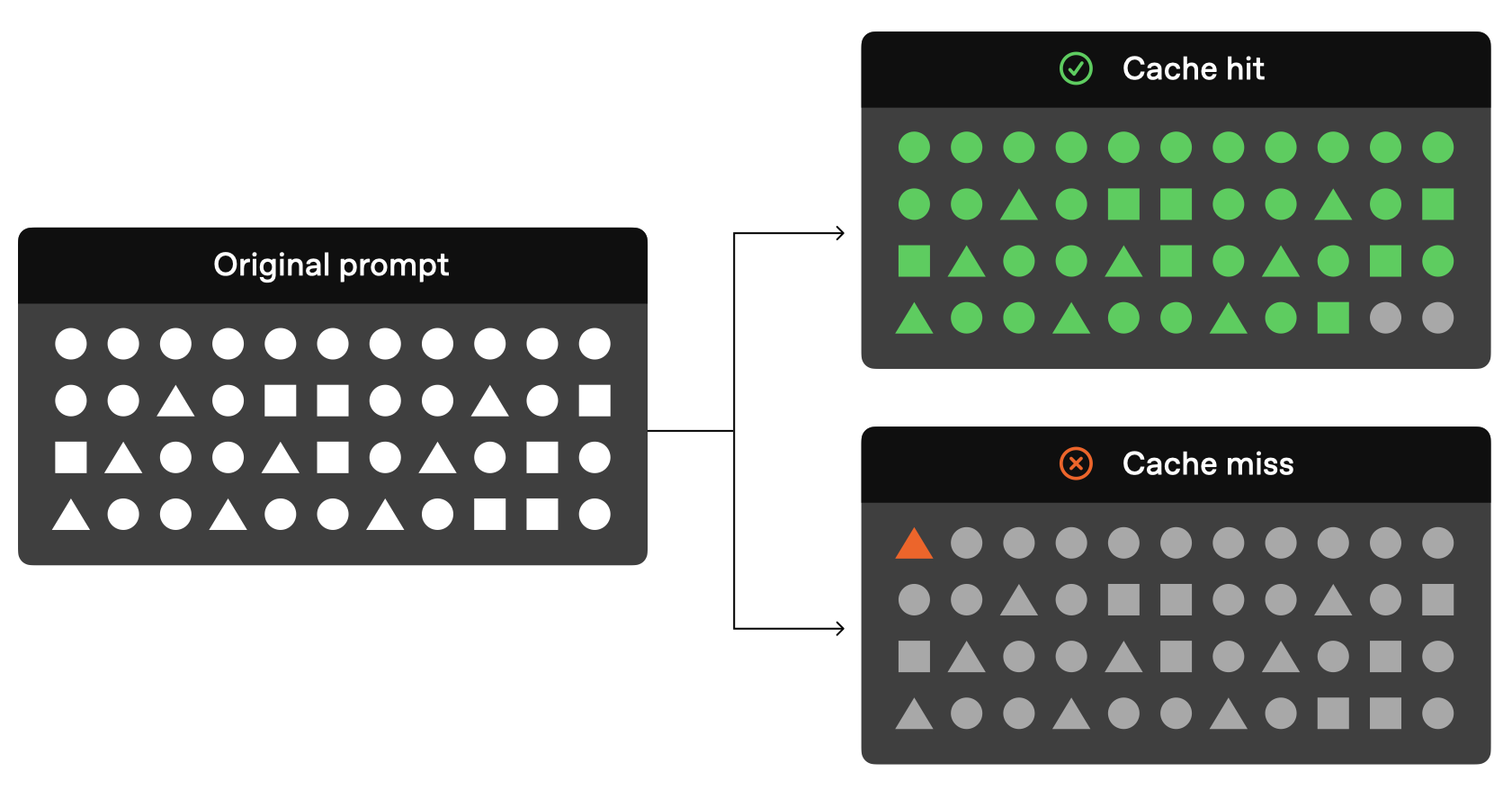}
    \caption{Prompt caching analogy used by OpenAI. Figure from \cite{openai_prompt_caching}.}
    \label{fig:openai_prompt_caching}
\end{figure}

OpenAI utilizes a prompt cache for very long prompts, exceeding 1,024 tokens, where they store/retrieve the prefill weights from the cache and do the decode stage independent of cache hit/cache miss; this approach is reported to have reduced latency by 80\% and costs by 50\%. In \Cref{fig:openai_prompt_caching} we showcase a figure taken from OpenAI's official blog on prefix caching~\cite{openai_prompt_caching}. On the left-hand side is a user prompt. If the first $n$ tokens match and $n$ exceeds the minimum threshold for the number of tokens in the matching prefix, then there is a cache hit (top right). However, if there is even one token in the prefix that doesn't match, the system gives a cache miss, even if the other tokens perfectly match (bottom right).

\textit{Design Choices:} We identify two main design choices for designing a prompt prefix caching system. First, we consider the exact-match approach, which OpenAI uses, where any mismatch in the cached prefix results in a cache miss (i.e., if there is at least one token which does not match, there is a cache miss). Second, we consider an approximate-match approach that allows minor mismatches, configurable by a user (i.e., allows for cache hit even if at least one of the e.g., 1,024 prefix tokens does not match). We identify the exact-match approach as better suited for Kavier's initial design and better suited for simulation-driven experiments from this work, in which we evaluate various prefix caching policies against OpenAI system. Specifically, the exact-match approach allows for keeping a similar experimental setup with the real-world ecosystem used by OpenAI. Still, we envision future research into further designign flexible caching strategies with user-adjustable tolerance levels for prefix mismatches.

To simulate prefix caching in Kavier, we design a simple representation of a cache store. The user configures a minimum length parameter; if the prompt length is higher than the user-configured length (i.e., if the user-configured length is $n$, and the prompt length is at least $n+1$), then the prompt input is cached. As Kavier iterates through the input trace of requests, we check for each prompt if the first $n$ tokens from the respective prompt are stored in the cache. In the case of a cache hit, the real-world system would skip the redundant prefill phase and retrieve the post-prefill data from memory, thereby only performing the decode stage.

To improve simulator performance, the caching system only contains the input tokens and does not store the output tokens. \Cref{lst:design:kv-pseudocode} shows a pseudocode of a system that simulates prompt caching:

\begin{lstlisting}[language=python, caption={Prompt caching pseudocode.}, float, label={lst:design:kv-pseudocode}]
PREFIX_CACHE = {}
PREFIX_CACHE_MIN_LEN = 256

for prompt in prompts:
    if len(prompt) > PREFIX_CACHE_MIN_LEN:
        prefix = prompt[:PREFIX_CACHE_MIN_LEN]
        if prefix in prompt_cache:
            handle_cache_hit()
            T_prefill = 0
            T_decode = simulate_decoding() # will be >0
            continue

    T_prefill, T_decode = simulate_decoding()
    if len(prompt) > PREFIX_CACHE_MIN_LEN:
        handle_cache_miss() # saves the prompt in the cache

\end{lstlisting}

We acknowledge the greedy yet powerful approach of this system. Although this design doesn't take into account the overhead of cache lookup, nor the overhead of retrieving the cache-stored response, we argue these actions are insignificant compared to the big model inference times, which would otherwise need to prefill and decode prompts of hundreds or thousands of tokens. However, albeit insignificant for individual prompts, in massive-scale operation scenarios, this overhead adds up. We leave the simulation of memory, memory levels, and networking for future research.

\section{Kavier Module for Performance Analysis}\label{sec:design:performance}

In this section, we describe the performance models Kavier uses to simulate LLM inference~\ref{fr:performance}. The inference process involves two core stages, prefill and decode, each with different performance particularities. 

\subsection{Performance simulation in the prefill stage}\label{sec:design:performance:prefill}
For the prefill stage, Kavier assumes a linear dependence between the number of input tokens and the decoding time, since the model does a full forward pass for each token in the user-given prompt~\cite{vaswani2017attention}. Specifically, the simulator computes the prefill based on the total floating-point operations (FLOPs) required for the prompt, divided by the GPU's throughput in FLOPs per second. According to~\cite{baseten_llm_inference_guide}, the number of FLOPs per token is estimated at twice the number of parameters in the model. The authors explain the processing of one token as involving a forward pass through all the layers of the model, including both the attention and the feed-forward networks, which is approximately equivalent to twice the size of the model. 

Kavier simulates the GPU's effective compute throughput by multiplying the amount of FLOPs per second by the efficiency factor, a hyperparameter that reflects the real-world limitations of GPUs; for example, empirical research conducted by \textit{Recasens et al.} shows that LLMs achieve only up to 30-35\% of the theoretical performance due to bottlenecks (e.g., memory, networking) or model (in)optimizations~\cite{recasens2025mind}. Similarly, systems may have overheads before each prompt execution; acknowledging this, we trace and measure real-world deployments and establish the prefill overhead as a hyperparameter, initially set to 25ms, but user-adjustable.  Transforming the paragraphs above into a formula, Kavier simulates prefill time through the formula defined in Equation \eqref{eq:time_prefill}.

\begin{align}
    T_{p} = \frac{n_\text{i} \times m_\text{p} \times 2}{F \times C_{e}} + O
    \label{eq:time_prefill}
\end{align}
\begin{small}
    \textit{$T_{p}$ is the simulated prefill time, $n_\text{i}$ is the number of input tokens, $m_\text{p}$ is the number of parameters in the model, $F$ is the theoretical throughput of the GPU, measured in FLOPs per second. $C_{e}$, the compute efficiency, and $O$, the system overhead, are hyperparameters system-dependent.}
\end{small}

\subsection{Performance simulation in the decode stage}\label{sec:design:performance:decode}

For the decode stage, Kavier models the time per output token and multiplies by the number of generated tokens, dependent on the presence or absence of KV-Caching. 

\textbf{KV on:} If KV-Caching is enabled, then the real-world LLM inference process executes in $O(n)$ time complexity; thus, computation of each token requires roughly the same amount of computation, leading to a decode time which grows linearly with the number of output tokens~\cite{vaswani2017attention, kwon2023efficient}. Kavier simulates the decode time of a model using KV-Caching using Equation~\eqref{eq:time_decoding_kv}.

\begin{align}
    T_{d, KV} = n_\text{o} \times T_{t}
    \label{eq:time_decoding_kv}
\end{align}
\begin{small}
    \textit{where $T_{d, KV}$ is the simulated decode time with KV-Caching enabled, $n_\text{o}$ is the number of output tokens, $T_\text{t}$ is the computed time per token.}
\end{small}

\textbf{KV off:} If KV-Caching is disabled, the real-world LLM inference process takes $O(n^2)$ time complexity; thus, computation per token grows as the LLM traverses the decode stage, leading to quadratic time complexity. This time complexity is due to the need to recompute attention, from scratch, over an ever-growing sequence, without the possibility of caching the previous computations. Kavier simulates the decode time of a model not using KV-Caching using the Equation~\eqref{eq:time_decoding_no_kv}:

\begin{align}
    T_{d, KV} = (n_\text{o} \times (n_\text{o} + 1) / 2) \times T_{t}
    \label{eq:time_decoding_no_kv}
\end{align}
\begin{small}
    \textit{where $T_{d}$ is the simulated decode time, $n_\text{o}$ is the number of output tokens, $T_\text{t}$ is the computed time per token.}
\end{small}

Equations \eqref{eq:time_decoding_kv} and \eqref{eq:time_decoding_no_kv} introduce a new variable, $T_{t}$, the time required to compute one token. \textit{Recasens et al.} empirically measure bottleneck in LLM inference, especially \textit{``unveiling GPU bottlenecks in large-batch LLM inference"}~\cite{recasens2025mind}; the time per token is either compute-bound or memory-bound. In our simulation approach we simulate the latency for both compute-bound and memory-bound, then select the highest latency between the two. \textit{``Minding the memory gap"}, and considering that \textit{``no model exceeds 35\% average (...) usage in either the prefill or decode phase"}~\cite{recasens2025mind}, we consider the same hyperparameter for compute efficiency set at 30\%.  Similarly, the memory-read efficiency is empirically measured and reported in Table 1, \cite{recasens2025mind}, averaging at 57.6\%, we thus implement a hyperparameter for memory-efficiency and set at 60\%.

We synthesise the above paragraphs in formulas; Equation \eqref{eq:compute_bound} shows the computation of compute-bound time per token, while Equation \eqref{eq:memory_bound} shows the computation of memory-bound time per token.

\begin{align}
    C = \frac{f_\text{tok}}{F \times C_{e}}
    \label{eq:compute_bound}
\end{align}
\begin{small}
    \textit{where $C$ is the compute-bound time per token, $f_\text{tok}$ is the number of FLOPs per token (estimated as $2 \times m_\text{p}$), $F$ is the theoretical throughput of the GPU in FLOPs per second, and $C_{e}$ is the compute efficiency hyperparameter.}
\end{small}

\begin{align}
    M = \frac{b \times m_\text{p}}{B \times M_{e}}
    \label{eq:memory_bound}
\end{align}
\begin{small}
    \textit{where $M$ is the memory-bound time per token, $b$ is the bytes per parameter, $m_\text{p}$ is the number of parameters in the model, $B$ is the memory bandwidth in bytes per second, and $M_{e}$ is the memory efficiency hyperparameter.}
\end{small}

Then, the final time of per-token computation is determined by taking the maximum between the compute-bound and the memory-bound, i.e., $max(C, M)$, computed in Equations \eqref{eq:compute_bound}, \eqref{eq:memory_bound}.

\subsection{GPU Utilization} \label{sec:design:performance:gpu_util}

Simulating the GPU utilization of the ecosystem under LLM inference is crucial for simulating sustainability metrics. From the amount of GPU utilization, we can estimate the amount of power used by the GPUs and further simulate the amount of CO2 emitted for running the workload, addressing~\ref{fr:sustainability}.

To the best of our knowledge, as of June 2025, there are no open-source traces showing the correlation between LLM inference and GPU utilization. Addressing this challenge, we decided to conduct our own ecosystem measurements. We deployed an LLM inference engine (vLLM 0.9.1, the latest version at the time of writing) and developed a tool for tracing LLM ecosystems, which we have released as open-source. We deployed the inference engine on clusters from two supercomputers: a cluster from SURF containing an NVIDIA A10 and a cluster from DAS-6 containing an NVIDIA A4000. We further detail and expand the tracing process in~\Cref{sec:evaluation:tracer}.

After empirical measurements, we observe an insignificant start-up time of $\approx$50-100~ms, when the GPU utilization grows from 4\% to the user-established maximum utilization e.g., 98\%; datacenter providers limit computing infrastructure to a certain cap, depending on the established SLOs and QoS. Then, throughout the inference process, the GPU utilization stays within the user-established cap, leading to an insignificant $\approx$50-100~ms when the GPU utilization decreases towards 0-10\%.

\textit{Design choices:} We identify two main design processes of simulating the GPU utilization, one observational-based, leveraging real-world traces, and one based on mathematical and statistical models. We identify the observation-based approach as superior, because GPU utilization remains largely constant throughout the inference, at the user-established maximum utilization, with negligible warm-up and cool-down periods. This design choice simplifies the computation complexity of the simulation process, while keeping a close-to-perfect simulation accuracy.

Hence, addressing the negligible variations in GPU utilization during inference, we simulate GPU utilization using the pseudocode presented in \Cref{lst:design:gpu_util}.

\begin{lstlisting}[language=python, float, caption={Prompt caching pseudocode.}, label=lst:design:gpu_util]
def get_gpu_utilization(t, t_prefill, t_decode, warm = 0.1, cool = 0.1):
    if t < warm: # if warming-up stage
        return 0.5 # i.e., 50% utilization

    if t < t_prefill + t_decode - cool: # if inference stage
        return MAX_GPU_UTILIZATION # i.e., user-established cap

    # if cooling stage
    return 0.5 # i.e., 50% utilization
\end{lstlisting} 
\section{Kavier Module for Sustainability Analysis}\label{sec:design:sustainability}

\begin{figure}[t]
    \centering
    \includegraphics[width=0.95\linewidth]{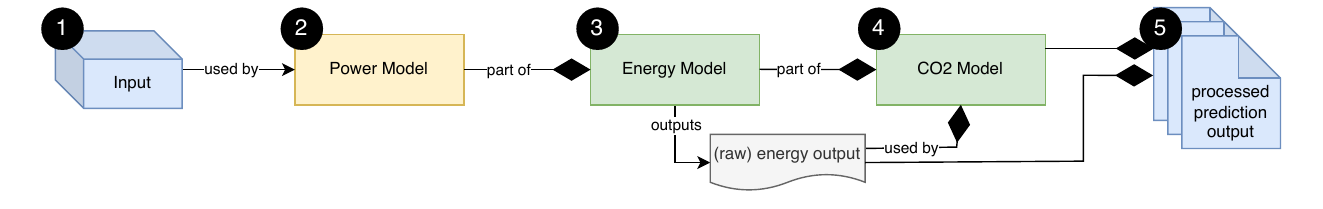}
    \caption{Relation between sustainability models in OpenDC.}
    \label{fig:design:sustainability-model-relation}
\end{figure}

In this section, we detail the sustainability component of the Kavier-OpenDC system \ref{fr:sustainability}. \Cref{fig:design:sustainability-model-relation} illustrates the relationship between the sustainability models OpenDC provides. The input is processed by a power model, which predicts energy consumption and generates the output trace; this output trace is then further leveraged by a CO2 model, which predicts CO2 emissions. In \Cref{sec:design:sustainability:energy}, we expand the energy simulation (component \circled{3}). In \Cref{sec:design:sustainability:co2}, we expand the simulation of CO2 emissions (component \circled{4}).

\subsection{Energy Simulation} \label{sec:design:sustainability:energy}

To simulate energy usage, we leverage the capabilities of OpenDC, a peer-reviewed and top-tier simulator, with which we are coupling Kavier. We argue that, since OpenDC is already a vetted simulator through publications in many peer-reviewed venues~\cite{DBLP:conf/ccgrid/MastenbroekAJLB21, DBLP:conf/ispdc/IosupABBENOTVV17, DBLP:journals/tpds/AndreadisMBI22, DBLP:journals/fgcs/MastenbroekMBI25, 2023-icpewip-scheduler-apis}, its models are reliable, accurate, and a robust model of reality. Furthermore, OpenDC implements capabilities of Multi- and Meta-Model simulation for energy models, which further increase the explainability and robustness of the simulation results~\cite{nicolae2025m3sa, radunicolae-hp-m3sa}. \Cref{table:power-draw-formulas} shows the simulation models OpenDC-Kavier simulation system uses to predict power draw and, further, energy usage; each model is leveraged from peer-reviewed literature on datacenter (energy) simulation.

OpenDC simulates energy usage and outputs the results into a discrete-event output format, which shows, at the user-selected granularity, the amount of power drawn at a certain timestamp (represented in Watts and derived units), as well as the total energy usage which a real-world infrastructure would consume in a real-world experimentation setup (represented in Watt-hour and derived units). OpenDC exports the results in Parquet format, due to scalability, efficient storage, and cross-system compatibility (and thus portability) of this storage format~\cite{nicolae2025m3sa, ApacheParquet}

\begin{table*}[t]
\centering
\begin{tblr}{
  hline{1-2,9-10} = {-}{},
}
Name                 & Formula                                                                                                                             & ~Source \\ 
Sqrt                 & $P(u)=P_{\text{idle}}+(P_{\text{max}}-P_{\text{idle}})\sqrt{u}$ \label{eq:sqrt_power}                                              & \cite{DBLP:conf/iccS/SilvaOCTDS19, DBLP:journals/spe/CalheirosRBRB11, DBLP:conf/ccgrid/MastenbroekAJLB21} \\ 
Linear               & $P(u)=P_{\text{idle}}+(P_{\text{max}}-P_{\text{idle}})\,u$ \label{eq:linear_power}                                                 & \cite{DBLP:conf/iccS/SilvaOCTDS19, DBLP:journals/spe/CalheirosRBRB11, DBLP:conf/ccgrid/MastenbroekAJLB21} \\ 
Square               & $P(u)=P_{\text{idle}}+(P_{\text{max}}-P_{\text{idle}})\,u^{2}$ \label{eq:square_power}                                             & \cite{DBLP:conf/iccS/SilvaOCTDS19, DBLP:journals/spe/CalheirosRBRB11, DBLP:conf/ccgrid/MastenbroekAJLB21} \\ 
Cubic                & $P(u)=P_{\text{idle}}+(P_{\text{max}}-P_{\text{idle}})\,u^{3}$ \label{eq:cubic_power}                                              & \cite{DBLP:conf/iccS/SilvaOCTDS19, DBLP:journals/spe/CalheirosRBRB11, DBLP:conf/ccgrid/MastenbroekAJLB21} \\ 
MSE                  & $P(u)=P_{\text{idle}}+(P_{\text{max}}-P_{\text{idle}})\,(2u-u^{r})$ \label{eq:mse_power}                                           & \cite{DBLP:conf/ccgrid/MastenbroekAJLB21, DBLP:conf/isca/FanWB07} \\ 
Asymptotic           & $P(u)=P_{\text{idle}}+\dfrac{P_{\text{max}}-P_{\text{idle}}}{2}\bigl(1+u-e^{-u/\alpha}\bigr)$ \label{eq:asymptotic_power}          & \cite{DBLP:conf/ccgrid/MastenbroekAJLB21} \\ 
Asymptotic DVFS      & $P(u)=P_{\text{idle}}+\dfrac{P_{\text{max}}-P_{\text{idle}}}{2}\bigl(1+u^{3}-e^{-u^{3}/\alpha}\bigr)$ \label{eq:asymptotic_power_dvfs} & \cite{DBLP:conf/ccgrid/MastenbroekAJLB21} \\ 
\end{tblr}
\caption{Formulas of the power models the peer-reviewed OpenDC uses to predict energy usage. $P_{\text{idle}}$, $P_{\text{max}}$ are the powers in idle and full-capacity states, $u$ is device utilization, $e$ is Euler’s number, $\alpha$ is the utilization fraction at which the host becomes asymptotic, and $r$ is a calibration parameter.}
\label{table:power-draw-formulas}

\end{table*}

\FloatBarrier


\subsection{CO2 Emissions Simulation} \label{sec:design:sustainability:co2}

Once the output of the energy predictions, OpenDC leverages the embedded carbon model to predict CO2 emissions. The CO2 model, specifically component \circled{4} from \Cref{fig:design:sustainability-model-relation}, predicts at the user-established granularity (same granularity as the prediction of energy consumption), leveraging the amount of power drawn and the carbon intensity at the specific granularity, as represented in Equation \eqref{eq:co2_model}. OpenDC implements the CO2 model explained in the peer-reviewed Footprinter~\cite{dniewenhuis_hotcloud_footprinter}. 

\textit{Niewenhuis et al.} show, using empirical measurements from ENTSO-E, the \textit{``Europe's most ambitious electricity data platform"}~\cite{hirth2018entso}, that the amount of CO2 emitted to produce one unit of energy varies significantly across location, time of the day, or weather conditions. In previous work, we showed that the same experiment run in locations with a high carbon footprint can emit up to 150-200 times more CO2 compared to low carbon footprint; for example, running the experiment in Germany emits a predicted 13.4~tCO2, while the same experiment run in Switzerland emits a predicted 0.081~tCO2 (Experiment 3 and Appendix C from~\cite{nicolae2025m3sa}).

OpenDC receives as input a carbon trace, which shows the carbon intensity per timestamp, in a discrete-event monitoring and reporting format. This trace is leveraged by \circled{4}, together with the power drawn, to simulate the CO2 emissions. OpenDC outputs CO2 predictions in Parquet format, matching the format of predictions of energy usage.

\begin{align}
    C_{e,{t=\alpha}} = P_{t=\alpha} \times C_{i, {t=\alpha}}
    \label{eq:co2_model}
\end{align}
\begin{small}
    \textit{Where $C_{e,{t=\alpha}}$ represents the CO2 emissions at timestamp alpha, $P_{t=\alpha}$ represents the power drawn at timestamp alpha, and $C_{i, {t=\alpha}}$ represents the carbon intensity at timestamp alpha. Alpha is identical for each variable.}
\end{small}

\section{Kavier Module for Efficiency Analysis}\label{sec:design:efficiency_simulation}

We now detail the efficiency models Kavier uses to compute the efficiency of the simulated experiment~\ref{fr:efficiency}. We note that the accuracy of the efficiency module is equal to the simulation accuracy; in other words, mathematics is never wrong, but simulation can be. If the simulations have 100\% accuracy, the efficiency predictions will also have 100\% accuracy. 

We further present two efficiency models, one for predicting financial efficiency and one for predicting sustainability efficiency. We argue that this component is crucial for datacenter and LLM operators in making informed decisions about potential deployments, as this component offers a homogeneous comparison metric for each type of efficiency, directly comparable, simple to understand, represent, and explain.

\subsection{Financial efficiency}\label{sec:design:efficiency_simulation:financial}
We express financial efficiency as the cost per token per second, essentially for the monetary aspect of running LLM ecosystems at scale, in profit-driven processes. Equation \eqref{eq:efficiency:financial} shows the formula Kavier uses for computing the financial efficiency across LLM prompts, containing the total cost (set up by the user for their own specific financial model), the total amount of tokens (derived from the trace), from both the prefill and decode phase, and the total time needed for the prefill and decode phase (simulated by Kavier).

$C$ from Equation \eqref{eq:efficiency:financial} represents the cost; we identify $C$ as a provider-dependent variable, as providers have distinct and highly diverse, some multi-dimensional financial models; for example, Microsoft Vidur shows a financial model in which users are charged at a GPU-hourly rate with $\approx$\$10 per hour~\cite{agrawal2024vidur}, while OpenAI API charges users by the amount of processed tokens~\cite{openai_api}. Thus, we provide the financial model as embodying an abstract variable, yet simple to implement a specific financial cost. 

The financial efficiency is, thus, represented in \textit{currency per token per second} e.g., \textit{€/t/s}, \textit{LEU/t/s}.

\begin{align}
    E_{f} = \frac{C}{T} = \frac{C}{\frac{T_p + T_d}{\Delta T_{P} + \Delta T_{D}}} = \frac{C \times (\Delta T_{P} + \Delta T_{D})}{T_P + T_D}
    \label{eq:efficiency:financial}
\end{align}
\begin{small}
    \textit{Where $E_{f}$ represents the financial efficiency, $C$ represents the operational cost, $T_P$ and $T_D$ represent the amount of prefill and decode tokens, respectively, and $\Delta T_{P}$ and $\Delta T_{D}$ represent the total inference time for prefill and decode stages, respectively.}
\end{small}

We further identify Equation \eqref{eq:efficiency:ratio} for comparing two systems through financial efficiency ratio. We note that the efficiency of both systems from Equation \eqref{eq:efficiency:ratio}, $s1$ and $s2$, should be quantified with the same metric.

\begin{align}
    R_f = \frac{E_\text{s1}}{E_\text{s2}}
    = \frac{\frac{P_{\text{s1}}}{\frac{T}{\Delta T_{\text{s1}}}} }{\frac{P_{\text{s2}}}{\frac{T}{\Delta T_{\text{s2}}}} }
    = \frac{P_{\text{s1}} \times \Delta T_{\text{s1}} \times \cancel{T}}{P_{\text{s2}} \times \Delta T_{\text{s2}} \times \cancel{T}}
   = \frac{\left( \cancel{P_{\text{A10}}} \times \Delta T_{\text{s1}}\right)  \times \Delta T_{\text{s1}}}{\left( \cancel{P_{\text{A10}}} \times \Delta T_{\text{s2}} \right) \times \Delta T_{\text{s2}}} 
    = \frac{\Delta T_{\text{s1}}^2} {\Delta T_{\text{s2}}^2}
    \label{eq:efficiency:ratio}
\end{align}
\begin{small}
    \textit{
    Where $R_f$ is the financial efficiency ratio between system $s1$ and $s2$, 
    $\text{s1}$ refers to the first system,
    $\text{s2}$ refers to the second system,
    $E$ is efficiency, 
    $P$ is price per hour, 
    $\Delta T$ is time, 
    $T$ is the amount of processed tokens.}
\end{small}

\subsection{Sustainability efficiency}\label{sec:design:efficiency_simulation:sustainability}
Following a similar approach as in Equation \eqref{eq:efficiency:financial}, we implement a sustainability model for computing the sustainability efficiency. Equation \eqref{eq:efficiency:sustainability} shows the formula Kavier uses for computing the sustainability efficiency across LLM prompts; in essence, the only difference between the formula for financial efficiency and sustainability efficiency is the cost element; in the former, the cost is monetary, in the latter, the cost is of sustainability. The proposed equation for computing sustainability efficiency, Equation \eqref{eq:efficiency:sustainability}, contains the sustainability cost, either in Energy Usage or CO2 emissions (simulated by the sustainability module of Kavier-OpenDC setup), the total amount of tokens (derived from the trace), from both the prefill and decode phase, and the total time needed for the prefill and decode phase (simulated by Kavier).

The sustainability efficiency is, thus, represented in \textit{Wh per token per second} (e.g., Wh/t/s) and \textit{CO2 per token per second} (e.g., CO2/t/s), or both.

\begin{align}
    E_{s} = \frac{S}{T} = \frac{S}{\frac{T_p + T_d}{\Delta T_{P} + \Delta T_{D}}} = \frac{S \times (\Delta T_{P} + \Delta T_{D})}{T_P + T_D}
    \label{eq:efficiency:sustainability}
\end{align}
\begin{small}
    \textit{Where $E_{s}$ represents the sustainability efficiency, $S$ represents the sustainability cost, $T_P$ and $T_D$ represent the amount of prefill and decode tokens, respectively, and $\Delta T_{P}$ and $\Delta T_{D}$ represent the total inference time for prefill and decode stages, respectively.}
\end{small}

To compare the sustainability efficiency ratio of two systems, $s1$ and $s2$, we propose the formula represented in Equation \eqref{eq:efficiency:ratio-sustainability}. We note that, similarly to the financial efficiency ratio, both efficiencies must be quantified using the same sustainability metric.

    

\begin{align}
R_s = \frac{E_{s1}}{E_{s2}} \
= \frac{\frac{S_{1} \times (\Delta T_{P_{1}} + \Delta T_{D_{1}})}{T_{P_{1}} + T_{D_{1}}}}{\frac{S_{2} \times (\Delta T_{P_{2}} + \Delta T_{D_{2}})}{T_{P_{2}} + T_{D_{2}}}} \
= \frac{S_{1} \times (\Delta T_{P_{1}} + \Delta T_{D_{1}}) \times (T_{P_{2}} + T_{D_{2}})}{S_{2} \times (\Delta T_{P_{2}} + \Delta T_{D_{2}}) \times (T_{P_{1}} + T_{D_{1}})}
\label{eq:efficiency:ratio-sustainability}
\end{align}
\begin{small}
\textit{
Where $R_s$ is the sustainability efficiency ratio between system $s1$ and $s2$,
$E_{s1}$ and $E_{s2}$ are the sustainability efficiencies of system $s1$ and $s2$, respectively,
$S_{s1}$ and $S_{s2}$ represent the sustainability cost (energy consumption or CO$2$ emissions),
$\Delta T{P}$ and $\Delta T_{D}$ represent the total inference time for prefill and decode stages, respectively,
and $T_{P}$ and $T_{D}$ represent the amount of prefill and decode tokens, respectively.
}
\end{small}
\section{Requirement Validation} \label{sec:design:requirement-addressal}

In this chapter, we presented a design of Kavier, a simulator for LLM ecosystems under inference, able to predict performance, sustainability, and multi-layer efficiency. We defined a set of requirements, both functional and non-functional, which guided our design process. We now evaluate the validity of our design against each requirement.

\begin{enumerate}[label=(\textbf{FR\arabic*}), leftmargin=*, align=left, labelwidth=1cm]
    \item \textbf{Support holistic simulation of the LLM inference process.} \\
    {Kavier models both inference stages of the LLM inference process, specifically the prefill stage and the decode stage, each with their own distinct characteristics and specific behaviours. Besides, Kavier follows a discrete event simulation paradigm, where the simulator predicts and exports predictions at a user-established granularity. Moreover, the Kavier-OpenDC system is designed to follow the same discrete event simulation model, reflecting real-world LLM performance and supporting detailed performance, sustainability, and efficiency reports at a user-set tradeoff between export granularity (with impact on performance) and available information (with impact of report detail).}

    \item \textbf{Simulate with cache awareness.} \\
    {We design Kavier as a cache-aware simulator, capable of predicting LLM ecosystems under inference with KV-Caching enabled or disabled, and under conditions where prefix matching follows various cache store and cache hit policies. Thus, Kavier allows for versatility in experimentation and exploration of the impact of various caching policies on system performance, environmental sustainability, and efficiency. Kavier models the different impacts of caching on the different execution stages within the inference process.
    }

    \item \textbf{Predict the performance of LLM ecosystems under workload.} \\
    {Kavier predicts system performance using community-vetted simulation models or models derived from these models, which the system then uses to simulate cache-awarely. Kavier predicts latency by determining the total amount of time required to answer a prompt, obtained by summing up the inference time for the prefill phase and the inference time of the decode phase. Throughput is further derived from the simulated latency (e.g., if a prompt contains, in total, $n$ tokens, and the simulated latency for that prompt is of $m$ seconds, then the throughput is $n/m$ tokens per second). Lastly, the results are exported in both task-based and fragment-based traces, ensuring compatibility with OpenDC, a top-tier datacenter simulation framework, also adhering to~\ref{sec:design:fr6}.}

    \item \textbf{Predict the sustainability of LLM ecosystems under workload.} \\
    {Kavier predicts sustainability as part of the Kavier-OpenDC simulation system. Leveraging the peer-reviewed simulation capabilities of OpenDC~\cite{DBLP:conf/ccgrid/MastenbroekAJLB21, dniewenhuis_hotcloud_footprinter, DBLP:conf/ispdc/IosupABBENOTVV17}, Kavier simulates power draw, measured in Watts, essential for discrete event simulation~\ref{fr:discrete}, energy usage, measured in Watt-Hours, essential for overall system predictions~\ref{fr:efficiency}, and, resulting from these, CO2 emissions, simulated in both discrete-event format~\ref{fr:discrete} and overall system sustainability~\ref{fr:sustainability}. For this, the simulation system uses real-world traces and user-defined granularity to provide accurate and detailed sustainability predictions. We design Kavier as modular and integrable with a peer-reviewed simulation framework. We select OpenDC and propose a detailed design of Kavier-OpenDC integration, emphasizing each party's role and how these two communicate with each other. Furthermore, the current design ensures a high degree of modularity and extensibility, allowing for a long software lifecycle and enabling the implementation of new functionality, modification of existing functionality, or deactivation of functionality.}

    \item \textbf{Predict the efficiency of LLM ecosystems under workload.} \\
    Kavier predicts efficiency as of LLM ecosystems through the Kavier Efficiency module, thus addressing~\ref{fr:efficiency}. Kavier can predict financial efficiency, represented in cost per token per second, and environmental efficiency, represented in CO2/Wh per token per second. This module of Kavier is crucial for operators to differentiate and compare systems with ease, by simply analysing one or more efficiency metrics of the ecosystem. 

    \item \textbf{Design Kavier compatible with other simulators and extensible.}
    We design Kavier as compatible with OpenDC, a peer-reviewed and state-of-the-art datacenter simulator. Kavier leverages the sustainability module of OpenDC to predict the energy consumption and CO2 emissions of LLM inference. We further validate this design component in \Cref{sec:prototype}, where we integrate an engineered prototype of Kavier with OpenDC. Addressing the extensibility aspect, we design Kavier as modular, where each core functionality can be modified, removed, or expanded, thus ensuring long lifetime of the simulator and also aligning with~\ref{nfr:software-art}.
\end{enumerate}

\begin{enumerate}[label=(\textbf{NFR\arabic*}), leftmargin=*, align=left, labelwidth=1cm]
    \item \textbf{Provide in-meeting, near-interactive, same-day simulation results.} \\
    {We design the architecture of Kavier to minimize redundancies and enable the engineering of the simulator according to best software engineering practices. In \Cref{sec:evaluation:experiment-1}, we analyze Kavier's performance through trace-based experimentation and observe that Kavier can simulate, at second-granularity, 500 GPU hours in under 10 seconds. Even more, Kavier meets \ref{nfr:speed} even when simulating at milisecond-granularity, and is able to simulate 500 GPU hours in about 150 minutes.
    }

    \item \textbf{Aim to provide adequate simulation accuracy.} \\
    {We design Kavier as modular, easily modifiable. Thus, every simulation model can be modified as "plug-and-play." We validate the accuracy of Kavier in predicting performance of LLM ecosystems using real-world traces in \Cref{sec:evaluation:experiment-1}. We obtain a MAPE error ratio of 7.39\% for prefill and 4.00\% for decode. Further addressing \ref{nfr:accuracy}, we argue that the prediction accuracy of the sustainability module of Kavier is already validated, by design, since Kavier leverages functionality from OpenDC, a peer-reviewed simulator within numerous venues~\cite{dniewenhuis_hotcloud_footprinter, DBLP:conf/ccgrid/MastenbroekAJLB21, DBLP:conf/ispdc/IosupABBENOTVV17, DBLP:journals/tpds/AndreadisMBI22, DBLP:journals/fgcs/MastenbroekMBI25} and used in national~\cite{DBLP:journals/corr/abs-2206-03259} and international scale projects~\cite{de2023boosting}. Lastly, we note that the validity of the efficiency component of Kavier is directly dependent on the validity of the sustainability component and the performance component.}

    \item \textbf{Facilitate reproducibility and open science.} \\
    {Addressing the major reproducibility and closed-science challenge in computer systems research, we release all the designs, prototypes, engineered tools and instruments, traces, experiments, and information to the community. In short, we perfectly adhere to concepts of open-(real-)science. We ensure experiment reproducibility by releasing a reproducibility capsule (expanded in \ref{sec:experiments}) which strengthens our experiments, claims, and findings.
    }

    \item \textbf{Adhere to modern software design and development standards.} \\
    {We present, in \Cref{sec:prototype}, the engineering process of Kavier, where we follow state-of-the-art standards of software architecture, design, development, and integration, some (also as) described in~\cite{rozanski2012software}. Matching the first stage of modern software design and development standards, we propose in this section high-level and detailed design of the Kavier simulation process, and detail each simulation model and component, and how they interact.}
\end{enumerate}

\section{Discussion} \label{sec:design:discussion}

We now summarize the contributions of this chapter, envision future research, and discuss potential threats to validity.

\textit{Summary:} In this chapter, we propose and validate a design for a discrete and cache-aware simulator for LLM ecosystems under inference. Leveraging the reference architecture proposed and validated in~\Cref{sec:refarch}, which matches a well-defined set of requirements, we provide a high-level overview of such a simulation instrument. We detail each simulation module: performance, sustainability, and efficiency. We also emphasize the simulation mechanisms our design uses for differentiating the specific behavior of the prefill and decode stages, and run cache-aware simulations.

\textit{Future Research:} We envision a future design of a simulator that predicts the performance, sustainability, and efficiency of LLM training, modeling specifics of the training process, similarly to how Kavier models the specifics of the inference process. 

We also envision integrating the Kavier-OpenDC simulation system into the first digital twin for datacenters, with a specific focus on measuring, simulating, and dynamically adjusting LLM ecosystems. 

\textit{Threats to Validity:} The design of Kavier, albeit robust for a first-of-its-kind tool, poses several limitations. Kavier assumes zero latency in processes of searching in and retrieving from the caching system, which is not applicable in real-world ecosystems. 

Furthermore, while the sustainability module of Kavier-OpenDC employs Multi-Model simulation, the performance component simulates using a single model, trained for general-purpose scenarios, and thus prone to errors when encountering edge-cases. We envision future research in simulating the performance component through Multi-Model simulation.

 \newpage
\thispagestyle{noheader}
\chapter{Prototype and integration of Kavier}\label{sec:prototype}

Designing an instrument for simulating LLM ecosystems and adhering to a robust reference architecture of the LLM inference Compute Continuum is a critical yet non-trivial challenge for the community. Addressing this challenge, in \Cref{sec:design}, we proposed a design for Kavier, a simulator for LLM inference that can predict the performance, sustainability, and efficiency of LLM ecosystems under inference. To fully validate this design and simulate LLM ecosystems, implementing a prototype is (also) a critical yet non-trivial challenge. We envision such a prototype as suitable for simulating LLM performance independently of other instruments (no such tools exist at the time of publication), suitable for simulating LLM sustainability when coupled with a peer-reviewed datacenter simulator, and suitable for predicting the efficiency of LLM ecosystems. This raises the research question: \textit{(RQ3) How to implement and integrate Kavier within a peer-reviewed, discrete-event datacenter simulator?}

In this chapter, we implement Kavier following state-of-the-art software engineering principles, aiming for performance, long-term codebase sustainability, and open science. Then, we integrate Kavier with OpenDC and release the instrument as open-source. Lastly, we evaluate our implementation against non-functional requirements established in~\Cref{sec:design}.

\section{Overview}\label{sec:prototype:overview}
We implement Kavier, matching the state-of-the-art AtLarge design, implementation, and valuation process of researching computer systems and ecosystems~\cite{DBLP:conf/icdcs/IosupVTETBFMT19}. Our contribution in this chapter is five-fold:

\begin{enumerate}

    \item We implement, in \Cref{sec:prototype:implementation}, a working prototype of Kavier engineering the core functionality of the simulator, and adhering to the design proposed in~\Cref{sec:design}. Kavier would, thus, be the first instrument for predicting performance, sustainability, and efficiency of LLM ecosystems under inference, following a discrete-event and cache-aware simulation approach.

    \item We integrate Kavier in OpenDC, thereby leveraging the peer-reviewed capabilities of OpenDC for predicting the sustainability of datacenters (\Cref{sec:prototype:integration}).

    \item We showcase the GPU and LLM library, as well as the input interface of Kavier in \Cref{sec:prototype:showcase}.
    
    \item We analyze, in \Cref{sec:prototype:requirements_addressal}, the engineered prototype against the requirements established in \Cref{sec:design}.

    \item We reflect on Kavier, its limitations, and envision future engineering work in \Cref{sec:prototype:discussion}.

\end{enumerate}

\section{Implementation of a Kavier Software Prototype}\label{sec:prototype:implementation}

\begin{figure}[t]
    \centering
    \includegraphics[width=0.95\linewidth]{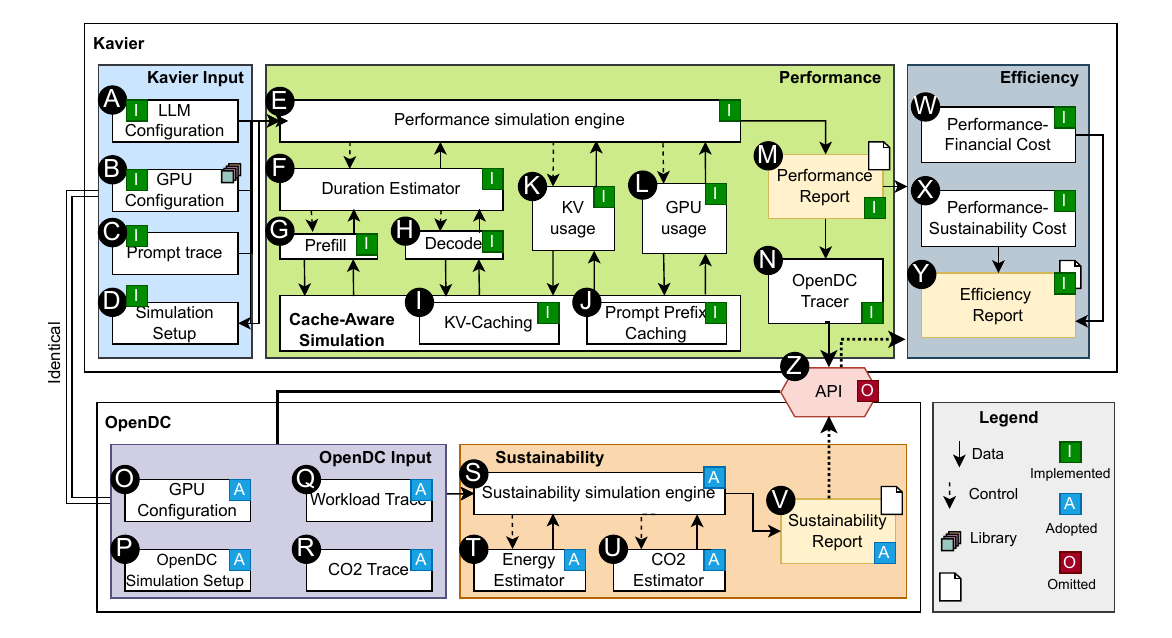}
    \caption{Kavier design from~\Cref{sec:design} showcasing specific components we implemented, adopted, or omitted in the engineered Kavier prototype.}
    \label{fig:implementation:kavier}
\end{figure}

In this section, we discuss the elements of Kavier that we implemented in our engineered prototype. \Cref{fig:implementation:kavier} represents the high-level design of Kavier and distinguishes the components of this design that we implemented in this work, adopted from the peer-reviewed OpenDC, or omitted.

\textit{(Implemented) Kavier input:} We implement and release to the community a prototype of Kavier, which receives input through CLI arguments, thus making Kavier easy to operate and easy to adopt in a simulation conglomerate, where external tools leverage the functionality of our prototype. We identify multiple ways to configure the experiment, including through an external configuration file, codebase tweaks, or visual interfaces. We argue that the CLI approach is the most versatile option, as it is simple to implement, operate, and adapt, and adheres best to our functional requirements, especially \ref{fr:modularity}, \ref{nfr:software-art}, and \ref{nfr:reproducibility}. Kavier can thus be configured either using default values or using one or more CLI arguments. We expand the input interface and how to set up experiments with Kavier in \Cref{sec:prototype:showcase}.

\textit{(Implemented) Performance:} We fully implement the performance component of Kavier, ensuring the prototype's ability to simulate discrete-event and KV-Cache aware. Kavier can mimic the distinct performance behavior of the prefill and decode stage, and can simulate various caching policies, such as autoregressive KV-Caching or Prompt-Prefix Caching. KV-Caching can be enabled or disabled through user input. Prompt-Prefix Caching can be configured by prefix length, where the minimum prefix length can be specified through user input and considered in the simulation. For example, the minimum sequence length should be at least $n$ tokens to be either cached or searched in the cache. This module exports a performance report, which is formatted to match the OpenDC input format.

\textit{(Leveraged) OpenDC:} We leverage OpenDC ``as-is," and, thus, leverage its status of community-reviewed and vetted simulator, which strengthens the validity of sustainability predictions. OpenDC outputs a sustainability report.

\textit{(Implemented) Efficiency}: We tailor the input format for the efficiency component of Kavier such that it matches the output format of OpenDC. 
\textit{(1) Performance-Financial Cost:} we implement a static financial model, where we consider a static price per hour for running LLM inference on a single GPU; we implement the default rate of 1.2\$ per hour which is consistent with prices of renting a GPU as NVIDIA A10 in July 2025~\cite{dilmegani2025cloud}. We use this specific GPU because it is the machine to which we have access, and we also utilize it in our tracing and experiments. The price per hour is simply adjustable by the user. The financial model, albeit not "one-command-away-adjustable," is relatively simple to modify without breaking external functionality of the Kavier, thanks to the modularity of the designed and engineered prototype~\ref{fr:modularity}. The output is reflected in the amount of money per million tokens.
\textit{(2) Performance-Sustainability Cost:} the sustainability efficiency module computes the total amount of CO2, the total amount of tokens, and computes the amount of CO2 per million tokens.
The results are ultimately packaged into a brief efficiency report.

\section{Integration of Kavier with OpenDC}\label{sec:prototype:integration}

In this section, we focus on component \circled{Z} from~\Cref{fig:implementation:kavier}, the \textit{(omitted) API}, then present a high-level overview of the software engineering processes used in the prototyping process. 

\subsection{The Human-in-the-loop}\label{sec:prototype:integration:human}

\textit{(Omitted) API:} We envision the API component as crucial for a digital-twinning system, with a human in the loop only for decision-making processes. However, in this prototype, Kavier acts as an LLM inference simulator decoupled from a digital twin (in fact, currently, there doesn't exist a digital twin for ICT ecosystems).

\begin{figure}[t]
    \centering
    \includegraphics[width=0.95\linewidth]{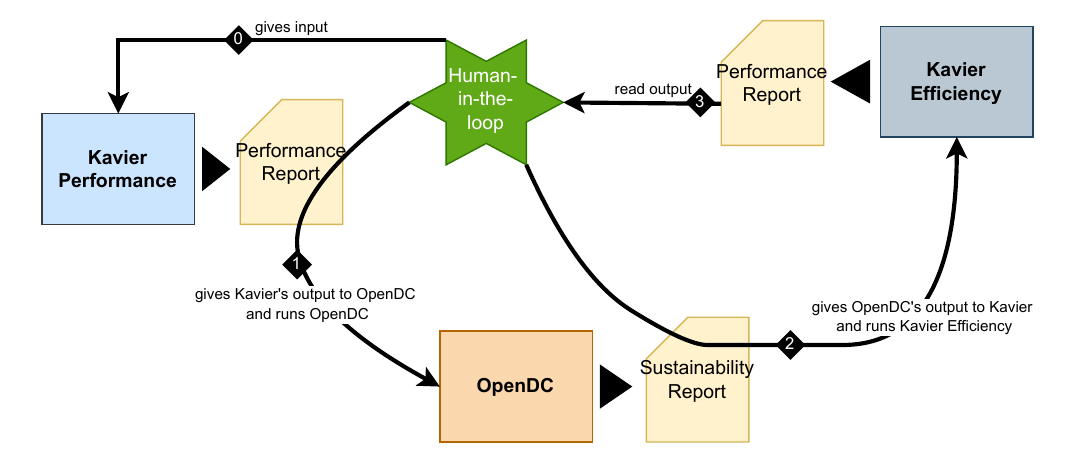}
    \caption{Kavier-OpenDC interaction with a human in the loop, who sets up the experiments, reads and analyzes outputs, and manipulates reports between simulators.}
    \label{fig:kavier-opendc-interaction}
\end{figure}

\Cref{fig:kavier-opendc-interaction} showcases the role of the “human-in-the-loop" in the simulation step. While a fully autonomous system would involve only steps~\circled{0} and~\circled{3}, our prototype involves two additional steps. The human gives input to Kavier~\circled{0} and waits until Kavier outputs a performance report; in \Cref{sec:evaluation:experiment-1} we show that this waiting period is usually a matter of seconds, and showcase that Kavier can simulate workloads of 500-GPU hours within 10 seconds, at second-granularity export rates. Then, in~\circled{1}, the human retrieves Kavier's input and gives it to OpenDC, then runs the OpenDC sustainability simulation part. After OpenDC's sustainability report~\circled{2}, the human forwards the predictions to Kavier to compute efficiency; the efficiency computation happens in a matter of milliseconds. Lastly, the user reads Kavier's prediction and analyzes the results.

\textit{Future work: } In this work, we identify the human-in-the-loop as necessary because many experiments are pioneering and exploratory in nature. We envision much of the human-in-the-loop's work could be automated once these experiments become de facto standards in the community. However, we regard this extra step as beyond the purpose this work, where we prioritize conceptual, experimental, and trace-based contributions over engineering contributions. We therefore prioritize the core engineering features (those without which a simulator would be unable to simulate and thus meet functional and non-functional requirements) and reserve the API component for future work.

\subsection{Software Engineering Processes}\label{sec:prototype:integration:human}

\begin{figure}[t]
    \centering
    \includegraphics[width=\linewidth]{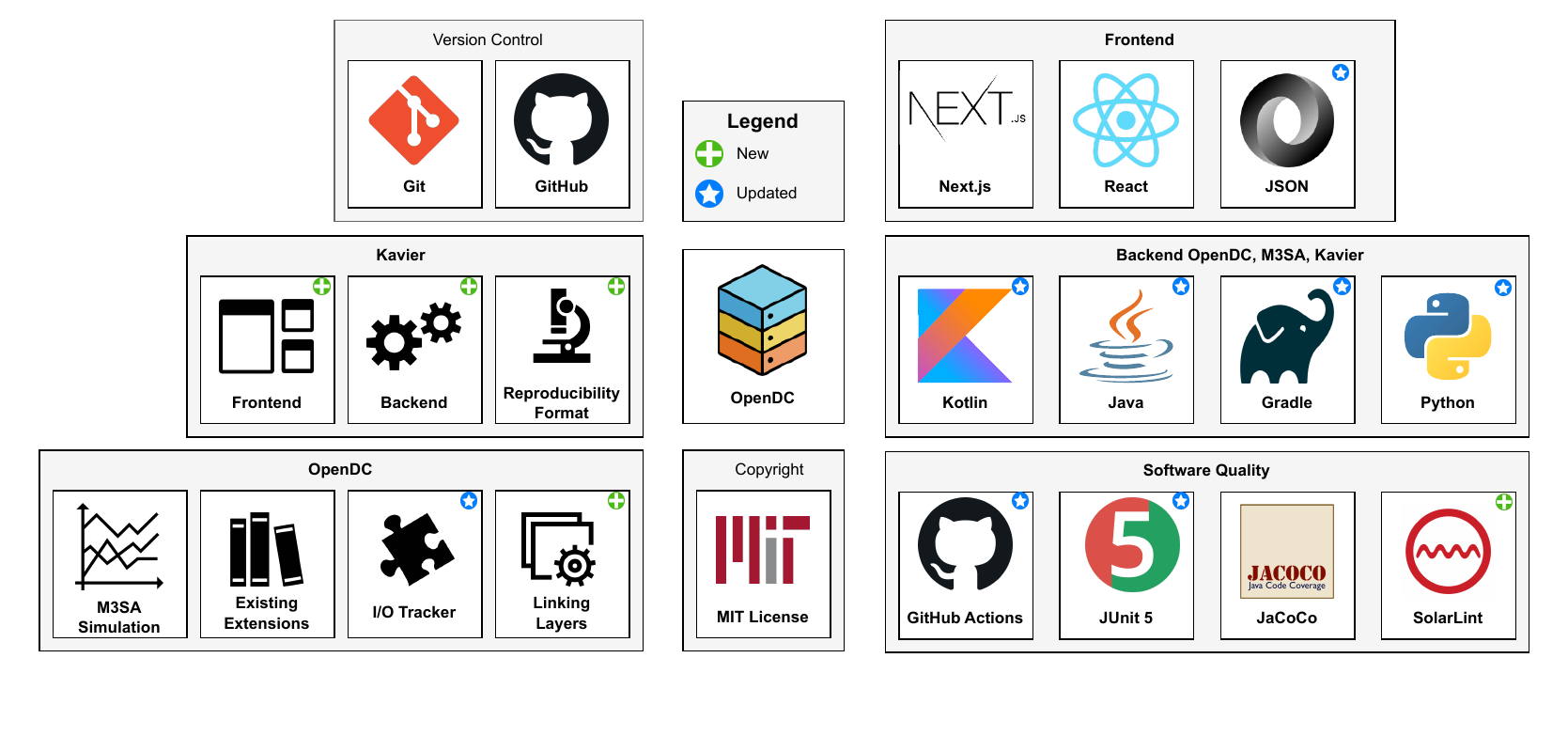}
    \vspace*{-1.5cm}
    \caption{Technologies Kavier and OpenDC use for simulating performance, sustainability, and efficiency.}
    \label{fig:prototype:technologies}
\end{figure}

We employ industry-best software development practices and technologies in the engineering process of Kavier. The main codebase of Kavier is written in Python, the second most used programming language, due to its extensive support for various libraries and frameworks~\cite{statista_programming_languages, radunicolae-hp-m3sa}. The main codebase of OpenDC, which represents the sustainability component of Kavier, is written in Kotlin, a modern and fast-growing programming language, fully interoperable with Java, and already widely adopted by large companies~\cite{jetbrains_kotlin, Mastenbroek2023RADICE, radunicolae-hp-m3sa}. \Cref{fig:prototype:technologies} shows the technologies Kavier prototype uses to simulate performance and efficiency of LLM ecosystems, and OpenDC and M3SA use to (multi-model-) simulate sustainability of such ecosystems.

We develop and envision future development of our simulator through industry-standard version control; we use Git and GitHub.
In the Kavier repository, development happens through branches and the main branch can be modified only through pull requests (in the future, assuming a large-scale adoption of the tool, pull requests would be reviewed and merged by authorized repository maintainers and techical leads). Commits messages and pull requests follow industry-standard formats employed in large-tech companies, such as Google~\cite{google_blockly_commits, google_blockly_pr}. 

We promote software quality through GitHub Actions, which run Continuous Integration (CI)~\cite{fowler_continuous_integration_CI} pipelines for each pull request. The CI pipeline consists of running automated test suites to spot code errors in functionality and simulation logic, and linting, to mandate adherence to best engineering/coding practices. 

Although these software engineering processes increase the overall burden of engineering and maintenance, they ensure high-quality implementations and integrations of simulation instruments, cross-component compatibility, and the long life of the system as Kavier and OpenDC evolve~\cite{softwareDesignPhilosophy, Mastenbroek2023RADICE, radunicolae-hp-m3sa}.
\section{Kavier Interface} \label{sec:prototype:showcase}

In this section, we detail the LLM and GPU library that Kavier provides, as well as the experimental setup and versatility that Kavier enables. Then, after establishing this operational background, we give two input examples to Kavier, the most simplistic and the most complex.

Kavier receives input through a CLI where the user can configure the specifics of the simulation, or use default setups. Unlike file-based setups, which add extra complexity to experiment configuration, or visual interfaces, which decrease the ease of adaptation or integration with other instruments, the CLI is simple to implement and maintain, easy to operate, and easy to integrate with third-party tools. 

\textit{LLM and GPU library:} To simplify the input process and setup of LLM and GPU properties, we provide an LLM library, from which the user would only select a specific LLM or GPU, instead of configuring from scratch. We include in the library 8 LLMs and 8 GPUs widely used in real-world LLM inference, where the LLMs vary in parameter size (e.g., 8B, 30B, 176B), architecture (e.g., LLama, OPT, Bloom), etc, and the GPUs vary in tensor core performance (e.g., 312 teraflops, 2,040 teraflops, 4,800 teraflops), memory (e.g., 24~GB, 80~GB, 141~GB), etc. However, if the user doesn't find the needed LLM or GPU in the library, they can append to this library by simply adding a new entry to the array.

\textit{Flags for setting up Kavier:} \Cref{table:kavier-flags-explained} describes the flags Kavier takes. For each flag, Kavier has a default value, which Kavier uses if the user leaves the field empty (e.g., the user doesn't need a specific simulation-driven experiment, but only wants to test the technical functionality of the setup).

\begin{table}[t]
\centering
\begin{tabular}{p{4cm}p{4cm}p{7cm}}
\toprule
\textbf{Flag} & \textbf{Default} & \textbf{Description}\\
\midrule
\verb!--llm! & \texttt{Llama-3-8B} & LLM prefab to simulate.\\
\verb!--gpu! & \texttt{A10} & GPU prefab to simulate.\\
\verb!--trace! & \texttt{N/A} & LLM workload trace to simulate.\\
\verb!--output_folder! & \texttt{data/output\_traces} & Output folder to save Kavier's predictions.\\
\verb!--kv_cache! & \texttt{on} & Toggles vLLM-style KV reuse.\\
\verb!--prefix_len! & \texttt{256} & Only prompts more than this many tokens populate the prefix cache (0 disables).\\
\verb!--export_rate! & \texttt{0.1} & Sets the simulation granularity, in seconds.\\
\verb!--flush_size! & \texttt{10,000} & Granularity of exporting to the output file, e.g., after 10,000 simulated prompts.\\
\bottomrule
\end{tabular}
\caption{Command-line flags to set up experiments for Kavier.}
\label{table:kavier-flags-explained}
\end{table}

We showcase in~\Cref{lst:kavier:simple-command} the simplest command Kavier takes, where the user only gives the workload trace. In contrast, we showcase in~\Cref{lst:kavier:complex-command} the most complex command Kavier takes, where the user exhaustively configures the simulation.

\begin{lstlisting}[language=bash,
                   float,
                   caption={Simplest input to Kavier.},
                   label={lst:kavier:simple-command}]
python -m kavier.main --trace name-of-the-trace.csv
\end{lstlisting}

\begin{lstlisting}[
language=bash,
float,
caption={Most detailed input to Kavier.}, 
label={lst:kavier:complex-command}]
python -m kavier.main \
  --llm Llama-3-8B \                  % LLM simulated in experiments
  --gpu A10 \                         % GPU simulated in experiments
  --trace input-workload.csv \        % the workload trace
  --outputfolder output-folder \      % the output folder
  --kv_cache on \                     % enable KV-Caching
  --prefix_cache_min_len 512 \        % prefix caching 512 tokens
  --export_rate 0.01 \                % second-granularity predictions
  --flush-size 100                    % flush granularity of 100 prompts
\end{lstlisting} 
\section{Requirement Validation}\label{sec:prototype:requirements_addressal}

We now evaluate our prototype against the requirements we established in~\Cref{sec:design:requirement-addressal}.

\subsection{Functional Requirements}
In this chapter, we implemented and integrated a prototype of Kavier that strictly matches the design proposed in~\Cref{sec:design}. The implemented prototype of Kavier (hereafter referred to as Kavier) is a discrete event simulator of LLM ecosystems under inference, with a user-configurable export rate~\ref{fr:discrete}, and can model the inference process based on the presence and absence of KV-Caching~\ref{fr:caching}. Kavier successfully predicts the performance~\ref{fr:performance}, sustainability~\ref{fr:sustainability}, and efficiency~\ref{fr:efficiency} of LLM ecosystems under inference, as we successfully validate in \Cref{sec:experiments}. Not only do we design Kavier as extensible and compatible with a peer-reviewed datacenter simulation framework, but we also implement and integrate Kavier with OpenDC~\ref{fr:modularity}, where a human-in-the-loop provides inputs, analyzes outputs, and manipulates intermediate files.

\subsection{Non-Functional Requirements}
Establishing and addressing FRs on Kavier, we answer \textit{``what it does"}~\cite{book-distributed-systems}. Now, with an engineered prototype of Kavier, we can successfully validate the design against non-functional requirements, and answer the \textit{``how well it does"}~\cite{book-distributed-systems}. In \Cref{sec:evaluation:experiment-1}, we measure the accuracy of the engineered prototype and observe an error rate (MAPE) of 7.39\% for the prefill stage and MAPE of 4.00\% for the decode stage, well below the NFR-established bar of ``under 10.00\%," thus successfully validating~\ref{nfr:accuracy}. Also in \Cref{sec:evaluation:experiment-1}, Kavier proves its efficiency by stimulating 500 GPU hours in a matter of seconds, and at second-granularity, thus meeting~\ref{nfr:speed}.

In this chapter, we engineered a prototype of Kavier, adhering to modern software design, development standards, and principles of open science~\ref{nfr:reproducibility},\ref{nfr:software-art}. We follow industry-standard technology, development, and version control pipelines, as well as software modularity. Matching \ref{nfr:software-art}, we ensure the long life potential of Kavier, an envisioned state-of-the-art component of a future digital twin for LLM ecosystems under inference.

\section{Discussion}\label{sec:prototype:discussion}

We now summarize the contributions of this chapter and envision future developments.

\textit{Summary:} In this chapter, we engineered Kavier, the first instrument capable of predicting the performance, sustainability, and efficiency of LLM ecosystems under inference. This aligns with the fifth step of the vetted AtLarge Design Process~\cite{DBLP:conf/icdcs/IosupVTETBFMT19} and addresses RQ3. 

\textit{Future work} We envision future work in maintaining and growing Kavier, from the current prototype, which serves core and basic functionality for simulating LLM ecosystems, to a tool able to mimic exhaustively end-to-end, planet-scale LLM ecosystems, simulating multi-user workloads, geo-distributed datacenters, heterogenous accelerators, multi-level caching systems, adapting scheduling, workload carbon-aware scheduling, migration, and distribution, and various caching policies.

\textit{Multi-Prompt, Multi-GPUs:} Currently, Kavier assumes one prompt running per GPU and simulates ecosystems with a single GPU. This is still valid for a prototype; inference engines such as vLLM keep the GPU usage close to maximum ($\approx$ 95-96\%) during the inference. However, for future versions, we envision scheduling as a crucial component of Kavier. This scheduler would enable operators of LLM ecosystems to analyze the impacts of different scheduling techniques (e.g., prioritizing jobs first, or batching small jobs on the same GPU) on performance, sustainability, and efficiency.

\textit{Multi-Level Caching:} Currently, Kavier assumes zero latency in the case of a prefill cache hit. This is valid for a prototype, as the process of cache searching and cache retrieval takes only tens of milliseconds, which is insignificant compared to the seconds, sometimes even minutes, taken by the LLM inference. However, we envision future work in exploring multi-level caching and exploring the tradeoffs between searching in the cache and running the inference workload. For example, if the latency of retrieving from the deepest cache level (e.g., a different datacenter) would take two seconds, while the inference itself takes one second, the system would choose inference instead of cache retrieval. 

\textit{Parallelism:} Currently, Kavier simulates sequentially, one prompt at a time. While this is already sufficient for a prototype, we envision future engineering research where Kavier would parallelize the simulation process. While simulating 500 GPU hours within 10 seconds, at second granularity, on a regular-user machine (current performance), it is even more impressive to simulate 5 GPU years within 10 seconds at second granularity, or 500 GPU hours within 10 seconds at millisecond granularity.
 \newpage
\thispagestyle{noheader}
\chapter{Trace-Based Experiments with Kavier}\label{sec:experiments} \label{sec:evaluation}

 Anticipating LLM ecosystems under inference is a critical, yet non-trivial, simulation challenge. In \Cref{sec:design}, we design Kavier, a KV-Caching-aware simulator, capable of predicting the performance, sustainability, and efficiency of LLM ecosystems under inference. Then, in \Cref{sec:prototype} we propose an engineered prototype of Kavier, which we implemented and integrated with a state-of-the-art datacenter simulator. This build-up raises the research question: \textit{(RQ4) How to evaluate a Kavier prototype with trace-based realistic scenarios?}

In this chapter, we address RQ4 by evaluating the engineered Kavier prototype against the requirements defined in \Cref{sec:design}. Then, we use Kavier's capabilities, many of which are novelties for the field, and analyze the impact of various caching policies on LLM inference performance, sustainability, and efficiency.

\section{Overview} \label{sec:design:overview}
We evaluate Kavier matching the seventh stage of the state-of-the-art AtLarge design, implementation, and valuation process of researching computer systems and ecosystems~\cite{DBLP:conf/icdcs/IosupVTETBFMT19}. Our contribution in this chapter is five-fold:

\begin{enumerate}

    \item We deploy a state-of-the-art inference engine on real-world infrastructure and engineer a tracing instrument, which we subsequently use to trace the real-world infrastructure. We leverage traces that map the relationship between the amount of prefill and decode tokens and the time required for prefill and decode across various infrastructures. We release all the obtained traces, as well as the tracing instrument, as open-source and open-science. (\Cref{sec:evaluation:tracer}).

    \item We present the experimental setup in \Cref{sec:evaluation:setup}. We run experiments through discrete-event simulation, where we use Kavier for simulating real-world setups.

    \item We analyze the impact of the prefill and decode length on the ecosystem performance. Then, we successfully validate Kavier's performance module against real-world measurements (\Cref{sec:evaluation:experiment-1}).

    \item We analyze how the presence and absence of KV-Caching affects the performance of various-sized, state-of-the-art models (\Cref{sec:evaluation:experiment-2}).

    \item Lastly, we analyze the impact of different prefix matching and caching policies on performance, and compare our simulation-driven, trace-based results with performance reports from OpenAI (\Cref{sec:evaluation:experiment-3}).

\end{enumerate}

\section{Deploying and tracing LLM ecosystems}\label{sec:evaluation:tracer}

In this section, we present our approach to measuring LLM ecosystems deployed on real-world clusters. 

\subsection{Context: what traces do we need}\label{sec:evaluation:tracer:context}

To validate Kavier's accuracy in simulating the performance of LLM ecosystems, we need traces showing, for a given infrastructure setup, the relationship between the prefill length (i.e., the number of tokens) and the time required by the ecosystem to perform the decode (i.e., the number of seconds). We need a similar trace for the decode phase. These measurements enable us to derive performance metrics, including latency and throughput. 

We present in \Cref{table:trace-example} an example of a trace that matches the content needs for our experiments and for validating Kavier's predictions. While the format structure is flexible, and columns such as \textit{latency} and \textit{throughput} are optional (they can be derived from the rest of the data), the trace should contain information on ecosystem setup, number of prefill and decode tokens, and the time needed for prefill and decode stage.

\begin{table}[t]
\centering
\begin{tblr}{
  cells = {c},
  hline{1-2,5} = {-}{},
}
Ecosystem Setup & NPT & NDT & Prefill Time & Decode Time & Latency & Throughput \\
s1                    & npt1             & ndt2            & t11           & t12          & l1       & t1          \\
s2                    & npt2            & ndt2           & t21          & t22         & l2     & t2        \\
...                  & ...            & ...           & ...          & ...         & ...     & ...        
\end{tblr}
\caption{Sample chunk of a trace containing the needed data for validating the performance tier of Kavier. NPT is the number of prefill tokens and NDT is the number of decode tokens.}
\label{table:trace-example}
\end{table}

\subsection{Context: existent traces}\label{sec:evaluation:tracer:existent-traces}

We identify a large bank of traces released as open science and with significant contributions to the community. However, none of these traces match the information setup we described in \Cref{sec:evaluation:tracer:context}.

\textit{Stojkovic et al.} release the \textit{Azure LLM inference trace 2024}, which contains three fields: timestamp, ContextTokens, equivalent to \textit{NPT} from \Cref{table:trace-example}, and GeneratedTokens, equivalent to \textit{NDT}. We present a sample from the Azure trace in \Cref{table:trace-azure-example}. We identify that this trace does not contain information about the amount of time needed per inference phase. While useful for their work published in HPCA 2025~\cite{stojkovic2025dynamollm}, this trace fails to present performance-related details.

\begin{table}[t]
\centering
\begin{tblr}{
  cells = {r},
  row{1} = {c},
  hline{1-2,5} = {-}{},
}
Timestamp                        & ContextTokens & GeneratedTokens \\
2024-05-10 00:00:00.009930+00:00 & 2,162          & 5               \\
2024-05-10 00:00:00.017335+00:00 & 2,399          & 6               \\
2024-05-10 00:00:00.022314+00:00 & 76            & 15              
\end{tblr}
\caption{Sample chunk from the \textit{Azure LLM inference trace 2024} showing context and generated token counts per request.}
\label{table:trace-azure-example}
\end{table}

\textit{Wang et al.} open the trace used in BurstGPT~\cite{wang2024burstgpt} to the public. The released traces contain six columns: timestamp, model (e.g., GPT-4), request tokens (i.e., prefill tokens), response tokens (i.e., decode tokens), total tokens, and log type (e.g., conversation log, API log). Their trace also reveals failures in the LLM inference process, which are caused by various operational phenomena. We present a sample from the BurstGPT trace in \Cref{table:trace-burst-example}. However, similarly to the Azure trace, their trace does not present performance-related details. 

\begin{table}[t]
\centering
\begin{tblr}{
  row{1} = {c},
  cell{2}{1} = {r},
  cell{2}{2} = {r},
  cell{2}{3} = {r},
  cell{2}{4} = {r},
  cell{2}{5} = {r},
  cell{3}{1} = {r},
  cell{3}{2} = {r},
  cell{3}{3} = {r},
  cell{3}{4} = {r},
  cell{3}{5} = {r},
  cell{4}{1} = {r},
  cell{4}{2} = {r},
  cell{4}{3} = {r},
  cell{4}{4} = {r},
  cell{4}{5} = {r},
  hline{1-2,5} = {-}{},
}
Timestamp & Model   & Request Tokens & Response Tokens & Total Tokens & Log Type         \\
5         & ChatGPT & 472            & 18              & 480          & Conversation log \\
825735    & ChatGPT & 94             & 11              & 105          & API log          \\
825731    & ChatGPT & 3,090           & 160             & 3,250         & API log          
\end{tblr}
\caption{Sample chunk from the \textit{BurstGPT} trace with request/response token counts and log-type metadata.}
\label{table:trace-burst-example}
\end{table}

\textit{Pan et al.} researched prefix caching techniques and released the traces they use in the Marconi paper~\cite{pan2024marconi}; in their work, the authors leveraged traces from peer-reviewed articles~\cite{zheng2023judging, sharegpt2024, yang2024sweagent, jimenez2024swebench}, used for the experiment, and then released them as part of their reproducibility capsule. These traces contain crucial information for prefix matching, which we utilize in our experimentation as input traces to evaluate the multi-tier impacts of prefix caching. We present a sample from the Marconi trace in \Cref{table:trace-marconi}. However, while this trace contains useful information for prefix caching, it does not contain performance-related tracing.

\begin{table}[t]
\centering
\begin{tblr}{
  row{1} = {c},
  cell{2}{1} = {r},
  cell{2}{2} = {r},
  cell{2}{3} = {r},
  cell{2}{4} = {r},
  cell{2}{5} = {r},
  cell{3}{1} = {r},
  cell{3}{2} = {r},
  cell{3}{3} = {r},
  cell{3}{4} = {r},
  cell{3}{5} = {r},
  cell{4}{1} = {r},
  cell{4}{2} = {r},
  cell{4}{3} = {r},
  cell{4}{4} = {r},
  cell{4}{5} = {r},
  cell{4}{7} = {fg=CodGray},
  hline{1-2,5} = {-}{},
}
session\_id & turn\_id & ts  & num\_in\_t & num\_out\_t & input\_tokens                                                            & output\_tokens                                                              \\
0           & 0        & 0.0 & 158        & 528         &
{{[}1, 8853, 3051, 1115, 376,\\~12148, 6773, 445, 5828, ...\\~1792, 9092]} &
{{[}1, 8853, 3051, 1115, 376,\\~2887, 27085, 29918, 29896, ...\\~22137, 9092]} \\
1           & 0        & 4.0 & 99         & 189         &
{{[}1, 8853, 3051, 1115, 376,\\~22550, 278, 1494, 2323, ...\\~1792, 9092]} &
{{[}1, 8853, 3051, 1115, 376,\\~1576, 1959, 1234, 338, ...\\~22137, 9092]}    \\
2           & 0        & 8.0 & 22         & 137         &
{{[}1, 8853, 3051, 1115, 376,\\~5816, 338, 278, 19087, ...\\~1792, 9092]}  &
{{[}1, 8853, 3051, 1115, 376,\\~1576, 19087, 4234, 491, ...\\~22137, 9092]}   
\end{tblr}
\caption{Excerpt of a token-level trace capturing session and turn identifiers together with the full input and output token sequences. num\_in\_t is the number of input tokens, num\_out\_t is the number of output tokens.}
\label{table:trace-marconi}
\end{table}

\subsection{Deploying  on real-world clusters}\label{sec:evaluation:tracer:deployment}
After analyzing existing traces, we conclude that, as of May 2025, no publicly available trace contains the necessary information for validation, as summarized in \Cref{table:trace-example}.

Thus, we conduct our own tracings.

\textbf{SURF:} We obtain access to real-world infrastructure from SURF, the largest datacenter provider in the Netherlands\footnote{Many thanks to my team from the Network Institute, especially to Radu Apșan and Ivano Malavolta, who helped gain access to the SURF infrastructure.}. Specifically, the offered infrastructure comprises a cluster with an NVIDIA GPU A10~\cite{nvidia_a10_2025}, on which we deploy the latest version of vLLM at the time of tracing (v0.9.0), serving Llama-3-8B~\cite{meta2024llama31}, and maintain the default settings~\cite{kwon2023efficient}.

\textbf{DAS-6:} Further, we obtain access to real-world infrastructure from the DAS-6~\cite{bal2016medium}, the set of clusters at Vrije Universiteit Amsterdam, which contains 32 nodes, and provides access to GPUs as NVIDIA A4000~\cite{nvidia_a4000_2025}, NVIDIA A6000~\cite{nvidia_a6000_2025}, and NVIDIA A100~\cite{nvidia_a100_2025}. We deployed vLLM (v0.9.0) and kept the settings default~\cite{kwon2023efficient}.

\subsection{Tracer and the tracing process}\label{sec:evaluation:tracer:tool}
We then engineer Tracer, an instrument for tracing real-world LLM deployments, tailored to the LLM serving infrastructure we deployed in~\Cref{sec:evaluation:tracer:deployment}. Similarly to the rest of our contributions, we release Tracer as open science.

\begin{figure}[t]
    \centering
    \includegraphics[width=0.95\linewidth]{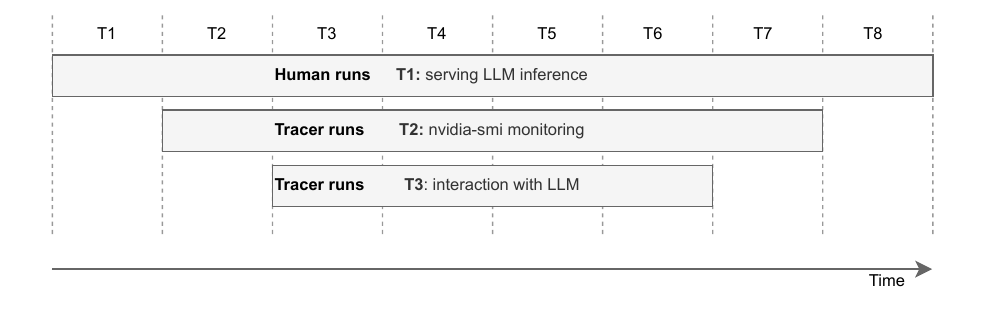}
    \caption{Threads for starting the inference engine (T1), running the measurement with NVIDIA-SMI (T2), and sending the prompt/receiving the answer (T3). Time progresses horizontally to the right, and the intervals between timestamps are considered equal, for the sake of exemplification and clarity.}
    \label{fig:tracer-timeframe}
\end{figure}

Tracer is a utility instrument that we use for automating the tracing process. Three main threads run in parallel: one for starting the inference engine, one for monitoring the cluster, particularly the GPU, and one for sending the input and receiving the output. However, the ``order of operations" matters: first, we need to run thread 1, then thread 2, then thread 3. The human (me!) runs thread 1, and Tracer runs threads 2 and 3. We illustrate in \Cref{fig:tracer-timeframe} the time frame of running and starting the threads.

\textbf{Thread 1: Running the inference} \\
Firstly, the infrastructure needs to run vLLM. To connect on DAS-6, for example, we run the following set of commands from \Cref{lst:tracer:thread1}:

\begin{lstlisting}[language=bash,
                   float,
                   caption={Thread 1 – set-up commands on DAS-6.},
                   label={lst:tracer:thread1}]
srun -p defq --gres=gpu:A6000:1 --time=00:15:00 --pty bash -i
source /var/scratch/$USER/conda/etc/profile.d/conda.sh
conda activate vllm
module load cuda12.3/toolkit
export HF_HOME=/var/scratch/$USER/hf
\end{lstlisting}

This set of commands from~\Cref{lst:tracer:thread1} selects a node by GPU, in this case, for exemplification purposes, an A6000, then reserves the node for 15 minutes (line 1). Then, the script loads Conda (line 2), vLLM (line 3), and the NVIDIA CUDA toolkit drivers (line 4). Then, to serve, e.g., LLama-3.1-8B, we run the command from \Cref{lst:tracer:thread1-serve}; we note that this command assumes the models are already loaded on the cluster, e.g., from HuggingFace~\cite{meta2024llama31}. 

\begin{lstlisting}[language=bash,
                   float,
                   caption={Thread 1 command to serve LLama-3.1-8B with vLLM.},
                   label={lst:tracer:thread1-serve}]
vllm serve meta-llama/Llama-3.1-8B
\end{lstlisting}


\textbf{Thread 2: NVIDIA-SMI sampler}

Tracer starts a remote sampler, through which it connects to the cluster and analyzes GPU utilization using NVIDIA-SMI~\cite{nvidia_smi_2025}. Tracer runs the command shown in \Cref{lst:tracer:thread2}. This command starts just before thread 3 begins, and ends just after thread 3 completes; in other words, NVIDIA-SMI measurements span the period between sending the prompt and receiving the system response.

\begin{lstlisting}[language=python,
                   float,
                   caption={Thread 2 – NVIDIA-SMI sampler.},
                   label={lst:tracer:thread2}]
cmd = [
        "ssh",
        "-i", SURF_KEY_PATH,
        "-o", "StrictHostKeyChecking=no",
        f"{SURF_USER}@{SURF_HOST}",
        (
            "nvidia-smi "
            "--query-gpu=timestamp,utilization.gpu,utilization.memory "
            "--format=csv,noheader,nounits "
            f"--loop-ms={loop_ms}"
        )
    ]
\end{lstlisting}

\textbf{Thread 3: Sending the prompt and receiving the system's output}

Lastly, Tracer sends the prompt to the system using the function presented in \Cref{lst:tracer:thread3}. This function returns the system's text response. Immediately after that, the monitoring with NVIDIA-SMI is closed, and the results are stored in files.

\begin{lstlisting}[language=python,
                   float,
                   caption={Thread 3 – sending prompt and collecting reply.},
                   label={lst:tracer:thread3}]
HEADERS = {
    "Content-Type": "application/json",
    **({"Authorization": f"Bearer {SURF_API_KEY}"})
}

def send_surf_prompt(model: str, prompt: str, max_tokens, temperature) -> str:
    payload = {
        "model": model,
        "prompt": prompt,
        "max_tokens": max_tokens,
        "temperature": temperature,
    }
    resp = requests.post(SURF_URL, headers=HEADERS, json=payload)
    ...
\end{lstlisting} 

\subsection{Traces}\label{sec:evaluation:tracer:results}
In this section, we present the traces we obtained. To ensure consistency and minimize system-dependent performance biases, we ran each measurement 10 times, then selected the median value. After each run, caches were deleted; between runs, the inference setup was kept identical. All measurements were run on the SURF infrastructure, which contains a cluster with an NVIDIA A10, serving vLLM, with a temperature of 0.8 and KV-Caching enabled.

We first measured the system's performance for the prefill phase. We sent prompts growing logarithmically in length, each of them asking \textit{"Which is the most common word in the following text? Answer in exactly \_1 word\_: LOREM IPSUM DOLOR SIT AMET..."}. We used Lorem Ipsum text, generated using~\cite{loremipsum}. We present results in \Cref{table:trace-surf-prefill}.

\begin{table}[t]
\centering
\begin{tblr}{
  cells = {r},
  row{1} = {c},
  hline{1-2,11} = {-}{},
}
Setup & PS & ML [s] & MT [tokens/s] \\
SURF, A10, LLama-3.1-8B, vLLM default & 64    & 0.054 & 1,192  \\
SURF, A10, LLama-3.1-8B, vLLM default & 128   & 0.072 & 1,776\\
SURF, A10, LLama-3.1-8B, vLLM default & 256   & 0.123 & 2,095 \\
SURF, A10, LLama-3.1-8B, vLLM default & 512   & 0.213 & 2,408  \\
SURF, A10, LLama-3.1-8B, vLLM default & 1,024  & 0.436 & 2,349 \\
SURF, A10, LLama-3.1-8B, vLLM default & 2,048  & 0.819 & 2,501  \\
SURF, A10, LLama-3.1-8B, vLLM default & 4,096  & 1.749 & 2,354   \\
SURF, A10, LLama-3.1-8B, vLLM default & 8,192  & 3.860 & 2,127  \\
SURF, A10, LLama-3.1-8B, vLLM default & 16,384 & 7.347 & 2,230  \\
\end{tblr}
\caption{Prefill performance trace. PS represents the prompt size, in tokens, ML represents the median latency, and MT represents the median throughput.}
\label{table:trace-surf-prefill}
\end{table}


Then, we measured the system's performance for the decode stage. We send prompts requesting increasingly large responses \textit{"Generate an exactly \{size\} word story about computers"}. We present results \Cref{table:trace-surf-decode}.

\begin{table}[t]
\centering
\begin{tblr}{
  cells = {r},
  row{1} = {c},
  hline{1-2,11} = {-}{},
}
Setup & RRS & MRS & ML [s] & MT [tokens/s] \\
{SURF, A10, LLama-3.1-8B, vLLM default} & 64    & 53    & 2.3   & 22.5  \\
SURF, A10, LLama-3.1-8B, vLLM default & 128   & 106   & 4.5   & 23.1  \\
SURF, A10, LLama-3.1-8B, vLLM default & 256   & 206   & 9.0   & 22.8 \\
SURF, A10, LLama-3.1-8B, vLLM default & 512   & 409   & 18.1  & 23.0 \\
SURF, A10, LLama-3.1-8B, vLLM default & 1,024  & 769   & 36.3  & 21.9 \\
SURF, A10, LLama-3.1-8B, vLLM default & 2,048  & 1,838  & 73.0  & 25.1 \\
SURF, A10, LLama-3.1-8B, vLLM default & 4,096  & 3,109  & 147.2 & 20.7 \\
SURF, A10, LLama-3.1-8B, vLLM default & 8,192  & 6,585  & 299.5 & 21.9 \\
SURF, A10, LLama-3.1-8B, vLLM default & 16,384 & 13,940 & 617.6 & 22.5 \\

\end{tblr}
\caption{Decode performance trace. RRS represents the requested response size, MRS represents the median response size, ML represents the median latency, and MT represents the median throughput.}
\label{table:trace-surf-decode}
\end{table}

\subsection{The LLM Trace Archive}
We release all the traces used in this research as FAIR dataset~\cite{GOFAIR_FAIRPrinciples}, which includes both traces leveraged from peer-reviewed scientific articles and the traces we obtained in this work, by deploying and measuring real-world LLM ecosystems. We name this archive \textit{the LLM Trace Archive.}

\textit{Societal impact:} We envision the LLM Trace Archive as a main contribution of our work; this FAIR dataset can significantly alleviate future research efforts, otherwise spent on data collection or system measurement. Furthermore, the LLM Trace Archive contains unique tracing, the first FAIR dataset in the community to map the relationship between the amount of tokens (in both prefill and decode phases) and the corresponding execution times. These traces are essential for accurately simulating performance and for validating predictions, as we show in this chapter. Lastly, the archive enables researches who don't have direct access to datacenter infrastructure to conduct experiments, thereby making a step towards equal scientific opportunities for everybody.

\textit{Future work:} Albeit already highly impactful on the community, we envision future work on the LLM Trace Archive, aided by Tracer, which would add traces for various ecosystems configurations run on various ecosystems deployments. These new traces could map the relationship between the system workload and system performance of e.g., various models (e.g., Llama, Granite), of different sizes (e.g., 8B parameter, 32B parameter), run on different GPUs (e.g., A10, A4000, A6000, A100), and with different vLLM configurations (e.g., KV enabled/disabled, different temperature varying from 0.0 to 1.0).

\section{Experimental setup}\label{sec:evaluation:setup}

We validate and run experiments matching step seven of the state-of-the-art and community-vetted AtLarge methodology on design and validation of computer ecosystems~\cite{DBLP:conf/icdcs/IosupVTETBFMT19}. To facilitate \textit{reproducibility} and \textit{consistency} among results, we run all experiments on the same physical infrastructure: an off-the-shelf MacBook Pro M3 Max, without other user programs running in the background. We run each non-deterministic experiment 10 times (e.g., performance validation) and report the standard deviation where applicable.

\textit{Marconi traces (public):} We simulate using traces from Marconi~\cite{pan2024marconi}, which leverages a set of traces from various peer-reviewed publicaitons, each of them containing real-world data anonymized. Marconi~\cite{pan2024marconi} release a set of traces which we aggregate into a singular, very-large LLM trace. We use the trace obtained in~\Cref{sec:evaluation:tracer:results} as ground truth. Marconi used these traces in their peer-reviewed paper on prefix caching on hybrid LLMs; thus, we regard this trace useful also for our work when evaluating various prefix caching policies. Furthermore, we select the Marconi trace for its volume of data, which is crucial for simulating operation of LLM ecosystems at scale. Specifically, we aggregate all traces from~\cite{pan2024marconi}, into a large trace which contains 96,870 entries, where each entry includes the user prompt and the system's response, as tokenized, the session ID, the turn ID, and the timestamp. With 3,000 sessions (i.e., 3,000 organizations, matching the terminology from~\cite{openai_prompt_caching}), we compute an average of 32.29 prompt-response pairs per session. 

\textit{CO2 trace (public):} To simulate CO2 emissions, we use a trace from ENTSO-E, leveraged and used also in our previous work~\cite{nicolae2025m3sa}. This trace was collected from ENTSO-E Transparency Platform~\cite{site:entso-e}, \textit{``an association representing 40 electricity transmission system operators from 36 countries across Europe"}~\cite{site:entso-e}. In this work, we use a trace from July 2023 monitoring the amount of CO2 emissions per Wh of energy, at 15-minute intervals, in the Netherlands. 

\begin{table}[t]
\centering
\begin{tblr}{r r r r r r r r}
\hline
Model & Source & P & L & H & $d_{h}$ & $d_{m}$ & B \\ \hline
Llama-3-8B      & Meta~\cite{meta2024llama31}      & 8  & 32 & 32 & 128 & 4,096 & 2 \\
Llama-2-13B     & Meta~\cite{touvron2023llama2}    & 13 & 40 & 40 & 128 & 5,120 & 2 \\
Granite-20B     & IBM~\cite{ibm2024granite20b}     & 20 & 52 & 48 & 128 & 6,144 & 2 \\
MPT-30B         & Mosaic~\cite{mosaicml2023mpt30b} & 30 & 48 & 64 & 112 & 7,168 & 2 \\ \hline
\end{tblr}
\caption{Configuration of the LLMs used in our experiments.  
P = parameters (billions), L = Transformer layers, H = attention heads, $d_{h}$ = dimension per head, $d_{m}$ = hidden dimension, B = precision in bytes (2 = FP16).}
\label{table:experiment-setup:llm-setup}
\end{table}

\textit{LLM Models:} We validate Kavier against real-world measurements and traces, with an identical experimental setup: Llama-3-8B, vLLM (default settings, v0.9.0), A10. Then, throughout the experimentation process, we consider that the LLMs are deployed via vLLM. For the experimentation process, we select four state-of-the-art LLMs from Kavier's LLM library, from different industry leaders, and with various configurations. We represent the specifications of each LLM we use in \Cref{table:experiment-setup:llm-setup}.

\begin{table}[t]
\centering
\begin{tblr}{r l r r r r r r r}
\hline
GPU & Vendor & M & B & FP16 & C & F & $P_\text{min}$ & $P_\text{max}$ \\ \hline
A10-24GB   & NVIDIA~\cite{nvidia_a10_2025} & 24  & 600   & 125 & 9,216 & 1,695 & 20 & 150 \\
A100-80GB  & NVIDIA~\cite{nvidia_a100_2025} & 80  & 2,039 & 312 & 6,912 & 1,410 & 50 & 400 \\ \hline
\end{tblr}
\caption{Configuration of the GPUs used in our experiments. M = memory (GB), B = bandwidth (GB/s), FP16 = tensor-core throughput (TFLOPS/s), C = CUDA cores, F = boost frequency (MHz), $P_\text{min}$/$P_\text{max}$ = power draw (W).}
\label{table:experiment-setup:gpu-setup}
\end{table}

\textit{GPU Units:} Throughout the experimentation process, we consider GPUs running vLLM, keeping default settings, as in v0.9.0. Throughout the experimentation process, we use A10 and A100, matching the real-world configurations, also represented in~\Cref{table:experiment-setup:gpu-setup}.

\section[Exploring Kavier accuracy and performance when simulating real-world LLM-inference processes]{Exploring Kavier accuracy and performance when simulating real-world LLM-inference processes}\label{sec:evaluation:experiment-1}

In this experiment, we investigate through discrete-event simulation the impact of input length (prompt size) and output length (LLM response) on performance. We first analyze the exponentially growing prefill and decode sizes and compare them with tracings of real-world deployments from \Cref{sec:evaluation:tracer:results}, thus successfully validating against the established accuracy~\ref {nfr:accuracy}. Then, we explore the performance of Kavier through large-trace experiments~\ref{nfr:speed}, and showcase the efficiency superiority of the simulation approach compared to running real-world experimentation.

\begin{figure}[t]
    \centering
    \includegraphics[width=0.75\linewidth]{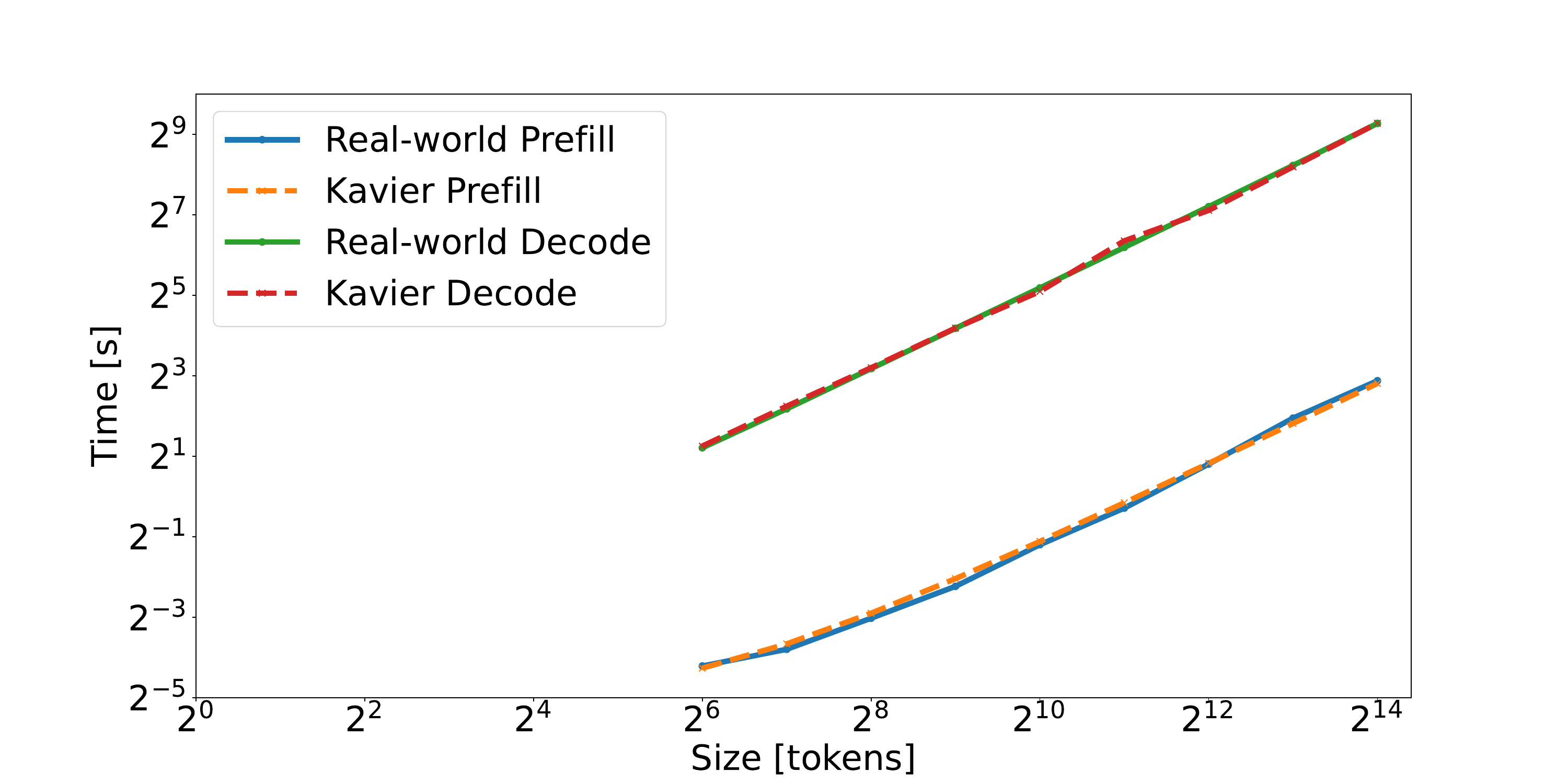}
    \caption{Kavier's predictions on prefill and decode time compared to the measured reality. The vertical axis depicts time, while the horizontal axis depicts the size/amount of tokens; specifically, the horizontal axis shows, for prefill, the amount of prefill tokens and, for decode, the amount of decode tokens. The MAPE for prefill time is 7.39\%, and the MAPE for decode time is 4.00\%.}
    \label{fig:experiment1:prefill-decode-reality}
\end{figure}

We run simulations of various prompts, which vary in prefill size exponentially between $2^{6}$ and $2^{14}$, and range in decode size logarithmically between $ 2^ {6} $ and $2^{14}$. \Cref{fig:experiment1:prefill-decode-reality} shows, on logarithmic scales (both vertical and horizontal), the simulated prefill and decode time against real-world measurements (\ref{capabilities}). 

We identify a constant gap of 1-2 orders of magnitude between the prefill time and decode time, thus emphasizing the heavy computation involved in the decoding stage and the lightweight computation from the prefill stage. Even for very-large prefill lengths of 16,384 tokens (i.e., $2^{14}$), the elapsed prefill time is under 10 seconds; in contrast, for the same very-large length, the elapsed decode time is over 500 seconds, approximately 9-10 minutes.

    Addressing \ref{nfr:accuracy}, we quantify the accuracy of our simulation instrument by measuring the MAPE error ratio against ground-truth. According to \ref{nfr:accuracy}, Kavier should model reality with an error rate of at most 10\%. However, in this experiment, Kavier achieves an MAPE of 7.39\% for the prefill time and a MAPE of 4.00\% for the decode time, successfully fulfilling \ref{nfr:accuracy}, and leading to \ref{errorrates}.

\begin{custombox}
  \finding{capabilities}{Kavier can simulate both stages of inference and model-specific behaviour.}
  \finding{errorrates}{Kavier simulates prefill with an error rate of 7.39~\% and decoding with an error rate of 4.00~\%.}
\end{custombox}

\begin{figure}[t]
    \centering
    \includegraphics[width=0.75\linewidth]{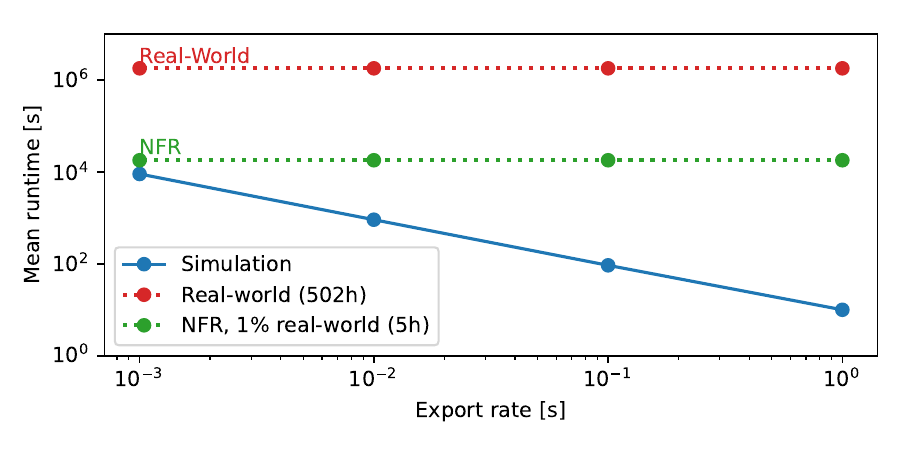}
    \vspace*{-0.5cm}
    \caption{Measurements of Kavier's performance across various export rates, compared to \ref{nfr:speed} requirements, and to the equivalent of running the experiment in a real-world setup. The vertical axis depicts time, and the horizontal axis depicts the export rate set for Kavier.}
    \label{fig:experiment1:performance}
\end{figure}

\begin{table}[t]
  \centering
  \begin{tabular}{|r|r|r|r|}
    \hline
    Export Rate\,[s] & Mean Time\,[s] & $\sigma$\,[s] & $\sigma$\,[\%] \\
    \hline
    1     & 9.9     & 0.1  & 1.0 \\
    0.1   & 92.8    & 0.8  & 0.9 \\
    0.01  & 914.8   & 4.8  & 0.5 \\
    0.001 & 9,039.7 & 21.7 & 0.2 \\
    \hline
  \end{tabular}
  \caption{Raw data from to Figure~\ref{fig:experiment1:performance}; $\sigma$ is the standard deviation over 10 runs, $s$ represents seconds.}
\label{fig:experiment1:stats}
\end{table}

Addressing \ref{nfr:speed}, we quantify Kavier's performance through a real-world trace, which aggregates all the traces released from Marconi~\cite{pan2024marconi}. This trace contains 96,869 tasks spanning over 502.1 GPU hours. According to \ref{nfr:speed}, Kavier should simulate in less than 1\% of the equivalent of a real-world experiment; in this case, Kavier should simulate 502.1~GPU-hours in less than 5 hours, on a regular user machine. We identify a trade-off between simulation granularity and performance; the higher the granularity, the longer it takes to simulate (and vice versa). Thus, we evaluate Kavier's performance for export rates of 1~second, 100~ms, 10~ms, and 1~ms, and present the results in \Cref{fig:experiment1:performance}. We observe that Kavier simulates the workload in under 10 seconds, at second-granularity~(\ref{mf:sim500h}), and even meets the established \ref{nfr:speed}, for millisecond granularity~(\ref{mf:sim_ms}).

\begin{custombox}
  \finding{mf:sim500h}{Kavier can simulate 500 GPU hours in 10 seconds, at second-granularity.}
  \finding{mf:sim_ms}{Kavier can simulate at millisecond granularity (2.5 hours), and still run in under 1\% of the real-world equivalent (500~GPU-hours)}
\end{custombox}
\section[Analyzing the Impact of KV-Caching on LLM-Inference Performance]{Analyzing the Impact of KV-Caching on \\ LLM-Inference Performance}\label{sec:evaluation:experiment-2}

In this experiment, we investigate the impact of KV-Caching presence and absence on ecosystem performance through simulation aided by Kavier, and analyze how KV on/off affects various 8B-parameter LLMs.

\textit{1,000$\times$Marconi Trace:} For this experiment, we use the Marconi trace, one thousand times, to simulate massive-scale, real-world operation. While the original Marconi trace contains 3,000 sessions (i.e., 3,000 users each with one session), and an average of 32.29 prompt-response pairs per session, in this experiment, we up-scale the input trace to 3 million sessions, each with the same average of 32.29 prompt-response pairs per session.

\textit{Caching:} In this experiment, we equip Kavier with a no-prompt prefix caching policy. We focus only on the impacts of token KV-Caching, where \textit{``computation of a new token depends on interactions between its embedding and the previously stored intermediate KV-Cache tensors"}~\cite{DBLP:conf/sosp/ZhangDLKMWLYLLZ25}. Since \textit{Vaswani et al.} introduced KV-Caching in \textit{``Attention Is All You Need''}, in 2017~\cite{vaswani2017attention}, KV-Caching became an industry standard and is widely used in LLM ecosystems~\cite{huggingface-kv-cache, vaswani2017attention, DBLP:conf/sosp/ZhangDLKMWLYLLZ25}. While KV-Caching has little impact on the prefill time, it reduces the time complexity for the decode phase from quadratic to linear; thus, in this experiment, we evaluate only the decode phase.

\textit{Simulated Infrastructure:} We evaluate the impacts of KV-Caching on and off policy as run on an NVIDIA A100-80GB~\cite{nvidia_a100_2025}. We simulate state-of-the-art LLMs, widely used in real-world setups, and growing in parameter sizes; we simulate Meta's LLama-3-8B~\cite{meta2024llama31}, Meta's Llama-2-13B~\cite{touvron2023llama2}, IBM's Granite-20B~\cite{ibm2024granite20b}, and Databricks' (MosaicML's) MPT-30B~\cite{mosaicml2023mpt30b}, all of them open-source and included in the LLM Library of Kavier. All the experiments run in this section total approximately 58 GPU years, as run on an NVIDIA A100. Although we do not have access, nor the physical time, to run these massive-scale experiments on a real-world NVIDIA A100-80GB, Kavier aids in predicting and anticipating how such real-world ecosystems would operate, in an availability-, time-, and cost-efficient way, in under 1 hour. 

\begin{figure}[t]
    \centering
    \includegraphics[width=0.75\linewidth]{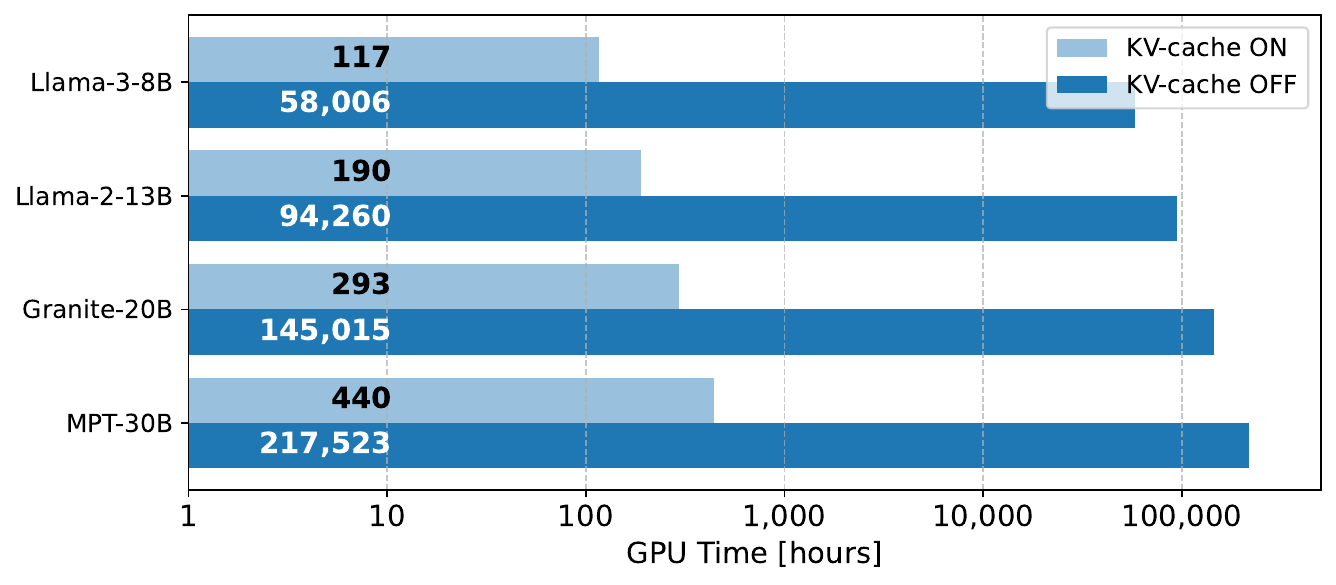}
    \caption{Impact of the presence and absence of KV-Caching on decode performance on industry state-of-the-art models.}
    \label{fig:evaluation:exp2:kv_cache_prefill_time}
\end{figure}

\Cref{fig:evaluation:exp2:kv_cache_prefill_time} shows the performance of the four different models, using and not using KV-Caching. On the vertical axis, we represent the used models, of 8, 13, 20, and 30 billion parameters, and the presence or absence of KV-Caching. On the horizontal axis, we represent the total GPU time, represented on a logarithmic scale, required to run the given workload. We identify a 2 to 3 order magnitude gap between the presence and the absence of KV-Caching~(\ref{mf:kv-caching-497}). Thus, while the absence of KV-Caching would lead to a total of 58.76 GPU years, the adoption of KV-Caching reduces the computation time by a factor of 497x, to only 0.11 GPU years~(\ref{mf:kv-caching-497}). Moreover, we observe a direct relationship, converging to a linear growth, between the model size (number of parameters) and the decode time, for both policies of using and not using KV-Caching.

We identify the 2 to 3 orders of magnitude difference between KV-Caching enabled and disabled as a direct consequence of their fundamentally different time complexities. Specifically, the absence of KV-Caching in autoregressive token generation (decode phase) has a time complexity of $O(n^2)$ ($n$ is the number of tokens in the decode sequence), and each new token needs to recompute attention over the entire sequence generated so far~\cite{vaswani2017attention}; to recompute attention, the attention mechanism repeatedly performs computations over a continuously growing set of previously generated tokens. However, the presence of KV-Caching allows for keeping cached previously conducted computations (i.e., token matrices), and computing only new, unseen, and not caches token multiplications.  The presence of KV-Caching reduces time complexity from quadratic ($O(n^2)$) to linear ($O(n)$).

This experiment thus emphasizes the need for a detailed analysis of the impacts of caches on performance. In general, caches offer significant performance benefits when workloads grow; for example, processing units (e.g., CPU, GPU) use multi-level caches to reduce latency when accessing frequently-used instructions and data, and adopt principles of temporal and spatial locality. However, caching can affect performance if workloads exhibit low locality, such as in cache-trashing scenarios in these processing units where frequent and incorrect cache misses leads to worsen performance, instead of improved performance. 

We argue that caching in LLM ecosystems can reflect similar behaviour to caching in processing units. However, there is currently a gap in understanding the degree, sometimes magnitude, to which caching can help or ``dishelp". We envision significant future work, both by us and by the community, in analyzing the impact of caches on the performance and subsequent aspects of LLM ecosystems.

\begin{custombox}
  \finding{mf:kv-caching-497}{KV-Caching can improve ecosystem performance by 2 to 3 orders of magnitude; in this experiment, KV-Caching improves performance by 497x.}
  \finding{mf:kv-on-off-time}{Simulation enables prediction of 59 GPU years in under 1 hour.}
\end{custombox}

\section[Analyzing the Impact of Prompt-Prefix Caching on Performance, Sustainability, and Efficiency]{Analyzing the Impact of Prompt-Prefix Caching Policies on LLM-Inference Performance, Sustainability, and Efficiency}\label{sec:evaluation:experiment-3}

In this experiment, we analyze the impact of prefix caching on prefill performance through discrete-event simulation aided by Kavier.

\textit{Prompt prefix matching - experiment setup:} In this experiment, we equip Kavier with a \texttt{Least Recently Used} \texttt{(LRU)} cache eviction policy. We simulate \textit{session caches} where users can benefit only from their own prefill caches, and caches are not shared between users. We also identify the existence of \textit{global caches}, which are a large, global, and shared pool of caches among all users; in this experiment, we do not simulate global caches. Lastly, in our experiment, we set various maximum capacities of this cache, between 2 and 64 prompts, and we analyze cache hit ratios.

\textit{Prompt prefix matching - OpenAI setup:} OpenAI acknowledges they use a prompt-prefix caching technique, available for 1,024 tokens or more, where \textit{``only the prompt itself is cached, while the actual response is computed anew each time based on the cached prompt"}~\cite{openai_prompt_caching}. OpenAI uses a \texttt{system-load}-based eviction policy, where cached prefixes remain active for 5-10 minutes or up to one hour during off-peak hours~\cite{openai_prompt_caching}. OpenAI claims to be using the equivalent of what we define as \textit{session caches}, which helps them reduce latency by up to 80\% and costs by up to 75\%~\cite{openai_prompt_caching}.

\subsection{Exploring matching prompt prefix length and size caches}

We analyze various prefix caching policies, specifically disabling prefix caching and setting the minimum matching tokens to 1,024 (used by OpenAI~\cite{openai_prompt_caching}), 2,048, and 4,096. We select the baseline at 1,024, and consider this number as the industry standard, and the minimum matching size of the prefix matching for which the cache hits are still helpful for accuracy; we select twice and four times higher prefix matching sizes, thus higher caching strictness, which in theory should ensure better accuracy when cache hits occur. 

\begin{figure}[t]
    \centering
    \begin{minipage}[t]{0.49\linewidth}
        \centering
        \includegraphics[width=\linewidth]{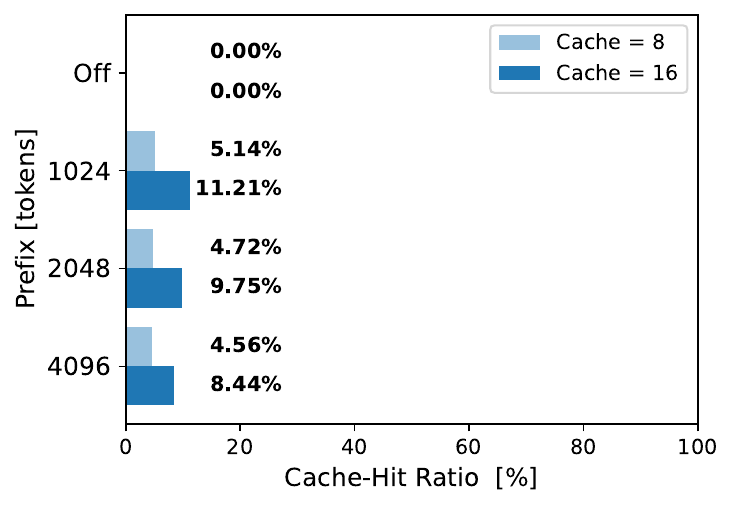}
        \caption{Prefix matching of various sizes against cache hit ratio. We measure with a cache size of 8 and 16 prompts, and an \texttt{LRU} eviction policy.}
        \label{fig:exp3_hit_ratio}
    \end{minipage}
    \hfill
    \begin{minipage}[t]{0.49\linewidth}
        \centering
        \includegraphics[width=\linewidth]{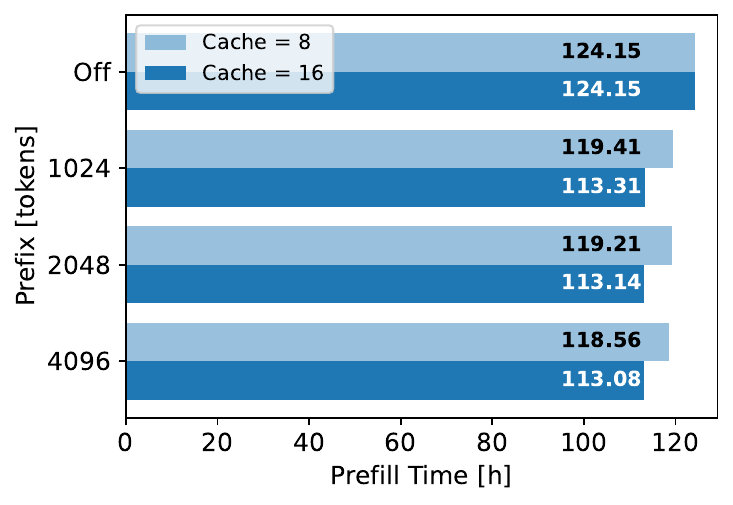}
        \caption{Prefill latency vs. prefix matching sizes. We measure with a cache size of 8 and 16 prompts, and an \texttt{LRU} eviction policy.}
        \label{fig:exp3_prefill_time}
    \end{minipage}
\end{figure}

We evaluated with caches of up to 8 prompts and 16 prompts. Considering the in-session scope of caches, caches of 8 and 16 prompts should be already sufficient, as the average conversation with an LLM has an average of \textit{``8.95 turns (n.b., prompts) per dialogue"}~\cite{shim2025tooldial} and \textit{``65.5\% of conversations finish within 10 turns"}~\cite{deng2023early}. We measured the impact of these setups on the Cache-Hit Ratio and Prefill time and show the results in~\Cref{fig:exp3_hit_ratio} and~\Cref{fig:exp3_prefill_time}. 

In~\Cref{fig:exp3_hit_ratio}, we represent the impact of the size of prefix matching on the cache-hit ratio and observe that a cache size of 16 prompts has approximately a double cache-hit ratio compared to a cache size of 8 prompts. We also identify a slight decreasing trend in the cache hit ratio as the prefix tokens increase and, thus, the caching policies become stricter. For 1,024 prefix tokens, the industry-standard (OpenAI) prefix caching policy yields a cache hit rate of 5.14\% and 11.21\%, for caches of maximum 8 and 16 prompts, respectively~(\ref{mf:hitratio16}, \ref{mf:openai_hitratio}).

\begin{custombox}
  \finding{mf:hitratio16}{In this experiment, we identify cache sizes of 16 prompts as having a twice higher cache-hit ratio than cache sizes of 8 prompts).}
  \finding{mf:openai_hitratio}{For minimum prefix caching size of 1,024 tokens the cache-hit ratio is 5.14\% for caches of 8 prompts and 11.21\% for caches of 16 prompts.}
\end{custombox}

In~\Cref{fig:exp3_prefill_time}, we illustrate the equivalent real-world prefill time for running the Marconi aggregated trace, which complements~\Cref {fig:exp3_hit_ratio}. Simulating the \texttt{no-caching} policy, we observe a total prefill time of 124.15~GPU-hours, while caching can reduce latency by up to 11 GPU-hours. For cache sizes of up to 8 prompts, we observe an average improvement of approximately 5.1 hours, equivalent to a 4.0\% improvement relative to no caching~(\ref{mf:cache-of-8}). For a cache size of up to 16 prompts, the average relative improvement is 8.8\%~(\ref{mf:latency16}), which is already substantial for SLOs and QoS when running LLM ecosystems at a societal scale. Lastly, we identify a relative improvement for the 1,024 tokens policy of 8.7\%.

\begin{custombox}
  \finding{mf:cache-of-8}{Prefix caching can reduce latency by 4.0\%, over caching of 8 prompts.}
  \finding{mf:latency16}{Prefix caching can reduce latency by 8.8\%, over caching of 16 prompts.}  
\end{custombox}

OpenAI reports that, using prompt prefix caching of 1,024 tokens, and a cache eviction policy based on system load, \textit{``can reduce latency by up to 80\% and cost by up to 75\%"}~\cite{openai_prompt_caching}. We observe a one-order-of-magnitude gap between our findings in this experiment and the improvements reported by OpenAI~\cite{openai_prompt_caching}. We identify three main possible causes for this finding.

\textit{Potential cause 1:} OpenAI's experimental setup and our experimental setup differ in the cache eviction policy (we use \texttt{LRU}, while OpenAI uses \texttt{system-load}), in cache size (we use various sizes for the cache, while OpenAI does not report the size), and in the input trace (our trace and their trace most probably don't coincide). Despite attempting to reproduce their experiments, OpenAI does not release the experimental setup or the used traces as open-source. Thus, we cannot investigate the eviction policy or the impact of the input trace further (we can further explore the effect of the cache size, which we do in \Cref{sec:evaluation:experiment-2:cache-size}). However, it seems unlikely that this specific difference in setup leads to such a large performance gap as identified in~\ref{mf:latency16}.

\textit{Potential cause 2:} OpenAI could be using \textit{global caches}, instead of \textit{session caches}. This is a large difference that goes beyond our setup and, because of the much higher potential to optimize when using the much larger global cache and its superior oversight on all prompts, it appears a likely explanation of the performance gap between our MF8 and OpenAI's reported performance. However, due to the closed-source nature of OpenAI's operational pipelines, we are unable to investigate this aspect further.

\textit{Potential cause 3:} Our measurements and reports, or OpenAI's measurements and reports, or both, may contain core errors that would affect the final reported results. For potential external validation of results and future research, we release all traces and codebase as open science. However, we cannot investigate the experimental process of OpenAI's measurements since this information is not publicly available.

\subsection{Exploring the implications of cache size}\label{sec:evaluation:experiment-2:cache-size}

Further exploring \textit{potential cause 1}, we analyze how the size of the cache, measured by the number of prompts it can hold, impacts the cache-hit ratio and the total GPU time. 

\begin{figure}[t]
    \centering
    \includegraphics[width=0.75\linewidth]{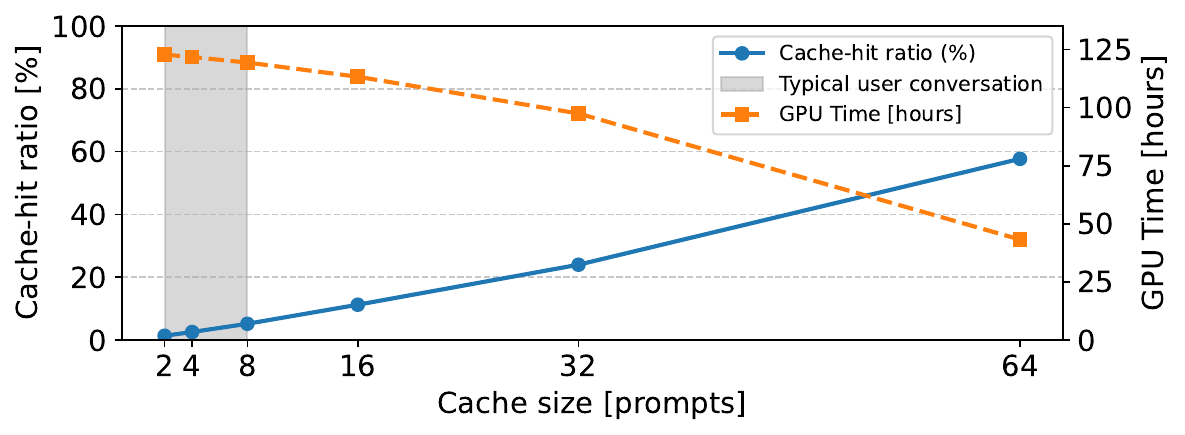}
    \caption{Impact of the size of in-session caches on cache hit ratio and total GPU time.}
    \label{fig:exp3_hit_ratio_cache_size}
\end{figure}

\Cref{fig:exp3_hit_ratio_cache_size} illustrates our findings. On the horizontal axis, we represent the size of the cache, growing exponentially from 2 to 64 prompts; on the left vertical axis and with the blue straight line, we illustrate the cache-hit ratio; on the right vertical axis and with the orange discontinuous line, we illustrate the GPU time measured in hours. 

We identify an increasing trend in the cache-hit ratio and a decreasing trend in GPU time, both of which are expected trends since the size of the cache increases and, thus, more prompts can be cached. However, the growing and diminishing trends are unexpectedly large, where the cache hit rate increases from 1.3\% for caches of 2 prompts to 57.7\% for caches of 64 prompts~(\ref{mf:cache64}), and the GPU total time (i.e., total latency) decreases from 122.83 hours to 43.21 hours.

Focusing on the cache size of 64 prompts, we observe a 65.2\% improvement in latency (relative to the absence of caching and 124.15 GPU hours), and expect a similar scale improvement in costs, both financial and environmental. While these findings match, at least in scale, the number reported by OpenAI~\cite{openai_prompt_caching}, we argue that caches of 64 prompts are challenging to operate from both computational and usability perspectives. 

From a computational perspective, storing individual caches of 64 prompts for each session would rapidly overwhelm the ecosystem's resources as the number of users scales to millions. LMSYS-Chat-1M, a large dataset containing one million real-world conversations from 25 LLMs, reports an average of 69.5 tokens per prompt~\cite{zheng2023lmsys}; for only 1 million concurrent users, all using GPT-4o-mini (an 8 billion parameter model, the smallest LLM OpenAI provides), 64 prompts per session, and 69.5 tokens per prompt, results in 2.33~GB per user, and 2.33~PB for hosting 1 million users at once. \eqref{eq:memory-caching-openai-1}-\eqref{eq:memory-caching-openai-6} show our computations.

\begin{align}
T &= M_t \times M_p \times M_s \times U \label{eq:memory-caching-openai-1} \\
M_t &= 2 \times 2 \times 32 \times 4096 = 524,288 \text{ bytes} = 0.52 \text{ MB per token} \\
M_p &= 524,288 \times 69.5 = 36,438,016 \text{ bytes} = 36.44 \text{ MB per prompt} \\
M_s &= 36.44 \times 64 = 2,332.16 \text{ MB} = 2.33 \text{ GB per user (session)} \\
U   & = 10^6 \text{ users (sessions)} \\
T &= 2.33 \times 10^6 \text{ GB} = 2.33 \text{ PB}
\label{eq:memory-caching-openai-6}
\end{align}
\begin{small}
\textit{where $T$=total memory, $M_t$=memory per token, $M_p$=memory per prompt, $M_s$=memory per session, $U$=users.}
\end{small}

From a usability perspective, the average user session does not even reach 64 interactions; according to~\cite{mcnichols2025studychat, deng2023early, shim2025tooldial}, the average chat contains less than 10 interactions (\cite{mcnichols2025studychat} reports an average of 3.5 interactions, \cite{deng2023early} \textit{``reports that over 65.5\% of conversations finish within 10 turns"}, \cite{shim2025tooldial} report an average of \textit{"8.95 turns per dialogue"}). This means that, usability-wise, the expected cache-hit ratio would be that corresponding to between 3.5 and 10 prompts; so, at most 10\% cache-hit ratio (see shaded area in Figure 6.7), far off the maximum of above 65\% for 64 or more prompts per dialogue.

We thus conclude that, given our experimental setup, storing in-session caches and reducing latency by costs by 75\%, respectively 80\%, is computationally- and usability- wise challenging~(\ref{mf:computationally}).

\begin{custombox}
  \finding{mf:cache64}{Session caches of 64 prompts can lead to 57\% cache hit ratios, and improve lantecy by 65\%. }
  \finding{mf:computationally}{Session caches of 64 prompts, with small models (8B parameters, e.g., GPT 4o-mini), and 1 million concurrent users, would constantly occupy 2.33~PB of caches. This is computationally- and usability- wise challenging.}
\end{custombox}

\subsection{Performance, Sustainability, Efficiency}

Throughout this experiment, we leveraged the capabilities of the Kavier prototype for predicting performance~\ref{fr:performance}, sustainability~\ref{fr:sustainability}, and efficiency~\ref{fr:efficiency}. In total, \Cref{sec:evaluation:experiment-3} contains measurements spanning over 2,500 GPU (A10) hours in a real-world setup, and only 0.6 simulation hours on a regular-user machine (i.e., not a supercomputer)~\ref{fr:performance},~(\ref{mf:exp3-performance}). Similarly, the simulation approach consumed approximately 7,075x less energy than the real-world simulation equivalent~\ref{fr:sustainability}~(\ref{mf:exp3-sustainability}). 

Lastly, through the Kavier performance module, we computed the financial efficiency and ratio between real-world-based and simulation-based experimentation~(\ref{mf:exp3-efficiency}). We consider the hourly running cost of the personal machine to be equal to the cost of renting an A10.

\begin{align}
    R = \frac{E_\text{sim}}{E_\text{real}}
    = \frac{\frac{P_{\text{sim}}}{\frac{T}{\Delta T_{\text{sim}}}} }{\frac{P_{\text{real}}}{\frac{T}{\Delta T_{\text{real}}}} }
    = \frac{P_{\text{sim}} \times \Delta T_{\text{sim}} \times \cancel{T}}{P_{\text{real}} \times \Delta T_{\text{real}} \times \cancel{T}}
   = \frac{\left( \cancel{P_{\text{A10}}} \times \Delta T_{\text{sim}}\right)  \times \Delta T_{\text{sim}}}{\left( \cancel{P_{\text{A10}}} \times \Delta T_{\text{real}} \right) \times \Delta T_{\text{real}}} 
    = \frac{\Delta T_{\text{sim}}^2} {\Delta T_{\text{real}}^2}
    = \frac{0.6^2} {2500^2} \approx 1:173,000,000
\end{align}
\begin{small}
    \textit{
    where $R$ is the financial efficiency ratio between simulation and reality, 
    $\text{sim}$ refers to simulation-driven experimentation, 
    $\text{real}$ refers to reality-driven experimentation,
    $E$ is efficiency, 
    $P$ is price, 
    $\Delta T$ is time, 
    $T$ is the amount of processed tokens.}
\end{small}

\ref{mf:simulation} - \ref{mf:exp3-efficiency} successfully validate the main function requirement \textit{MFR:} \textit{"Simulate performance, sustainability, and efficiency of LLM ecosystems under inference"}, and prove the superiority of our proposed approach, simulation-based experiments, over real-world-based experiments \ref{mf:exp3-efficiency}.

\begin{custombox}
    \finding{mf:simulation}{Kavier enables conducting real-world experiments in a time and cost-efficient way, through discrete-event simulation.}
    
    \finding{mf:exp3-performance}{Performance -- Kavier: 0.6~h, Reality: 2,500~GPU~h (via Kavier performance).}

    \finding{mf:exp3-sustainability}{Energy -- Kavier: 0.054~KWh.  Reality: 375~KWh (via Kavier Sustainability).}

    \finding{mf:exp3-efficiency}{Financial efficiency improvement --  1:173,000,000 Kavier : reality (via Kavier Efficiency).}
\end{custombox}




\section{Discussion}\label{sec:evaluation:discussion}

We now summarize the contributions of this chapter, the final content chapter of this thesis, and envision future experimentation. Many thanks, reader, if you have reached the page of our work, we hope you enjoyed the journey, 'cause we surely did!

\textit{Summary:} In this chapter, we conducted the first open-science tracings of LLM ecosystems, which show the relationship between the prefill length, the decode length, and the time required to run the prefill and decode stages. We engineered Tracer, a utility tool for tracing LLM infrastructure. We release Tracer as open-science, and we release all the traces as an open-science LLM Trace Archive.

Then, using the ground-truth measurements, we validated Kavier against the established non-functional requirements for accuracy and performance, and validated Kavier against the remaining requirements through experimentation. We explored the massive impact (Token) KV-Caching has on LLM ecosystem performance, and identified differences of 2-3 orders of magnitude between its KV-on and KV-off. Lastly, we explored prompt prefix caching and reproduced results from OpenAI through experimentation aided by Kavier; we identified discrepancies between our findings and their reports, and identified three possible causes. We then explored in depth one of these causes (the other two could not be explored because OpenAI's operational pipelines are closed source). We then validated Kavier-aided experimentation versus real-world experimentation and identified orders of magnitude improvements of the simulation approach, the largest being of 1:173,000,000 in financial efficiency improvement.

\textit{Future exploration:} Kavier enables the exploration of large-scale systems in a time- and cost-efficient manner, without requiring access to real-world ecosystems or incurring the financial, time, and configuration burdens. We envision Kavier as aiding in exploring future aspects of LLM inference, with the current prototype. 

We envision future research and exploration on how different prefix caching policies and cache eviction policies (e.g., least recently used, least frequently used, random) can impact system metrics. We also envision future research in exploring the most energy-efficient configurations while still meeting performance-sustainability real or synthetic SLOs. 
 \newpage
\thispagestyle{noheader}
\chapter{Conclusion and Future Work}\label{sec:conclusion}

In this chapter, we summarize the contributions of our work and envision future research and exploration of LLM ecosystems through simulation.

\section{Conclusion}\label{sec:conclusion:conclusion}
We investigated in this work \textit{how to enable analysis of LLM ecosystems through discrete-event simulation (MRQ)}. We identified and addressed three research questions, methodologically matching the state-of-the-art AtLarge vision on design of distributed systems and ecosystems~\cite{DBLP:conf/icdcs/IosupVTETBFMT19}. 
In \Cref{sec:intro}, we described the societal impact of LLM ecosystems and described the potential benefits of a simulation instrument of LLM ecosystems under inference. We identified two main problems: the lack of such a scientific instrument and the lack of a (robust) reference architecture of LLM ecosystems under inference, on which the simulator would map against. In \Cref{sec:background}, we provided relevant background and analyzed existing reference architectures, prior to this work. In \Cref{sec:refarch}, we designed a reference architecture for LLM ecosystems under inference and validated the architecture against real-world ecosystems. In \Cref{sec:design}, we proposed Kavier, a scientific instrument for simulating the performance, sustainability, and efficiency of LLM ecosystems under inference. We then prototyped Kavier~\Cref{sec:prototype}. In the absence of real-world traces necessary for validating Kavier, we deployed LLM ecosystems and deployed real-world infrastructure. We then successfully validated Kavier in~\Cref{sec:evaluation}, and analyzed the impact of various caching policies on ecosystem, performance, sustainability, and efficiency. We now answer each research question punctually:

\noindent
\begin{tabular}{@{}p{1cm} p{15.2cm}@{}}
    \textbf{RQ1} & \textbf{How to synthesize and validate a reference architecture of LLM ecosystems?} \\
    & In \Cref{sec:refarch}, we have conducted a literature review and analyzed existent reference architectures of LLM ecosystems under inference. We detailed positives and negatives and identified that none of the existent reference architectures are sufficient to map an LLM ecosystems simulator upon. Existent architectures are either incomplete~\cite{lu2023towards}, non-inference oriented~\cite{lu2023towards, bucaioni2025functional, mahr2024reference}, assume a (too) high degree of homogeneity of LLM ecosystems~\cite{bucaioni2025functional}, vetted, following state-of-the-art approaches in distributed systems, but too universal~\cite{DBLP:conf/ccgrid/JansenAPTI23}, or do not follow a distributed systems approach~\cite{bucaioni2025functional}. To design a robust reference architecture for LLM ecosystems under inference, we defined a set of design requirements and design principles which guide our design process. We then proposed a reference architecture of the current continuum of LLM ecosystems, mapping to real-world deployments. To validate our reference architecture, we explicitly mapped our model to four real-world ecosystems, out of which two in-detail (IBM, OpenAI) and two high-level (Ubicloud, Databricks), and against a state-of-the-art reference architecture from the scientific community.
\end{tabular}

\noindent
\begin{tabular}{@{}p{1cm} p{15.2cm}@{}}
    \textbf{RQ2} & \textbf{How to design Kavier, a scientific instrument for cache-aware simulation analysis of the performance, sustainability, and efficiency of LLM ecosystems under inference?} \\
    & In \Cref{sec:design}, we have designed Kavier adhering to stage 1 of AtLarge Design Process~\cite{DBLP:conf/icdcs/IosupVTETBFMT19}. We established a set of requirements which guide our design process, then propose a high-level design of Kavier, the first scientific instrument for predicting the performance, sustainability, and efficiency of LLM ecosystems under inference, through discrete-event simulation and cache-awareness. We designed Kavier as modular and leveraging peer-reviewed capabilities of predicting sustainability of OpenDC~\cite{DBLP:conf/ccgrid/MastenbroekAJLB21, dniewenhuis_hotcloud_footprinter}. We also designed Kavier as able to predict the continuum cache-aware, and model the impacts various caching policies (e.g., prompt prefix caching, KV-Caching) have on LLM ecosystems. We then detailed each main module of Kavier, specifically the performance module, the sustainability module, and the efficiency module. Lastly, we systematically evaluated our proposed design against established functional and non-functional requirements.
\end{tabular}

\noindent
\begin{tabular}{@{}p{1cm} p{15.2cm}@{}}
    \textbf{RQ3} & \textbf{How to implement and integrate Kavier within a peer-reviewed, discrete-event datacenter simulator?} \\
    & In \Cref{sec:prototype}, we implemented a prototype of Kavier and integrated with OpenDC. We developed Kavier following state-of-the-art software engineering practices and so industry-standard software engineering processes. We integrated Kavier and OpenDC, thus allowing Kavier to leverage the peer-reviewed capabilities of OpenDC to simulate sustainability~\cite{DBLP:conf/ccgrid/MastenbroekAJLB21, dniewenhuis_hotcloud_footprinter}.
\end{tabular}

\noindent
\begin{tabular}{@{}p{1cm} p{15.2cm}@{}}
    \textbf{RQ4} & \textbf{How to evaluate a Kavier prototype with trace-based realistic scenarios? } \\
    & In \Cref{sec:evaluation}, we collected traces for validation of Kavier and simulation-driven experimentation aided by Kavier. We identify a gap in open-source traces, as none of them was revealing the relationship between the prefill/decode size and the prefill/decode performance. To address this challenge, we deployed LLM ecosystems on real world infrastructure from SURF, engineered a utility tracing tool, and obtained traces matching the validation needs for Kavier. We then validated Kavier and identify its ability to simulate hundreds of GPU hours within seconds, with at-second-granularity predictions, and with error rates of less than 8\%. With a validated prototype, we analyzed impacts of the presence and absence of KV-Caching on massive-scale LLM inference. Laslty, we analyzed impacts of prompt prefix caching on system performance and contrasted our findings with real-world reports from OpenAI.
\end{tabular}

We released all instruments, tools, and traces as open-source and open-science. Specifically, we release a parent-repository containing:

\begin{enumerate}
    \item Kavier, the scientific instrument we designed, engineered, and validated in this work.

    \item The LLM Trace Archive, containing all the traces used in this work, both leveraged from peer-reviewed articles and obtained by us, by tracing real-world deployments.

    \item Tracer, the utility tool we built for tracing LLM ecosystems.

    \item A reproducibility capsule of all our experiments, with a guide on how to reproduce our findings.

    \item This thesis.
\end{enumerate}

The parent-repository can be found on GitHub, via \\\url{https://github.com/Radu-Nicolae/On-Simulating-LLM-Ecosystems-under-Inference}.

\section{Future work}\label{sec:conclusion:future-work}

We envision four main areas of future research, building upon our contributions from this work.

\begin{enumerate}
    \item \textit{Simulation of heterogeneous, highly-distributed LLM ecosystems:} We plan to broaden the fidelity of the models and expand the current prototype of Kavier to predicting high-heterogeneity infrastructure: multi-GPU, TPU, and NPU parallelism, network, memory contention, geo-distributed and multi-layered caches, and multi-level metric report card. By supporting simulation of highly heterogeneous and distributed infrastructure, Kavier can evolve from the current simulator status to an LLM ecosystem simulation twin, able to close to perfectly mimic reality.

    \item \textit{The LLM ecosystem Digital Twin:} Digital twins are simulation ecosystems, where simulation instruments (e.g., Kavier) are connected to the operational ecosystem. There is a continuous feedback loop between \textit{1) the LLM ecosystem} running LLM workloads (e.g., inference, fine-tuning, training), \textit{2) the metrics reported by the ecosystem} (operational data analytics, live telemetry), which are transmitted to the simulator, \textit{3) the simulator's predictions} to adjust the infrastructure such that the ecosystem operates as efficient and performant, while still meeting the SLOs and QoS, predictions which are transmitted to the LLM ecosystem and \textit{4) the LLM ecosystem} which reacts based on the simulator's predictions. There is currently no such digital twin for LLM ecosystems, nor for ICT infrastructure. We envision Kavier as taking the role of the simulator within a potential digital twin of LLM ecosystems under inference.

    \item \textit{LLM ecosystems under training workloads:} We identify simulation of the training stage as potential future work and future capabilities of Kavier. While inference represents the largest proportion of an LLM's lifetime, the training of \textit{large} language models also raises high performance, sustainability, and efficiency challenges. We envision a future version of Kavier as capable of simulating the holistic lifecycle of an LLM ecosystem, from the initial deployment in the training pipeline until the last prompt of the last user interacting with the respective LLM ecosystem. Similarly, to the inference process, we envision digital twins as crucial also for LLM training.

    \item \textit{Educating future generations:} We plan to develop educative material around Kavier and OpenDC, and deliver as a series of interactive workshops, seminars, and assignments to educate future generation of scientists and engineers, on how to responsibly use, deploy, and monitor LLM ecosystems, focusing on the model inference aspect. Thanks to the open-source nature of all our contributions, such material can be developed both by us or by other researchers and educators from the community, and can be in-depth explored by the students who would engage in these educational activities. We envision various difficulty educational materials, matching various academic ages, from highschool, to Bachelor's, Master's, and Doctorate levels. 
    Such educational activities, both emerging from this work and from other work, are essential for training future generations on systematically, in-depth, and ethically exploring, researching, and engineering LLM ecosystems, and, overall, on responsibly \textit{Massivizing Computer Systems}.
\end{enumerate}

\clearpage

\clearpage
\thispagestyle{noheader}
\bibliographystyle{ieeetr}
\bibliography{main}

@article{patterson2022carbon,
  title={Carbon emissions and large neural network training},
  author={Patterson, David and Gonzalez, Joseph and Le, Quoc and others},
  journal={arXiv preprint arXiv:2104.10350},
  year={2022}
}

@techreport{epa2023emissions,
  title={Greenhouse Gas Emissions from Transportation},
  author={Environmental Protection Agency},
  year={2023},
  institution={EPA}
}

@inproceedings{DBLP:conf/mlsys/WuRGAAMCBHBGGOM22,
  author       = {Carole{-}Jean Wu and
                  Ramya Raghavendra and
                  Udit Gupta and
                  Bilge Acun and
                  Newsha Ardalani and
                  Kiwan Maeng and
                  Gloria Chang and
                  Fiona Aga Behram and
                  Jinshi Huang and
                  Charles Bai and
                  Michael Gschwind and
                  Anurag Gupta and
                  Myle Ott and
                  Anastasia Melnikov and
                  Salvatore Candido and
                  David Brooks and
                  Geeta Chauhan and
                  Benjamin Lee and
                  Hsien{-}Hsin S. Lee and
                  Bugra Akyildiz and
                  Maximilian Balandat and
                  Joe Spisak and
                  Ravi Jain and
                  Mike Rabbat and
                  Kim M. Hazelwood},
  editor       = {Diana Marculescu and
                  Yuejie Chi and
                  Carole{-}Jean Wu},
  title        = {Sustainable {AI:} Environmental Implications, Challenges and Opportunities},
  booktitle    = {Proceedings of the Fifth Conference on Machine Learning and Systems,
                  MLSys 2022, Santa Clara, CA, USA, August 29 - September 1, 2022},
  publisher    = {mlsys.org},
  year         = {2022},
  url          = {https://proceedings.mlsys.org/paper\_files/paper/2022/hash/462211f67c7d858f663355eff93b745e-Abstract.html},
  bibsource    = {dblp computer science bibliography, https://dblp.org}
}

@techreport{iea2023netherlands,
  title={Netherlands 2023 Energy Policy Review},
  author={International Energy Agency},
  year={2023},
  institution={IEA}
}

@article{DBLP:journals/jmlr/LuccioniVL23,
  author       = {Alexandra Sasha Luccioni and
                  Sylvain Viguier and
                  Anne{-}Laure Ligozat},
  title        = {Estimating the Carbon Footprint of BLOOM, a 176B Parameter Language
                  Model},
  journal      = {J. Mach. Learn. Res.},
  volume       = {24},
  pages        = {253:1--253:15},
  year         = {2023},
  url          = {https://jmlr.org/papers/v24/23-0069.html},
  bibsource    = {dblp computer science bibliography, https://dblp.org}
}

@article{masanet2020recalibrating,
  title={Recalibrating global data center energy-use estimates},
  author={Masanet, Eric and Shehabi, Arman and Lei, Nuoa and others},
  journal={Science},
  volume={367},
  number={6481},
  pages={984--986},
  year={2020}
}

@article{andrae2015global,
  title={On global electricity usage of communication technology: trends to 2030},
  author={Andrae, Anders SG and Edler, Tomas},
  journal={Challenges},
  volume={6},
  number={1},
  pages={117--157},
  year={2015}
}

@inproceedings{DBLP:conf/sosp/ZhangDLKMWLYLLZ25,
  author       = {Chen Zhang and
                  Kuntai Du and
                  Shu Liu and
                  Woosuk Kwon and
                  Xiangxi Mo and
                  Yufeng Wang and
                  Xiaoxuan Liu and
                  Kaichao You and
                  Zhuohan Li and
                  Mingsheng Long and
                  Jidong Zhai and
                  Joseph Gonzalez and
                  Ion Stoica},
  editor       = {Youjip Won and
                  Youngjin Kwon and
                  Ding Yuan and
                  Rebecca Isaacs},
  title        = {Jenga: Effective Memory Management for Serving {LLM} with Heterogeneity},
  booktitle    = {Proceedings of the {ACM} {SIGOPS} 31st Symposium on Operating Systems
                  Principles, {SOSP} 2025, Lotte Hotel World, Seoul, Republic of Korea,
                  October 13-16, 2025},
  pages        = {446--461},
  publisher    = {{ACM}},
  year         = {2025},
  url          = {https://doi.org/10.1145/3731569.3764823},
  doi          = {10.1145/3731569.3764823},
  bibsource    = {dblp computer science bibliography, https://dblp.org}
}

@article{DBLP:journals/corr/abs-2206-03259,
  author       = {Alexandru Iosup and
                  Fernando Kuipers and
                  Ana Lucia Varbanescu and
                  Paola Grosso and
                  Animesh Trivedi and
                  Jan S. Rellermeyer and
                  Lin Wang and
                  Alexandru Uta and
                  Francesco Regazzoni},
  title        = {Future Computer Systems and Networking Research in the Netherlands:
                  {A} Manifesto},
  journal      = {CoRR},
  volume       = {abs/2206.03259},
  year         = {2022},
  url          = {https://doi.org/10.48550/arXiv.2206.03259},
  doi          = {10.48550/ARXIV.2206.03259},
  eprinttype    = {arXiv},
  eprint       = {2206.03259},
  bibsource    = {dblp computer science bibliography, https://dblp.org}
}

@article{DBLP:journals/corr/abs-2408-07326,
  author       = {Seungjae Moon and
                  Jung{-}Hoon Kim and
                  Junsoo Kim and
                  Seongmin Hong and
                  Junseo Cha and
                  Minsu Kim and
                  Sukbin Lim and
                  Gyubin Choi and
                  Dongjin Seo and
                  Jongho Kim and
                  Hunjong Lee and
                  Hyunjun Park and
                  Ryeowook Ko and
                  Soongyu Choi and
                  Jongse Park and
                  Jinwon Lee and
                  Joo{-}Young Kim},
  title        = {{LPU:} {A} Latency-Optimized and Highly Scalable Processor for Large
                  Language Model Inference},
  journal      = {CoRR},
  volume       = {abs/2408.07326},
  year         = {2024},
  url          = {https://doi.org/10.48550/arXiv.2408.07326},
  doi          = {10.48550/ARXIV.2408.07326},
  eprinttype    = {arXiv},
  eprint       = {2408.07326},
  bibsource    = {dblp computer science bibliography, https://dblp.org}
}

@article{DBLP:journals/corr/abs-2501-11006,
  author       = {Shashikant Ilager and
                  Lukas Florian Briem and
                  Ivona Brandic},
  title        = {{GREEN-CODE:} Optimizing Energy Efficiency in Large Language Models
                  for Code Generation},
  journal      = {CoRR},
  volume       = {abs/2501.11006},
  year         = {2025},
  url          = {https://doi.org/10.48550/arXiv.2501.11006},
  doi          = {10.48550/ARXIV.2501.11006},
  eprinttype    = {arXiv},
  eprint       = {2501.11006},
  bibsource    = {dblp computer science bibliography, https://dblp.org}
}

@article{alzaabi2023chatgpt,
  title={ChatGPT applications in academic research: a review of benefits, concerns, and recommendations},
  author={AlZaabi, Adhari and ALamri, Amira and Albalushi, Halima and Aljabri, Ruqaya and AalAbdulsalam, AbdulRahman},
  journal={Biorxiv},
  pages={2023--08},
  year={2023},
  publisher={Cold Spring Harbor Laboratory}
}

@article{kim2023chatgpt,
  title={ChatGPT and large language model (LLM) chatbots: The current state of acceptability and a proposal for guidelines on utilization in academic medicine},
  author={Kim, Jin K and Chua, Michael and Rickard, Mandy and Lorenzo, Armando},
  journal={Journal of Pediatric Urology},
  volume={19},
  number={5},
  pages={598--604},
  year={2023},
  publisher={Elsevier}
}

@article{meyer2023chatgpt,
  title={ChatGPT and large language models in academia: opportunities and challenges},
  author={Meyer, Jesse G and Urbanowicz, Ryan J and Martin, Patrick CN and O’Connor, Karen and Li, Ruowang and Peng, Pei-Chen and Bright, Tiffani J and Tatonetti, Nicholas and Won, Kyoung Jae and Gonzalez-Hernandez, Graciela and others},
  journal={BioData mining},
  volume={16},
  number={1},
  pages={20},
  year={2023},
  publisher={Springer}
}

@article{liang2024mapping,
  title={Mapping the increasing use of LLMs in scientific papers},
  author={Liang, Weixin and Zhang, Yaohui and Wu, Zhengxuan and Lepp, Haley and Ji, Wenlong and Zhao, Xuandong and Cao, Hancheng and Liu, Sheng and He, Siyu and Huang, Zhi and others},
  journal={arXiv preprint arXiv:2404.01268},
  year={2024}
}

@article{qin2024study,
  title={A Study on Enhancing Government Efficiency and Public Trust: The Transformative Role of Artificial Intelligence and Large Language Models},
  author={Qin, Hao and Li, Zhi},
  journal={International Journal of Engineering and Management Research},
  volume={14},
  number={3},
  pages={57--61},
  year={2024}
}

@article{safaei2024end,
  title={The end of the policy analyst? testing the capability of artificial intelligence to generate plausible, persuasive, and useful policy analysis},
  author={Safaei, Mehrdad and Longo, Justin},
  journal={Digital Government: Research and Practice},
  volume={5},
  number={1},
  pages={1--35},
  year={2024},
  publisher={ACM New York, NY}
}

@inproceedings{dai2024applications,
  title={Applications and Challenges of Large Language Models in Smart Government-From technological Advances to Regulated Applications},
  author={Dai, Ziqing},
  booktitle={Proceedings of the 2024 3rd International Conference on Frontiers of Artificial Intelligence and Machine Learning},
  pages={275--280},
  year={2024}
}

@article{Agrawal2025EnergyEL,
  title={Energy Efficient Large Language Models: Advancements and Challenges},
  author={Vishakha Agrawal},
  journal={INTERANTIONAL JOURNAL OF SCIENTIFIC RESEARCH IN ENGINEERING AND MANAGEMENT},
  year={2025},
  howpublished={\url{https://api.semanticscholar.org/CorpusID:275730745}}
}

@inproceedings{DBLP:conf/mlsys/KorthikantiCLMA23,
  author       = {Vijay Anand Korthikanti and
                  Jared Casper and
                  Sangkug Lym and
                  Lawrence McAfee and
                  Michael Andersch and
                  Mohammad Shoeybi and
                  Bryan Catanzaro},
  editor       = {Dawn Song and
                  Michael Carbin and
                  Tianqi Chen},
  title        = {Reducing Activation Recomputation in Large Transformer Models},
  booktitle    = {Proceedings of the Sixth Conference on Machine Learning and Systems,
                  MLSys 2023, Miami, FL, USA, June 4-8, 2023},
  publisher    = {mlsys.org},
  year         = {2023},
  url          = {https://proceedings.mlsys.org/paper\_files/paper/2023/hash/80083951326cf5b35e5100260d64ed81-Abstract-mlsys2023.html},
  bibsource    = {dblp computer science bibliography, https://dblp.org}
}

@article{DBLP:journals/cacm/Chien23a,
  author       = {Andrew A. Chien},
  title        = {GenAI: Giga{\textdollar}{\textdollar}{\textdollar}, TeraWatt-Hours,
                  and GigaTons of CO\({}_{\mbox{2}}\)},
  journal      = {Commun. {ACM}},
  volume       = {66},
  number       = {8},
  pages        = {5},
  year         = {2023},
  url          = {https://doi.org/10.1145/3606254},
  doi          = {10.1145/3606254},
  bibsource    = {dblp computer science bibliography, https://dblp.org}
}

@misc{nvidia_h100,
  author = {NVIDIA Corporation},
  title = {NVIDIA H100 Tensor Core GPU},
  year = {2023},
  howpublished = {\url{https://www.nvidia.com/en-us/data-center/h100/}},
  note = {Accessed: 2025}
}

@article{DBLP:journals/corr/abs-2405-21015,
  author       = {Ben Cottier and
                  Robi Rahman and
                  Loredana Fattorini and
                  Nestor Maslej and
                  David Owen},
  title        = {The rising costs of training frontier {AI} models},
  journal      = {CoRR},
  volume       = {abs/2405.21015},
  year         = {2024},
  url          = {https://doi.org/10.48550/arXiv.2405.21015},
  doi          = {10.48550/ARXIV.2405.21015},
  eprinttype    = {arXiv},
  eprint       = {2405.21015},
  bibsource    = {dblp computer science bibliography, https://dblp.org}
}

@inproceedings{DBLP:conf/ccgrid/MastenbroekAJLB21,
  author       = {Fabian Mastenbroek and
                  Georgios Andreadis and
                  Soufiane Jounaid and
                  Wenchen Lai and
                  Jacob Burley and
                  Jaro Bosch and
                  Erwin Van Eyk and
                  Laurens Versluis and
                  Vincent van Beek and
                  Alexandru Iosup},
  editor       = {Laurent Lef{\`{e}}vre and
                  Stacy Patterson and
                  Young Choon Lee and
                  Haiying Shen and
                  Shashikant Ilager and
                  Mohammad Goudarzi and
                  Adel Nadjaran Toosi and
                  Rajkumar Buyya},
  title        = {OpenDC 2.0: Convenient Modeling and Simulation of Emerging Technologies
                  in Cloud Datacenters},
  booktitle    = {21st {IEEE/ACM} International Symposium on Cluster, Cloud and Internet
                  Computing, CCGrid 2021, Melbourne, Australia, May 10-13, 2021},
  pages        = {455--464},
  publisher    = {{IEEE}},
  year         = {2021},
  url          = {https://doi.org/10.1109/CCGrid51090.2021.00055},
  doi          = {10.1109/CCGRID51090.2021.00055},
  bibsource    = {dblp computer science bibliography, https://dblp.org}
}

@mastersthesis{radunicolae-hp-m3sa,
  author       = {Radu Nicolae},
  title        = {M3SA: Exploring the Performance and Climate Impact of Datacenters by Multi-Model Simulation and Analysis},
  school       = {Vrije Universiteit Amsterdam},
  year         = {2024},
  address      = {Amsterdam, The Netherlands},
  type         = {Honours Program Thesis},
  note         = {Submitted in partial fulfillment of the requirements for the Honors Program.}
}

@inproceedings{DBLP:conf/wosp/ChowTLRB24,
  author       = {Kingsum Chow and
                  Yu Tang and
                  Zhiheng Lyu and
                  Anil Rajput and
                  Khun Ban},
  editor       = {Simonetta Balsamo and
                  William J. Knottenbelt and
                  Cristina L. Abad and
                  Weiyi Shang},
  title        = {Performance Optimization in the {LLM} World 2024},
  booktitle    = {Companion of the 15th {ACM/SPEC} International Conference on Performance
                  Engineering, {ICPE} 2024, London, United Kingdom, May 7-11, 2024},
  pages        = {156--157},
  publisher    = {{ACM}},
  year         = {2024},
  url          = {https://doi.org/10.1145/3629527.3651436},
  doi          = {10.1145/3629527.3651436},
  bibsource    = {dblp computer science bibliography, https://dblp.org}
}

@inproceedings{lazuka2024llm,
  title={LLM-Pilot: Characterize and Optimize Performance of your LLM Inference Services},
  author={Lazuka, Malgorzata and Anghel, Andreea and Parnell, Thomas},
  booktitle={SC24: International Conference for High Performance Computing, Networking, Storage and Analysis},
  pages={1--18},
  year={2024},
  organization={IEEE}
}

@article{bommasani2021opportunities,
  title={On the opportunities and risks of foundation models},
  author={Bommasani, Rishi and Hudson, Drew A and Adeli, Ehsan and Altman, Russ and Arora, Simran and von Arx, Sydney and Bernstein, Michael S and Bohg, Jeannette and Bosselut, Antoine and Brunskill, Emma and others},
  journal={arXiv preprint arXiv:2108.07258},
  year={2021}
}

@misc{simon2024llm-mooreslaw,
  author       = {Julien Simon},
  title        = {Large Language Models: A New Moore’s Law?},
  year         = {2024},
  howpublished = {\url{https://huggingface.co/blog/large-language-models}},
  note         = {Accessed: 2025}
}

@inproceedings{narayanan2021efficient,
  title={Efficient large-scale language model training on gpu clusters using megatron-lm},
  author={Narayanan, Deepak and Shoeybi, Mohammad and Casper, Jared and LeGresley, Patrick and Patwary, Mostofa and Korthikanti, Vijay and Vainbrand, Dmitri and Kashinkunti, Prethvi and Bernauer, Julie and Catanzaro, Bryan and others},
  booktitle={Proceedings of the international conference for high performance computing, networking, storage and analysis},
  pages={1--15},
  year={2021}
}

@misc{ramponi2024llm,
  author       = {Marco Ramponi},
  title        = {Why Language Models Became Large Language Models and the Hurdles in Developing LLM-Based Applications},
  year         = {2024},
  howpublished = {\url{https://www.assemblyai.com/blog/why-language-models-became-large-language-models}},
  note         = {Accessed: 2025}
}

@misc{long_context2025,
  title={The Long Context Conundrum: Challenges and Innovations in Scaling LLM Memory},
  author={Anjanava Biswas},
  year={2025},
  howpublished = {https://www.semanticscholar.org/paper/613706b04c5cd3f3cb1b35aba73977ae2c5b6f64}
}

@misc{shoham2024longcontext,
  author       = {Yoav Shoham},
  title        = {Why Language Models Became Large Language Models and the Hurdles in Developing LLM-Based Applications},
  year         = {2024},
  howpublished = {\url{https://www.ai21.com/blog/long-context-yoav-shoham/}},
  note         = {Accessed: 2025}
}

@article{achiam2023gpt,
  title={Gpt-4 technical report},
  author={Achiam, Josh and Adler, Steven and Agarwal, Sandhini and Ahmad, Lama and Akkaya, Ilge and Aleman, Florencia Leoni and Almeida, Diogo and Altenschmidt, Janko and Altman, Sam and Anadkat, Shyamal and others},
  journal={arXiv preprint arXiv:2303.08774},
  year={2023}
}

@article{xu2024hethub,
  title={HETHUB: A Distributed Training System with Heterogeneous Cluster for Large-Scale Models},
  author={Xu, Si and Huang, Zixiao and Zeng, Yan and Yan, Shengen and Ning, Xuefei and Zhang, Quanlu and Ye, Haolin and Gu, Sipei and Shui, Chunsheng and Lin, Zhezheng and others},
  journal={arXiv preprint arXiv:2405.16256},
  year={2024}
}

@article{cheng2009ice,
  title={Ice age terminations},
  author={Cheng, Hai and Edwards, R Lawrence and Broecker, Wallace S and Denton, George H and Kong, Xinggong and Wang, Yongjin and Zhang, Rong and Wang, Xianfeng},
  journal={science},
  volume={326},
  number={5950},
  pages={248--252},
  year={2009},
  publisher={American Association for the Advancement of Science}
}

@article{DBLP:journals/fgcs/MastenbroekMBI25,
  author       = {Fabian Mastenbroek and
                  Tiziano De Matteis and
                  Vincent van Beek and
                  Alexandru Iosup},
  title        = {RADiCe: {A} Risk Analysis Framework for Data Centers},
  journal      = {Future Gener. Comput. Syst.},
  volume       = {166},
  pages        = {107702},
  year         = {2025},
  url          = {https://doi.org/10.1016/j.future.2024.107702},
  doi          = {10.1016/J.FUTURE.2024.107702},
  bibsource    = {dblp computer science bibliography, https://dblp.org}
}

@article{DBLP:journals/tpds/AndreadisMBI22,
  author       = {Georgios Andreadis and
                  Fabian Mastenbroek and
                  Vincent van Beek and
                  Alexandru Iosup},
  title        = {Capelin: Data-Driven Compute Capacity Procurement for Cloud Datacenters
                  Using Portfolios of Scenarios},
  journal      = {{IEEE} Trans. Parallel Distributed Syst.},
  volume       = {33},
  number       = {1},
  pages        = {26--39},
  year         = {2022},
  url          = {https://doi.org/10.1109/TPDS.2021.3084816},
  doi          = {10.1109/TPDS.2021.3084816},
  bibsource    = {dblp computer science bibliography, https://dblp.org}
}

@inproceedings{DBLP:conf/hpec/SamsiZMLMJBKTG23,
  author       = {Siddharth Samsi and
                  Dan Zhao and
                  Joseph McDonald and
                  Baolin Li and
                  Adam Michaleas and
                  Michael Jones and
                  William Bergeron and
                  Jeremy Kepner and
                  Devesh Tiwari and
                  Vijay Gadepally},
  title        = {From Words to Watts: Benchmarking the Energy Costs of Large Language
                  Model Inference},
  booktitle    = {{IEEE} High Performance Extreme Computing Conference, {HPEC} 2023,
                  Boston, MA, USA, September 25-29, 2023},
  pages        = {1--9},
  publisher    = {{IEEE}},
  year         = {2023},
  url          = {https://doi.org/10.1109/HPEC58863.2023.10363447},
  doi          = {10.1109/HPEC58863.2023.10363447},
  bibsource    = {dblp computer science bibliography, https://dblp.org}
}

@misc{iosup2021massivizingkeynote,
  author       = {A. Iosup},
  title        = {Massivizing Computer Systems},
  howpublished = {Keynote Presentation, February 4, 2021},
  url          = {https://atlarge-research.com/pdfs/pres-20210204-aiosup-massivizing.pdf},
}

@article{DBLP:journals/simpra/JawaddiI24,
  author       = {Siti Nuraishah Agos Jawaddi and
                  Azlan B. Ismail},
  title        = {Integrating OpenAI Gym and CloudSim Plus: {A} simulation environment
                  for {DRL} Agent training in energy-driven cloud scaling},
  journal      = {Simul. Model. Pract. Theory},
  volume       = {130},
  pages        = {102858},
  year         = {2024},
  url          = {https://doi.org/10.1016/j.simpat.2023.102858},
  doi          = {10.1016/J.SIMPAT.2023.102858},
  bibsource    = {dblp computer science bibliography, https://dblp.org}
}

@article{DBLP:journals/corr/abs-2408-13386,
  author       = {Remo Andreoli and
                  Jie Zhao and
                  Tommaso Cucinotta and
                  Rajkumar Buyya},
  title        = {CloudSim 7G: An Integrated Toolkit for Modeling and Simulation of
                  Future Generation Cloud Computing Environments},
  journal      = {CoRR},
  volume       = {abs/2408.13386},
  year         = {2024},
  url          = {https://doi.org/10.48550/arXiv.2408.13386},
  doi          = {10.48550/ARXIV.2408.13386},
  eprinttype    = {arXiv},
  eprint       = {2408.13386},
  bibsource    = {dblp computer science bibliography, https://dblp.org}
}

@article{DBLP:journals/tomacs/GuptaBAVJFGM14,
  author       = {Sandeep K. S. Gupta and
                  Ayan Banerjee and
                  Zahra Abbasi and
                  Georgios Varsamopoulos and
                  Michael Jonas and
                  Joshua Ferguson and
                  Rose Robin Gilbert and
                  Tridib Mukherjee},
  title        = {GDCSim: {A} simulator for green data center design and analysis},
  journal      = {{ACM} Trans. Model. Comput. Simul.},
  volume       = {24},
  number       = {1},
  pages        = {3:1--3:27},
  year         = {2014},
  url          = {https://doi.org/10.1145/2553083},
  doi          = {10.1145/2553083},
  bibsource    = {dblp computer science bibliography, https://dblp.org}
}

@article{spence2000ancient,
  title={Ancient Egyptian chronology and the astronomical orientation of pyramids},
  author={Spence, Kate},
  journal={Nature},
  volume={408},
  number={6810},
  pages={320--324},
  year={2000},
  publisher={Nature Publishing Group UK London}
}

@article{microsoft-openai-10billion,
  author  = {Q.ai},
  title   = {Microsoft Confirms Its \$10 Billion Investment Into {ChatGPT}, Changing How Microsoft Competes With Google, Apple And Other Tech Giants},
  journal = {Forbes},
  year    = {2023},
  month   = {January},
  url     = {https://www.forbes.com/sites/qai/2023/01/27/microsoft-confirms-its-10-billion-investment-into-chatgpt-changing-how-microsoft-competes-with-google-apple-and-other-tech-giants/},
  note    = {Accessed: 2025}
}

@article{threemileisland,
  author  = {The Guardian},
  title   = {{Three Mile Island} Nuclear Plant to Reopen Under {Microsoft} Initiative},
  journal = {The Guardian},
  year    = {2024},
  month   = {September},
  url     = {https://www.theguardian.com/environment/2024/sep/20/three-mile-island-nuclear-plant-reopen-microsoft},
  note    = {Accessed: 2025}
}

@article{harrison2018brief,
  author  = {Harrison, X. A. and Donaldson, L. and Correa-Cano, M. E. and Evans, J. and Fisher, D. N. and Goodwin, C. E. and Robinson, B. S. and Hodgson, D. J. and Inger, R.},
  title   = {A brief introduction to mixed effects modelling and multi-model inference in ecology},
  journal = {PeerJ},
  year    = {2018},
  volume  = {6},
  pages   = {e4794},
  doi     = {10.7717/peerj.4794}
}

@article{myhre2017multi,
  author  = {Myhre, G. and Aas, W. and Cherian, R. and Collins, W. and Faluvegi, G. and Flanner, M. and Forster, P. and Hodnebrog, O. and Klimont, Z. and Lund, M. T. and others},
  title   = {Multi-model simulations of aerosol and ozone radiative forcing due to anthropogenic emission changes during the period 1990–2015},
  journal = {Atmospheric Chemistry and Physics},
  year    = {2017},
  volume  = {17},
  number  = {4},
  pages   = {2709–2720},
  doi     = {10.5194/acp-17-2709-2017}
}

@article{nicolae2025m3sa,
  author       = {Radu Nicolae and
                  Dante Niewenhuis and
                  Sacheendra Talluri and
                  Alexandru Iosup},
  title        = {{M3SA:} Exploring Datacenter Performance and Climate-Impact with Multi-
                  and Meta-Model Simulation and Analysis},
  journal      = {CoRR},
  volume       = {abs/2603.29778},
  year         = {2026},
  url          = {https://doi.org/10.48550/arXiv.2603.29778},
  doi          = {10.48550/ARXIV.2603.29778},
  eprinttype   = {arXiv},
  eprint       = {2603.29778},
  bibsource    = {dblp computer science bibliography, https://dblp.org}
}

@article{bucaioni2025functional,
  title={A Functional Software Reference Architecture for LLM-Integrated Systems},
  author={Bucaioni, Alessio and Weyssow, Martin and He, Junda and Lyu, Yunbo and Lo, David},
  journal={arXiv preprint arXiv:2501.12904},
  year={2025}
}

@inproceedings{mahr2024reference,
  title={A Reference Architecture for Deploying Large Language Model Applications in Industrial Environments},
  author={Mahr, Felix and Angeli, Giulia and Sindel, Till and Schmidt, Konstantin and Franke, J{\"o}rg},
  booktitle={2024 IEEE 30th International Symposium for Design and Technology in Electronic Packaging (SIITME)},
  pages={19--23},
  year={2024},
  organization={IEEE}
}

@inproceedings{DBLP:conf/acl/YangHGHZ024,
  author       = {Dongjie Yang and
                  Xiaodong Han and
                  Yan Gao and
                  Yao Hu and
                  Shilin Zhang and
                  Hai Zhao},
  editor       = {Lun{-}Wei Ku and
                  Andre Martins and
                  Vivek Srikumar},
  title        = {PyramidInfer: Pyramid {KV} Cache Compression for High-throughput {LLM}
                  Inference},
  booktitle    = {Findings of the Association for Computational Linguistics, {ACL} 2024,
                  Bangkok, Thailand and virtual meeting, August 11-16, 2024},
  pages        = {3258--3270},
  publisher    = {Association for Computational Linguistics},
  year         = {2024},
  url          = {https://doi.org/10.18653/v1/2024.findings-acl.195},
  doi          = {10.18653/V1/2024.FINDINGS-ACL.195},
  bibsource    = {dblp computer science bibliography, https://dblp.org}
}

@article{team2025aibrix,
  title={AIBrix: Towards Scalable, Cost-Effective Large Language Model Inference Infrastructure},
  author={Team, The AIBrix and Shan, Jiaxin and Gupta, Varun and Xu, Le and Shi, Haiyang and Zhang, Jingyuan and Wang, Ning and Xu, Linhui and Kang, Rong and Liu, Tongping and others},
  journal={arXiv preprint arXiv:2504.03648},
  year={2025}
}

@inproceedings{DBLP:conf/icdcs/IosupVTETBFMT19,
  author       = {Alexandru Iosup and
                  Laurens Versluis and
                  Animesh Trivedi and
                  Erwin Van Eyk and
                  Lucian Toader and
                  Vincent van Beek and
                  Giulia Frascaria and
                  Ahmed Musaafir and
                  Sacheendra Talluri},
  title        = {The AtLarge Vision on the Design of Distributed Systems and Ecosystems},
  booktitle    = {39th {IEEE} International Conference on Distributed Computing Systems,
                  {ICDCS} 2019, Dallas, TX, USA, July 7-10, 2019},
  pages        = {1765--1776},
  publisher    = {{IEEE}},
  year         = {2019},
  url          = {https://doi.org/10.1109/ICDCS.2019.00175},
  doi          = {10.1109/ICDCS.2019.00175},
  bibsource    = {dblp computer science bibliography, https://dblp.org}
}

@inproceedings{DBLP:conf/ccgrid/JansenAPTI23,
  author       = {Matthijs Jansen and
                  Auday Al{-}Dulaimy and
                  Alessandro V. Papadopoulos and
                  Animesh Trivedi and
                  Alexandru Iosup},
  editor       = {Yogesh Simmhan and
                  Ilkay Altintas and
                  Ana Lucia Varbanescu and
                  Pavan Balaji and
                  Abhinandan S. Prasad and
                  Lorenzo Carnevale},
  title        = {The {SPEC-RG} Reference Architecture for The Compute Continuum},
  booktitle    = {23rd {IEEE/ACM} International Symposium on Cluster, Cloud and Internet
                  Computing, CCGrid 2023, Bangalore, India, May 1-4, 2023},
  pages        = {469--484},
  publisher    = {{IEEE}},
  year         = {2023},
  url          = {https://doi.org/10.1109/CCGrid57682.2023.00051},
  doi          = {10.1109/CCGRID57682.2023.00051},
  bibsource    = {dblp computer science bibliography, https://dblp.org}
}

@book{DBLP:journals/sigsoft/Herzog15,
  title={Software architecture in practice},
  author={Bass, Len and Clements, Paul and Kazman, Rick},
  year={2021},
  publisher={Addison-Wesley Professional}
}

@article{malone2006proceedings-pue,
  title={Proceedings of 2006 Digital Power Forum Richardson TX},
  author={Malone, C and Belady, C},
  journal={Metrics to Characterize Data Center IT Equipment Energy Use},
  year={2006}
}

@masterthesis{hongyuhe2021hpreport,
  author       = {Hongyu He},
  title        = {Modelling Energy Consumption in the OpenDC Datacenter Simulator for Analyzing Energy-Aware Cloud Infrastructure},
  school       = {Vrije Universiteit Amsterdam},
  year         = {2021},
  type         = {Research Thesis},
  address      = {Amsterdam},
  month        = {5},
  note         = {Honours Programme, Research Thesis},
  supervisor   = {Prof. Alexandru Iosup},
  daily-supervisor = {Fabian Mastenbroek},
  third-supervisor = {Georgios Andreadis},
  second-reader = {Vincent van Beek}
}

@article{avelar2012pue,
  title={PUE: a comprehensive examination of the metric},
  author={Avelar, Victor and Azevedo, Dan and French, Alan and Power, Emerson Network},
  journal={White paper},
  volume={49},
  year={2012}
}

@misc{ClimateNeutralDataCentre,
  author = {{Climate Neutral Data Centre}},
  title = {{Home}},
  year = {2023},  
  howpublished = {\url{https://www.climateneutraldatacentre.net/}},
  note = {Accessed: 2024}
}

@inproceedings{dniewenhuis_hotcloud_footprinter,
  author    = {Dante Niewenhuis
              and Sacheendra Talluri
              and Alexandru Iosup
              and Tiziano de Matteis},
  title     = {FootPrinter: Quantifying Data Center Carbon Footprint},
  booktitle = {Companion of the 15th ACM/SPEC International Conference on Performance Engineering (ICPE ’24 Companion)},
  publisher = {{ACM}},
  year      = {2024},
}

@misc{google2023efficiency,
  title={Data Center Efficiency},
  author={Google},
  year={2023},
  howpublished={\url{https://www.google.com/about/datacenters/efficiency/}},
  note={Accessed: 2024}
}

@misc{statistaPUE2023,
  author = {Statista},
  title = {Data center average annual PUE worldwide},
  year = 2023,
  howpublished = {\url{https://www.statista.com/statistics/1229367/data-center-average-annual-pue-worldwide/}},
  note = {Accessed: 2024}
}

@article{Mastenbroek2023RADICE,
  title={RADICE: A Risk Analysis Framework for DataCenters},
  author={Fabian Mastenbroek and Tiziano De Matteis and Vincent van Beek and Alexandru Iosup},
  journal={IEEE Transactions on Cloud Computing},
  year={2023},
  publisher={IEEE}
}

@article{lowestPUE-website,
  title={Holistic cooling at the world's most efficient data center},
  author={Summers, Jon and Kozma, Alan and others},
  journal={Data Center Dynamics},
  howpublished={\url{https://www.datacenterdynamics.com/en/analysis/holistic-cooling-at-the-worlds-most-efficient-data-center/}},
  year={2019},
  month={Oct}
}

@article{lowestPUE-article,
  title={A realistic view on heat reuse from direct free air-cooled data centres},
  author={Ljungqvist, Hampus Markeby and Risberg, Mikael and Toffolo, Andrea and Vesterlund, Mattias},
  journal={Energy Conversion and Management: X},
  volume={20},
  pages={100473},
  year={2023},
  publisher={Elsevier}
}

@misc{StatistaElectricityPrices2024,
  author = {Statista},
  title = {Electricity Prices in Selected Countries},
  year = {2024},
  howpublished = {\url{https://www.statista.com/statistics/263492/electricity-prices-in-selected-countries/}},
  note = {Accessed: 2024}
}

@inproceedings{gupta2011gdcsim,
  title={Gdcsim: A tool for analyzing green data center design and resource management techniques},
  author={Gupta, Sandeep KS and Gilbert, Rose Robin and Banerjee, Ayan and Abbasi, Zahra and Mukherjee, Tridib and Varsamopoulos, Georgios},
  booktitle={2011 International Green Computing Conference and Workshops},
  pages={1--8},
  year={2011},
  organization={IEEE}
}

@misc{eea_energy_consumption,
    author = {European Environment Agency},
    title = {Share of energy consumption from renewable sources},
    year = {2023},
    howpublished = {\url{https://www.eea.europa.eu/en/analysis/indicators/share-of-energy-consumption-from}},
    note = {Accessed: 2024}
}

@inproceedings{DBLP:conf/ispdc/IosupABBENOTVV17,
  author       = {Alexandru Iosup and
                  Georgios Andreadis and
                  Vincent van Beek and
                  Matthijs Bijman and
                  Erwin Van Eyk and
                  Mihai Neacsu and
                  Leon Overweel and
                  Sacheendra Talluri and
                  Laurens Versluis and
                  Maaike Visser},
  editor       = {Radu Prodan and
                  Florin Pop and
                  Ralf{-}Peter Mundani},
  title        = {The OpenDC Vision: Towards Collaborative Datacenter Simulation and
                  Exploration for Everybody},
  booktitle    = {16th International Symposium on Parallel and Distributed Computing,
                  {ISPDC} 2017, Innsbruck, Austria, July 3-6, 2017},
  pages        = {85--94},
  publisher    = {{IEEE}},
  year         = {2017},
  url          = {https://doi.org/10.1109/ISPDC.2017.25},
  doi          = {10.1109/ISPDC.2017.25},
  bibsource    = {dblp computer science bibliography, https://dblp.org}
}

@inproceedings{zhou2013axpue,
title={AxPUE: Application level metrics for power usage effectiveness in data centers},
author={Zhou, Runlin and Shi, Yingjie and Zhu, Chunge},
booktitle={2013 IEEE International Conference on Big Data},
pages={110--117},
year={2013},
organization={IEEE}
}

@article{seagateData,
title={Data age 2025: The evolution of data to life-critical. Don’t Focus on Big Data},
author={Reinsel, David and Gantz, John and Rydning, John},
journal={2},
year={2017}
}

@misc{IEADataCentresNetworks,
title = {Data Centres and Networks},
author = {{International Energy Agency}},
howpublished = {\url{[https://www.iea.org/fuels-and-technologies/data-centres-networks](https://www.iea.org/fuels-and-technologies/data-centres-networks)}},
note = {Accessed: 2023}
}

@inproceedings{DBLP:conf/wsc/LeeMS03,
  author       = {Y. Tina Lee and
                  Charles R. McLean and
                  Guodong Shao},
  editor       = {Stephen E. Chick and
                  Paul J. Sanchez and
                  David M. Ferrin and
                  Douglas J. Morrice},
  title        = {Neutral information structure for manufacturing simulations: a neutral
                  information model for simulating machine shop operations},
  booktitle    = {Proceedings of the 35th Winter Simulation Conference: Driving Innovation,
                  New Orleans, Louisiana, USA, December 7-10, 2003},
  pages        = {1296--1304},
  publisher    = {{IEEE} Computer Society},
  year         = {2003},
  url          = {https://doi.org/10.1109/WSC.2003.1261565},
  doi          = {10.1109/WSC.2003.1261565},
  bibsource    = {dblp computer science bibliography, https://dblp.org}
}

@book{modsim:book/ZaraiN15:orig,
 editor = {Faouzi Zarai and Petros Nicopolitidis},
 nolocation = {The Netherlands},
 publisher = {Elsevier},
 subtitle = {Methodologies and Applications},
 title = {Modeling and Simulation of Computer Networks and Systems},
 year = {2015},
isbnpaperback = {9780128008874},
isbnebook = {9780128011584}
}

@article{vaswani2017attention,
  title={Attention is all you need},
  author={Vaswani, Ashish and Shazeer, Noam and Parmar, Niki and Uszkoreit, Jakob and Jones, Llion and Gomez, Aidan N and Kaiser, {\L}ukasz and Polosukhin, Illia},
  journal={Advances in neural information processing systems},
  volume={30},
  year={2017}
}

@article{DBLP:journals/corr/abs-2102-08606,
  author       = {Lemeng Wu and
                  Xingchao Liu and
                  Qiang Liu},
  title        = {Centroid Transformers: Learning to Abstract with Attention},
  journal      = {CoRR},
  volume       = {abs/2102.08606},
  year         = {2021},
  url          = {https://arxiv.org/abs/2102.08606},
  eprinttype    = {arXiv},
  eprint       = {2102.08606},
  bibsource    = {dblp computer science bibliography, https://dblp.org}
}

@misc{huggingface-kv-cache,
  author = {Hugging Face},
  title = {KV Cache Strategies},
  year = {2024},
  howpublished = {\url{https://huggingface.co/docs/transformers/en/kv_cache}},
  note = {Accessed: 2025}
}

@misc{efficientnlp2023kv,
  title        = {The KV Cache: Memory Usage in Transformers},
  author       = {Efficient NLP},
  year         = {2023},
  howpublished = {YouTube video},
  note         = {Available at: https://www.youtube.com/watch?v=Jt8Xc3pG6cg},
  month        = {July},
  url          = {https://www.youtube.com/watch?v=Jt8Xc3pG6cg},
  urldate      = {2025}
}

@article{zhang2022opt,
  title={Opt: Open pre-trained transformer language models},
  author={Zhang, Susan and Roller, Stephen and Goyal, Naman and Artetxe, Mikel and Chen, Moya and Chen, Shuohui and Dewan, Christopher and Diab, Mona and Li, Xian and Lin, Xi Victoria and others},
  journal={arXiv preprint arXiv:2205.01068},
  year={2022}
}

@inproceedings{DBLP:conf/wmcsa/SatyanarayananG19,
  author       = {Mahadev Satyanarayanan and
                  Wei Gao and
                  Brandon Lucia},
  editor       = {Alec Wolman and
                  Lin Zhong},
  title        = {The Computing Landscape of the 21st Century},
  booktitle    = {Proceedings of the 20th International Workshop on Mobile Computing
                  Systems and Applications, HotMobile 2019, Santa Cruz, CA, USA, February
                  27-28, 2019},
  pages        = {45--50},
  publisher    = {{ACM}},
  year         = {2019},
  url          = {https://doi.org/10.1145/3301293.3302357},
  doi          = {10.1145/3301293.3302357},
  bibsource    = {dblp computer science bibliography, https://dblp.org}
}

@inproceedings{DBLP:conf/edge/MortazaviSGPL17,
  author       = {Seyed Hossein Mortazavi and
                  Mohammad Salehe and
                  Carolina Simoes Gomes and
                  Caleb Phillips and
                  Eyal de Lara},
  editor       = {Junshan Zhang and
                  Mung Chiang and
                  Bruce M. Maggs},
  title        = {Cloudpath: a multi-tier cloud computing framework},
  booktitle    = {Proceedings of the Second {ACM/IEEE} Symposium on Edge Computing,
                  San Jose / Silicon Valley, {SEC} 2017, CA, USA, October 12-14, 2017},
  pages        = {20:1--20:13},
  publisher    = {{ACM}},
  year         = {2017},
  url          = {https://doi.org/10.1145/3132211.3134464},
  doi          = {10.1145/3132211.3134464},
  bibsource    = {dblp computer science bibliography, https://dblp.org}
}

@article{lu2023towards,
  author       = {Qinghua Lu and
                  Liming Zhu and
                  Xiwei Xu and
                  Zhenchang Xing and
                  Jon Whittle},
  title        = {Toward Responsible {AI} in the Era of Generative {AI:} {A} Reference
                  Architecture for Designing Foundation Model-Based Systems},
  journal      = {{IEEE} Softw.},
  volume       = {41},
  number       = {6},
  pages        = {91--100},
  year         = {2024},
  url          = {https://doi.org/10.1109/MS.2024.3406333},
  doi          = {10.1109/MS.2024.3406333},
  bibsource    = {dblp computer science bibliography, https://dblp.org}
}

@inproceedings{DBLP:conf/sc/AndreadisVMI18,
  author       = {Georgios Andreadis and
                  Laurens Versluis and
                  Fabian Mastenbroek and
                  Alexandru Iosup},
  title        = {A reference architecture for datacenter scheduling: design, validation,
                  and experiments},
  booktitle    = {Proceedings of the International Conference for High Performance Computing,
                  Networking, Storage, and Analysis, {SC} 2018, Dallas, TX, USA, November
                  11-16, 2018},
  pages        = {37:1--37:15},
  publisher    = {{IEEE} / {ACM}},
  year         = {2018},
  url          = {http://dl.acm.org/citation.cfm?id=3291706},
  bibsource    = {dblp computer science bibliography, https://dblp.org}
}

@book{rozanski2012software,
  title={Software systems architecture: working with stakeholders using viewpoints and perspectives},
  author={Rozanski, Nick and Woods, Eoin},
  year={2012},
  publisher={Addison-Wesley}
}

@article{DBLP:journals/corr/abs-2404-14294,
  author       = {Zixuan Zhou and
                  Xuefei Ning and
                  Ke Hong and
                  Tianyu Fu and
                  Jiaming Xu and
                  Shiyao Li and
                  Yuming Lou and
                  Luning Wang and
                  Zhihang Yuan and
                  Xiuhong Li and
                  Shengen Yan and
                  Guohao Dai and
                  Xiao{-}Ping Zhang and
                  Yuhan Dong and
                  Yu Wang},
  title        = {A Survey on Efficient Inference for Large Language Models},
  journal      = {CoRR},
  volume       = {abs/2404.14294},
  year         = {2024},
  url          = {https://doi.org/10.48550/arXiv.2404.14294},
  doi          = {10.48550/ARXIV.2404.14294},
  eprinttype    = {arXiv},
  eprint       = {2404.14294},
  bibsource    = {dblp computer science bibliography, https://dblp.org}
}

@article{yan2025we,
  title={Are We There Yet? A Measurement Study of Efficiency for LLM Applications on Mobile Devices},
  author={Yan, Xiao and Ding, Yi},
  journal={arXiv preprint arXiv:2504.00002},
  year={2025}
}

@article{DBLP:journals/corr/abs-2501-14205,
  author       = {Minrui Xu and
                  Dusit Niyato and
                  Christopher G. Brinton},
  title        = {Serving Long-Context LLMs at the Mobile Edge: Test-Time Reinforcement
                  Learning-based Model Caching and Inference Offloading},
  journal      = {CoRR},
  volume       = {abs/2501.14205},
  year         = {2025},
  url          = {https://doi.org/10.48550/arXiv.2501.14205},
  doi          = {10.48550/ARXIV.2501.14205},
  eprinttype    = {arXiv},
  eprint       = {2501.14205},
  bibsource    = {dblp computer science bibliography, https://dblp.org}
}

@article{DBLP:journals/tmc/HeFYL24,
  author       = {Ying He and
                  Jingcheng Fang and
                  F. Richard Yu and
                  Victor C. M. Leung},
  title        = {Large Language Models (LLMs) Inference Offloading and Resource Allocation
                  in Cloud-Edge Computing: An Active Inference Approach},
  journal      = {{IEEE} Trans. Mob. Comput.},
  volume       = {23},
  number       = {12},
  pages        = {11253--11264},
  year         = {2024},
  url          = {https://doi.org/10.1109/TMC.2024.3415661},
  doi          = {10.1109/TMC.2024.3415661},
  bibsource    = {dblp computer science bibliography, https://dblp.org}
}

@misc{googlegemini,
  author = {{Google}},
  title = {Gemini {AI}},
  year = {2025},
  howpublished = {\url{https://gemini.google.com/}},
  note = {Accessed: 2025}
}

@misc{openaichatgpt,
  author = {{OpenAI}},
  title = {ChatGPT},
  year = {2025},
  howpublished = {\url{https://chatgpt.com/}},
  note = {Accessed: 2025}
}

@misc{openai2025deepresearch,
  author       = {{OpenAI}},
  title        = {Introducing Deep Research},
  howpublished = {\url{https://openai.com/index/introducing-deep-research/}},
  year         = {2025},
  month        = {Feb},
  day          = {2},
  note         = {Accessed: 2025}
}

@inproceedings{2023-icpewip-scheduler-apis,
  author    = {Aratz Manterola Lasa and
              Sacheendra Talluri and
              Alexandru Iosup
            },
  title     = {A Reference Architecture for Datacenter Scheduler Programming Abstractions: Design and Experiments (Work In Progress Paper)},
  booktitle = {Proceedings of the International Conference on Performance Engineering, Coimbra, Portugal, April, 2023},
  year      = {2023},
  doi       = {},
}

@misc{openai_prompt_caching,
  author = {{OpenAI}},
  title = {Prompt caching},
  year = {2025},
  howpublished = {\url     {https://platform.openai.com/docs/guides/prompt-caching}},
  urldate = {2025-05-22},
  organization = {OpenAI},
  note = {Accessed: 2025}
}

@article{ouyang2022training,
  title={Training language models to follow instructions with human feedback},
  author={Ouyang, Long and Wu, Jeffrey and Jiang, Xu and Almeida, Diogo and Wainwright, Carroll and Mishkin, Pamela and Zhang, Chong and Agarwal, Sandhini and Slama, Katarina and Ray, Alex and others},
  journal={Advances in neural information processing systems},
  volume={35},
  pages={27730--27744},
  year={2022}
}

@article{dai2023safe,
  title={Safe rlhf: Safe reinforcement learning from human feedback},
  author={Dai, Josef and Pan, Xuehai and Sun, Ruiyang and Ji, Jiaming and Xu, Xinbo and Liu, Mickel and Wang, Yizhou and Yang, Yaodong},
  journal={arXiv preprint arXiv:2310.12773},
  year={2023}
}

@article{dong2024rlhf,
  title={Rlhf workflow: From reward modeling to online rlhf},
  author={Dong, Hanze and Xiong, Wei and Pang, Bo and Wang, Haoxiang and Zhao, Han and Zhou, Yingbo and Jiang, Nan and Sahoo, Doyen and Xiong, Caiming and Zhang, Tong},
  journal={arXiv preprint arXiv:2405.07863},
  year={2024}
}

@misc{ibm_milvus_2024,
    author = {{IBM}},
    title = {Milvus overview},
    howpublished = {IBM Documentation for watsonx.data, \url{https://www.ibm.com/docs/en/watsonx/watsonxdata/2.0.x?topic=overview-milvus}},
    note = {IBM watsonx.data version 2.0.x documentation},
    year = {2024}
}

@misc{ibm_elasticsearch_2024,
    author = {{IBM}},
    title = {IBM Cloud Databases for Elasticsearch},
    howpublished = {IBM Product Documentation \url{https://www.ibm.com/products/databases-for-elasticsearch}},
    note = {Fully managed Elasticsearch Service offering on IBM Cloud},
    year = {2024}
}

@misc{openai_api,
  author = {OpenAI},
  title = {OpenAI {API} Reference},
  year = {2024},
  howpublished = {\url{https://platform.openai.com/docs/api-reference/introduction}},
  note = {Online documentation for the OpenAI API}
}

@misc{openai_deep_learning_infrastructure,
  author = {OpenAI},
  title = {Infrastructure for deep learning},
  year = {2016},
  howpublished = {\url{https://openai.com/index/infrastructure-for-deep-learning/}},
  note = {Technical blog post detailing infrastructure approaches}
}

@misc{microsoft_azure_openai_data,
  author = {Microsoft},
  title = {Azure OpenAI On Your Data},
  howpublished = {\url{https://learn.microsoft.com/en-us/azure/ai-services/openai/concepts/use-your-data}},
  year = {2025},
  note = {Accessed: 2025}
}

@article{patke2025hierarchical,
  title={Hierarchical Autoscaling for Large Language Model Serving with Chiron},
  author={Patke, Archit and Reddy, Dhemath and Jha, Saurabh and Narayanaswami, Chandra and Kalbarczyk, Zbigniew and Iyer, Ravishankar},
  journal={arXiv preprint arXiv:2501.08090},
  year={2025}
}

@misc{openai2021kubernetes,
  author = {OpenAI},
  title = {Scaling {Kubernetes} to 7,500 Nodes},
  howpublished = {\url{https://openai.com/index/scaling-kubernetes-to-7500-nodes/}},
  year = {2021},
  note = {Accessed: 2025}
}

@misc{openai2025stargate,
  author = {OpenAI},
  title = {Announcing The {Stargate} Project},
  howpublished = {\url{https://openai.com/index/announcing-the-stargate-project/}},
  year = {2025},
  note = {Accessed: 2025}
}

@misc{openai2025funding,
  author = {OpenAI},
  title = {New funding to build towards {AGI}},
  howpublished = {\url{https://openai.com/index/march-funding-updates/}},
  year = {2025},
  note = {Accessed: 2025}
}

@misc{nvidia_tensorrt_2023,
  author = {NVIDIA},
  title = {NVIDIA TensorRT},
  year = {2023},
  howpublished = {\url{https://developer.nvidia.com/tensorrt}},
  note = {Accessed: 2025}
}

@misc{microsoft_azure_cosmosdb_2024,
  author = {Microsoft},
  title = {Data source - Azure Cosmos DB for MongoDB vCore},
  year = {2024},
  howpublished = {\url{https://learn.microsoft.com/en-us/azure/ai-services/openai/references/cosmos-db?tabs=python}},
  note = {Accessed: 2025}
}

@misc{ibm_watson_assistant_getting_started,
  author = {{IBM}},
  title = {{Getting Started with watsonx Assistant - IBM Cloud Docs}},
  year = {2025},
  howpublished = {\url{https://cloud.ibm.com/docs/watson-assistant}},
  note = {Accessed: 2025. Official IBM Cloud documentation for Watson Assistant.}
}

@misc{ibm_genai_rag_2024,
  title = {Gen AI Pattern for Watsonx on IBM Cloud},
  author = {{IBM Cloud Documentation}},
  year = {2024},
  howpublished = {\url{https://cloud.ibm.com/docs/pattern-genai-rag?topic=pattern-genai-rag-genai-pattern}},
  note = {Accessed: 2025}
}

@misc{ibm_workload_placement_2025,
  title = {Workload Placement Planning for IBM Cloud Pak for Integration},
  author = {{IBM Cloud Documentation}},
  year = {2025},
  howpublished = {\url{https://www.ibm.com/docs/en/cloud-paks/cp-integration/16.1.0?topic=planning-workload-placement}},
  note = {Accessed: 2025}
}

@misc{ibm_elastic_partnership_2024,
  title = {IBM Partners with Elasticsearch to Deliver Conversational Search with Watsonx Assistant},
  author = {Hemant Malik and Serena Chou and Eser Saydam},
  year = {2024},
  howpublished = {\url{https://www.elastic.co/blog/ibm-elasticsearch-partnership-conversational-search-watsonx-assistant}},
  note = {Accessed: 2025}
}

@misc{ibm_watsonx_milvus_2025,
  title = {Milvus Overview for IBM watsonx.data},
  author = {{IBM Cloud Documentation}},
  year = {2025},
  howpublished = {\url{https://www.ibm.com/docs/en/watsonx/watsonxdata/2.0.x?topic=overview-milvus}},
  note = {Accessed: 2025}
}

@misc{ibm_think_milvus_2025,
  title = {What is Milvus? - IBM},
  author = {{IBM Cloud Documentation}},
  year = {2025},
  howpublished = {\url{https://www.ibm.com/think/topics/milvus}},
  note = {Accessed: 2025}
}

@misc{ibm_watsonx_apis,
  title = {Watsonx APIs},
  author = {{IBM}},
  howpublished = {\url{https://www.ibm.com/docs/en/watsonx/saas?topic=tutorials-watsonx-apis}},
  note = {Accessed: 2025},
  year = {2025},
  organization = {IBM}
}

@inproceedings{kwon2023efficient,
  title={Efficient memory management for large language model serving with pagedattention},
  author={Kwon, Woosuk and Li, Zhuohan and Zhuang, Siyuan and Sheng, Ying and Zheng, Lianmin and Yu, Cody Hao and Gonzalez, Joseph and Zhang, Hao and Stoica, Ion},
  booktitle={Proceedings of the 29th Symposium on Operating Systems Principles},
  pages={611--626},
  year={2023}
}

@misc{ibm_watsonx_governance,
  author = {IBM},
  title = {Governing assets with watsonx.governance},
  year = {2023},
  howpublished = {\url{https://www.ibm.com/docs/en/watsonx/saas?topic=governing-ai}},
  note = {Accessed: 2025}
}

@misc{ibm_watson_machine_learning,
  author = {IBM},
  title = {Watson Machine Learning},
  year = {2023},
  howpublished = {\url{https://www.ibm.com/docs/en/software-hub/5.1.x?topic=services-watson-machine-learning}},
  note = {Version: 5.1.2; Accessed: 2025}
}

@misc{baseten_llm_inference_guide,
  author    = {{Baseten}},
  title     = {A Guide to LLM Inference and Performance},
  howpublished = {\url{https://www.baseten.co/blog/llm-transformer-inference-guide/}},
  urldate   = {2024-05-24}
}

@article{recasens2025mind,
  title={Mind the memory gap: Unveiling gpu bottlenecks in large-batch llm inference},
  author={Recasens, Pol G and Agullo, Ferran and Zhu, Yue and Wang, Chen and Lee, Eun Kyung and Tardieu, Olivier and Torres, Jordi and Berral, Josep Ll},
  journal={arXiv preprint arXiv:2503.08311},
  year={2025}
}

@article{agrawal2024vidur,
  title={Vidur: A large-scale simulation framework for llm inference},
  author={Agrawal, Amey and Kedia, Nitin and Mohan, Jayashree and Panwar, Ashish and Kwatra, Nipun and Gulavani, Bhargav S and Ramjee, Ramachandran and Tumanov, Alexey},
  journal={Proceedings of Machine Learning and Systems},
  volume={6},
  pages={351--366},
  year={2024}
}

@inproceedings{DBLP:conf/iccS/SilvaOCTDS19,
  author       = {Rafael Ferreira da Silva and
                  Anne{-}C{\'{e}}cile Orgerie and
                  Henri Casanova and
                  Ryan Tanaka and
                  Ewa Deelman and
                  Fr{\'{e}}d{\'{e}}ric Suter},
  editor       = {Jo{\~{a}}o M. F. Rodrigues and
                  Pedro J. S. Cardoso and
                  J{\^{a}}nio M. Monteiro and
                  Roberto Lam and
                  Valeria V. Krzhizhanovskaya and
                  Michael Harold Lees and
                  Jack J. Dongarra and
                  Peter M. A. Sloot},
  title        = {Accurately Simulating Energy Consumption of I/O-Intensive Scientific
                  Workflows},
  booktitle    = {Computational Science - {ICCS} 2019 - 19th International Conference,
                  Faro, Portugal, June 12-14, 2019, Proceedings, Part {I}},
  series       = {Lecture Notes in Computer Science},
  volume       = {11536},
  pages        = {138--152},
  publisher    = {Springer},
  year         = {2019},
  url          = {https://doi.org/10.1007/978-3-030-22734-0\_11},
  doi          = {10.1007/978-3-030-22734-0\_11},
  bibsource    = {dblp computer science bibliography, https://dblp.org}
}

@article{DBLP:journals/spe/CalheirosRBRB11,
  author       = {Rodrigo N. Calheiros and
                  Rajiv Ranjan and
                  Anton Beloglazov and
                  C{\'{e}}sar A. F. De Rose and
                  Rajkumar Buyya},
  title        = {CloudSim: a toolkit for modeling and simulation of cloud computing
                  environments and evaluation of resource provisioning algorithms},
  journal      = {Softw. Pract. Exp.},
  volume       = {41},
  number       = {1},
  pages        = {23--50},
  year         = {2011},
  url          = {https://doi.org/10.1002/spe.995},
  doi          = {10.1002/SPE.995},
  bibsource    = {dblp computer science bibliography, https://dblp.org}
}

@inproceedings{DBLP:conf/isca/FanWB07,
  author       = {Xiaobo Fan and
                  Wolf{-}Dietrich Weber and
                  Luiz Andr{\'{e}} Barroso},
  editor       = {Dean M. Tullsen and
                  Brad Calder},
  title        = {Power provisioning for a warehouse-sized computer},
  booktitle    = {34th International Symposium on Computer Architecture {(ISCA} 2007),
                  June 9-13, 2007, San Diego, California, {USA}},
  pages        = {13--23},
  publisher    = {{ACM}},
  year         = {2007},
  url          = {https://doi.org/10.1145/1250662.1250665},
  doi          = {10.1145/1250662.1250665},
  bibsource    = {dblp computer science bibliography, https://dblp.org}
}

@misc{ApacheParquet,
  title = {Apache Parquet},
  author = {{The Apache Software Foundation}},
  year = {2024},
  howpublished = {\url{https://parquet.apache.org/}}
}

@article{hirth2018entso,
  title={The ENTSO-E Transparency Platform--A review of Europe’s most ambitious electricity data platform},
  author={Hirth, Lion and M{\"u}hlenpfordt, Jonathan and Bulkeley, Marisa},
  journal={Applied energy},
  volume={225},
  pages={1054--1067},
  year={2018},
  publisher={Elsevier}
}

@inproceedings{stojkovic2025dynamollm,
  title={Dynamollm: Designing llm inference clusters for performance and energy efficiency},
  author={Stojkovic, Jovan and Zhang, Chaojie and Goiri, {\'I}{\~n}igo and Torrellas, Josep and Choukse, Esha},
  booktitle={2025 IEEE International Symposium on High Performance Computer Architecture (HPCA)},
  pages={1348--1362},
  year={2025},
  organization={IEEE}
}

@misc{wang2024burstgpt,
      title={BurstGPT: A Real-world Workload Dataset to Optimize LLM Serving Systems}, 
      author={Yuxin Wang and Yuhan Chen and Zeyu Li and Xueze Kang and Zhenheng Tang and Xin He and Rui Guo and Xin Wang and Qiang Wang and Amelie Chi Zhou and Xiaowen Chu},
      year={2024},
      eprint={2401.17644},
      archivePrefix={arXiv},
      primaryClass={id='cs.DC' full_name='Distributed, Parallel, and Cluster Computing' is_active=True alt_name=None in_archive='cs' is_general=False description='Covers fault-tolerance, distributed algorithms, stabilility, parallel computation, and cluster computing. Roughly includes material in ACM Subject Classes C.1.2, C.1.4, C.2.4, D.1.3, D.4.5, D.4.7, E.1.'}
}

@article{pan2024marconi,
  title={Marconi: Prefix Caching for the Era of Hybrid LLMs},
  author={Pan, Rui and Wang, Zhuang and Jia, Zhen and Karakus, Can and Zancato, Luca and Dao, Tri and Netravali, Ravi and Wang, Yida},
  journal={arXiv preprint arXiv:2411.19379},
  year={2024}
}

@inproceedings{zheng2023judging,
  title={Judging LLM-as-a-Judge with MT-Bench and Chatbot Arena},
  author={Zheng, Lianmin and Chiang, Wei-Lin and Sheng, Ying and Zhuang, Siyuan and Wu, Zhanghao and Zhuang, Yonghao and Lin, Zi and Li, Zhuohan and Li, Dacheng and Xing, Eric and Zhang, Hao and Gonzalez, Joseph E. and Stoica, Ion},
  booktitle={Advances in Neural Information Processing Systems},
  volume={36},
  pages={46595--46623},
  year={2023}
}

@misc{sharegpt2024,
  title={ShareGPT: Share your wildest ChatGPT conversations with one click},
  author={{ShareGPT Team}},
  year={2024},
  howpublished={\url{https://sharegpt.com}},
  note={Accessed: 2025}
}

@article{yang2024sweagent,
  title={SWE-agent: Agent-Computer Interfaces Enable Automated Software Engineering},
  author={Yang, John and Jimenez, Carlos E. and Wettig, Alexander and Lieret, Kilian and Yao, Shunyu and Narasimhan, Karthik and Press, Ofir},
  journal={arXiv preprint arXiv:2405.15793},
  year={2024}
}

@inproceedings{jimenez2024swebench,
  title={SWE-bench: Can Language Models Resolve Real-World GitHub Issues?},
  author={Jimenez, Carlos E. and Yang, John and Wettig, Alexander and Yao, Shunyu and Pei, Kexin and Press, Ofir and Narasimhan, Karthik R.},
  booktitle={International Conference on Learning Representations},
  year={2024}
}

@misc{nvidia_a10_2025,
  author       = {{NVIDIA Corporation}},
  title        = {NVIDIA A10 Tensor Core GPU},
  howpublished = {\url{https://www.nvidia.com/en-us/data-center/products/a10-gpu/}},
  note         = {Accelerated graphics and video with AI for mainstream enterprise servers. Accessed 2025-07-03},
  year         = {2025}
}

@article{bal2016medium,
  title={A medium-scale distributed system for computer science research: Infrastructure for the long term},
  author={Bal, Henri and Epema, Dick and De Laat, Cees and Van Nieuwpoort, Rob and Romein, John and Seinstra, Frank and Snoek, Cees and Wijshoff, Harry},
  journal={Computer},
  volume={49},
  number={5},
  pages={54--63},
  year={2016},
  publisher={IEEE}
}

@misc{nvidia_a4000_2025,
  author       = {{NVIDIA Corporation}},
  title        = {NVIDIA RTX A4000 Graphics Card},
  howpublished = {\url{https://www.nvidia.com/en-us/products/workstations/rtx-a4000/}},
  note         = {Single-slot professional GPU with real-time ray tracing and AI acceleration. Accessed 2025-07-03},
  year         = {2025}
}

@misc{nvidia_a6000_2025,
  author       = {{NVIDIA Corporation}},
  title        = {NVIDIA RTX A6000 Graphics Card},
  howpublished = {\url{https://www.nvidia.com/en-us/products/workstations/rtx-a6000/}},
  note         = {48 GB GDDR6, third-generation NVLink, for advanced visualization \& AI workloads. Accessed 2025-07-03},
  year         = {2025}
}

@misc{nvidia_a100_2025,
  author       = {{NVIDIA Corporation}},
  title        = {NVIDIA A100 Tensor Core GPU},
  howpublished = {\url{https://www.nvidia.com/en-us/data-center/a100/}},
  note         = {Ampere-architecture accelerator for AI, HPC, and data analytics. Accessed 2025-07-03},
  year         = {2025}
}

@misc{nvidia_smi_2025,
  author       = {{NVIDIA Corporation}},
  title        = {NVIDIA System Management Interface (nvidia-smi) Documentation},
  howpublished = {\url{https://docs.nvidia.com/deploy/nvidia-smi/index.html}},
  note         = {Accessed 2025-07-03},
  year         = {2025}
}

@misc{meta2024llama31,
  title        = {Meta Llama 3.1 8B},
  author       = {Meta AI},
  year         = {2024},
  howpublished = {\url{https://huggingface.co/meta-llama/Llama-3.1-8B}},
  note         = {Llama 3.1 is licensed under the Llama 3.1 Community License, Copyright © Meta Platforms, Inc. All Rights Reserved. Model release date: July 23, 2024.}
}

@misc{loremipsum,
  title        = {loremipsum: A Lorem Ipsum text generator},
  author       = {Luca De Vitis},
  year         = {2011--2014},
  howpublished = {\url{https://loremipsum.readthedocs.io/}},
  note         = {Version 1.0.4. GNU General Public License v3 or later.}
}

@inproceedings{de2023boosting,
  title={Boosting the Impact of Extreme and Sustainable Graph Processing for Urgent Societal Challenges in Europe Graph-Massivizer: A Horizon Europe Project},
  author={de Lama Sanchez, Nuria and Haase, Peter and Roman, Dumitru and Prodan, Radu},
  booktitle={Companion of the 2023 ACM/SPEC International Conference on Performance Engineering},
  pages={233--238},
  year={2023}
}

@misc{li2025life,
  author    = {Junhao Li},
  title     = {Life of an inference request (vLLM V1): How LLMs are served efficiently at scale},
  date      = {2025-06-27},
  year      = {2025},
  url       = {https://www.ubicloud.com/blog/life-of-an-inference-request-vllm-v1},
  urldate   = {2025-07-03},
  publisher = {Ubicloud Blog}
}

@book{book-distributed-systems,
  title  = {Distributed Systems: Lecture Notes 2019--2020},
  author = {Iosup, Alexandru and Trivedi, Animesh and Donkervliet, Jesse and
            Versluis, Laurens and Talluri, Sacheendra},
  year   = {2019},
  date   = {2019-12-03},
  note   = {Lecture notes, compiled 3~Dec~2019},
  publisher = {Vrije Universiteit Amsterdam}
}

@misc{flinders2024ai,
  author    = {Mesh Flinders and Ian Smalley},
  title     = {What is AI inference?},
  date      = {2024-06-18},
  year      = {2024},
  howpublished  = {\url{https://www.ibm.com/think/topics/ai-inference},
  urldate   = {2025-07-03}},
  publisher = {IBM Think}
}

@misc{site:entso-e,
  title = {ENTSO-E Transparency Platform},
  author = {{European Network of Transmission System Operators for Electricity (ENTSO-E)}},
  howpublished = {Official Website, \url{https://transparency.entsoe.eu/}},
  year = {2025}
}

@misc{ibm_tokens,
  author       = {IBM},
  title        = {Tokens and tokenization},
  howpublished = {\url{https://www.ibm.com/docs/en/watsonx/saas?topic=solutions-tokens}},
  note         = {IBM watsonx documentation, accessed 2025-07-04}
}

@article{de2016mean,
  title={Mean absolute percentage error for regression models},
  author={De Myttenaere, Arnaud and Golden, Boris and Le Grand, B{\'e}n{\'e}dicte and Rossi, Fabrice},
  journal={Neurocomputing},
  volume={192},
  pages={38--48},
  year={2016},
  publisher={Elsevier}
}

@misc{oracle2014mape,
  author = {{Oracle}},
  title = {MAPE (Mean Absolute Percentage Error) Documentation},
  howpublished = {Oracle Cloud Infrastructure Documentation, \url{https://docs.oracle.com/en/cloud/saas/planning-budgeting-cloud/pfusu/insights_metrics_MAPE.html}},
  year = {2024},
  note = {Accessed: 2025}
}

@article{moreno2013using-mape,
  title={Using the R-MAPE index as a resistant measure of forecast accuracy},
  author={Moreno, Juan Jos{\'e} Monta{\~n}o and others},
  journal={Psicothema},
  year={2013}
}

@article{dilmegani2025cloud,
  title={Cloud GPUs for Deep Learning: Availability \& Price / Performance},
  author={Dilmegani, Cem},
  journal={AIMultiple Research},
  year={2025},
  month={July},
  day={3},
  howpublished={\url{https://research.aimultiple.com/cloud-gpu/},
  urldate={2025-07-05}},
  note={Accessed: 2025}
}

@misc{statista_programming_languages,
  author       = {{Statista}},
  title        = {Most Used Programming Languages among Developers Worldwide},
  howpublished = {\url{https://www.statista.com/statistics/793628/worldwide-developer-survey-most-used-languages/}},
  year         = {2025},
  note         = {Accessed: 2025}
}

@misc{jetbrains_kotlin,
  author       = {{JetBrains}},
  title        = {Kotlin Programming Language},
  howpublished = {\url{https://www.jetbrains.com/opensource/kotlin/}},
  year         = {2024},
  note         = {Accessed: 2025}
}

@misc{fowler_continuous_integration_CI,
  author       = {Martin Fowler},
  title        = {Continuous Integration},
  howpublished = {\url{https://martinfowler.com/articles/continuousIntegration.html}},
  year         = {2006},
  note         = {Accessed: 2025}
}

@misc{google_blockly_commits,
  author       = {{Google Developers}},
  title        = {Contributing to Blockly: Getting Started with Commits},
  howpublished = {\url{https://developers.google.com/blockly/guides/contribute/get-started/commits}},
  year         = {2024},
  note         = {Accessed: 2025}
}

@misc{google_blockly_pr,
    title = {Contributing to Blockly: Writing a Good Pull Request},
    author = {{Google Developers}},
    year = {2024},
    howpublished = {\url{https://developers.google.com/blockly/guides/contribute/get-started/write_a_good_pr}},
    note = {Accessed: 2025}
}

@book{softwareDesignPhilosophy,
  title={A Philosophy of Software Design},
  author={Ousterhout, John},
  year={2018},
  edition={1},
  publisher={Yaknyam Press},
  isbn={978-1732102200}
}

@misc{Cho2023AICarbon,
  author       = {Cho, Ren{\'e}e},
  title        = {{AI’s Growing Carbon Footprint}},
  year         = {2023},
  date         = {2023-06-09},
  url          = {https://news.climate.columbia.edu/2023/06/09/ais-growing-carbon-footprint/},
  organization = {Columbia Climate School – State of the Planet},
  note         = {Accessed 2025-07-06}
}

@misc{Patterson2022GoodNews,
  author       = {Patterson, David},
  title        = {{Good News About the Carbon Footprint of Machine Learning Training}},
  year         = {2022},
  date         = {2022-02-15},
  url          = {https://research.google/blog/good-news-about-the-carbon-footprint-of-machine-learning-training/},
  organization = {Google Research Blog},
  note         = {Accessed 2025-07-06}
}

@inproceedings{chien2023reducing,
  title={Reducing the Carbon Impact of Generative AI Inference (today and in 2035)},
  author={Chien, Andrew A and Lin, Liuzixuan and Nguyen, Hai and Rao, Varsha and Sharma, Tristan and Wijayawardana, Rajini},
  booktitle={Proceedings of the 2nd workshop on sustainable computer systems},
  pages={1--7},
  year={2023}
}

@misc{ibm-watson,
  author       = {{IBM}},
  title        = {IBM Watson},
  howpublished = {\url{https://www.ibm.com/watson}},
  year         = {2025},
  note         = {[Online; accessed 6 July 2025]}
}

@book{banks2005discrete,
  title={Discrete event system simulation},
  author={Banks, Jerry},
  year={2005},
  publisher={Pearson Education India}
}

@inproceedings{deng2023early,
  title={Early chatgpt user portrait through the lens of data},
  author={Deng, Yuyang and Zhao, Ni and Huang, Xin},
  booktitle={2023 IEEE International Conference on Big Data (BigData)},
  pages={4770--4775},
  year={2023},
  organization={IEEE}
}

@article{mcnichols2025studychat,
  title={The StudyChat Dataset: Student Dialogues With ChatGPT in an Artificial Intelligence Course},
  author={McNichols, Hunter and Lan, Andrew},
  journal={arXiv preprint arXiv:2503.07928},
  year={2025}
}

@article{shim2025tooldial,
  title={Tooldial: Multi-turn dialogue generation method for tool-augmented language models},
  author={Shim, Jeonghoon and Seo, Gyuhyeon and Lim, Cheongsu and Jo, Yohan},
  journal={arXiv preprint arXiv:2503.00564},
  year={2025}
}

@article{zheng2023lmsys,
  title={Lmsys-chat-1m: A large-scale real-world llm conversation dataset},
  author={Zheng, Lianmin and Chiang, Wei-Lin and Sheng, Ying and Li, Tianle and Zhuang, Siyuan and Wu, Zhanghao and Zhuang, Yonghao and Li, Zhuohan and Lin, Zi and Xing, Eric P and others},
  journal={arXiv preprint arXiv:2309.11998},
  year={2023}
}

@misc{Li2025,
  author       = {Li, Junhao},
  title        = {Life of an inference request (vLLM V1): How LLMs are served efficiently at scale},
  howpublished = {\url{https://www.ubicloud.com/blog/life-of-an-inference-request-vllm-v1}},
  year         = {2025},
  note         = {Accessed: 2025}
}

@misc{Erdogan2024,
  author       = {Erdogan, Ozgun},
  title        = {EuroGPT: Open source and privacy-conscious alternative to ChatGPT Enterprise},
  howpublished = {\url{https://www.ubicloud.com/blog/eurogpt-open-source-and-privacy-conscious-alternative-to-chatgpt-enterprise}},
  year         = {2024},
  note         = {Accessed: 2025}
}

@misc{Cubukcu2024,
  author       = {Cubukcu, Umur},
  title        = {Lantern on Ubicloud: Build AI applications with PostgreSQL},
  howpublished = {\url{https://www.ubicloud.com/blog/build-ai-apps-with-postgresql}},
  year         = {2024},
  note         = {Accessed: 2025}
}

@article{touvron2023llama2,
  title        = {{Llama 2}: Open Foundation and Fine-Tuned Chat Models},
  author       = {Hugo Touvron and Louis Martin and Kevin Stone and Peter Albert and Amjad Almahairi
                  and Yasmine Babaei and Sergey Edunov and Thomas Scialom and et al.},
  journal      = {arXiv preprint arXiv:2307.09288},
  year         = {2023},
  url          = {https://arxiv.org/abs/2307.09288}
}

@misc{ibm2024granite20b,
  title        = {Granite-20B Foundation Model},
  author       = {{IBM Research}},
  howpublished = {\url{https://huggingface.co/ibm-granite/granite-20b-code-base-8k}},
  note         = {Version granite-20b-code-base-8k},
  year         = {2024}
}

@misc{mosaicml2023mpt30b,
  title        = {{MPT-30B}: A 30-Billion-Parameter Open-Source Transformer},
  author       = {{MosaicML}},
  howpublished = {\url{https://huggingface.co/mosaicml/mpt-30b}},
  year         = {2023}
}

@misc{meta_llama31_2024,
  title        = {Introducing Llama 3.1: Our Most Capable Models to Date},
  author       = {{Meta AI}},
  year         = {2024},
  howpublished = {\url{https://ai.meta.com/blog/meta-llama-3-1/}},
  note         = {Accessed 2025}
}

@misc{databricks_resnet_tensorrt_2025,
  title  = {Model inference using TensorFlow and TensorRT},
  author = {{Databricks Documentation}},
  year   = {2025},
  url    = {https://docs.databricks.com/aws/en/machine-learning/model-inference/resnet-model-inference-tensorrt},
  note   = {Accessed 2025}
}

@misc{databricks_notebooks_doc,
  title  = {Introduction to Databricks notebooks},
  author = {{Databricks Documentation}},
  year   = {2025},
  url    = {https://docs.databricks.com/aws/en/notebooks/},
  note   = {Accessed 2025}
}

@misc{databricks_mosaic_serving_2025,
  title  = {Deploy models using Mosaic AI Model Serving},
  author = {{Databricks Documentation}},
  year   = {2025},
  url    = {https://docs.databricks.com/aws/en/machine-learning/model-serving},
  note   = {Accessed 2025}
}

@misc{databricks_guardrails_2025,
  title  = {Implementing LLM Guardrails for Safe and Responsible Generative AI Deployment on Databricks},
  author = {{Databricks}},
  year   = {2025},
  url    = {https://www.databricks.com/blog/implementing-llm-guardrails-safe-and-responsible-generative-ai-deployment-databricks},
  note   = {Accessed 2025}
}

@misc{databricks_k8s_upgrade_2024,
  title  = {Scalable Kubernetes Upgrade Using Operators},
  author = {{Databricks}},
  year   = {2024},
  url    = {https://www.databricks.com/blog/scalable-kubernetes-upgrade-using-operators},
  note   = {Accessed 2025}
}

@misc{databricks_vector_search_2025,
  title  = {Vector Search},
  author = {{Databricks}},
  year   = {2025},
  url    = {https://www.databricks.com/product/machine-learning/vector-search},
  note   = {Accessed 2025}
}

@misc{databricks_docs_2025,
  title  = {Databricks documentation},
  author = {{Databricks Documentation}},
  year   = {2025},
  url    = {https://docs.databricks.com/aws/en/},
  note   = {Accessed 2025}
}

@misc{databricks_mosaic_capabilities_2025,
  title  = {Mosaic AI capabilities for generative AI apps},
  author = {{Databricks Documentation}},
  year   = {2025},
  url    = {https://learn.microsoft.com/en-us/azure/databricks/generative-ai/guide/mosaic-ai-gen-ai-capabilities},
  note   = {Accessed 2025}
}

@inproceedings{cho2024llmservingsim,
  title={Llmservingsim: A hw/sw co-simulation infrastructure for llm inference serving at scale},
  author={Cho, Jaehong and Kim, Minsu and Choi, Hyunmin and Heo, Guseul and Park, Jongse},
  booktitle={2024 IEEE International Symposium on Workload Characterization (IISWC)},
  pages={15--29},
  year={2024},
  organization={IEEE}
}

@misc{GOFAIR_FAIRPrinciples,
  title        = {FAIR Principles},
  organization = {GO FAIR},
  year         = {2016},
  url          = {https://www.go-fair.org/fair-principles/},
  urldate      = {2025}
}

@article{arpaci2018operating,
  title={Operating systems: Three easy pieces},
  author={Arpaci-Dusseau, Remzi H and Arpaci-Dusseau, Andrea C},
  year={2018},
  publisher={Arpaci-Dusseau Books, LLC Madison, WI, USA}
}

\end{document}